\documentclass[aps,prx,superscriptaddress,twocolumn,floatfix,nofootinbib,nobibnotes,10pt,longbibliography]{revtex4-2}
\usepackage[utf8]{inputenc}
\makeatletter
\renewcommand{\bibinfo}[2]{%
  \ifstrequal{#1}{title}{\textit{#2}}{#2}}
\makeatother
\usepackage{mathpazo}

\usepackage[titletoc,toc,title,page]{appendix}
\usepackage{titlesec}

\usepackage{enumitem}

\usepackage{csquotes}
\usepackage[italian,english]{babel}

\usepackage{bm}

\usepackage{dsfont} 
\usepackage{amsmath,amssymb,amsthm,thmtools}
\usepackage{mathtools}
\usepackage{cases}
\usepackage{calc}
\usepackage{flafter}
\usepackage{mathrsfs} 
\usepackage[normalem]{ulem} 

\usepackage{parskip}
\usepackage[colorlinks=true]{hyperref}
\usepackage[nameinlink]{cleveref}
\crefalias{appsec}{section}
\crefalias{endsec}{section}
\crefname{appsec}{Appendix}{Appendices}
\Crefname{appsec}{Appendix}{Appendices}
\crefname{endsec}{Section}{Sections}
\Crefname{endsec}{Section}{Sections}
\crefname{conjecture}{conjecture}{conjectures}
\crefname{box}{Box}{Box}
\hypersetup{
  pdftitle    = {Triple descent in quantum linear regression},
  pdfauthor   = {Gabriele Lo Monaco},
  colorlinks   = true, 
  urlcolor     = green!80!black, 
  linkcolor    = blue, 
  citecolor    = red!80!black 
}

\usepackage{physics} 
\usepackage[dvipsnames,table]{xcolor}
\usepackage{orcidlink}
\usepackage{graphicx}
\usepackage{float}        

\usepackage{placeins}
\usepackage{multirow,booktabs}
\usepackage{tcolorbox}
\tcbuselibrary{theorems,breakable}

\makeatletter
\newcommand{\columntoc}{%
  \begingroup
    \setlength{\parskip}{0pt}%
    \small
    \def\l@f@section{\addvspace{0.3em}}%
    \def\l@f@subsection{}%
    \let\columntoc@numberline\numberline@@sections
    \def\numberline@@sections##1{%
      \if\relax\detokenize{##1}\relax
        \leavevmode\ignorespaces
      \else
        \columntoc@numberline{##1}%
      \fi}%
    \par\noindent\rule{\columnwidth}{0.4pt}\par\smallskip
    {\centering\bfseries Contents\par}\smallskip
    \let\appendix\appendix@toc\@starttoc{toc}%
    \par\smallskip\noindent\rule{\columnwidth}{0.4pt}\par
  \endgroup
  \bigskip
}
\makeatother

\graphicspath{{./figures/}}

\newcommand{\RR}{\mathbb{R}}
\newcommand{\CC}{\mathbb{C}}

\newcommand{\EE}{\mathbb{E}}
\newcommand{\PP}{\mathbb{P}}

\newcommand{\calC}{\mathcal{C}}

\newcommand{\calE}{\mathcal{E}}

\newcommand{\calH}{\mathcal{H}}
\newcommand{\calL}{\mathcal{L}}

\newcommand{\calN}{\mathcal{N}}
\newcommand{\calO}{\mathcal{O}}

\newcommand{\bs}[1]{\boldsymbol{#1}}
\newcommand{\on}[1]{\operatorname{#1}}

\newcommand{\parTitle}[1]{\textit{#1 ---}}

\newcommand{\MSE}{\text{MSE}}

\newcommand{\hatbs}[1]{\bs{\hat{#1}}}

\newcommand{\bfI}{\mathbf{I}}

\newcommand{\im}{{\rm im}}

\newcommand{\diag}{\on{diag}}

\newcommand{\Var}{\on{Var}}

\newcommand{\nout}{n_{\rm out}}

\newcommand{\ntr}{n_{\rm tr}}

\newcommand{\hatbsp}{\hat{\bs p}}
\newcommand{\bsp}{\bs{p}}

\newcommand{\vect}[1]{\mathrm{vec}(#1)}

\newcommand{\frameOp}{\mathcal{F}}

\newcommand{\barSigma}{\overline{\bs{\Sigma}}}

\newtheorem{theorem}{Theorem}

\newtheorem{lemma}{Lemma}
\newtheorem{proposition}{Proposition}
\newtheorem{corollary}{Corollary}

\theoremstyle{definition}
\newtheorem{remark}{Remark}
\newtheorem{definition}{Definition}
\newtcbtheorem[crefname={example}{examples}]{example}{Example}%
{
  parbox=false,
  breakable,
  colback=red!0,          
  colframe=gray!10,         
  fonttitle=\bfseries,      
  arc=2mm,                  
  boxrule=0.2pt,           
  left=3mm,                
  right=3mm,               
  top=2mm,                 
  bottom=2mm,
  fonttitle=\bfseries\color{black}
}{}

\NewCommandCopy{\rawcite}{\cite}
\NewCommandCopy{\rawcref}{\cref}
\NewDocumentCommand{\safecite}{om}{%
  \mbox{\IfValueTF{#1}{\rawcite[#1]{#2}}{\rawcite{#2}}}%
}
\NewDocumentCommand{\safecref}{m}{%
  \mbox{\rawcref{#1}}%
}
\NewDocumentCommand{\addLI}{+m +g}{%
  {\color{RoyalBlue}#1}%
  \IfNoValueF{#2}{%
    \begingroup
      \let\cite\safecite
      \let\cref\safecref
      \color{red}\small\sout{#2}%
    \endgroup
  }%
}

\newif\ifshowrevisions
\showrevisionstrue
\ifdefined\cleanversion\showrevisionsfalse\fi
\definecolor{revnewcolor}{RGB}{0,118,140}
\definecolor{revoldcolor}{RGB}{196,30,58}
\definecolor{revnotecolor}{RGB}{176,92,0}
\ifshowrevisions
  \NewDocumentCommand{\revadd}{+m}{\ifvmode\leavevmode\fi{\color{revnewcolor}\relax#1}}
  \NewDocumentCommand{\revdel}{+m}{\ifvmode\leavevmode\fi{\let\cite\safecite\let\cref\safecref\color{revoldcolor}\sout{#1}}}
  \NewDocumentCommand{\revnote}{+m}{\ifvmode\leavevmode\fi{\color{revnotecolor}\footnotesize\sffamily[#1]}}
\else
  \NewDocumentCommand{\revadd}{+m}{#1}
  \NewDocumentCommand{\revdel}{+m}{}
  \NewDocumentCommand{\revnote}{+m}{}
\fi

\begin{document}

\title{Triple descent in quantum linear regression}

\author{Gabriele Lo Monaco\,\orcidlink{0000-0002-3594-3477}}
\email{gabriele.lomonaco@unipa.it}
\let\comma,
\affiliation{Universit\`a degli Studi di Palermo\comma{} Dipartimento di Fisica e Chimica - Emilio Segr\`e\comma{} via Archirafi 36\comma{} I-90123 Palermo\comma{} Italy}

\author{Salvatore Lorenzo\,\orcidlink{0000-0002-0827-5549}}
\let\comma,
\affiliation{Universit\`a degli Studi di Palermo\comma{} Dipartimento di Fisica e Chimica - Emilio Segr\`e\comma{} via Archirafi 36\comma{} I-90123 Palermo\comma{} Italy}

\author{Alessandro Ferraro\,\orcidlink{0000-0002-7579-6336}}
\affiliation{Quantum Technology Lab\comma{} Dipartimento di Fisica Aldo Pontremoli\comma{} Universit\`a degli Studi di Milano\comma{} I-20133 Milano\comma{} Italy}

\author{Mauro Paternostro\,\orcidlink{0000-0001-8870-9134}}
\affiliation{Universit\`a degli Studi di Palermo\comma{} Dipartimento di Fisica e Chimica - Emilio Segr\`e\comma{} via Archirafi 36\comma{} I-90123 Palermo\comma{} Italy}
\affiliation{Centre for Quantum Materials and Technologies\comma{} School of Mathematics and Physics\comma{} Queen's University Belfast\comma{} BT7 1NN\comma{} United Kingdom}

\author{G. Massimo Palma\,\orcidlink{0000-0001-7009-4573}}
\let\comma,
\affiliation{Universit\`a degli Studi di Palermo\comma{} Dipartimento di Fisica e Chimica - Emilio Segr\`e\comma{} via Archirafi 36\comma{} I-90123 Palermo\comma{} Italy}

\author{Luca Innocenti\,\orcidlink{0000-0002-7678-1128}}
\let\comma,
\affiliation{Universit\`a degli Studi di Palermo\comma{} Dipartimento di Fisica e Chimica - Emilio Segr\`e\comma{} via Archirafi 36\comma{} I-90123 Palermo\comma{} Italy}

\begin{abstract}\noindent
Quantum linear regression models, such as quantum extreme learning machines and quantum fidelity kernels, are trained on features that can only be estimated from finitely many measurement shots. Sampling noise is usually treated as a nuisance that more shots suppress; we show instead that it can fundamentally reshape the instabilities of the learning problem. As the number of measurement outcomes grows, the prediction error displays a \emph{triple-descent} profile, with two peaks of fundamentally different origins. The first, \emph{classical}, peak occurs when the measurement first becomes informationally complete, is a direct analogue of double descent in classical linear regression, and is caused by ill-conditioning of the measurement itself. This peak affects predictions only through their variance, and is thus suppressed by increasing the number of shots at inference. The second, \emph{quantum}, peak occurs beyond informational completeness when the number of outcomes matches the number of training states: shot noise populates unphysical outcome combinations and the regression fits these pure-noise features with large weights, corrupting the learned model itself. More training or test shots cannot remove this quantum peak, which disappears only in the ideal limit of exact training features. Using random-matrix theory, we derive the shape of both peaks analytically, accounting also for ridge regularization. Our results show that treating quantum features as exactly known overestimates classical instabilities while missing genuinely quantum ones.
\end{abstract}

\maketitle

\columntoc

\section{Introduction}
\label{sec:intro}

Most quantum machine learning (QML) models share a common structure: an input quantum state is measured after evolution through some quantum dynamics, and a classical model is then trained on features extracted from the measurement outcomes~\cite{biamonte2017quantum,schuld2019quantum,havlivcek2019supervised,huang2021power}. These features are never known exactly, since expectation values, outcome probabilities, and state overlaps can only be estimated from finitely many shots. While the resulting sampling noise is usually treated as a nuisance that more shots can make arbitrarily small, this picture is incomplete: sampling noise can produce effects that do not vanish with more shots, and reducing it can even \textit{worsen} performance in some cases. In a companion paper~\cite{lomonaco2026WhoCanSample}, for example, we showed that standard training methods can end up fitting the sampling noise itself, a form of \textit{quantum overfitting} that worsens as the training statistics increase.

\parTitle{Quantum linear regression} We study these effects in \textit{quantum linear regression} (QLR) models~\cite{lomonaco2026WhoCanSample}, a class of QML protocols that includes quantum extreme learning machines~\cite{innocenti2023potential,xiong2025fundamental,vetrano2025state,lomonaco2024quantum,suprano2024experimental,zia2025quantum}, quantum reservoir computing~\cite{fujii2017HarnessingDisorderedQuantum,ghosh2019reservoir,mujal2021OpportunitiesQuantumReservoir}, and quantum fidelity kernels~\cite{havlivcek2019supervised,schuld2021SupervisedQuantumMachine,huang2021power,jerbi2023QuantumMachineLearning}. In QLR, a quantum device produces the features as expectation values of fixed measurement operators, and the prediction is a linear combination of these features whose coefficients are fitted classically by standard regression~\cite{scholkopf2001learning,hastie2009Elements,huang2006extreme,jaeger2004harnessing}.

\parTitle{Spectral origins of double descent} Fitting these coefficients generally involves inverting the matrix of features measured on the training states, an operation which can amplify errors along its nearly singular directions. In ridgeless and random-feature regression, this mechanism underlies \emph{double descent} phenomena~\cite{hastie2020Surprises,belkin2019reconciling,mei2022double}: as features are added, the test error peaks when their number matches the number of training examples, and then decreases again. The peak arises because the feature matrix is square at that point, and random matrix theory shows that square random matrices typically have singular values close to zero~\cite{vershynin2018high,marvcenko1967distribution,bai2010spectral}. In shadow tomography, the same mechanism makes observable estimation from poorly conditioned measurements particularly sensitive to shot noise~\cite{innocenti2023shadow}. Spectral analyses have recently extended this picture to QML~\cite{thanasilp2024exponential,kempkes2026double,kamisoyama2026double,belis2026spectral}, revealing double-descent phenomena that do not originate from shot noise and persist even when the features are known exactly.

\parTitle{Triple descent from finite sampling} Here, we show that sampling noise adds a second peak to this picture, producing a \textit{triple} descent as the number of measurement outcomes is varied. Unlike previously identified double-descent peaks, this new peak has an intrinsically stochastic origin: more training statistics do not suppress it --- in the test variance, its height actually grows and then saturates to a value depending on the dimension --- and it disappears only in the theoretical limit of exactly known training features. Unlike the peak at informational completeness, which affects only the test variance when the target is reconstructed exactly, the new peak can also distort the learned observable itself, and is therefore not removed by increasing the statistics at inference. We focus our analysis on the simplest type of quantum features: the outcome probabilities of a fixed measurement performed after a fixed quantum evolution. This choice loses little generality, since the mechanisms we identify rely only on two properties shared by all QLR models: the exact features depend linearly on the input state, and they are estimated from finite samples. Learning a target observable then means reconstructing its expectation value from the outcome probabilities, using $\ntr$ training states labeled by their target values.

\begin{figure}[ht!]
    \centering
    \includegraphics[width=\linewidth]{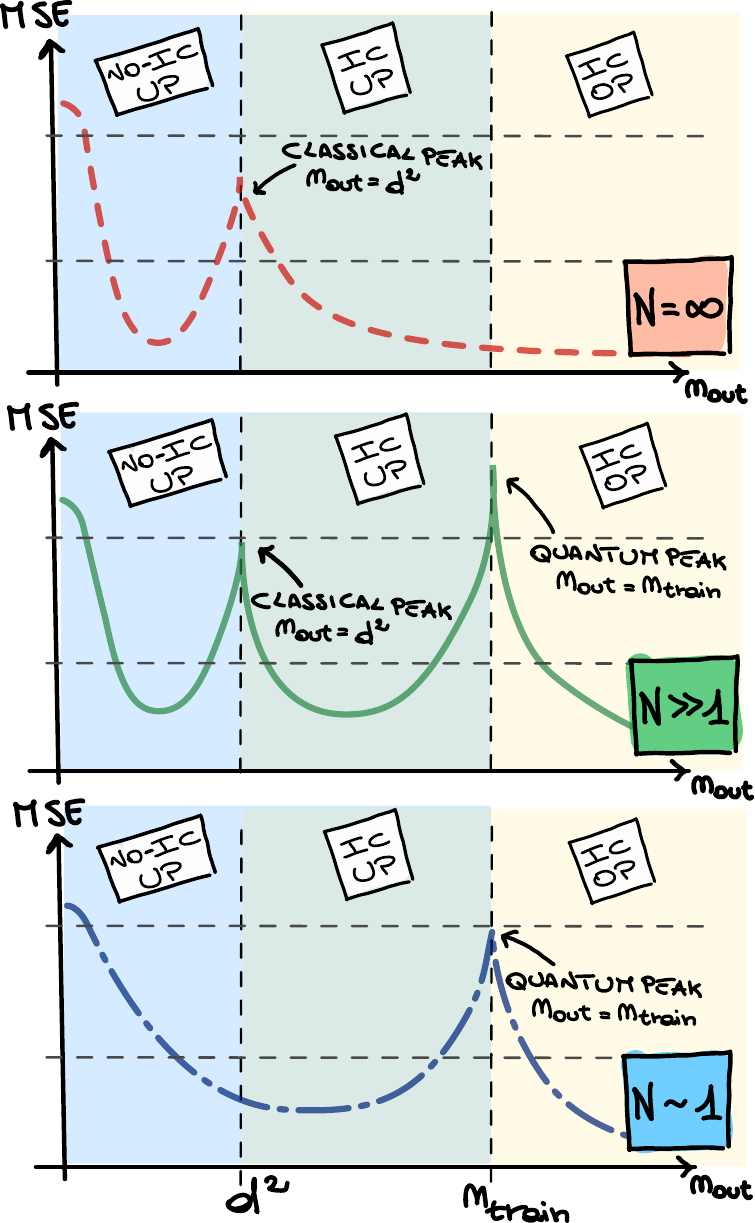}
    \caption{
    \textbf{Triple descent in quantum linear regression.}
    Pictorial representation of the MSE versus the number of measurement outcomes $n_{\mathrm{out}}$, for $n_{\mathrm{tr}}>d^2$ training states and various test statistics $N$.
    \textit{Top:} With exact training probabilities ($N=\infty$), the error exhibits a classical peak at the threshold for informational completeness, $n_{\mathrm{out}}=d^2$, caused by ill-conditioning of the measurement frame.
    \textit{Middle:} With large but finite training statistics ($N\gg1$ shots per training state), sampling fluctuations populate additional feature directions absent from the exact probabilities. These generate a second, quantum peak at the interpolation threshold, $n_{\mathrm{out}}=n_{\mathrm{tr}}$, producing the triple-descent profile.
    \textit{Bottom:} At low training statistics, shot noise suppresses the classical peak through effective regularization, and the quantum peak becomes dominant.
    The labels indicate regions where the measurement is informationally complete (IC), not informationally complete (NO-IC), underparametrized (UP), and overparametrized (OP).}
    \label{fig:variance}
\end{figure}

\begin{figure}[ht!]
\centering
\includegraphics[width=\linewidth]{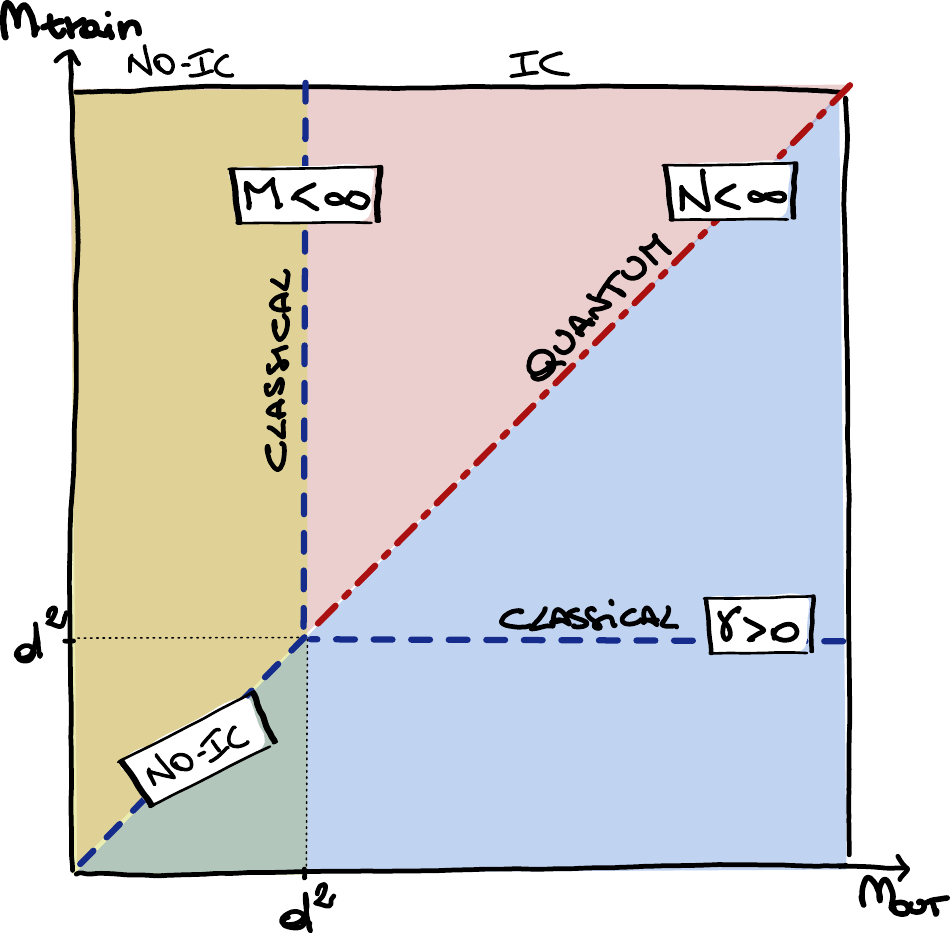}
\caption{\textbf{Overview of classical and quantum peaks.} Schematic representation of the regression regimes and the instabilities that separate them, for generic random POVMs and training states. With high probability, the measurement is informationally complete for $\nout \ge d^2$, while the training states span operator space for $\ntr \ge d^2$. For $\ntr>d^2$, increasing $\nout$ we cross we cross two peak: a classical one at $\nout=d^2$, caused by ill-conditioning of the measurement and affecting the test variance even with exact training features (\cref{sec:classical}), and a quantum peak at $\nout=\ntr$, caused by finite training shot-noise (\cref{sec:quantum}). Increasing $\nout$ at $\ntr>d^2$ thus causes a \textit{triple descent} phenomenon, as also illustrated in~\cref{fig:variance}. Similarly, for $\nout>d^2$, increasing $\ntr$ crosses a classical interpolation threshold at $\ntr=d^2$, where label noise can produce a peak even with exact features, followed by the quantum peak at $\ntr=\nout$. When both $\nout,\ntr<d^2$, the interpolation peak $\nout=\ntr$ changes nature and becomes \textit{classical}, and remains even with exact training and test features and no label noise, being caused by inaccessible components of the target observable (\cref{sec:uc_main}). This is analogous to the peak observed in~\cite{meyer2026shaping}. The classical peak at $\ntr>d^2$ is caused by a divergence in the test variance and is thus visible only accounting for finite test statistics; the classical peak at $\ntr=d^2$ is caused by label noise and is thus visible only for $\gamma>0$; the quantum peak at $\nout=\ntr$ is instead caused by stochastic instabilities in the estimated regression weights, and is thus visible only accounting for finite training statistics $N<\infty$.}
\label{fig:square}
\end{figure}

\parTitle{Two spectral instabilities} More precisely, as the number of measurement outcomes $\nout$ grows, the prediction error peaks twice, as illustrated schematically in~\cref{fig:variance}, and the two peaks have different origins. We call a peak \emph{classical} if it is caused by ill-conditioning of the exact feature matrix, typically inherited from the measurement itself, and \emph{quantum} if it is caused by nearly singular directions that sampling fluctuations create. In both cases, the peak appears because a random matrix that is close to square is likely to be nearly singular, as described by the Marchenko--Pastur law. We also map out the full landscape of possible peaks arising at the various completeness and interpolation thresholds in~\cref{fig:square}.

\parTitle{Peak at informational completeness} The first peak occurs at informational completeness, $\nout=d^2$ with $d$ the Hilbert-space dimension, the point at which exact outcome probabilities first suffice to reconstruct the input state. This peak is classical and is the direct counterpart of the double descent observed in classical linear regression~\cite{hastie2020Surprises} and quantum kernel methods~\cite{kempkes2026double}. Formally, it originates from the near-singularity of the measurement frame operator~\cite{innocenti2023shadow}, a superoperator built from the POVM that describes which operator directions the measurement resolves, and how well.
For the random POVMs we consider, the POVM elements form an effectively random $\nout\times d^2$ matrix, so the frame-operator spectrum follows the Marchenko--Pastur law, and becomes poorly conditioned exactly where the POVM first becomes informationally complete.
Because this ill-conditioning belongs to the measurement, the peak persists with exact training probabilities, and would appear even if the measurement were known exactly and the target were reconstructed with the optimal unbiased linear estimator~\cite{innocenti2023shadow}; suitably chosen symmetric measurements, however, can avoid it: for instance symmetric informationally complete POVMs, whose measurement frame operator is proportional to the identity on traceless operators~\cite{scott2006TightInformationallyComplete}. At low training statistics, shot noise acts as a regularizer and dampens the peak~\cite{lomonaco2026WhoCanSample,meyer2026shaping}, while at high training statistics its height saturates to a level set by the exact features. Because this peak stems from large exact regression coefficients, its effect on predictions is confined to their variance, and more shots at inference suppress it.

\parTitle{Classical peak from label noise}
For $\nout>d^2$, a further classical peak can occur at
$\ntr=d^2$ when the training labels are noisy, even
with exact measurement probabilities. Near this threshold,
the training states become nearly linearly dependent:
certain combinations $\sum_i c_i\rho_i$ almost vanish,
while the corresponding combinations of noisy labels
generally do not. Least-squares regression attempts to fit
these fluctuations by introducing large components in the
reconstructed observable along the poorly constrained
operator directions. The resulting amplification of label
noise produces a peak in the prediction error that
persists even for $N=M=\infty$. In the absence of label
noise, the near-dependencies are exactly reflected in the
labels, and this instability does not occur.

\parTitle{Peak at interpolation} At the interpolation threshold $\nout=\ntr$, the classification depends on whether the measurement is informationally complete. For $\ntr<d^2$, the exact training-feature matrix can itself become poorly conditioned and the resulting peak is classical, as also observed in~\cite{meyer2026shaping}. For $\ntr>d^2$, the peak is quantum: exact probabilities span at most $d^2$ dimensions, but training-shot fluctuations populate the remaining directions. The noise-only part of the empirical feature matrix becomes square at $\nout=\ntr$; its small singular values make the least-squares fit amplify test-shot noise and perturb the reconstructed observable. Consequently, adding outcomes beyond informational completeness can worsen predictions even as the measurement frame becomes better conditioned.

\parTitle{Finite-size behavior and limits} At finite Hilbert-space dimension, the peaks are rounded and their heights fluctuate across measurements and training records. The divergences arise in distinct high-dimensional limits, studied below. In particular, the typical quantum test-variance peak saturates as the training-shot budget $N$ grows at fixed dimension, although the peak is absent when the training probabilities are supplied exactly. The order of these limiting operations is therefore essential.

\parTitle{Analytical approach and implications} We make all these observations quantitative by analyzing the frame operator with random-matrix methods and by separating, in the empirical feature matrix, the directions carried by the training states from those generated only by sampling. The resulting analysis proves that both peaks exist and yields their shape analytically, both with and without ridge regularization, in good agreement with extensive numerical simulations. More broadly, these results show that the spectrum of an ideal quantum feature map is only part of the learning problem: treating the features as exactly known misses the quantum peak entirely and overestimates the classical one. A reliable analysis must therefore treat training and test statistics separately, and account both for the constraints that quantum states and measurements impose on the features and for the finite records from which they are estimated. In practice, measurement design, training-set size, and shot budget cannot be chosen independently: whether adding outcomes helps or hurts depends on how $\nout$ compares with both $d^2$ and $\ntr$.

We first define the regression protocol and separate prediction error into its ideal-test and test-variance contributions in~\cref{sec:setup}. \Cref{sec:classical} identifies the classical peak at informational completeness. \Cref{sec:quantum} analyzes the quantum peak at interpolation above informational completeness, including its training-shot and ridge dependence. The classical interpolation peak below informational completeness is treated separately in~\cref{sec:uc_main}. We close with implications and open questions in~\cref{sec:conclusions}. For readability, theorem statements in the main text are informal; precise formulations and proofs are in the Supplemental Material.

\section{Quantum linear regression and prediction error}
\label{sec:setup}

\begin{figure}[th]
    \centering
    \includegraphics[width=\linewidth]{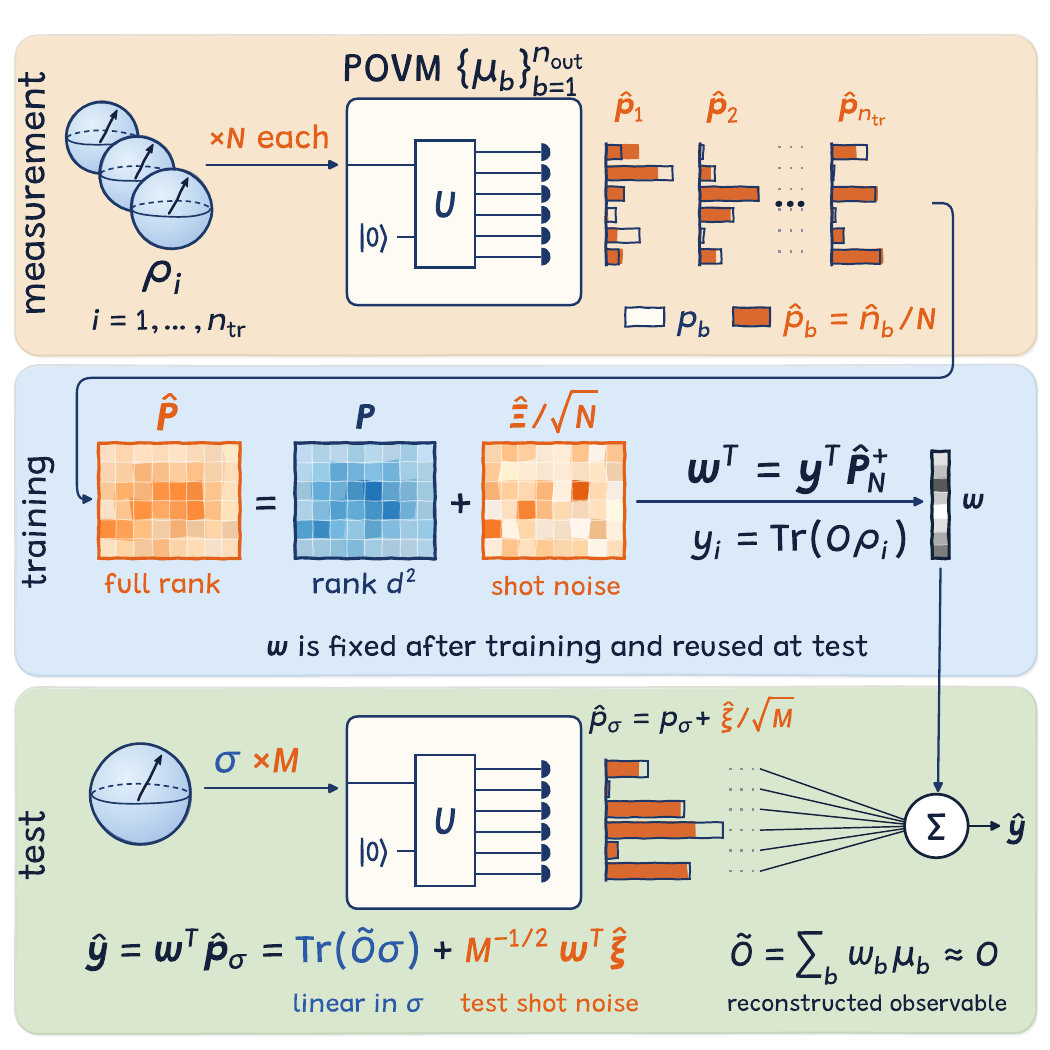}
    \caption{\textbf{Learning protocol.} Each training state $\rho_i$ is measured $N$ times with the POVM $\bs\mu$, and the resulting outcome frequencies are used to train a linear least-squares regressor with weights $\bs w$. The regressor then predicts the target value for a new test state $\sigma$ from the outcome frequencies of $M$ measurements of $\sigma$. For any $\bs{w}$, the prediction $\hat y$ averaged over test shot noise is $\Trace(\widetilde\calO\sigma)$, with the effective observable $\widetilde\calO=\sum_b w_b\mu_b$. When the training labels are $y_i=\Trace(\calO\rho_i)$ and $\calO\in\on{span}(\{\mu_b\})$, which always holds for informationally complete POVMs, and the training states span the operator space, the learned weights give $\widetilde\calO\approx\calO$ for large $N$.}
    \label{fig:qlr_1col}
\end{figure}

In this section we set up the learning problem and isolate the contributions of training and test measurement statistics to the prediction error. This separation will let us identify the two thresholds characterizing the triple-descent phenomenon and the nature of the quantum and classical peaks. A schematic of the measurement protocol we consider is given in~\cref{fig:qlr_1col}.

\subsection{Quantum features and empirical regression}
\label{sec:setup_features}

We consider $d$-dimensional states $\rho$ measured with a fixed positive-operator-valued measure (POVM) $\bs\mu=(\mu_a)_{a=1}^{\nout}$ with $\nout$ outcomes, where $\mu_a\ge0$ and $\sum_a\mu_a=I_d$. Here $\bs\mu$ is the \emph{effective POVM}~\cite{innocenti2023potential}, which accounts for both the physical measurement and any quantum evolution applied before it. The outcome probabilities $\bs p(\rho)$, with components $p_a(\rho)=\Tr(\mu_a\rho)$, serve as features for a classical linear model trained to predict the expectation value $y(\rho)=\Tr(\calO\rho)$ of a Hermitian target observable $\calO$. We assume that the training labels are known exactly, so that finite statistics affect learning only through the features.

\parTitle{Rank constraint and informational completeness} For a training set of $\ntr$ states $\bs\rho=(\rho_i)_{i=1}^{\ntr}$, we collect the labels $y_i=y(\rho_i)$ in the vector $\bs y$ and the exact probabilities in the matrix $\bs P\in\RR^{\nout\times\ntr}$, whose $i$-th column is $\bs p(\rho_i)$. The probability matrix factorizes as
\begin{equation}
\label{eq:main_probability_factorization}
 \bs P=\bs M_{\bs\mu}^T\bs M_{\bs\rho},
 \qquad
 \rank\bs P\le d^2,
\end{equation}
where the columns of $\bs M_{\bs\mu}\in\RR^{d^2\times\nout}$ and $\bs M_{\bs\rho}\in\RR^{d^2\times\ntr}$ are the effects $\mu_a$ and the states $\rho_i$, vectorized in an orthonormal basis of the $d^2$-dimensional real vector space of Hermitian operators on $\CC^d$; this rank is bounded because all effects and states live in this $d^2$-dimensional space. Unless stated otherwise, we assume that the training states span the space of Hermitian operators, which requires $\ntr\ge d^2$. The bound is then saturated, $\rank\bs P=d^2$, precisely when $\bs\mu$ is informationally complete, i.e., when its effects also span the space. For an informationally complete POVM, every target observable lies in $\on{span}(\{\mu_b\})$ and can therefore be reconstructed from the measurement data. For other POVMs, the components of $\calO$ outside $\on{span}(\{\mu_b\})$ are inaccessible: whenever such components are present, the prediction error remains nonzero even with unlimited training resources.

\parTitle{Finite-shot feature matrix} In an experiment, however, $\bs P$ is not directly available: its $i$-th column is estimated as $\hatbs n_i/N$, where the count vector $\hatbs n_i$ records the outcomes of $N$ measurements of $\rho_i$ and is multinomially distributed,
\begin{equation}
\label{eq:main_counts}
 \hatbs n_i\mid\rho_i
 \sim\operatorname{Mult}(N,\bs p_i),
 \qquad
 \bs\Sigma_i=
 \diag\bs p_i-\bs p_i\bs p_i^T,
\end{equation}
with $\bs p_i=\bs p(\rho_i)$, and $\bs\Sigma_i$ the single-shot covariance. We collect the observed frequencies in the empirical feature matrix $\hatbs P_N=[\hatbs n_1/N\ \cdots\ \hatbs n_{\ntr}/N]$ and write $\hatbs P_N=\bs P+N^{-1/2}\hatbs{\Xi}_N$, which defines the normalized fluctuation matrix $\hatbs{\Xi}_N$. For fixed training states, the columns of $\hatbs{\Xi}_N$ are independent, with zero mean and covariances $\bs\Sigma_i$. The mean and covariance of $\hatbs\Xi_N$ are thus independent of $N$, which enters the distribution of $\hatbs\Xi_N$ only through higher-order cumulants that vanish as $N\to\infty$; when only these first two moments matter, we thus drop the subscript and write $\hatbs\Xi$. Shot noise also changes the spectrum of the features: it adds singular directions absent from the exact features, with singular values of order $N^{-1/2}$, making $\hatbs P_N$ full rank with high probability.

\parTitle{Least-squares and ridge estimators} Training the resulting classical model amounts to ordinary least squares with noisy features. The learner assigns a weight to each measurement outcome so as to minimize the training error $\|\hatbs P_N^T\bs w-\bs y\|^2$, and takes the minimum-norm solution via the Moore--Penrose pseudoinverse when the minimizer is not unique. The learned weights and the resulting prediction for a test state $\sigma$ are
\begin{equation}
\label{eq:qelm_exact_linear_rule}
 \hatbs{w}_N=(\hatbs P_N^T)^+\bs y,\qquad
 \hat y=\hatbs{w}_N^T\hatbs p_M(\sigma),
\end{equation}
where $(\hatbs P_N^T)^+$ is the Moore--Penrose pseudoinverse, which inverts all nonzero singular values of $\hatbs P_N^T$, including the small ones generated by shot noise, and $\hatbs p_M(\sigma)$ is the vector of outcome frequencies obtained from $M$ measurements of $\sigma$. Thus $N$ sets the precision of the training features and $M$ that of the test features. We also consider regularization via ridge regression, which gives weights $\hatbs{w}_{N,\tau}$ solving
\begin{equation}\label{eq:main_ridge_estimator}
    \min_{\bs{w}}\left[
    \ntr^{-1}\|\hatbs{P}_N^T\bs w-\bs y\|^2
    + \tau \|\bs{w}\|^2\right].
\end{equation}
Explicitly, this minimization has solution $\hatbs w_{N,\tau}=(\hatbs P_N\hatbs P_N^T+\ntr\tau I_{\nout})^{-1}\hatbs P_N\bs y$. Ridge regression filters each singular value $s$ of $\hatbs{P}_N$ as $s/(s^2+\ntr \tau)$ instead of inverting it as $1/s$, thus suppressing the contribution of singular values below $\sqrt{\ntr\tau}$, and $\hatbs w_{N,\tau}\to\hatbs w_N$ as $\tau\to0^+$.

\subsection{Prediction error and observable reconstruction}
\label{sec:setup_error}

\parTitle{MSE decomposition averaging over test shot noise}
We quantify the performance of the learned predictor by its mean squared error (MSE). For a fixed training record, a fixed target $\calO$, and a fixed test state $\sigma$, averaging over the test shot noise gives
\begin{equation}
\label{eq:hat_MSE}
\begin{gathered}
    \EE_{\hat y}[(\hat y-y_\sigma)^2]
    =\calE_N(\sigma,\calO) + \frac{V_N(\sigma,\calO)}{M},
    \\
    \calE_N(\sigma,\calO)\equiv (\hatbs{w}_N^T\bs p_\sigma-y_\sigma)^2,
    \quad
    V_N(\sigma,\calO) \equiv \hatbs{w}_N^T\bs\Sigma_\sigma\hatbs{w}_N,
\end{gathered}
\end{equation}
where $\bs p_\sigma=\bs p(\sigma)$, $y_\sigma=y(\sigma)$, and $\bs\Sigma_\sigma=\diag(\bs p_\sigma)-\bs p_\sigma\bs p_\sigma^T$ is the single-shot covariance of the test outcomes; the target enters through the training labels from which $\hatbs w_N$ is learned. The first term, $\calE_N(\sigma,\calO)$, is the error that remains when the test probabilities are known exactly ($M\to\infty$), which depends on $N$ because $\hatbs{w}_N$ was learned from noisy features. The second term is the additional error caused by amplification of shot noise at test, with $V_N(\sigma,\calO)$ the single-shot test variance. We call $\calE$ the \emph{ideal-test error} and $V$ the \emph{test variance}, which measures how sensitive the learned predictor is to finite test statistics.

\parTitle{Averages over test states and targets} The arguments of $\calE_N$ and $V_N$ list what is held fixed, and we drop an argument when averaging over it. Averaging over Haar-random pure test states gives
$\calE_N(\calO)\equiv\EE_\sigma[\calE_N(\sigma,\calO)]$ and $V_N(\calO)\equiv\EE_\sigma[V_N(\sigma,\calO)]=\hatbs w_N^T\barSigma_{\rm te}\hatbs w_N$, with $\barSigma_{\rm te}\equiv\EE_\sigma[\bs\Sigma_\sigma]$, so that $\MSE_{N,M}(\calO)\equiv\EE_\sigma\EE_{\hat y}[(\hat y-y_\sigma)^2]=\calE_N(\calO)+V_N(\calO)/M$. Averaging further over the random targets introduced below defines $\calE_N\equiv\EE_\calO[\calE_N(\calO)]$, $V_N\equiv\EE_\calO[V_N(\calO)]$, and $\MSE_{N,M}=\calE_N+V_N/M$. All these quantities still depend on the measurement, the training states, and the training record; averages over these are taken only where stated. Replacing $\hatbs w_N$ with $\hatbs w_{N,\tau}$ gives the ridge counterparts $\calE_{N,\tau}$, $V_{N,\tau}$, and $\MSE_{N,M,\tau}$, with the same arguments.

\parTitle{Ideal-test error and the reconstructed observable} Averaging over the test shot noise gives $\EE_{\hat y}[\hat y]=\hatbs w_N^T\bs p_\sigma=\Trace(\widetilde\calO_N\sigma)$, where $\widetilde\calO_N=\sum_b(\hatbs w_N)_b\,\mu_b$ is the observable reconstructed by the learned (noisy) weights $\hatbs{w}_N$; for a general $\bs w$, we write $\widetilde\calO_{\bs w}=\sum_bw_b\mu_b$. Denoting by $A_0\equiv A-(\Trace(A)/d)I_d$ the traceless part of an operator $A$, the Haar average over test states gives (\cref{lem:haar_error_split})
\begin{equation}
\label{eq:ideal_test_error_operator}
\calE_N(\calO)=\frac{\Trace[(\widetilde\calO_N-\calO)_0^2]}{d(d+1)}+\left[\frac{\Trace(\widetilde\calO_N-\calO)}{d}\right]^2.
\end{equation}
The first term is the squared Hilbert--Schmidt error of the state-dependent part of the reconstruction. The factor $1/[d(d+1)]$ is the variance of $\Tr(A\sigma)$ over Haar-random $\sigma$ for any traceless $A$ with $\Trace(A^2)=1$, i.e., the typical size of the state-to-state variations of an expectation value. Thus $\calE_N(\calO)$ vanishes only when the learned weights reproduce the target exactly, $\widetilde\calO_N=\calO$, which is impossible when $\calO\notin\on{span}(\{\mu_b\})$.

\parTitle{Reference coefficients} We compare the learned weights with the minimum-norm reconstruction coefficients $\bs w_{\rm ex}\equiv\bs M_{\bs\mu}^+\bs v_\calO$, where $\bs v_\calO$ is the vectorization of $\calO$ in the basis of~\cref{eq:main_probability_factorization}. Their effective observable $\widetilde\calO_{\bs w_{\rm ex}}$ is the Hilbert--Schmidt projection of $\calO$ onto $\on{span}(\{\mu_b\})$, i.e., $\calO$ itself for an informationally complete POVM, and they are the weights that least squares returns from exact training probabilities when the POVM is informationally complete and the training states span the operator space (\cref{sec:classical_frame}). Replacing $\hatbs w_N$ with $\bs w_{\rm ex}$ in~\cref{eq:hat_MSE} defines the reference test variance $V_{\rm ex}(\sigma,\calO)$ and its averages $V_{\rm ex}(\calO)$ and $V_{\rm ex}$.

\parTitle{Target averaging and normalization} Finally, to describe a typical target rather than a specific one, we average over the unitarily rotated targets $U^\dagger\calO U$, with $U$ a Haar-random unitary independent of the measurement and of the training data. As anticipated, these target averages are denoted by dropping the argument $\calO$, e.g., $V_N=\EE_U[V_N(U^\dagger\calO U)]$; they depend on $\calO$ only through its spectrum. Decomposing $\calO=(\Trace(\calO)/d)I_d+\calO_0$, we observe that only the traceless part $\calO_0$ distinguishes input states, and its size can be quantified by \begin{equation}
\label{eq:def_Delta_squared_main}
\Delta_{\calO}^2=\Trace(\calO_0^2)/d=\Trace(\calO^2)/d-(\Trace(\calO)/d)^2\,,
\end{equation} which equals the variance of the eigenvalues of $\calO$ (equivalently, of $\calO$ in the maximally mixed state). Note that $\Delta_{\calO}^2>0$ iff $\calO$ is not a multiple of the identity and thus represents a nontrivial target; we will assume throughout the rest of the paper that this is the case. The rotation leaves the identity component unchanged, whereas the rotated traceless part $U^\dagger\calO_0U$ has zero mean and a second moment proportional to the identity on traceless operators (\cref{eq:moment_2_operator}), so that no operator direction is preferred. Since $\hatbs w_N$ is linear in $\bs y$, and hence in $\calO$, both $\calE_N(\calO)$ and $V_N(\calO)$ are quadratic in the target, and their target averages are linear combinations of $d\Delta_{\calO}^2$ and $(\Trace(\calO)/d)^2$ with coefficients set by the measurement and the training data. We therefore report the normalized averages $V_N/(d\Delta_\calO^2)$ and $d(d+1)\calE_N/(d\Delta_\calO^2)$, which do not depend on the scale of the target. By~\cref{eq:ideal_test_error_operator}, the second is the mean squared Hilbert--Schmidt error of the reconstructed state-dependent part relative to $\Tr(\calO_0^2)$, plus the contribution of the identity component; it equals one for the trivial predictor that assigns the constant value $\Tr(\calO)/d$ to every input.
The same averages hold for any random target, drawn independently of the measurement and training data, whose traceless part has these two moments (\cref{def:target_ensembles}), such as a target drawn uniformly from the unit sphere of traceless operators. Unless stated otherwise, our simulations draw one such target per realization, so that $d\Delta_\calO^2=\Tr(\calO^2)=1$, and report the realized costs $V_N(\calO)$ and $\calE_N(\calO)$; at large $d$ and fixed ridge parameter, these concentrate around their target averages (\cref{lem:centered_full_sphere_sign_concentration}).

\section{Classical peak at informational completeness}
\label{sec:classical}

The first instability occurs at informational completeness, $\nout=d^2$, persists even with exact training probabilities, and is caused by a property of the measurement itself: near this threshold, the POVM frame operator has small positive eigenvalues with high probability; this causes a blow up in the learned exact weights, which strongly amplify shot noise at inference.

In~\cref{sec:classical_frame} we show why weakly resolved operator directions cause large reconstruction weights. In~\cref{sec:classical_spectrum} we determine when these directions occur for random POVMs. We combine these two ingredients in~\cref{sec:classical_peak} to derive the classical variance peak and interpret its finite-dimensional behavior. Finally, in~\cref{sec:classical_training} we return to regression on finite training data and explain when shot noise suppresses the peak rather than amplifying it.

\subsection{Prediction variance and the POVM frame}
\label{sec:classical_frame}

\parTitle{Training with exact features} We start from the idealized case where training features, i.e. the elements of the probability matrix $\bs{P}$, are known exactly. In this scenario the learner has perfect knowledge of how the effective measurement interacts with the training states. With labels $y_i=\Tr(\calO\rho_i)$, i.e., $\bs y=\bs M_{\bs\rho}^T\bs v_\calO$, least squares returns $(\bs P^T)^+\bs y$. When the POVM is informationally complete and the training states span the operator space, both $\bs M_{\bs\mu}$ and $\bs M_{\bs\rho}$ have rank $d^2$, and~\cref{eq:main_probability_factorization} gives (\cref{cor:classical_training})
\begin{equation}
\label{eq:var_epsilon_0}
    (\bs{P}^T)^+\bs{y}
    =\bs M_{\bs\mu}^{+}\bs v_\calO=\bs{w}_{\rm ex}.
\end{equation}
The reconstruction is then exact, $\widetilde\calO_{\bs w_{\rm ex}}=\calO$~\cite{innocenti2023potential}, and the error vanishes if the test probabilities are also known exactly. Below informational completeness, $(\bs P^T)^+\bs y$ generally differs from $\bs w_{\rm ex}$ for a finite training set, but approaches it for large Haar training sets (\cref{sec:classical_training}); we therefore use $\bs w_{\rm ex}$, which reconstructs the projection of $\calO$ onto the span of the effects, in both regimes. Even with exact training features, however, the weights $w_{{\rm ex},a}$ can be large, so that their test variance $V_{\rm ex}(\sigma,\calO)$ strongly amplifies shot noise at inference.

\parTitle{Weakly resolved measurement directions} The size of these coefficients is controlled by the \emph{POVM frame operator} $\frameOp_{\bs\mu}=\bs M_{\bs\mu}\bs M_{\bs\mu}^{T}$, which acts on operators as $\frameOp_{\bs\mu}(A)=\sum_a\Tr(\mu_aA)\,\mu_a$. The quadratic form induced by this operator has a direct operational meaning: for a Hermitian $A$ with vectorization $\bs v_A$, $
    \bs v_A^{T}\frameOp_{\bs\mu}\bs v_A=\Tr[A\frameOp_{\bs\mu}(A)]=\sum_a[\Tr(\mu_aA)]^2.
$
If a state is shifted along a traceless direction, $\rho\to\rho+\epsilon A$, the outcome probabilities change by $\delta p_a=\epsilon\Tr(\mu_aA)$, so this quadratic form equals the squared change of the outcome distribution per unit shift. In particular, if $\Trace(A^2)=1$ and $\frameOp_{\bs\mu}(A)=\lambda A$, the outcome distribution moves only by $\epsilon\sqrt\lambda$. A small eigenvalue $\lambda$ therefore marks an operator direction that the measurement barely resolves: states that differ along it produce nearly identical statistics. Reconstructing an observable along such a direction requires correspondingly large coefficients, since the minimum-norm coefficients satisfy $\|\bs M_{\bs\mu}^+\bs v_A\|^2=\bs v_A^T\frameOp_{\bs\mu}^+\bs v_A=1/\lambda$. Large coefficients in turn amplify the statistical fluctuations of the test frequencies, so the test variance is governed by the inverse eigenvalues of $\frameOp_{\bs\mu}$; specifically, by the largest eigenvalue, $\|\frameOp_{\bs\mu}^+\|$, for the worst-case target, and by their average, $\Trace(\frameOp_{\bs\mu}^+)$, for a typical one~\cite{innocenti2023shadow}.
For effects of equal trace, $\Trace(\mu_a)=d/\nout$, this connection is exact. By~\cref{eq:barSigma_exact}, every weight vector that reconstructs the target of an informationally complete POVM, $\widetilde\calO_{\bs w}=\calO$, has Haar-averaged test variance $\|\bs w\|^2/\nout-[\Trace(\calO^2)+(\Trace\calO)^2]/[d(d+1)]$, because the remaining terms of $\barSigma_{\rm te}$ depend on $\bs w$ only through $\widetilde\calO_{\bs w}$. The minimum-norm coefficients $\bs w_{\rm ex}$ therefore also minimize the test variance among all unbiased linear readouts~\cite{innocenti2023shadow}. Averaging $\|\bs w_{\rm ex}\|^2=\bs v_\calO^T\frameOp_{\bs\mu}^+\bs v_\calO$ over the random targets of~\cref{sec:setup_error} gives, for equal effect traces and with or without informational completeness (\cref{eq:exact_classical_average}),
\begin{equation}
\label{eq:main_classical_inverse_trace}
\frac{V_{\rm ex}}{d\Delta_\calO^2}=\frac{\Trace(\frameOp_{\bs\mu}^+)}{\nout(d^2-1)}+O(d^{-2}).
\end{equation}
For random POVMs, whose effect traces are only approximately equal, the same relation holds up to relative corrections that vanish as $d\to\infty$ (\cref{thm:classical_peak}).

\subsection{Spectrum of a random POVM}
\label{sec:classical_spectrum}

To describe the frame spectrum of a typical measurement, we consider Haar rank-one POVMs. Their effects are $\mu_a=\bs v_a\bs v_a^\dagger$, where the $\bs v_a$ are the rows of a Haar-random isometry $\bs V_{\bs\mu}\in\CC^{\nout\times d}$~\cite{mezzadri2006generate}. The isometry condition guarantees $\sum_a\mu_a=I$, but also makes the effects mutually dependent, so it is not obvious that the frame behaves like a standard random covariance matrix. The following result shows that, asymptotically, it does.

\begin{theorem}[Spectrum of random POVM (informal)]
\label{thm:informal_spectrum}
Consider a normalized Haar rank-one POVM with $\nout$ outcomes on a $d$-dimensional Hilbert space. In the limit $d\to\infty$ with $d^2/\nout\to\gamma_C\in(0,\infty)$, the empirical eigenvalue distribution of the rescaled frame $\nout\frameOp_{\bs\mu}$ converges in probability to the Marchenko--Pastur law with aspect ratio $\gamma_C$, whose continuous part is supported between
\begin{equation}
\label{eq:main_frame_edges}
    x_\pm=(1\pm\sqrt{\gamma_C})^2.
\end{equation}
For $\nout<d^2$, the frame also has $d^2-\nout$ exact zero eigenvalues, corresponding to inaccessible operator directions. A single further eigenvalue, asymptotically associated with the identity, lies on a parametrically larger scale than the bulk.
\end{theorem}

\parTitle{Informational completeness and the spectral gap} The precise statement and proof are given in \cref{thm:main,cor:outlier_additive}. \Cref{thm:informal_spectrum} identifies informational completeness as a critical point of the measurement spectrum. For $\nout<d^2$ the POVM cannot reconstruct every observable, and the frame has exact zero modes. As $\nout\to d^2$, the lower edge $x_-$ of the nonzero spectrum also approaches zero, and at $\nout=d^2$ it reaches the origin, so that some operator directions are resolved arbitrarily weakly. For $\nout>d^2$ a spectral gap opens, and each additional outcome improves the conditioning.

\begin{figure*}[t]
    \centering
    \includegraphics[width=\linewidth]
    {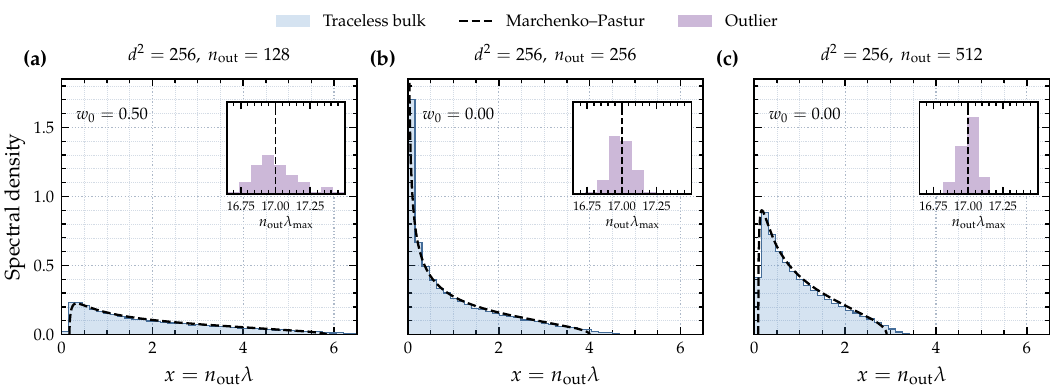}
    \caption{\textbf{Spectral shape of the measurement frame operator.}
    Eigenvalues $x=\nout\lambda$ of the rescaled frame operator $\nout\frameOp_{\bs\mu}$, restricted to the subspace of traceless Hermitian operators, for normalized Haar rank-one POVMs (\cref{sec:classical_spectrum}) with $d^2=256$ and $\nout=128,256,512$. Each panel aggregates 64 independent POVMs. Dashed curves show the limiting Marchenko--Pastur density at $\gamma_C=d^2/\nout$ (\cref{thm:informal_spectrum}). For $\nout<d^2$, exact zero eigenvalues are omitted without renormalizing the positive weights; $w_0=\max(d^2-\nout,0)/d^2$ is the fraction of zero eigenvalues. Insets show the largest eigenvalue of the full rescaled frame, compared with its asymptotic value $d+1$ (\cref{cor:outlier_additive}); the corresponding eigenvector is asymptotically the identity.}
    \label{fig:spectrum}
\end{figure*}

\parTitle{Numerical spectrum and identity outlier} \Cref{fig:spectrum} shows that the Marchenko--Pastur law already describes random POVMs at moderate dimension, both below and above informational completeness. The insets show the isolated eigenvalue associated with the identity, well separated from the bulk. This outlier plays no role in the instability: since $\sum_a\mu_a=I$, the identity component of a target can be reconstructed by uniform outcome weights, whose prediction does not depend on the state and is therefore not affected by test shot noise. What matters is the lower edge of the spectrum.

\subsection{Classical variance peak}
\label{sec:classical_peak}

Here we combine the relation between the POVM frame and test variance from~\cref{sec:classical_frame} with the random-POVM spectrum of~\cref{sec:classical_spectrum}. Our aim is to predict how the classical peak depends on the number of measurement outcomes, and to distinguish the asymptotic divergence from the finite peak observed in numerical experiments.

\parTitle{Why small frame eigenvalues dominate} With exact training probabilities,~\cref{eq:main_classical_inverse_trace} expresses the target-averaged test variance through the inverse of the POVM frame operator. Writing $x=\nout\lambda$ for an eigenvalue $\lambda$ of the frame, the identity $\frameOp_{\bs\mu}^+=\nout(\nout\frameOp_{\bs\mu})^+$ shows that the normalized variance is, to leading order, the sum of $1/x$ over the positive rescaled eigenvalues, divided by $d^2-1$. The inverse gives greatest weight to the operator directions that the POVM resolves most weakly: reconstructing a target along such a direction requires large outcome weights, which in turn amplify fluctuations in the test frequencies. Exact zero eigenvalues do not contribute to this sum, because the pseudoinverse discards them. Below informational completeness, the corresponding inaccessible target components instead contribute to the ideal-test error defined in~\cref{eq:hat_MSE}.

For Haar-random POVMs,~\cref{thm:informal_spectrum} gives a Marchenko--Pastur spectrum with aspect ratio $\gamma_C=d^2/\nout$. As $\nout$ approaches $d^2$ from either side, the smallest \emph{positive} frame eigenvalues approach zero. At $\gamma_C=1$ the spectral density behaves as $x^{-1/2}$ near the origin, so its inverse-eigenvalue average becomes unbounded in the large-$d$ limit. The resulting instability is already present in the measurement frame, without any noise in the training features.

\begin{figure}[t]
    \centering
    \includegraphics[width=\linewidth]{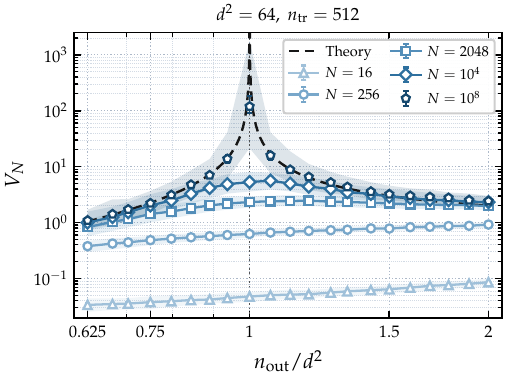}
    \caption{\textbf{Classical variance peak and its suppression by finite training statistics.} Realized test variance $V_N(\calO)$ at $d=8$ and $\ntr=512$. Colored curves correspond to multinomial training records with $N=16,256,2048,10^4,10^8$ shots, with darker blue for larger $N$. Each point aggregates 192 independent realizations of the normalized Haar rank-one POVM, Haar training ensemble, traceless unit-sphere target, and training record. Markers and curves show medians; shading shows pointwise 10th--90th percentiles. The $N=10^8$ median is drawn with markers only, to distinguish it from the theory. The black dashed curve is the asymptotic prediction of~\cref{eq:classical_MP_comparison} at $d\Delta_{\calO}^2=1$, which diverges at $\nout=d^2=64$ (vertical line). Records for which $\hatbs P_N$ is rank deficient, e.g.\ because some outcome is never observed, are retained; test states are averaged analytically.}
    \label{fig:classical}
\end{figure}

\begin{theorem}[Classical test-variance peak (informal)]
\label{thm:informal_classical}
Consider the random POVM ensemble of~\cref{thm:informal_spectrum} and the minimum-norm reconstruction, with perfect device knowledge, of a target observable with $\Delta_{\calO}^2>0$ and bounded $\lvert\Trace(\calO)\rvert/(d\sqrt{d\Delta_{\calO}^2})$. In the limit $d\to\infty$ with $d^2/\nout\to\gamma_C\in(0,\infty)$, the normalized test variance, averaged over Haar test states and unitary rotations of the target, converges in probability to
\begin{equation}
\label{eq:classical_MP_comparison}
    \frac{V_{\rm ex}}{d\Delta_{\calO}^2}
    \xrightarrow{\PP}
    \frac{\min\{1,\gamma_C^{-1}\}}
    {|1-\gamma_C|},
    \qquad \gamma_C\ne1.
\end{equation}
At $\gamma_C=1$, i.e.\ $\nout=d^2$, the normalized variance diverges in probability.
\end{theorem}

\parTitle{Classical double-descent profile} \Cref{thm:informal_classical} makes this mechanism quantitative. For $\nout<d^2$ ($\gamma_C>1$), only a fraction of operator directions is accessible and the exact zero modes are omitted from the variance; the missing target components nevertheless leave an ideal-test error. As $\nout$ increases toward $d^2$, the smallest \emph{nonzero} frame eigenvalues shrink, and the normalized variance grows as $1/|1-\gamma_C|$, with the prefactor specified in~\cref{eq:classical_MP_comparison}. For $\nout>d^2$ ($\gamma_C<1$), all operator directions are accessible and additional outcomes open a spectral gap, reducing the variance again. The classical double-descent peak thus measures the difficulty of reconstructing an observable from a nearly informationally complete measurement, not the difficulty of estimating its training probabilities.

The theorem concerns the minimum-norm coefficients obtained from the POVM itself, before introducing finite training-shot noise. Its bound on the trace of the target prevents a growing state-independent component from dominating the normalization; such a component is reconstructed by nearly uniform outcome weights and does not cause the peak (\cref{sec:classical_spectrum}). The precise control of small frame eigenvalues used to establish~\cref{eq:classical_MP_comparison} is given in~\cref{thm:classical_peak}.

\parTitle{Finite-dimensional peak} The divergence at $\gamma_C=1$ refers to the limit $d\to\infty$ with $d^2/\nout\to1$. At any fixed dimension, each informationally complete POVM has finite test variance, but a few exceptionally small positive eigenvalues can make it very large. Their values fluctuate substantially between random POVMs, so the finite-dimensional peak has a broad distribution even though its typical height remains finite. We therefore compare the asymptotic curve with sample medians and percentile bands in~\cref{fig:classical}, rather than relying on averages that are sensitive to nearly singular realizations. The large-$N$ results already reflect the predicted peak at $d=8$. The progressive suppression of the peak at lower training-shot budgets is the subject of the next subsection.

\subsection{Finite training data and noise-induced regularization}
\label{sec:classical_training}

So far, the coefficients have been obtained from exact knowledge of the measurement apparatus. We now ask how finite training resources affect the classical peak: first, which coefficients regression on exact training probabilities returns; then, how training-shot noise changes the least-squares problem.

\parTitle{What exact-feature regression reconstructs} With exact training probabilities, an informationally complete POVM and training states spanning the operator space give $\widetilde\calO_{\bs w_{\rm ex}}=\calO$ (\cref{eq:var_epsilon_0}): the labels determine the target observable without reconstruction error. Below informational completeness, instead, only the component of $\calO$ in the span of the POVM effects can be reconstructed. Write $\calO=\calO_\parallel+\calO_\perp$, where $\calO_\parallel=\sum_a w_{{\rm ex},a}\mu_a$ is the Hilbert--Schmidt projection onto that span, with the minimum-norm coefficients $\bs w_{\rm ex}$ of~\cref{sec:setup_error}, and $\calO_\perp$ is orthogonal to every effect. Because the POVM effects sum to $I_d$, the inaccessible part is traceless: $\Trace(\calO_\perp)=0$.

To see why a large Haar training ensemble selects $\calO_\parallel$, consider any observable $A$ in the POVM span and put $D=\calO-A$. For a Haar-random pure training state $\rho$, the second-moment identity (\cref{lem:haar_error_split}) gives
\begin{equation}
\label{eq:main_haar_training_loss}
    \EE_\rho[\Trace(D\rho)^2]
    =\frac{\Trace(D^2)+[\Trace(D)]^2}{d(d+1)}.
\end{equation}
The difference $A-\calO_\parallel$ lies in the POVM span and is therefore Hilbert--Schmidt orthogonal to $\calO_\perp$. The squared error in~\cref{eq:main_haar_training_loss} consequently equals the unavoidable contribution from $\calO_\perp$ plus a nonnegative contribution from $A-\calO_\parallel$, minimized at $A=\calO_\parallel$. Among coefficient vectors producing that observable, the pseudoinverse chooses $\bs w_{\rm ex}$ with minimum norm.

For a \emph{finite} Haar training set the same orthogonality need not hold after summing over the particular sampled states. An accessible operator $A$ and an inaccessible one $B$ satisfy $\Trace(AB)=0$ and $\EE_\rho[\Trace(A\rho)\Trace(B\rho)]=0$, but the sample sum $\sum_i\Trace(A\rho_i)\Trace(B\rho_i)$ is generally nonzero. Least squares can exploit this accidental overlap by changing the accessible observable to reproduce part of the inaccessible labels on the training states. Such a change can reduce the \emph{training} error while increasing the mean squared error on independently sampled Haar test states. For Haar training ensembles these sample overlaps become negligible in the regime $d^2\log d/\ntr\to0$, under the assumptions of~\cref{cor:classical_training}; the fitted variance then approaches the frame-based variance of~\cref{sec:classical_frame}.

\parTitle{Finite shots suppress the classical peak} With $N$ shots per training state, least squares uses $\hatbs P_N$ rather than the exact matrix $\bs P$. \Cref{fig:classical} shows that large $N$ gives a sharp variance peak near $\nout=d^2$, whereas reducing $N$ progressively flattens it. The reason is that the learned weights act on noisy frequencies both at training and at test. Reconstructing a target component along a weakly resolved operator direction requires large weights (\cref{sec:classical_frame}), and these make the prediction for a test state fluctuate with variance $V(\sigma,\calO)/M$. On each training state, the same weights make the prediction $\bs w^T\hatbs n_i/N$ fluctuate around its exact value with variance $\bs w^T\bs\Sigma_i\bs w/N$. When there are many more training states than outcomes, the weights cannot follow these fluctuations, and since the labels are exact, the fluctuations add to the training error. Least squares therefore trades reconstruction accuracy against weight size, and shrinks the weights along directions whose variation across training states is weaker than their sampling noise. These are precisely the directions responsible for the classical peak.

\parTitle{The shot-averaged training loss} To make this trade-off precise, fix the training states and their exact labels, and consider any weight vector $\bs w$ that does not depend on the measured counts. We denote by $\EE_{\hatbs n}$ the average over the multinomial count records $\hatbs n_1,\ldots,\hatbs n_{\ntr}$ of~\cref{eq:main_counts}, with the training states held fixed. On state $\rho_i$, the predicted label $\bs w^T\hatbs n_i/N$ has mean $\bs w^T\bs p_i$ and variance $\bs w^T\bs\Sigma_i\bs w/N$, so its mean squared deviation from $y_i$ is the squared bias $(\bs w^T\bs p_i-y_i)^2$ plus this variance. Averaging over the training set the loss that enters~\cref{eq:main_ridge_estimator} therefore gives the exact identity (\cref{lem:shot_averaged_ls})
\begin{equation}
\label{eq:noise_induced_regularization}
    \frac1{\ntr}\EE_{\hatbs n}
    \left[\|\hatbs P_N^T\bs{w}-\bs y\|^2\right]
    =\frac1{\ntr}\|\bs P^T\bs{w}-\bs y\|^2
    +\frac{\bs w^T\barSigma_{\rm tr}\bs w}{N},
\end{equation}
where $\barSigma_{\rm tr}\equiv\ntr^{-1}\sum_i\bs\Sigma_i$ is the single-shot covariance averaged over the training states. The first term is the loss obtained with exact features. The second measures the shot noise that the weights feed into the training predictions. Only the counts are averaged: the training states are the given ones, not samples from an ensemble.

\parTitle{Averaged objective and fitted weights} Least squares minimizes the loss of the record actually observed, not the average in~\cref{eq:noise_induced_regularization}, which holds for a fixed $\bs w$. The two are related through the normal equations $\hatbs P_N\hatbs P_N^T\bs w=\hatbs P_N\bs y$ satisfied by the fitted weights. Their coefficient matrices have shot averages $\EE_{\hatbs n}[\hatbs P_N\hatbs P_N^T]/\ntr=\bs P\bs P^T/\ntr+\barSigma_{\rm tr}/N$ and $\EE_{\hatbs n}[\hatbs P_N\bs y]=\bs P\bs y$, which are exactly the normal equations of the averaged objective (\cref{lem:shot_averaged_ls}). Both matrices are sums of $\ntr$ independent contributions, one per training state. When $\ntr\gg\nout$ they concentrate around their averages, and the fitted weights approach the minimizer of~\cref{eq:noise_induced_regularization}. We use this concentration as a heuristic and do not quantify it; near $\nout=\ntr$ it fails, as discussed at the end of this subsection.

\parTitle{When the penalty resembles ridge} The extra term in~\cref{eq:noise_induced_regularization} suppresses large weights in directions sensitive to shot noise, much as explicit ridge regularization does in classical regression~\cite{bishop1995training,fuller1987MeasurementErrorModels,carroll2006MeasurementErrorNonlinear}. There is one exception: for weights proportional to the all-ones vector $\mathbf1_{\nout}$, the prediction is constant because the frequencies sum to one, and $\bs\Sigma_i\mathbf1_{\nout}=0$, so shot noise leaves the identity component unpenalized. On the remaining, nonconstant outcome directions the penalty is nearly isotropic for balanced measurements. If the mean training probabilities are nearly uniform, $\ntr^{-1}\sum_i\bs p_i\approx\mathbf1_{\nout}/\nout$, as for Haar training states and effects of nearly equal trace, then $\nout\barSigma_{\rm tr}$ is close to the projector $\bs\Pi_{\nout}=I_{\nout}-\mathbf1_{\nout}\mathbf1_{\nout}^T/\nout$ onto these directions, up to corrections of order $d^{-2}$ from the variation of the exact probabilities across training states (\cref{lem:shot_averaged_ls}). The averaged objective is then the ridge objective of~\cref{eq:main_ridge_estimator} applied to exact features, with an unpenalized uniform weight and ridge parameter $\tau_N=1/(N\nout)$.

\parTitle{Shot noise sets a spectral resolution floor} The scale $\tau_N$ tells us which weakly resolved measurement directions are damped. For Haar-random pure training states, a traceless operator $A$ with $\Trace(A^2)=1$ has mean squared expectation value $1/[d(d+1)]$ (\cref{lem:haar_error_split}). Accordingly, a traceless frame direction with eigenvalue $\lambda$ contributes an eigenvalue approximately $\lambda/[d(d+1)]$ to the normalized exact-feature matrix $\bs P\bs P^T/\ntr$, and ridge reduces its coefficient when this eigenvalue is smaller than $\tau_N$. We define the training-shot ratio
\begin{equation}
\alpha\equiv\frac{N}{d(d+1)}.
\label{eq:main_alpha}
\end{equation}
In terms of the rescaled frame eigenvalue $x=\nout\lambda$ of~\cref{thm:informal_spectrum}, the damping condition then reads $\alpha x\lesssim1$. The product $\alpha x$ is the signal-to-noise ratio of the direction: the variance of its exact expectation value across Haar-random states divided by the variance of its $N$-shot estimate (\cref{lem:snr_direction}). Finite training shots therefore impose a resolution floor: regression tends not to invert measurement directions whose variation across training states is weaker than their sampling noise. Since the classical peak comes from frame eigenvalues approaching zero, the floor suppresses the peak. Because the bulk of the Marchenko--Pastur spectrum has $x$ of order one, the crossover occurs for $\alpha$ of order one, i.e., for shot budgets of order $d^2$; recovering the contribution of the very smallest eigenvalues of a finite-dimensional frame can require substantially more shots. At fixed $d$ and for a sufficiently large training set, increasing $N$ eventually removes this floor and restores the exact-feature classical peak.

\parTitle{Why the quantum peak survives} The effective-ridge picture requires the measured normal equations to stay close to their shot averages. Their noise part $\hatbs\Xi_N\hatbs\Xi_N^T/(N\ntr)$ has average $\barSigma_{\rm tr}/N\approx\bs\Pi_{\nout}/(N\nout)$, but it is a sample covariance of $\ntr$ independent vectors in $\nout$ dimensions. Its eigenvalues therefore spread around this average over a range set by $\nout/\ntr$, and for $\nout$ close to $\ntr$ some of them approach zero, as for any nearly square random matrix~\cite{marvcenko1967distribution,bai2010spectral}. On the outcome directions that carry no exact signal this noise part is all there is (\cref{sec:quantum_mechanism}). Near $\nout=\ntr$, the measured feature matrix may thus have extremely small singular values in an individual record even though its averaged loss is regularized, and least squares strongly amplifies the fluctuations of that particular record. Finite-shot noise can therefore suppress the \emph{classical} peak of the measurement frame while creating the distinct \emph{quantum} interpolation peak analyzed in~\cref{sec:quantum}.

\section{Quantum peak at the interpolation threshold}
\label{sec:quantum}

We now analyze interpolation above informational completeness, $\nout=\ntr>d^2$, where the peak originates in the empirical feature matrix rather than in the exact measurement frame. Although the POVM can reconstruct every target observable, the exact probability matrix has rank at most $d^2$; finite-shot fluctuations generate additional outcome directions without exact signal. Near interpolation, least squares can fit these directions with large weights even for exact training labels. We first explain how this mechanism amplifies test variance and distorts the reconstructed observable, then quantify its singularity and its dependence on training shots and ridge regularization. Interpolation below informational completeness has a distinct, classical mechanism and is considered in~\cref{sec:uc_main}.

\subsection{Noise-only features and the quantum mechanism}
\label{sec:quantum_mechanism}

The origin of the quantum peak can be seen by separating the exact quantum subspace from the directions created by noise. When the POVM and training states span operator space, let the orthonormal columns of $\bs U_1,\bs U_2$ span $\im(\bs P)$ and $\ker(\bs P^T)$ in $\RR^{\nout}$, and let those of $\bs V_1,\bs V_2$ span $\im(\bs P^T)$ and $\ker(\bs P)$ in $\RR^{\ntr}$, respectively. Because the POVM is informationally complete, $\ker(\bs P)$ consists of the combinations $\bs c\in\RR^{\ntr}$ of training states that cancel as operators, $\sum_ic_i\rho_i=0$, whereas $\ker(\bs P^T)$ consists of the outcome weights $\bs w \in \RR^{\nout}$ that reconstruct the zero observable, $\sum_aw_a\mu_a=0$. Since $\rank(\bs{P})=d^2$, the matrices $\bs U_1$ and $\bs V_1$ have $d^2$ columns, whereas $\bs U_2$ and $\bs V_2$ have $\nout-d^2$ and $\ntr-d^2$ columns. The measured matrix takes the block form
\begin{equation}
 \begin{pmatrix}\bs U_1^T\\\bs U_2^T\end{pmatrix}\!
 \hatbs P_N[\bs V_1\ \bs V_2]
 \!=\!\begin{pmatrix}
 \bs P_{11}+N^{-\frac12}\hatbs\Xi_{11} & N^{-\frac12}\hatbs\Xi_{12}\\
 N^{-\frac12}\hatbs\Xi_{21} & N^{-\frac12}\hatbs\Xi_{22}
 \end{pmatrix},
 \label{eq:alt_companion_blocks}
\end{equation}
where $\bs{P}_{11}=\bs U_1^T\bs{P} \bs V_1$ and $\hatbs\Xi_{kl}=\bs U_k^T\hatbs\Xi_N\bs V_l$; we drop the subscript $N$ from the blocks of $\hatbs\Xi$ for brevity.
Only the upper-left block contains exact probabilities; the other three blocks consist entirely of measurement fluctuations and vanish for large $N$. In particular the noise-only block $\hatbs\Xi_{22}$ has size $(\nout-d^2)\times(\ntr-d^2)$ and thus becomes square precisely at $\nout=\ntr$, where, as for any nearly square random matrix, its smallest singular values are much smaller than its typical ones with high probability~\cite{marvcenko1967distribution,bai2010spectral,vershynin2018high}.

\parTitle{Weights along noise-only directions} At interpolation, the effect of this block on the learned weights can be made explicit. Whenever $\nout\ge\ntr$ and $\hatbs P_N$ has full column rank, least squares fits the training labels exactly, $\hatbs P_N^T\hatbs w_N=\bs y$; at $\nout=\ntr>d^2$ this solution is unique and can be obtained block by block. Write $\hatbs w_1=\bs U_1^T\hatbs w_N$ and $\hatbs w_2=\bs U_2^T\hatbs w_N$ for the learned weights along the signal and noise-only outcome directions. Because the labels are exact and the target lies in the span of the effects, $\bs y=\bs P^T\bs w_{\rm ex}$ has no component along $\ker(\bs P)$: the label combinations $\sum_ic_iy_i=\Tr(\calO\sum_ic_i\rho_i)$ vanish on the cancelling combinations of training states. By~\cref{eq:alt_companion_blocks}, the interpolation conditions along $\ker(\bs P)$ therefore read $N^{-1/2}(\hatbs\Xi_{12}^T\hatbs w_1+\hatbs\Xi_{22}^T\hatbs w_2)=0$, so that
\begin{equation}
 \hatbs w_2=-(\hatbs\Xi_{22}^T)^{-1}\hatbs\Xi_{12}^T\hatbs w_1.
 \label{eq:main_null_weights}
\end{equation}
The factor $N^{-1/2}$ has cancelled. The signal weights $\hatbs w_1$ are of order one, since they reconstruct the target, and through the noisy training features they add a spurious contribution $N^{-1/2}\hatbs\Xi_{12}^T\hatbs w_1$ to the predictions on the cancelling combinations of training states, whose labels vanish. The noise-only weights $\hatbs w_2$ must cancel this contribution using the only features available along $\ker(\bs P)$, which are themselves of order $N^{-1/2}$. The noise-only weights are therefore of order one as well, independently of $N$, and they become large when $\hatbs\Xi_{22}$ is nearly singular. Inserting~\cref{eq:main_null_weights} into the conditions along $\im(\bs P^T)$ shows that the signal weights solve the exact equations $\bs P_{11}^T\bs U_1^T\bs w_{\rm ex}=\bs V_1^T\bs y$ with $\bs P_{11}$ replaced by $\bs P_{11}+N^{-1/2}\bigl(\hatbs\Xi_{11}-\hatbs\Xi_{12}\hatbs\Xi_{22}^{-1}\hatbs\Xi_{21}\bigr)$ (\cref{eq:main_signal_weights} of~\cref{prop:block_interpolation}).

\parTitle{Two contributions to prediction error} Two distinct effects follow, matching the two error contributions in~\cref{eq:hat_MSE}. First, the reconstructed observable $\widetilde\calO_N$ does not depend on the noise-only weights: every vector $\bs U_2\hatbs w_2\in\ker(\bs P^T)$ reconstructs the zero observable, so that $\widetilde\calO_N=\sum_a(\bs U_1\hatbs w_1)_a\mu_a$. The noise-only weights still respond to the shot noise of the test frequencies, however, and large $\hatbs w_2$ inflate the test variance.
Second, the signal weights solve a perturbed version of the exact equations, in which the Schur complement $\hatbs\Xi_{12}\hatbs\Xi_{22}^{-1}\hatbs\Xi_{21}$ can be large near interpolation. The weights $\hatbs w_1$ then deviate from the exact coefficients $\bs U_1^T\bs w_{\rm ex}$, and the reconstructed observable acquires an error that persists with exact test probabilities and is measured by the ideal-test error. The block identities are exact whenever $\hatbs P_N$ and $\hatbs\Xi_{22}$ are invertible (\cref{prop:block_interpolation}). The heuristic step is the size estimate, which treats the blocks of $\hatbs\Xi_N$ as generic random matrices, although they are projections of the same multinomial fluctuations, whose covariances $\bs\Sigma_i$ depend on the training state. The results below do not rely on this estimate; they are proved for the minimum-norm and ridge estimators computed from the actual multinomial counts, including records for which $\hatbs P_N$ is rank deficient, for instance because some outcome is never observed.

\subsection{Quantum peak: divergence and ridge-rounded profiles}
\label{sec:quantum_ic}

To make these statements quantitative, we consider large Hilbert-space dimensions, $d\to\infty$, with $\nout,\ntr$, and $N$ also increasing proportionally to $d^2$. More precisely, we generally consider the limit with fixed values of the ratios
\begin{equation}
 \gamma_Q=\frac{\nout}{\ntr},\qquad
 \kappa=\frac{d^2-1}{\nout-1},\qquad
 \alpha=\frac{N}{d(d+1)},
 \label{eq:main_quantum_scaling}
\end{equation}
and consider the scaled ridge parameter $\eta=\nout N\tau$. We study $\gamma_Q$ near one and take $0<\kappa<1$, so that the measurement is informationally complete. The ratio $\kappa$ is the fraction of nonconstant outcome directions that can carry exact signal: probability normalization removes one of the $\nout$ outcome directions, and the trace removes one of the $d^2$ operator directions. In the limit, $\kappa$ coincides with the aspect ratio $\gamma_C=d^2/\nout$ of~\cref{sec:classical}. The training-shot ratio $\alpha$ of~\cref{eq:main_alpha} sets the training signal-to-noise ratio: an operator direction with rescaled frame eigenvalue $x$ is resolved by $N$ shots with signal-to-noise ratio $\alpha x$ across Haar-random states (\cref{lem:snr_direction}). Since the Marchenko--Pastur law of~\cref{thm:informal_spectrum} has unit mean, $\alpha$ is the signal-to-noise ratio of a typical direction, and $N\propto d^2$ is precisely the regime in which the training shots resolve the state-dependent signal only partially.
The scaled ridge parameter $\eta$ compares $\tau$ with the shot-noise scale of the measured features: a fixed $\eta>0$ corresponds to $\tau=\eta/(\nout N)$, which decreases as the training statistics improve, and $\eta=1$ reproduces the parameter $\tau_N$ of the regularization that shot noise induces on average (\cref{sec:classical_training}).

\parTitle{Divergence at interpolation}
Our first result concerns the \emph{physical, unregularized} estimator~$\hatbs w_N$ of~\cref{eq:qelm_exact_linear_rule}. Its divergence does not follow from the analysis of ridge regression below, because the limits $\tau\to0$ and $d\to\infty$ need not commute.

\begin{theorem}[Divergence of the ridgeless test variance (informal)]
\label{thm:informal_quantum_peak}
Consider the proportional regime of~\cref{eq:main_quantum_scaling} with $\nout/\ntr\to1$, $0<\kappa<1$ and $\alpha>0$, independent multinomial training records, and a target with $\Delta_{\calO}^2>0$ and bounded $\lvert\Tr(\calO)\rvert/(d\sqrt{d\Delta_{\calO}^2})$. Assume that (i) the training-state and measurement frames are bounded, so that no operator direction carries an anomalously large weight; (ii) shot noise is nondegenerate, with strength of order $1/\nout$ along every nonconstant outcome direction, for both training and test states; and (iii) the training states remain spread around their mean, so that the labels vary across the training set. Then, as $d\to\infty$, the target-averaged ridgeless test variance diverges,
\begin{equation}
 \frac{V_N}{d\Delta_{\calO}^2}\xrightarrow{\PP}+\infty.
 \label{eq:main_central_growth}
\end{equation}
The same holds for the test variance $V_N(\calO)$ of a single random target and for its median. Records for which $\hatbs P_N$ is rank deficient are included.
\end{theorem}
The precise conditions (i)--(iii) are~\cref{eq:alt_geometry,eq:general_noise_bounds,eq:general_noncollapse}, and the formal statement is~\cref{thm:alt_central}.

\begin{figure}[tbh]
 \centering
 \includegraphics[width=\linewidth]{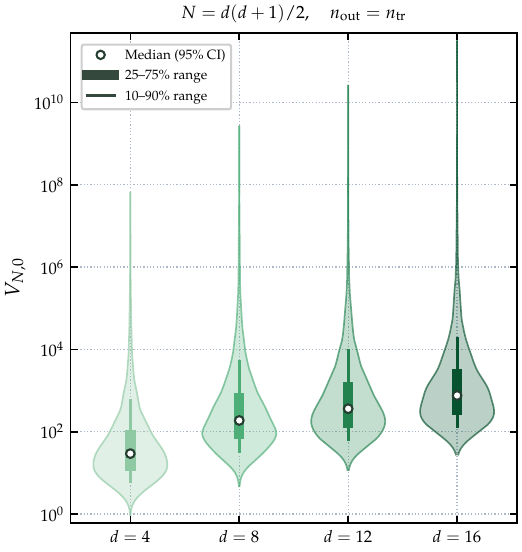}
 \caption{\textbf{Distribution of the ridgeless test variance at interpolation.} Distributions of the ridgeless test variance $V_N(\calO)$ at interpolation, $\nout=\ntr$, for $d=4,8,12,16$ and $\nout=64,256,512,1024$, respectively, with $N=d(d+1)/2$. These choices give $\nout\approx4d^2$, so that $\kappa=(d^2-1)/(\nout-1)\approx0.24$--$0.28$ and $\alpha=N/[d(d+1)]=1/2$ stay nearly fixed while $d$ grows (\cref{eq:main_quantum_scaling}). Each violin contains 10\,000 independent realizations of a normalized Haar rank-one POVM, a Haar training ensemble, a traceless target drawn uniformly from the unit sphere ($d\Delta_\calO^2=1$), and multinomial training records. Widths show densities on a logarithmic scale; markers and bars indicate medians, 95\% confidence intervals and percentile ranges. Realizations with rank-deficient $\hatbs P_N$ or extremely large $V_N$ are retained.}
 \label{fig:quantum_distribution}
\end{figure}

\parTitle{Generality of the interpolation divergence} \Cref{thm:informal_quantum_peak} does not require the training states to span the full operator space: whenever $\nout>d^2$, the deviations $\bsp_i-\bar{\bsp}$ of the exact probability vectors from their training mean $\bar{\bsp}$ span at most $d^2-1$ of the $\nout-1$ nonconstant outcome directions, independently of the training ensemble, leaving at least $\nout-d^2$ directions without exact signal. Finite-shot noise populates these directions, creating spurious features that can be strongly amplified near interpolation. Conditions (ii) and (iii) ensure that these fluctuations are nondegenerate and that the training labels vary enough to require large weights. Since the divergence holds with probability tending to one over the shot noise in the training records, it describes typical records rather than rare ill-conditioned realizations. The theorem does not specify the limiting distribution of the peak height, which depends on the smallest singular values of $\hatbs{P}_N$;~\cref{fig:quantum_distribution} shows the finite-size distributions at $\gamma_Q=1$ and their growth with $d$ at nearly fixed $\kappa$ and $\alpha$.

\parTitle{Shape of the peak}
To describe the shape of the peak, we analyze ridge regression at a fixed scaled ridge parameter $\eta>0$ and use the normalized target averages $V_{N,\tau}/(d\Delta_{\calO}^2)$ and $d(d+1)\calE_{N,\tau}/(d\Delta_{\calO}^2)$ of~\cref{sec:setup}; the second is the relative Hilbert--Schmidt error of the reconstructed state-dependent observable (\cref{eq:ideal_test_error_operator}). The analysis is simplest for least squares with an unpenalized intercept, which is equivalent to subtracting the training-set averages from labels and features before fitting; we call this the \emph{centered} fit. The intercept absorbs the constant part of the prediction, so that the centered fit isolates the state-dependent part of the problem, where the peak arises. We denote by $F_\eta(\gamma)$ and $G_\eta(\gamma)$ the large-$d$ limits, at $\gamma_Q\to\gamma$, of the two normalized error contributions of the centered fit, defined in~\cref{eq:general_centered_costs,eq:general_centered_ideal_cost}. The physical fit has the same leading singular test variance. Its ideal-test error can contain additional contributions from the empirical training mean and from the identity component of the target (\cref{thm:alt_ridge_profiles}). At every fixed $\eta>0$, these contributions vanish for traceless targets whenever the training states are pure and satisfy conditions (i) and (iii), and for all targets when, in addition, the training states are Haar random and the effects have nearly equal traces (\cref{app:proof_haar_centered_profiles}); training sets of mixed states can retain them.

\begin{figure}[t]
 \centering
 \includegraphics[width=\linewidth]{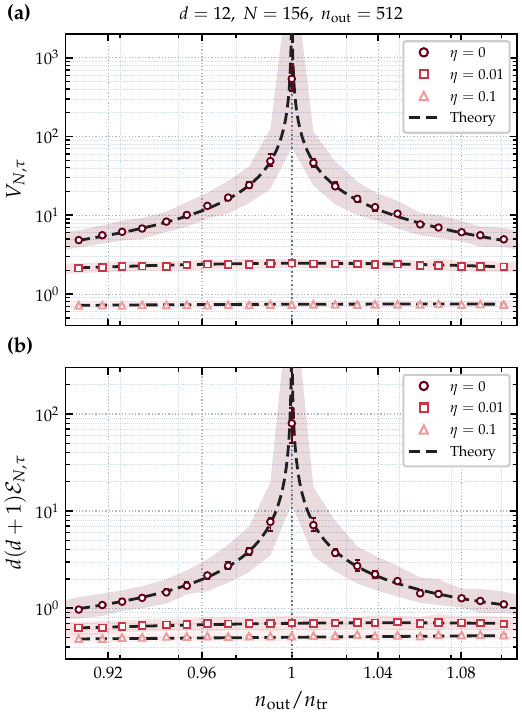}
 \caption{\textbf{Quantum peak and ridge rounding.} Haar prediction-error profiles near the quantum interpolation threshold, as $\ntr$ varies at fixed $d=12$, $\nout=512$ and $N=d(d+1)=156$, i.e., $\kappa\approx0.28$ and $\alpha=1$. (a) Test variance $V_{N,\tau}(\calO)$. (b) Normalized ideal-test error $d(d+1)\calE_{N,\tau}(\calO)$. The three red tones indicate scaled ridge parameters $\eta=\nout N\tau=0,0.01,0.1$, consistently in both panels. Open red markers show simulated medians; charcoal dashed curves show the complete analytical predictions for Haar training states and Haar POVMs, \cref{eq:alt_F,eq:alt_G,eq:alt_left,eq:alt_right}, including the nonsingular background, evaluated without fitting at the stated $d$, $N$, and $\nout$. Each point contains 128 independent realizations of a normalized Haar rank-one POVM, a Haar pure training ensemble, a traceless unit-sphere target ($d\Delta_\calO^2=1$), and multinomial records; within each realization, all values of $\eta$ share the same apparatus, states, target, and training counts. Shading shows pointwise 10th--90th percentiles.}
 \label{fig:quantum}
\end{figure}

\begin{theorem}[Quantum peak and ridge rounding (informal)]
\label{thm:informal_quantum_profiles}
Under the assumptions of~\cref{thm:informal_quantum_peak}, together with a regularity condition on the large-$d$ limits of the normalized traces entering the calculation (\cref{cor:alt_rounding}), $F_\eta(\gamma)$ and $G_\eta(\gamma)$ are deterministic functions for every $\eta>0$. Taking the dimensional limit first and then $(\gamma,\eta)\to(1,0)$ gives
\begin{equation}
 \begin{aligned}
 F_\eta(\gamma)&\sim\frac{C_V}{\sqrt{(\gamma-1)^2+4h\eta}},\\
 G_\eta(\gamma)&\sim\frac{C_G}{\sqrt{(\gamma-1)^2+4h\eta}}
 \qquad (C_G>0),
 \end{aligned}
 \label{eq:main_ridge_rounding}
\end{equation}
where $C_V>0$, $C_G\ge0$ and $h>0$ depend on the training and measurement geometry at interpolation, as given in~\cref{eq:general_peak_coefficients,eq:general_ridge_width}.
\end{theorem}
\parTitle{Pole and ridge-rounded profile} The square-root denominator in~\cref{eq:main_ridge_rounding} interpolates between two regimes. For $2\sqrt{h\eta}\ll|\gamma-1|\ll1$, both contributions decay as $1/|\gamma-1|$; the upper bound on $|\gamma-1|$ is needed because~\cref{eq:main_ridge_rounding} describes only the singular part of the profiles, which dominates near interpolation. This pole is the profile of the ridgeless fit: \cref{cor:alt_fixed_gap} shows that the centered ridgeless costs obey $F_0(\gamma)\sim C_V/|\gamma-1|$ and $G_0(\gamma)\sim C_G/|\gamma-1|$ (\cref{eq:general_poles}), provided the smallest measured singular values carry a vanishing share of the coefficient norm. For $|\gamma-1|\lesssim2\sqrt{h\eta}$, any fixed $\eta>0$ rounds off the pole into a peak of finite height. The two amplitudes $C_V$ and $C_G$ correspond to the two effects identified at the beginning of this section: $C_V$ quantifies the amplification of finite test-shot fluctuations by large weights, while $C_G$ quantifies the error of the reconstructed observable, which persists even with exact test probabilities. Their dependence on $N$ need not be the same (\cref{sec:ridge_risk}). The simulations in~\cref{fig:quantum} compare both contributions with the complete Haar profiles as $\gamma_Q$ crosses interpolation.

\parTitle{Frame-spectrum control of the amplitudes} For Haar pure training states and effects of nearly equal trace, $\Trace(\mu_a)\approx d/\nout$, shot noise has the same strength along every nonconstant outcome direction, and the coefficients $C_V,C_G$ and $h$ have an intuitive spectral form. If $\nu$ is the limiting distribution of the rescaled traceless POVM-frame eigenvalues (the Marchenko--Pastur law of~\cref{thm:informal_spectrum} for Haar POVMs), then
\begin{equation}
 \begin{gathered}
 \phi=\int\frac{\dd\nu(x)}{1+\alpha x},\qquad
 h=1-\kappa+\kappa\phi,\\
 C_V=\alpha\phi h,\qquad C_G=\kappa\phi(1-\phi).
 \end{gathered}
 \label{eq:main_phi_general}
\end{equation}
The quantity $\phi$ has a direct operational meaning: it is the relative error of the best linear estimate of a Haar-random state from $N$ shots of the measurement, i.e., the minimum mean squared Hilbert--Schmidt error divided by the mean squared distance of the state from $I_d/d$ (\cref{eq:weak_haar_training_moments}). Along each frame eigendirection, the traceless coordinate of a Haar-random state has variance $1/[d(d+1)]$ and is measured with the signal-to-noise ratio $\alpha x$ (\cref{lem:snr_direction}). The best linear estimate of such a coordinate retains the fraction $1/(1+\alpha x)$ of its variance, and $\phi$ averages this fraction over the frame spectrum: well-resolved directions ($\alpha x\gg1$) contribute little, and inaccessible ones ($x=0$) contribute fully. Both amplitudes vanish when the training shots resolve nothing, $\phi\to1$, but they behave differently as the resolution improves: $C_G$ also vanishes for $\phi\to0$, whereas $C_V$ does not.

\begin{figure}[t]
    \centering
    \includegraphics[width=\linewidth]{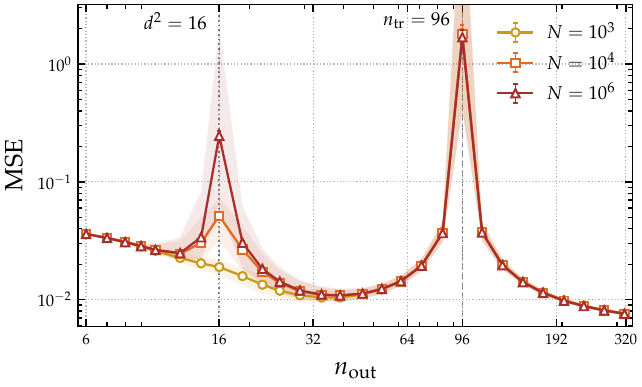}
    \caption{\textbf{Training-shot dependence of triple descent.} Full test MSE. We fix $d=4$, $\ntr=96$, and $M=256$ test shots, and vary the number of training shots per state: $N=10^3$ (yellow), $10^4$ (orange), and $10^6$ (dark red). The minimum-norm readout is trained without ridge regularization, and its full test error is $\MSE_{N,M}=\calE_N+V_N/M$.
    The outcome count varies crossing $\nout=d^2=16$ (dotted line) and $\nout=\ntr=96$ (dashed line). Decreasing $N$ strongly rounds the first peak, while the interpolation peak remains pronounced.
    Markers show medians over 1024 independent Haar-POVM, training-state, and measurement-record realizations per point; apparatus and training states are paired across budgets. Shading shows the 10th--90th percentiles, and error bars show pointwise 95\% confidence intervals for the median. Target-direction and Haar test-state averages are evaluated exactly.}
 \label{fig:quantum_mse}
\end{figure}

\subsection{Training-shot dependence and noncommuting limits}
\label{sec:ridge_risk}

The effect of increasing training precision depends on how the shot budget scales with dimension. For Haar-random measurements the peak amplitudes can be evaluated explicitly; fixed-shot and fixed-dimension limits then reveal behaviors that are not apparent from proportional scaling alone.

\parTitle{Haar-random measurements}
For independent Haar pure training states, the expectation value $\Tr(A\rho_i)$ of every traceless observable $A$ with $\Tr A^2=1$ has the same variance $1/[d(d+1)]$ over the training states, and the measurement enters the coefficients of~\cref{eq:main_phi_general} only through the eigenvalue distribution $\nu$ of the rescaled traceless POVM frame. For normalized Haar rank-one POVMs, $\nu$ is the limiting law of~\cref{thm:informal_spectrum} restricted to traceless operators (\cref{thm:main}), i.e., the Marchenko--Pastur law $\mu_{{\rm MP},\kappa}$, because $\kappa$ and $\gamma_C$ have the same limit. Its support is $[(1-\sqrt{\kappa})^2,(1+\sqrt{\kappa})^2]$ in the informationally complete regime $\kappa<1$.

\parTitle{Closed-form Haar amplitudes} For this law, $\phi$ is available in closed form: $\phi$ equals $\alpha^{-1}$ times the Stieltjes transform $\int\dd\nu(x)/(x-z)$ evaluated at $z=-1/\alpha$, and this transform obeys a quadratic self-consistent equation~\cite{bai2010spectral} (\cref{app:proof_peak_amplitudes}). For $\phi$, the equation reads $1-\phi=\alpha\phi(1-\kappa+\kappa\phi)$, whose positive solution is
\begin{equation}
 \phi=\frac{2}{1+\alpha(1-\kappa)+\sqrt{[1+\alpha(1-\kappa)]^2+4\kappa\alpha}}.
 \label{eq:main_phi_mp}
\end{equation}
The right-hand side of this equation is $\alpha\phi h$, by~\cref{eq:main_phi_general}, so that $C_V=\alpha\phi h=1-\phi$, and the two amplitudes depend only on $\phi$ and $\kappa$:
\begin{equation}
 C_V=1-\phi,\qquad
 C_G=\kappa\phi(1-\phi).
 \label{eq:main_random_amplitudes}
\end{equation}

\begin{figure}[t]
 \centering
 \includegraphics[width=\linewidth]{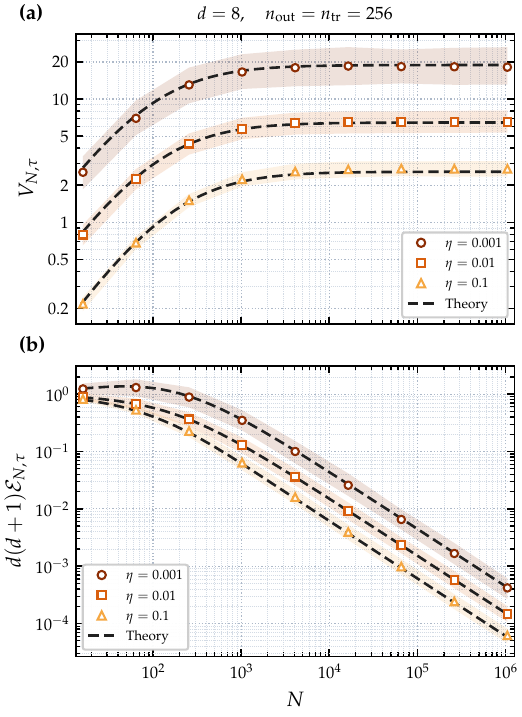}
 \caption{\textbf{Training-shot dependence at interpolation.} Parameters: $d=8$, $\nout=\ntr=256$ ($\kappa\approx0.25$). Panels (a),(b) show $V_{N,\tau}(\calO)$ and $d(d+1)\calE_{N,\tau}(\calO)$ as $N=4^k$, $k=2,\ldots,10$, varies at $\eta=10^{-3},10^{-2},10^{-1}$. Open markers are medians and shading gives 10th--90th percentiles over 1024 independent Haar and multinomial realizations. Dashed curves are the complete analytical Haar predictions, evaluated at the actual budgets.}
 \label{fig:shot_profiles}
\end{figure}

\parTitle{Contrasting dependence on training shots} The two amplitudes respond differently to training statistics. As $\alpha$ increases, $\phi$ decreases, reflecting the improved resolution of the exact probability features. The variance amplitude $C_V$ therefore increases monotonically towards $1$, whereas the ideal-test amplitude $C_G$ is nonmonotonic: it reaches its maximum $C_G=\kappa/4$ at $\phi=1/2$, i.e., at $\alpha=2/(2-\kappa)$, and then decays. For fixed $0<\kappa<1$ and large $\alpha$,
\begin{equation}
 \begin{aligned}
 C_V&=1-\frac{1}{(1-\kappa)\alpha}+O(\alpha^{-2}),\\
 C_G&=\frac{\kappa}{(1-\kappa)\alpha}+O(\alpha^{-2}).
 \end{aligned}
 \label{eq:main_ic_shot_contrast}
\end{equation}
Increasing $N$ thus improves the reconstructed observable but strengthens the variance pole, whose amplitude saturates at $C_V=1$ rather than diverging. The target-averaged MSE of the centered fit is approximately $d\Delta_\calO^2[G_\eta/(d(d+1))+F_\eta/M]$, so that its singular coefficient is $d\Delta_\calO^2[C_G/(d(d+1))+C_V/M]$.
With finite test statistics, additional training shots therefore do not suppress the peak: the $V/M$ contribution approaches the nonzero amplitude $d\Delta_\calO^2/M$, while the ideal-test contribution decays as $1/\alpha$. Because the two contributions are weighted by $1/[d(d+1)]$ and $1/M$, the variance dominates the peak unless $M\gtrsim d(d+1)C_V/C_G$. The complete Haar profiles and multinomial simulations in~\cref{fig:shot_profiles} show this contrast.

\parTitle{Fixed shot number or fixed dimension} The proportional limit fixes $\alpha=N/[d(d+1)]$, so the amplitudes above describe training budgets that grow with the dimension. Two other ways of varying $N$ lead to different behaviors.

\begin{figure}[t]
 \centering
 \includegraphics[width=\linewidth]{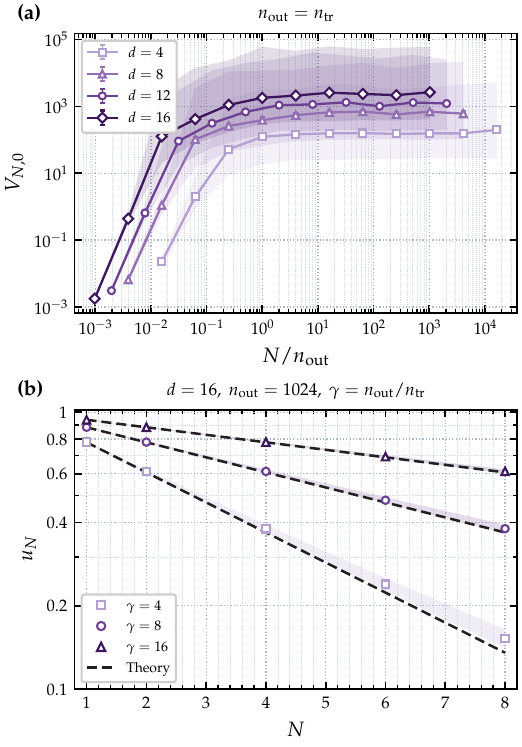}
 \caption{\textbf{Finite-record effects.} (a) Ridgeless test variance $V_N(\calO)$ at interpolation for $d=4,8,12,16$, with $\nout=\ntr=64,256,512,1024$ and $N=4^k$, $k=0,\ldots,10$. (b) Fraction $u_N$ of outcomes never observed in any training record (\cref{sec:ridge_risk}) at $d=16$, $\nout=1024$, $\gamma_Q=\nout/\ntr=4,8,16$ and $N=1,2,4,6,8$. Markers show medians in (a) and ensemble means in (b), using 1024 independent realizations per point. Dashed lines in (b) show the large-$d$ prediction $e^{-N/\gamma_Q}$ of~\cref{prop:alt_finite_N}; bars and shading indicate uncertainty and percentile ranges. Realizations with rank-deficient $\hatbs P_N$ are included.}
 \label{fig:quantum_shots}
\end{figure}

\parTitle{Sparse records at fixed shot number} At fixed $N$ and growing dimension, $\alpha\to0$ and the training records become sparse. For approximately balanced outcome probabilities, a fraction $e^{-N/\gamma_Q}$ of the outcomes is never observed in any training record (\cref{prop:alt_finite_N}). Their rows in $\hatbs P_N$ vanish, so that $\hatbs P_N$ becomes rank deficient and the minimum-norm fit assigns these outcomes zero weight. Discarding these outcomes removes directions that could otherwise carry large weights and thus limits the test variance: at $N=1$ and $\Tr(\calO^2)=1$, every record gives $V_1(\calO)\le1$ and $\MSE_{1,M}(\calO)\le4$. Predictions do not become accurate, however, since unseen outcomes still occur at test time, where they contribute nothing to the prediction. \Cref{fig:quantum_shots} shows the finite-shot test variance and the fraction of unseen outcomes.

\parTitle{Saturation at fixed dimension} At fixed dimension and growing $N$, instead, the peak saturates, as can be understood from~\cref{eq:main_null_weights}. As $N\to\infty$, $\hatbs w_1$ converges to the exact coefficients and $\hatbs\Xi_N$ to a Gaussian matrix, by the multivariate central limit theorem, so $\hatbs w_2$ converges in distribution to a nonzero random vector that does not depend on $N$. The test variance at the peak, $V_N(\calO)=\hatbs w_N^T\barSigma_{\rm te}\hatbs w_N$, is a continuous function of $\hatbs w_N$ and therefore converges in distribution to a finite random variable, whose typical size is set by the smallest singular values of the square noise block $\hatbs\Xi_{22}$, whereas the ideal-test error, to which $\hatbs w_2$ does not contribute, decreases as $1/N$. Away from interpolation, an analogous argument with a pseudoinverse gives smaller but still $N$-independent weights $\hatbs w_2$, because the noise block is rectangular and better conditioned. With exact features ($N=\infty$), in contrast, the noise-only directions do not exist, and the variance reduces to the shadow variance of~\cref{sec:classical}. The limits $N\to\infty$ and $N=\infty$ thus do not commute, consistently with the proportional-limit amplitudes $C_V\to1$ and $C_G=O(1/\alpha)$; \cref{fig:quantum_shots}(a) shows the resulting plateau of the test variance at large $N$. This argument is heuristic: it assumes that $\hatbs P_N$ and the limiting Gaussian block remain invertible with probability one, and we do not prove the convergence in distribution.

\subsection{Explicit ridge regularization}
\label{sec:quantum_ridge}

Explicit ridge regularization, \cref{eq:main_ridge_estimator}, acts on the measured matrix itself: in every record, it replaces each inverse singular value $1/s$ of $\hatbs P_N$ by $s/(s^2+\ntr\tau)$ (\cref{sec:setup}), and thus bounds the weights along the small singular values of the particular noise realization. The noise-induced regularization of~\cref{sec:classical_training}, by contrast, acts only on the loss averaged over training records, and near interpolation the fit follows the particular record instead. Since $\eta=\nout N\tau$ sets the natural scale of the ridge parameter, varying $N$ at fixed $\eta$ is not the same experiment as varying $N$ at fixed $\tau$.

\parTitle{Ridge scaling at interpolation} At $\gamma=1$, \cref{eq:main_ridge_rounding} of~\cref{thm:informal_quantum_profiles} gives peak heights that grow as the ridge parameter decreases,
\begin{equation}
 F_\eta(1)\sim\frac{C_V}{2\sqrt{h\eta}},\qquad
 G_\eta(1)\sim\frac{C_G}{2\sqrt{h\eta}}\quad(C_G>0).
 \label{eq:main_central_ridge}
\end{equation}
The exponent $-1/2$ is fixed by the square-root denominator and is therefore the same for every training set and measurement satisfying the assumptions of~\cref{thm:informal_quantum_profiles}, whereas the prefactors $C_V,C_G$ and the width $2\sqrt{h\eta}$ depend on the training and measurement geometry, through $\nu$ in the Haar case, where $h=1-\kappa+\kappa\phi$. Moreover, \cref{eq:main_central_ridge} holds when $d\to\infty$ is taken \emph{before} $\eta\to0$, and does not apply to arbitrarily weak ridge at finite dimension: at fixed $d$, the values of $V_{N,\tau}$ and $\calE_{N,\tau}$ at $\gamma_Q=1$ saturate at their finite ridgeless values as $\eta\to0$.

\begin{figure*}[t]
 \centering
 \includegraphics[width=\linewidth]{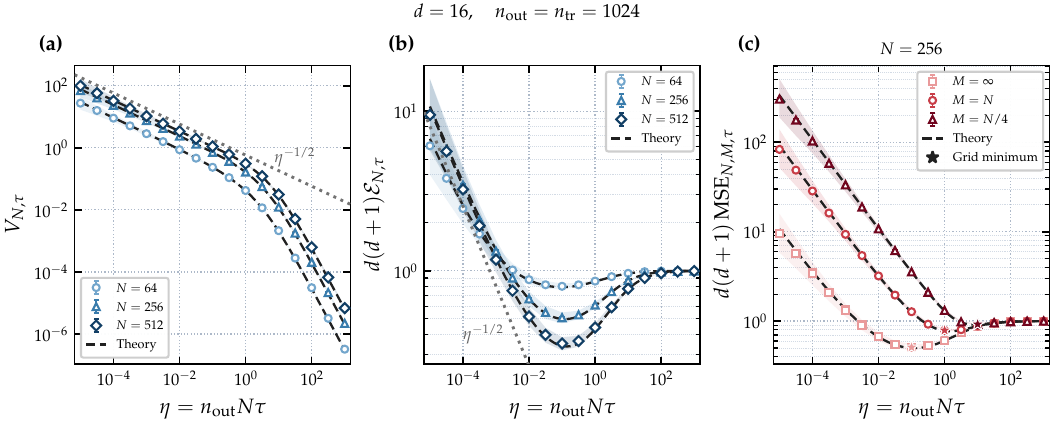}
 \caption{\textbf{Ridge rounding at interpolation.} Parameters: $d=16$, $\nout=\ntr=1024$. (a),(b) Test variance $V_{N,\tau}(\calO)$ and $d(d+1)\calE_{N,\tau}(\calO)$ at interpolation versus $\eta$, for $N=64,256,512$; (c) $d(d+1)\MSE_{N,M,\tau}(\calO)$ at $N=256$ for $M=\infty,N,N/4$. The scaled ridge parameter is $\eta=\nout N\tau$. Markers show medians of 1024 Haar and multinomial realizations, each shared by all values of $\eta$, shading gives 10th--90th percentiles, and dashed curves are the full analytical Haar predictions. Dotted guides in (a),(b) illustrate the $\eta^{-1/2}$ laws of~\cref{eq:main_central_ridge} (rescaled for visibility); stars in (c) mark minima on the displayed grid of $\eta$.}
 \label{fig:ridge_rounding}
 \label{fig:ridge_mse}
\end{figure*}

\parTitle{Regularization and prediction accuracy} As shown in~\cref{fig:ridge_rounding}(a),(b), ridge rounds both error contributions, while the remaining deviations at extremely small $\eta$ reflect the finite-dimensional saturation discussed above. In panel (c), increasing ridge first reduces the MSE by suppressing overfitting and then increases it by attenuating the target signal; the minimum depends on $M$. For Haar training states, $d(d+1)\MSE_{N,M,\tau}/(d\Delta_\calO^2)$ tends to $G_\eta+[d(d+1)/M]F_\eta$; for general training sets, the additional terms of~\cref{thm:alt_ridge_profiles} must be added. Ridge offers a controlled way to reduce the peak, although its useful strength depends on the POVM, the training states, and both shot budgets.

\section{Classical peak below informational completeness}
\label{sec:uc_main}

When $\nout=\ntr<d^2$, i.e., $\kappa>1$, the interpolation peak is classical: it is present even with exact features, and a peak of this type was also recently reported in~\cite{meyer2026shaping}. For generic measurements and training states, the exact feature matrix now has full rank, so no outcome direction is guaranteed to carry pure noise. Instead, the target $\calO$ generally contains a component outside the POVM span, which the measurement cannot access. Let $\bs w_{\rm ex}=\bs M_{\bs\mu}^+\bs v_\calO$ be the minimum-norm reconstruction coefficients of~\cref{sec:setup_error}, which reconstruct the orthogonal projection of $\calO$ onto the span of the effects, and write $\calO_\perp=\calO-\sum_a w_{{\rm ex},a}\mu_a$ for the inaccessible part. Since $\bsp_i^T\bs w_{\rm ex}=\Tr[(\calO-\calO_\perp)\rho_i]=y_i-\Tr(\calO_\perp\rho_i)$, the labels differ from the predictions of $\bs w_{\rm ex}$ on the measured training frequencies by the residual
\begin{equation}
 r_{N,i}=y_i-\hatbsp_i^T\bs w_{\rm ex}
 =\Tr(\calO_\perp\rho_i)-(\hatbsp_i-\bsp_i)^T\bs w_{\rm ex}.
 \label{eq:uc_main_residual}
\end{equation}
The first contribution is the inaccessible part of the target and survives even for exact training probabilities; the second is measurement noise. Above informational completeness, $\calO_\perp=0$ and only the noise term remains.

\parTitle{Amplification of inaccessible residuals} Least squares corrects $\bs w_{\rm ex}$ so as to reproduce this residual on the training states: it seeks a correction $\hatbs w_N-\bs w_{\rm ex}$ with $\hatbs P_N^T(\hatbs w_N-\bs w_{\rm ex})\approx\bs r_N$. Let $s_j>0$ be a singular value of $\hatbs P_N$, with left (outcome) and right (training) singular vectors $\bs u_j\in\RR^{\nout}$ and $\bs v_j\in\RR^{\ntr}$. Writing $\bs y=\hatbs P_N^T\bs w_{\rm ex}+\bs r_N$ and using $(\hatbs P_N^T)^+=\sum_j\bs u_j\bs v_j^T/s_j$, the minimum-norm solution $\hatbs w_N=(\hatbs P_N^T)^+\bs y$ of~\cref{eq:qelm_exact_linear_rule} satisfies
\begin{equation}
 \bs u_j^T(\hatbs w_N-\bs w_{\rm ex})=\frac{\bs v_j^T\bs r_N}{s_j}.
 \label{eq:uc_main_amplification}
\end{equation}
Each right singular vector $\bs v_j$ defines a combination of training states whose measured frequency vectors add up to $\sum_i(\bs v_j)_i\hatbsp_i=\hatbs P_N\bs v_j=s_j\bs u_j$; for small $s_j$, these frequency vectors nearly cancel. To reproduce the component $\bs v_j^T\bs r_N=\sum_i(\bs v_j)_ir_{N,i}$ of the residual on such a combination, the weights must change by $\bs v_j^T\bs r_N/s_j$ along $\bs u_j$. Near $\nout=\ntr$, the matrix $\hatbs P_N$ is nearly square and has singular values close to zero, and the resulting large weights inflate the test variance. Because these weights also change the reconstructed observable, adapting it to the inaccessible values $\Tr(\calO_\perp\rho_i)$ on the training states but not on new states, they inflate the ideal-test error as well.

\parTitle{Small singular values without shot noise}
Unlike above informational completeness, these small singular values are already present in the exact feature matrix $\bs P$. Its columns $\bsp_i$ are random because the training states are random, and near $\nout=\ntr$ the matrix is nearly square, as in random-design regression~\cite{hastie2020Surprises}. The near-singularity thus belongs to the finite training set rather than to the measurement frame, and it occurs even if the POVM resolves all the operator directions it can access equally well. To isolate this effect, we remove the constant part absorbed by the intercept of the fit and consider the centered exact matrix $\bs P-\bar{\bsp}\mathbf1_{\ntr}^T$, whose columns are the deviations of the exact probability vectors from their training mean $\bar{\bsp}$, with $\mathbf1_{\ntr}$ the all-ones vector. For Haar pure training states, unitary invariance makes these columns nearly independent and isotropic on the $\nout-1$ nonconstant outcome directions, so the centered matrix behaves like a square random matrix with independent columns. Take, for instance, a POVM whose rescaled traceless frame has all its nonzero eigenvalues equal to some $x_0>0$. The squared singular values of the suitably rescaled centered matrix then follow a Marchenko--Pastur law with lower edge $x_0(1-\sqrt\gamma)^2$, where $\gamma$ is the limit of $(\nout-1)/\ntr$ (\cref{lem:uc_hard_edge}). The lower edge reaches zero at $\gamma=1$, where the limiting density diverges as $x^{-1/2}$, and the inverse moment of its positive part diverges as $|\gamma-1|^{-1}$, as for the measurement frame at $\nout=d^2$. A target whose residual has negligible projection on the small singular directions need not display the peak, so this divergence alone is not a statement about individual observables.

\begin{figure}[tp]
 \centering
 \includegraphics[width=0.9\columnwidth]{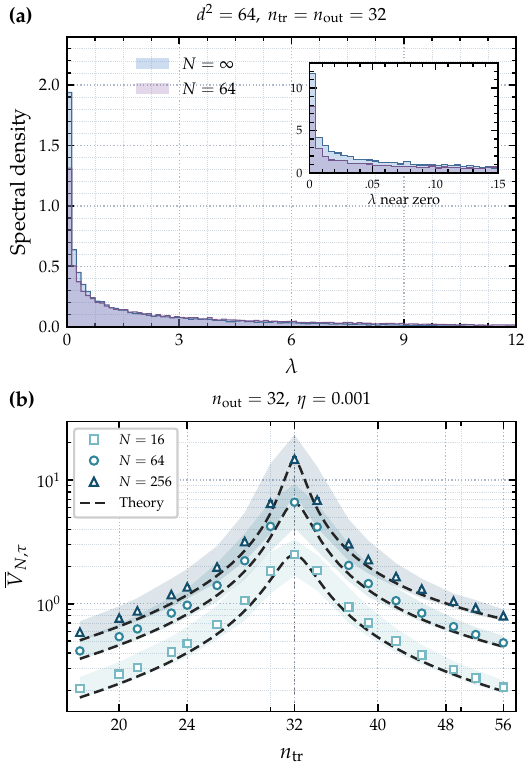}
 \caption{\textbf{Interpolation below informational completeness.} Parameters: $d=8$, $\nout=32<d^2$ ($\kappa\approx2.0$). (a) Small-eigenvalue spectra of centered exact and $N=64$-shot training matrices at $\ntr=32$. (b) Target-averaged physical test variance as $\ntr$ crosses interpolation, for $N=16,64,256$ and $\eta=10^{-3}$. Histograms and medians use 1024 independent Haar and multinomial realizations; shading indicates 10th--90th percentiles and dashed curves are the analytical Haar predictions.}
 \label{fig:undercomplete_interpolation}
\end{figure}

\parTitle{Effect of finite shots}
Finite-shot noise adds a random matrix to the centered features. This changes their singular values, but does not in general move the smallest ones away from zero at $\nout=\ntr$. Such a gap, i.e., a lower bound on the singular values that stays positive as $d\to\infty$, would keep the factors $1/s_j$ in~\cref{eq:uc_main_amplification} bounded and remove the peak. A gap does open for the classical peak when $\ntr\gg\nout$: there, the noise contribution to $\hatbs P_N\hatbs P_N^T$ concentrates around a multiple of the identity on the nonconstant outcome directions and lifts every eigenvalue (\cref{sec:classical_training}). At $\nout=\ntr$, instead, the noise is itself a nearly square random matrix. As an illustration, model the centered exact features and the shot noise by independent Gaussian matrices with independent entries, of variances $\alpha x_0/\ntr$ and $1/\ntr$, respectively, in the shot-noise units in which the scaled ridge parameter $\eta$ is defined (\cref{eq:uc_gaussian_surrogate}). Their sum is again such a Gaussian matrix, with entry variance $(1+\alpha x_0)/\ntr$, so its spectrum still reaches zero at $\gamma=1$. At $\gamma=1$, the squared norm of the ridge weights fitted to a residual with no preferred direction then grows as $[4(1+\alpha x_0)\eta]^{-1/2}$ as $\eta\to0$ (\cref{eq:uc_gaussian_inverse_moment}), rather than remaining bounded. Finite shots thus lower the peak without removing it, as~\cref{fig:undercomplete_interpolation} shows for actual multinomial records.

\parTitle{Condition for a singular profile} Whether this hard edge produces a singular profile at finite $\alpha$ is decided by comparing the state-dependent signal with shot noise. Two sources of variability compete in the measured features: the exact probability vectors differ from one training state to another, which is the signal to be learned, and each of these vectors is estimated from only $N$ shots. To compare them, we use the training-averaged single-shot covariance $\barSigma_{\rm tr}$ of~\cref{eq:noise_induced_regularization}. Define $\lambda_j$, $j=1,\ldots,\nout-1$, as the squared singular values, including zeros, of
\begin{equation}
 \sqrt{N\ntr^{-1}}\,(\barSigma_{\rm tr}^{+})^{1/2}
 (\bs P-\bar{\bsp}\mathbf1_{\ntr}^T).
\label{eq:main_whitened_probability_matrix}
\end{equation}
The factor $(\barSigma_{\rm tr}^{+})^{1/2}$ measures the deviations of the exact probability vectors from their training mean in units of shot noise. The $\lambda_j$ are therefore signal-to-shot-noise ratios: they measure the variation of an exact outcome combination between training states, relative to the fluctuations of its $N$-shot estimate~(\cref{eq:general_critical_signal_to_noise}), generalizing the ratios $\alpha x$ of~\cref{sec:quantum_ic}. The singular profile~\cref{eq:main_ridge_rounding} rests on a positive solution $\zeta_*$ of the critical equation (\cref{app:proof_quantum_peak,app:proof_uc_interpolation})
\begin{equation}
 \frac{1}{\nout-1}\sum_{j=1}^{\nout-1}\frac{1}{\zeta_*+\lambda_j}=1,
 \qquad \zeta_*>0.
 \label{eq:main_critical_inverse}
\end{equation}
Since the left-hand side decreases with $\zeta_*$, a positive solution exists if and only if
\begin{equation}
 \frac{1}{\nout-1}\sum_{j=1}^{\nout-1}\lambda_j^{-1}>1,
 \label{eq:main_positive_root_condition}
\end{equation}
where a zero eigenvalue contributes $+\infty$. Above informational completeness this condition holds automatically: the centered exact probabilities span at most $d^2-1$ of the $\nout-1$ nonconstant outcome directions, whereas shot noise acts on all of them, so that at least $\nout-d^2$ of the $\lambda_j$ vanish. These are the noise-only directions of~\cref{sec:quantum_ic}, and they are the reason why~\cref{thm:informal_quantum_peak,thm:informal_quantum_profiles} need no spectral assumption. Below informational completeness, the exact features can occupy all $\nout-1$ nonconstant directions, so no eigenvalue is forced to vanish, and the condition holds only if sufficiently many signal directions are weak compared with shot noise. Equivalently, multiplying~\cref{eq:main_critical_inverse} by $\zeta_*$ gives $\zeta_*=1-(\nout-1)^{-1}\sum_j\lambda_j/(\zeta_*+\lambda_j)$.
The normalized sum is the fraction of nonconstant outcome directions in which the state-dependent signal is resolved above the self-consistent noise level $\zeta_*$, and $\zeta_*$ is the fraction left unresolved. For the equal-strength POVM above, the hard edge of the exact features makes the left-hand side of~\cref{eq:main_positive_root_condition} diverge at $\gamma_Q=1$, so that the condition holds at every finite $\alpha$.

\begin{theorem}[Incomplete interpolation peak (informal)]
\label{thm:informal_uc_interpolation}
Assume $\kappa>1$, $\gamma\to1$, fixed $\alpha\in(0,\infty)$, conditions (i)--(iii) of~\cref{thm:informal_quantum_peak}, a POVM whose rescaled traceless frame has a positive average eigenvalue, a positive solution of~\cref{eq:main_critical_inverse} bounded away from zero, and the trace regularity of~\cref{thm:uc_interpolation}. For independent multinomial training records, traceless targets averaged as in~\cref{sec:setup} and Haar test states, with a training mean approaching the maximally mixed state, $\|\bar\rho-I_d/d\|_F\to0$, the target-averaged \emph{physical} variance and ideal-test error have the same ridge-rounded singular denominator as in~\cref{eq:main_ridge_rounding}, with positive amplitudes. At interpolation the target-averaged minimum-norm variance diverges in probability; the ideal-test error also diverges when the accessible traceless POVM frame is uniformly well conditioned.
\end{theorem}
Below informational completeness the interpolation peak is therefore conditional on~\cref{eq:main_positive_root_condition}, whereas above completeness the noise-only directions guarantee it.

\parTitle{Haar-random measurements} For Haar pure training states and normalized Haar rank-one POVMs, the amplitudes are again given by~\cref{eq:main_random_amplitudes}, now with $\kappa>1$, where $\nu=\mu_{{\rm MP},\kappa}$ has an atom of mass $1-1/\kappa$ at zero. At small $\alpha$, $C_V=\alpha+O(\alpha^2)$ and $C_G=\kappa\alpha+O(\alpha^2)$; at large $\alpha$,
\begin{equation}
 C_V\longrightarrow\frac{1}{\kappa},\qquad
 C_G\longrightarrow1-\frac{1}{\kappa},\qquad
 h\sim\frac{1}{(\kappa-1)\alpha}.
 \label{eq:uc_main_large_shot_amplitudes}
\end{equation}
The singular amplitude of the ideal-test error no longer vanishes with increasing training precision: the inaccessible part of the target persists even with exact features. For $1<\kappa<2$, $C_G$ is also nonmonotonic and reaches $\kappa/4$ before approaching its large-shot limit. The interpolation peaks above and below informational completeness, driven by shot noise and by inaccessible target components, respectively, thus differ both in mechanism and in their scaling with measurement resources.

\section{Conclusions and perspectives}
\label{sec:conclusions}

In this work, we have investigated how finite quantum measurement statistics affect the generalization properties of quantum linear regression, with quantum extreme learning machines as a particularly relevant setting. We have shown that the interplay between the geometry of quantum measurements and the statistical fluctuations of their outcomes can give rise to a \emph{triple-descent} behavior as the number of measurement outcomes increases. The triple descent originates from two distinct spectral instabilities, which occur at different thresholds and respond differently to finite measurement statistics.

\parTitle{Classical peak} The first peak arises near informational completeness, when the number of outcomes approaches the dimension $d^2$ of operator space. In this regime, small eigenvalues of the POVM frame make the reconstruction of certain observables increasingly sensitive to measurement fluctuations. This instability is already present when the training probabilities are known exactly, and its asymptotic behavior can be understood through the spectrum of random quantum measurements. Interestingly, finite training statistics can suppress this classical peak, illustrating how measurement noise can sometimes have an effective regularizing role.

\parTitle{Two interpolation mechanisms} Above informational completeness, the interpolation peak at $\nout=\ntr>d^2$ has a fundamentally different origin: statistical fluctuations populate directions absent from the exact quantum feature representation. Our analysis of multinomial measurement records establishes that the test variance grows at this threshold for typical training realizations, rather than only through rare events. Away from interpolation, we derive deterministic high-dimensional profiles for both the test variance and the error that remains when test probabilities are known exactly. Their singular behavior is governed by the distance from interpolation, while the measurement-frame spectrum and the training-shot budget determine its magnitude. At fixed dimension, the peak in the test variance saturates as the number of training shots grows, yet it is absent when the training features are exact: the limits $N\to\infty$ and $N=\infty$ do not commute, which is the clearest signature of the quantum origin of the peak. Below informational completeness, the interpolation peak is instead classical: it is driven by target components that the measurement cannot access, it is present with exact features, and finite statistics change its amplitude without generally removing it. 

\parTitle{Training shots and regularization} These results reveal a subtle relationship between measurement precision and learning stability. Increasing the number of shots improves the accuracy of individual quantum features, but does not necessarily reduce the variance of the resulting predictor. Conversely, very limited statistics may suppress unstable directions by leaving the empirical feature matrix rank deficient, without necessarily improving prediction accuracy. We have also shown that ridge regularization provides a controlled way to mitigate the quantum peak. A positive ridge parameter makes the limiting prediction-error profiles finite at interpolation, while the peak height grows as the inverse square root of the ridge parameter when regularization is weakened. The optimal choice depends on both training and test measurement resources, since suppressing statistical fluctuations must be balanced against the loss of predictive information.

\parTitle{Estimator and measurement design} Several questions remain open. One is to characterize more precisely the behavior at the quantum interpolation threshold, including the finite-size scaling and distribution of the peak height. Another is whether the quantum peak can be avoided by training methods that exploit the structure of quantum features. Because it originates in outcome directions that exact Born probabilities cannot occupy, the quantum peak is tied to the least-squares estimator rather than to the measurement, and we expect estimators that discard these directions, for instance by restricting the fit to a $d^2$-dimensional feature subspace, to suppress it. The classical peak, in contrast, is tied to the measurement: for informationally complete POVMs with equal effect traces, the minimum-norm reconstruction already minimizes the average test variance among unbiased linear estimators (\cref{sec:classical_frame}), so that the classical peak can only be traded against bias, as ridge regularization does, or avoided by redesigning the measurement. A related question concerns the design of quantum measurements for learning. Our results show that informational completeness alone does not guarantee stable predictions: the spectrum of the measurement frame and its interplay with finite sampling are equally important. This suggests exploring POVMs designed to optimize generalization performance, rather than reconstruction accuracy alone, and investigating how measurement geometry, shot allocation, and regularization can be jointly optimized under a fixed experimental budget. 

\parTitle{Beyond linear regression} A further direction is to understand whether the mechanisms identified here extend beyond linear regression. Particularly relevant settings include nonlinear classical readouts, quantum models with multiple data injections and quantum memory, and learning algorithms based on quantum fidelity kernels. In these architectures, the relationship between quantum measurements and the effective feature representation can be considerably more complex, and finite sampling may introduce different correlations and spectral instabilities. Understanding whether analogous double- or triple-descent phenomena arise, and how they depend on the structure of the model and the available quantum resources, remains an interesting open problem.

\parTitle{Broader implication} More broadly, our findings show that finite measurement statistics cannot always be regarded as a small perturbation of an ideal quantum learning model. They can fundamentally alter the geometry of the learning problem, with consequences for both generalization and the efficient use of quantum measurement resources.

\section*{Acknowledgements}
We are grateful to S.~Meyer, E.~Yilmaz, A.~Lucchi, F.~Tacchino, and F.~Scala for insightful discussions about their work~\cite{meyer2026shaping}, which addresses aspects complementary to those investigated here. We thank Alessandro Romancino for helpful discussions. GLM acknowledges funding from the European Union--NextGenerationEU through the Italian Ministry of University and Research under PNRR-M4C2-I1.3 Project PE-00000019 ``HEAL ITALIA'' (CUP B73C22001250006).
SL, AF, and GMP acknowledge support by MUR under PRIN Project No.~2022FEXLYB, Quantum Reservoir Computing (QuReCo).
LI, SL, MP, and GMP acknowledge funding from the ``National Centre for HPC, Big Data and Quantum Computing (HPC)'' Project CN00000013, HyQELM--Spoke 10.
MP is grateful for support from the Royal Society Wolfson Fellowship (RSWF/R3/183013), the Department for the Economy of Northern Ireland under the US-Ireland R\&D Partnership Programme, the PNRR PE Italian National Quantum Science and Technology Institute (PE0000023), and the EU Horizon Europe EIC Pathfinder project QuCoM (GA no.~10032223).


\clearpage
\onecolumngrid
\appendix

\setcounter{equation}{0}
\setcounter{figure}{0}
\setcounter{table}{0}
\setcounter{section}{0}

\makeatletter
\renewcommand{\thesection}{\Alph{section}}
\renewcommand{\theequation}{\thesection\arabic{equation}}
\renewcommand{\thefigure}{\thesection\arabic{figure}}
\renewcommand{\thetable}{\thesection\arabic{table}}
\@addtoreset{theorem}{section}
\@addtoreset{lemma}{section}
\@addtoreset{proposition}{section}
\@addtoreset{corollary}{section}
\@addtoreset{conjecture}{section}
\@addtoreset{result}{section}
\@addtoreset{remark}{section}
\renewcommand{\thetheorem}{\thesection\arabic{theorem}}
\renewcommand{\thelemma}{\thesection\arabic{lemma}}
\renewcommand{\theproposition}{\thesection\arabic{proposition}}
\renewcommand{\thecorollary}{\thesection\arabic{corollary}}
\renewcommand{\theconjecture}{\thesection\arabic{conjecture}}
\renewcommand{\theresult}{\thesection\arabic{result}}
\renewcommand{\theremark}{\thesection\arabic{remark}}

\renewcommand{\theHsection}{SM.\Alph{section}}
\renewcommand{\theHequation}{SM.\Alph{section}.\arabic{equation}}
\renewcommand{\theHfigure}{SM.\Alph{section}.\arabic{figure}}
\renewcommand{\theHtable}{SM.\Alph{section}.\arabic{table}}

\renewcommand{\theHtheorem}{SM.\Alph{section}.\arabic{theorem}}
\renewcommand{\theHlemma}{SM.\Alph{section}.\arabic{lemma}}
\renewcommand{\theHproposition}{SM.\Alph{section}.\arabic{proposition}}
\renewcommand{\theHcorollary}{SM.\Alph{section}.\arabic{corollary}}
\renewcommand{\theHconjecture}{SM.\Alph{section}.\arabic{conjecture}}
\renewcommand{\theHresult}{SM.\Alph{section}.\arabic{result}}
\renewcommand{\theHremark}{SM.\Alph{section}.\arabic{remark}}
\makeatother

\makeatletter
\let\SMoriginaladdtocontents\addtocontents
\long\def\addtocontents#1#2{%
  \def\SMtocfile{#1}%
  \def\SMstandardtoc{toc}%
  \ifx\SMtocfile\SMstandardtoc
    \SMoriginaladdtocontents{smt}{#2}%
  \else
    \SMoriginaladdtocontents{#1}{#2}%
  \fi
}
\newcommand{\SMtableofcontents}{%
  \begingroup
    \small
    \setlength{\parskip}{0pt}%
    \def\l@f@section{\addvspace{0.2em}}%
    \let\l@subsubsection\@gobbletwo
    \let\SM@numberline\numberline@@sections
    \def\numberline@@sections##1{%
      \if\relax\detokenize{##1}\relax
        \leavevmode\ignorespaces
      \else
        \SM@numberline{##1}%
      \fi}%
    \@starttoc{smt}%
  \endgroup
}
\newcommand{\SMtwocoltableofcontents}{%
  \begingroup
    \small
    \setbox\z@\vbox{%
      \hsize=\dimexpr(\textwidth-1cm)/2\relax 
      \linewidth\hsize \columnwidth\hsize
      \interlinepenalty\@M \clubpenalty\@M \widowpenalty\@M 
      \vbox{}
      \SMtableofcontents}%
    \splittopskip\baselineskip \vbadness\@M \vfuzz\maxdimen
    \setbox\tw@\vsplit\z@ to\z@ 
    \dimen@=.5\ht\z@
    \loop 
      \setbox\tw@\copy\z@
      \setbox4\vsplit\tw@ to\dimen@
      \ifdim\ht\tw@>\dimen@ \advance\dimen@ 1pt \repeat
    \par\nointerlineskip
    \noindent\hbox to\textwidth{\vtop{\unvbox4}\hfil\vtop{\unvbox\tw@}}\par
  \endgroup
}
\makeatother

\makeatletter
\begingroup
  \let\addtocontents\SMoriginaladdtocontents
  \def\toclevel@section{0}%
  \phantomsection
  \addcontentsline{toc}{section}{Supplemental Material}%
\endgroup
\makeatother

\begin{center}
    {\Large\bfseries Supplemental Material}
\end{center}
\vspace{1em}
\noindent\textbf{Contents}\par\smallskip
\SMtwocoltableofcontents
\bigskip

Here we discuss separately the two mechanisms discussed in the main text: weakly resolved measurement directions near informational completeness ($\nout=d^2$), and small singular directions of the measured training matrix near interpolation ($\nout=\ntr$).
\Cref{app:conventions} fixes operator coordinates, norms, asymptotic conventions, and the target and test-state averages. These conventions specify which error is being studied and which randomness remains in the main-text spectral and regression claims.
\Cref{app:concentration_POVM} proves the Haar-POVM spectrum used in~\cref{sec:classical_spectrum}: a bulk of state-dependent directions, an identity outlier, and exact zero modes when outcomes are insufficient. It also treats the finite-dimensional outlier correction and fixed-rank extensions.
\Cref{app:bounds} turns this spectrum into the classical test-variance peak of~\cref{sec:classical_peak}. Small eigenvalues are inverted by the shadow readout, so their total inverse weight requires control beyond a limiting spectral density. The final subsections relate the shadow result to exact-feature training and derive the shot-averaged training loss used in~\cref{sec:classical_training}.
\Cref{app:quantum_profiles} develops the common regression calculation behind~\cref{sec:quantum_ic,sec:ridge_risk,sec:uc_main}. It separates training signal from shot noise, evaluates a centered auxiliary fit at positive ridge, and restores the empirical mean needed for the physical readout and its ideal-test error.
\Cref{app:interpolation_mechanisms} derives the peak near $\nout=\ntr$. Above informational completeness, signal-free outcome directions force a critical solution. Below completeness, a positive-root condition identifies when weak training directions still produce a peak. The section gives amplitudes and ridge rounding, and distinguishes positive-ridge limits from claims about the actual minimum-norm estimator.
\Cref{app:complete_profiles_resources} adds the regular background to the Haar-training curves used in~\cref{sec:ridge_risk}, then treats finite records and fixed training-shot budgets. These limits explain why the proportional interpolation divergence does not imply infinite error at fixed resources.
\Cref{app:rm_prelim} collects the spectral comparisons, Gaussian-lift bounds, and covariance-resolvent calculations used in the proofs. The physical results and their assumptions can be read first, returning to these estimates at their cited uses.

\section{Conventions and preliminaries}
\label[appsec]{app:conventions}

The spectral and regression arguments use a common separation between the identity response and the state-dependent, traceless operator directions. \Cref{end:notation} fixes the coordinates, matrix operations, and probabilistic limits; \cref{end:haar} specifies the target and test-state averages used in the main-text error profiles.

\subsection{Coordinates and asymptotic notation}
\label[appsec]{end:notation}
\label[appsec]{end:little_big_o}

The learning problem of~\cref{sec:setup} uses outcome probabilities $p_a(\rho_i)=\Trace(\mu_a\rho_i)$ as features and exact expectation values $y_i=\Trace(\calO\rho_i)$ as labels. Each training state is measured $N$ times, and a fitted outcome-weight vector predicts a new state's label from $M$ test measurements. Operator coordinates turn the Born features into matrix products; the conventions below distinguish exact identities from large-dimension statements.

\parTitle{Two contributions to prediction error} For fixed training data, weights $\bs w$ reconstruct the observable $\widetilde\calO_{\bs w}=\sum_aw_a\mu_a$. As in~\cref{sec:setup_error}, for a fixed target, Haar test-state and test-shot averaging gives $\MSE_{N,M,\tau}(\calO)=\calE_{N,\tau}(\calO)+V_{N,\tau}(\calO)/M$: the ideal-test error is $\calE_{N,\tau}(\calO)=\EE_\sigma[\Trace((\widetilde\calO_{\bs w}-\calO)\sigma)^2]$, while $V_{N,\tau}(\calO)=\bs w^T\barSigma_{\rm te}\bs w$ measures sensitivity to test-shot fluctuations, with $\bs w=\hatbs w_{N,\tau}$ and $\barSigma_{\rm te}$ the single-shot outcome covariance averaged over Haar test states. The training-shot count $N$ and ridge penalty $\tau$ enter through the fitted weights. Dropping the argument $\calO$ denotes the average over a random target drawn from~\cref{def:target_ensembles}. All costs still depend on the training record; averages over the apparatus, the training states, or the record are taken only when stated.

\parTitle{Operator coordinates and traceless compression} Represent Hermitian operators on $\calH\simeq\CC^d$ in an orthonormal Hermitian basis $(\eta_j)_{j=1}^{d^2}$ for $\langle A,B\rangle=\Trace(A^\dagger B)$. Their real coordinates $(\bs v_A)_j=\Trace(\eta_jA)$ satisfy $\bs v_A^T\bs v_B=\langle A,B\rangle$; $\vect A$ denotes the same vector unless another vectorization is specified. For the $\ntr$ training states, $\bs M_{\bs\rho}=[\bs v_{\rho_1}\ \cdots\ \bs v_{\rho_{\ntr}}]$. The measurement matrix $\bs M_{\bs\mu}$ is defined analogously, and $\bs P=\langle\bs\mu,\bs\rho\rangle=\bs M_{\bs\mu}^T\bs M_{\bs\rho}$ is the $\nout\times\ntr$ Born probability matrix. The operator space has dimension $d^2$; its traceless sector and orthogonal projector are
\begin{equation}
\label{eq:common_projectors}
 \calL(\calH)_0=\{A:\Trace(A)=0\},\qquad
 \bs P_0=\bfI_{d^2}-\frac{\bs v_{I_d}\bs v_{I_d}^T}{d}.
\end{equation}
A compression to this sector means $\bs P_0\bs A\bs P_0$ restricted to its $(d^2-1)$-dimensional range. The identity and inverses on the sector are understood in this restricted sense.

\parTitle{Norms, inverses, and sampled matrices} Bold Latin symbols denote vectors and matrices. Unsubscripted vector norms are Euclidean. Matrix and operator norms carry an explicit subscript: $\|\cdot\|_{\rm op}$ is the spectral norm and $\|A\|_F^2=\Trace(A^\dagger A)$ defines the Frobenius norm. For a linear map on operators, $\|\cdot\|_{\rm op}$ is induced by the Frobenius norm, equivalently the spectral norm in orthonormal operator coordinates. The entrywise maximum norm is $\|\bs A\|_{\max}=\max_{ij}\lvert A_{ij}\rvert$. The Moore--Penrose inverse $\bs A^+$ inverts nonzero singular values and sets zero modes to zero; $\lambda_{\min}^+(\bs A)$ is the smallest positive eigenvalue, and $\Pi_{\bs\mu}=\bs M_{\bs\mu}\bs M_{\bs\mu}^+$ projects onto the measurement-visible operator subspace. Hats identify sampled quantities; removing a hat need not denote an expectation. In particular, $N$ training shots give $\hatbs P_N=\bs P+N^{-1/2}\hatbs\Xi$, while test records use $M$ shots. We use $\gamma_{\rm out}=d^2/\nout$, $\gamma_{\rm tr}=d^2/\ntr$, and $\gamma_Q=\nout/\ntr$; any finite-dimensional adjustment to these ratios is stated where needed.

\parTitle{Probabilistic limits and medians} For random $X_j$ and deterministic $a_j>0$, $X_j=o_{\PP}(a_j)$ means $X_j/a_j\to0$ in probability, and $X_j=O_{\PP}(a_j)$ means that $\lvert X_j\rvert/a_j$ is tight: for every $\varepsilon>0$, some finite $C_\varepsilon$ bounds $\PP(\lvert X_j\rvert>C_\varepsilon a_j)$ by $\varepsilon$ for all sufficiently large $j$. Almost-sure $o(a_j)$ means $X_j/a_j\to0$ almost surely, and almost-sure $O(a_j)$ means $\limsup_j\lvert X_j\rvert/a_j<\infty$ almost surely. For vectors or matrices, these definitions apply to the stated norm; $a_j\asymp b_j$ means their ratio is eventually bounded above and below by positive deterministic constants. The median is $\operatorname{med}X=\inf\{x:\PP(X\le x)\ge1/2\}$. If $X_j/b_j\to c$ in probability for deterministic $b_j>0$ and $c$, then $\operatorname{med}X_j/b_j\to c$, also under a conditional law with fixed apparatus.

\subsection{Target ensembles and Haar moments}
\label[endsec]{end:haar}

The main-text error profiles separate imperfect observable reconstruction from amplified test-shot fluctuations. Both costs are quadratic in the target and depend on the averages below; these are the moments used in~\cref{sec:classical_peak,sec:quantum_ic}. At fixed target trace and Hilbert--Schmidt norm, Haar unitary rotations, sphere directions, and sign directions have the same traceless second moment, but their concentration properties must be treated separately.

\parTitle{Unitary target rotations} Classical variance averages use $\calO_U=U^\dagger\calO U$ for Haar $U$ and a fixed Hermitian $\calO$. Its eigenvalue variance is $\Delta_{\calO}^2=\Trace(\calO^2)/d-(\Trace(\calO)/d)^2$. Haar integration gives $\EE_U\bs v_{\calO_U}=(\Trace(\calO)/d)\bs v_{I_d}$ and~\cite{mele2024introduction,collins2003moments}
\begin{equation}
\label{eq:moment_2_operator}
 \EE_U[\bs v_{\calO_U}\bs v_{\calO_U}^T]
 =\frac{(\Trace(\calO))^2}{d^2}\bs v_{I_d}\bs v_{I_d}^T
 +\frac{d\Delta_{\calO}^2}{d^2-1}\bs P_0.
\end{equation}

\parTitle{Haar test states} For an independent Haar pure state, $\EE_\rho\bs v_\rho=\bs v_{I_d}/d$ and
\begin{equation}
\label{eq:moment_2_state}
 \EE_\rho[\bs v_\rho\bs v_\rho^T]
 =\frac{\bfI_{d^2}+\bs v_{I_d}\bs v_{I_d}^T}{d(d+1)}.
\end{equation}

\parTitle{Averaged test-shot covariance} Write $\omega_a=\Trace(\mu_a)$ for the effect traces, $p_{I/d,a}=\omega_a/d$, $\bs D_\rho=\diag\bs p(\rho)$, and $\bs\Sigma(\rho)=\bs D_\rho-\bs p(\rho)\bs p(\rho)^T$. The state moments give $\EE p_a=\omega_a/d$ and $\EE[p_ap_b]=(\omega_a\omega_b+\Trace(\mu_a\mu_b))/[d(d+1)]$. Thus, for every POVM, the Haar-averaged categorical covariance is
\begin{equation}
\label{eq:barSigma_exact}
 \barSigma_{\rm te}=\bs D_{I/d}
 -\frac{\bs\omega\bs\omega^T+\mathfrak G_{\bs\mu}}{d(d+1)},
 \qquad \barSigma_{\rm te}\mathbf1_{\nout}=0,
\end{equation}
where $\mathfrak G_{\bs\mu}=\bs M_{\bs\mu}^T\bs M_{\bs\mu}$.

\parTitle{Haar average of a squared expectation value} The ideal-test error and the training loss with exact features are both Haar averages of squared expectation values. The following lemma gives their exact form.
\begin{lemma}[Haar average of a squared expectation value]
\label{lem:haar_error_split}
For every Hermitian operator $D$ and a Haar-random pure state $\sigma$,
\begin{equation}
\label{eq:haar_error_split}
\EE_\sigma[\Trace(D\sigma)^2]
=\frac{\Trace(D^2)+(\Trace D)^2}{d(d+1)}
=\frac{\Trace(D_0^2)}{d(d+1)}+\left(\frac{\Trace D}{d}\right)^2,
\end{equation}
where $D_0=D-(\Trace(D)/d)I_d$. In particular, for outcome weights $\bs w$ and a target $\calO$, the ideal-test error of~\cref{eq:hat_MSE} averaged over Haar test states is
\begin{equation}
d(d+1)\calE(\calO)
=\|\bs P_0\bs M_{\bs\mu}\bs w-\bs v_{\calO_0}\|^2
+d(d+1)\bigl(\bs p_{I/d}^T\bs w-c_\calO\bigr)^2,
\label{eq:haar_error_split_coordinates}
\end{equation}
with $\bs p_{I/d}=\bs p(I_d/d)$ and $c_\calO=\Trace(\calO)/d$.
\end{lemma}
\begin{proof}
Since $\Trace(D\sigma)=\bs v_D^T\bs v_\sigma$,~\cref{eq:moment_2_state} gives $\EE_\sigma[\Trace(D\sigma)^2]=[\|\bs v_D\|^2+(\bs v_{I_d}^T\bs v_D)^2]/[d(d+1)]=[\Trace(D^2)+(\Trace D)^2]/[d(d+1)]$. Inserting $\Trace(D^2)=\Trace(D_0^2)+(\Trace D)^2/d$ gives the second form, because $(1+1/d)(\Trace D)^2/[d(d+1)]=(\Trace D/d)^2$. For~\cref{eq:haar_error_split_coordinates}, take $D=\widetilde\calO_{\bs w}-\calO$, so that $(\bs w^T\bs p_\sigma-\Trace(\calO\sigma))^2=\Trace(D\sigma)^2$ and $\bs v_D=\bs M_{\bs\mu}\bs w-\bs v_\calO$. The traceless part has coordinates $\bs P_0\bs v_D=\bs P_0\bs M_{\bs\mu}\bs w-\bs v_{\calO_0}$, and $\Trace(D)/d=\sum_aw_a\Trace(\mu_a)/d-c_\calO=\bs p_{I/d}^T\bs w-c_\calO$.
\end{proof}

\parTitle{Random target ensembles} The error profiles describe a typical target observable rather than a specific one. We therefore draw the target at random, independently of the measurement, the training states, and the counts, and average the quadratic costs over it. The definition below fixes the admissible ensembles: their second moments are the only input to target-averaged costs, whereas their concentration properties control the realized cost of a single draw.
\begin{definition}[Target ensembles]
\label{def:target_ensembles}
Write a target as $\calO=c_\calO I_d+\calO_0$, with $c_\calO=\Trace(\calO)/d$ and $\Trace(\calO_0)=0$, and assume
\begin{equation}
\label{eq:full_target_class}
 \Trace(\calO_0^2)=d\Delta_{\calO}^2>0,\qquad
 \frac{\lvert c_\calO\rvert}{\sqrt{d\Delta_{\calO}^2}}\le C.
\end{equation}
A \emph{random target} is drawn independently of the apparatus, the training states, and the counts. Its invariants $c_\calO$ and $\Delta_\calO^2$ may be deterministic or random; in the latter case target averages $\EE_\calO$ are conditional on their values, and the bound in~\cref{eq:full_target_class} holds uniformly or with probability tending to one. Its normalized traceless direction $\bs v_{\calO_0}/\sqrt{d\Delta_{\calO}^2}$ follows one of three laws:
(i) \emph{sphere}: uniform on the real unit sphere of the traceless sector;
(ii) \emph{sign}: independent coordinates $\pm1/\sqrt{d^2-1}$ in a fixed orthonormal traceless basis;
(iii) \emph{unitary orbit}: $\bs v_{U^\dagger\Omega U}$ for a fixed traceless $\Omega$ with $\Trace(\Omega^2)=1$ and a Haar unitary $U$.
\end{definition}
By~\cref{eq:moment_2_operator} for law (iii), and by direct computation for (i) and (ii), all three laws share the isotropic moments
\begin{equation}
\EE_\calO\bs v_{\calO_0}=0,\qquad
\EE_\calO[\bs v_{\calO_0}\bs v_{\calO_0}^T]
=\frac{d\Delta_{\calO}^2}{d^2-1}\bs P_0.
\label{eq:centered_full_target_covariance}
\end{equation}
Law (iii) with $\Omega=\calO_0/\sqrt{d\Delta_\calO^2}$ describes the rotated targets $U^\dagger\calO U$ of~\cref{sec:setup_error}, and law (i) is the one used in our simulations. The laws differ in their concentration bounds (\cref{lem:centered_full_sphere_sign_concentration,lem:identity_target_concentration}). Pure identity targets have $\Delta_{\calO}^2=0$ and are treated in~\cref{lem:identity_readout}.

\section{Spectrum of Haar-random rank-one POVMs}
\label[appsec]{app:concentration_POVM}

The spectral claim in~\cref{sec:classical_spectrum} rests on two scales: the positive traces of the effects make their identity components add coherently, producing an outlier of order $d/\nout$, while their centered components produce a bulk on the scale $\nout^{-1}$. \Cref{nav:povm_construction} separates these contributions using independent Gaussian vectors, and~\cref{nav:povm_spectrum} establishes the bulk law and exact rank. \Cref{app:thm_POVM_extensions} refines the outlier at finite dimension; \cref{app:thm_random_states} supplies the fixed-rank covariance laws used in the generalized peak amplitudes.

\subsection{Gaussian construction and effect traces}
\label[appsec]{nav:povm_construction}

Normalizing a Gaussian matrix into a Haar isometry enforces $\sum_a\mu_a=I_d$ through a common inverse square root. The Gaussian lift below simply represents each rank-one matrix $gg^\dagger$ as an operator-coordinate vector. When $\nout$ grows in proportion to $d^2$, the common normalization approaches the identity, allowing the spectrum to be obtained from independent lifts. The trace estimate below also makes the mean outcome probabilities nearly uniform, as required to convert the spectrum into test variance.

\parTitle{Exact frame and Gram matrices} Sample a Haar isometry $\mathbf V_\mu\in\CC^{\nout\times d}$, equivalently the first $d$ columns of a Haar unitary in $U(\nout)$~\cite{mezzadri2006generate}. For canonical basis vectors $e_a$, put $\mathbf v_a=\mathbf V_\mu^\dagger e_a$ and $\mu_a=\mathbf v_a\mathbf v_a^\dagger$. Then $\sum_a\mu_a=I_d$, and
\begin{equation}
\label{eq:frame_def}
 \mathbf M_{\bs\mu}=[\vect{\mu_1}\ \cdots\ \vect{\mu_{\nout}}],\qquad
 \mathfrak G_{\bs\mu}=\mathbf M_{\bs\mu}^\dagger\mathbf M_{\bs\mu},\qquad
 \frameOp_{\bs\mu}=\mathbf M_{\bs\mu}\mathbf M_{\bs\mu}^\dagger.
\end{equation}
The Gram matrix compares effects with one another; the frame measures apparatus response along operator directions. For a Hermitian $A$, $\bs v_A^T\frameOp_{\bs\mu}\bs v_A=\sum_a[\Trace(A\mu_a)]^2$, so small frame eigenvalues identify weak probability responses. The two matrices are positive semidefinite and have the same nonzero spectrum. In particular, $(\mathfrak G_{\bs\mu})_{ab}=\Trace(\mu_a\mu_b)=\lvert(\mathbf V_\mu\mathbf V_\mu^\dagger)_{ab}\rvert^2$, or $\mathfrak G_{\bs\mu}=\lvert\mathbf V_\mu\mathbf V_\mu^\dagger\rvert^{\odot2}$, with entrywise squared modulus. We may use column-stacking coordinates for the following Gaussian calculations: the fixed unitary change from Hermitian coordinates preserves all spectra, norms, and inner products. The notation $\mathcal{CN}(0,1)$ denotes a zero-mean circular complex Gaussian variable with mean squared modulus one.

\begin{lemma}[Gaussian lift and rank]
\label{lem:gaussian_lift_rank}
Let $\nout\ge d$ and let $\hat{\mathbf G}\in\CC^{\nout\times d}$ have independent $\mathcal{CN}(0,1)$ entries, $\mathbf S=\hat{\mathbf G}^\dagger\hat{\mathbf G}/\nout$, and $\hat{\mathbf g}_a=\hat{\mathbf G}^\dagger e_a$. Define $\mathbf x_a=\vect{\hat{\mathbf g}_a\hat{\mathbf g}_a^\dagger}$, $\mathbf X=[\mathbf x_1,\ldots,\mathbf x_{\nout}]$, and $\mathbf T=\mathbf S^{-T/2}\otimes\mathbf S^{-1/2}$. The exact Haar construction and lifted covariance are
\begin{equation}
\label{eq:V_gauss_exact}
 \begin{aligned}
 \mathbf V_\mu&=\frac{\hat{\mathbf G}\mathbf S^{-1/2}}{\sqrt{\nout}},
 &\mathbf M_{\bs\mu}&=\frac{\mathbf T\mathbf X}{\nout},\\
 \mathbf A&=\frac{\mathbf X\mathbf X^\dagger}{\nout},
 &\nout\frameOp_{\bs\mu}&=\mathbf T\mathbf A\mathbf T^\dagger.
 \end{aligned}
\end{equation}
Almost surely, $\operatorname{rank}\mathbf X=\operatorname{rank}\mathbf M_{\bs\mu}=\min(d^2,\nout)$. If $\nout\asymp d^2$, then $\|\mathbf T-I_{d^2}\|_{\rm op}=O_\PP(d^{-1/2})$. The centered lifts $\mathbf m_a=\mathbf x_a-\bs v_{I_d}$ are independent and isotropic:
\begin{equation}
\label{eq:wicks_identity}
 \EE[\mathbf x_a\mathbf x_a^\dagger]=I_{d^2}+\bs v_{I_d}\bs v_{I_d}^\dagger,
 \qquad \EE\mathbf m_a=0,\qquad
 \EE[\mathbf m_a\mathbf m_a^\dagger]=I_{d^2}.
\end{equation}
\end{lemma}
\begin{proof}
\parTitle{Normalization and covariance} The Gaussian representation of a Haar isometry~\cite{mezzadri2006generate} and $\vect{AXB}=(B^T\otimes A)\vect X$ give~\cref{eq:V_gauss_exact}. The Wishart estimate~\cref{lem:wishart_conc,eq:S_inv_sqrt_bound}~\cite{tao2023topics} gives $\|\mathbf S^{-1/2}-I_d\|_{\rm op}=O_\PP(\sqrt{d/\nout})$, hence
$\|\mathbf T-I_{d^2}\|_{\rm op}\le2\|\mathbf S^{-1/2}-I_d\|_{\rm op}+\|\mathbf S^{-1/2}-I_d\|_{\rm op}^2=O_\PP(d^{-1/2})$ in the proportional regime. Wick's identity gives~\cref{eq:wicks_identity}.

\parTitle{Almost-sure rank} The $d^2$ rank-one matrices $e_ie_i^\dagger$ $(1\le i\le d)$, $(e_i+e_j)(e_i+e_j)^\dagger$, and $(e_i+\mathrm i e_j)(e_i+\mathrm i e_j)^\dagger$ $(i<j)$ form a real Hermitian basis: subtracting their diagonal parts recovers all real-symmetric and imaginary-antisymmetric off-diagonal units. Therefore, with $q=\min(d^2,\nout)$, some $q\times q$ minor of $[\vect{g_1g_1^\dagger},\ldots,\vect{g_qg_q^\dagger}]$ is a nonzero polynomial in the real and imaginary Gaussian coordinates. Its zero set has Lebesgue measure zero. Thus $\mathbf X$ has rank $q$ almost surely, and the invertibility of $\mathbf T$ transfers the rank to $\mathbf M_{\bs\mu}$.
\end{proof}

\parTitle{Why effect traces matter} The Haar-averaged test probability of outcome $a$ is $\omega_a/d$. Concentration of all $\omega_a$ around $d/\nout$ therefore makes these probabilities uniformly close to $1/\nout$; the same traces are the Gram-matrix row sums and will locate the identity outlier in~\cref{eq:gram_outlier_loc_simple}.
\begin{lemma}[Uniform concentration of effect traces]
\label{lem:trace_conc}
For a Haar-random rank-one POVM with $\nout\ge2d$, each $\omega_a=\Trace(\mu_a)$ has distribution $\mathrm{Beta}(d,\nout-d)$ and mean $d/\nout$. For $\varepsilon\in(0,1/2)$,
\begin{equation}
 \PP\!\left(\max_{a\le\nout}\lvert\omega_a-d/\nout\rvert\ge4\varepsilon d/\nout\right)
 \le4\nout\exp(-\varepsilon^2d/3).
\end{equation}
If $\log\nout=o(d)$, then
\begin{equation}
\label{eq:trace_rate}
 \max_a\lvert\nout\omega_a/d-1\rvert
 =\|\nout\bs D_{I/d}-I_{\nout}\|_{\rm op}
 =O_\PP\!\left(\sqrt{\frac{\log\nout}{d}}\right).
\end{equation}
\end{lemma}
\begin{proof}
\parTitle{Beta representation} A row of a Haar unitary is a normalized complex Gaussian vector, so $\omega_a=X/(X+Y)$ for independent $X\sim\Gamma(d,1)$ and $Y\sim\Gamma(\nout-d,1)$. This gives the Beta law and its mean. On $\{\lvert X-d\rvert\le\varepsilon d,\ \lvert Y-(\nout-d)\rvert\le\varepsilon(\nout-d)\}$, the ratio lies between $(d/\nout)(1-\varepsilon)/(1+\varepsilon)$ and $(d/\nout)(1+\varepsilon)/(1-\varepsilon)$, hence $\lvert\omega_a-d/\nout\rvert<4\varepsilon d/\nout$.

\parTitle{Uniform tail bound} For $S\sim\Gamma(k,1)$, $\EE e^{tS}=(1-t)^{-k}$. The Chernoff choice $t=\varepsilon/(1+\varepsilon)$ and $\varepsilon-\log(1+\varepsilon)\ge\varepsilon^2/3$, together with the analogous lower-tail estimate, give
\begin{equation}
\label{eq:gamma_two_sided}
 \PP(S\ge(1+\varepsilon)k)\le e^{-\varepsilon^2k/3},\qquad
 \PP(\lvert S-k\rvert>\varepsilon k)\le2e^{-\varepsilon^2k/3}.
\end{equation}
Apply these bounds to $X,Y$, use $\nout-d\ge d$, and take a union bound over outcomes. Choosing $\varepsilon$ proportional to $\sqrt{\log\nout/d}$ proves~\cref{eq:trace_rate}.
\end{proof}

\subsection{Bulk spectrum, exact rank, and identity outlier}
\label[appsec]{nav:povm_spectrum}

Subtracting the mean from the independent Gaussian lifts gives identity covariance: every unit operator direction has the same second-moment strength, although the coordinates of a lift remain dependent. Their sample covariance generates the Marchenko--Pastur (MP) bulk. Restoring the mean is a rank-at-most-two change and cannot alter the limiting eigenvalue histogram; Haar normalization changes positive eigenvalues by a factor tending to one. The theorem below makes the main-text spectrum claim in~\cref{sec:classical_spectrum} precise, including the zero modes below informational completeness and the separate identity outlier.

\parTitle{Limiting spectral measure} The empirical spectral measure is the eigenvalue histogram with equal weight assigned to each eigenvalue. Its large-dimension limit is the Marchenko--Pastur law $\mu_{\mathrm{MP},\gamma}$,
\begin{equation}
\label{eq:MP_density}
 d\mu_{\mathrm{MP},\gamma}(x)
 =\left(1-\frac1\gamma\right)_+\delta_0
 +\frac{\sqrt{(\lambda_+-x)(x-\lambda_-)}}{2\pi\gamma x}
 \mathbf1_{[\lambda_-,\lambda_+]}(x)\,dx,
 \qquad \lambda_\pm=(1\pm\sqrt\gamma)^2.
\end{equation}
Here $\delta_0$ is unit mass at zero and $(x)_+=\max(x,0)$. The atom at zero counts inaccessible operator directions. The lower endpoint of the continuous bulk reaches zero at $\gamma=1$, identifying the weakly resolved directions responsible for the classical peak. Weak convergence means convergence of averages of bounded continuous functions of the eigenvalues; it does not by itself bound the smallest positive eigenvalue.
\begin{theorem}[Spectrum of a Haar-random rank-one POVM]
\label{thm:main}
Let $d\to\infty$ with $\gamma_{\rm out}=d^2/\nout\to\gamma_\infty\in(0,\infty)$. For a Haar-random rank-one POVM, let $\lambda_1,\ldots,\lambda_{d^2}$ be the frame eigenvalues and $\widetilde\lambda_1,\ldots,\widetilde\lambda_{d^2-1}$ those of its traceless compression. For every bounded continuous $f$,
\begin{equation}
\label{eq:mp_convergence}
 \frac1{d^2}\sum_{i=1}^{d^2}f(\nout\lambda_i)
 \xrightarrow{\PP}\int f\,d\mu_{\mathrm{MP},\gamma_\infty},\qquad
 \frac1{d^2-1}\sum_{i=1}^{d^2-1}f(\nout\widetilde\lambda_i)
 \xrightarrow{\PP}\int f\,d\mu_{\mathrm{MP},\gamma_\infty}.
\end{equation}
The largest eigenvalue is $\lambda_{\max}(\frameOp_{\bs\mu})=(d/\nout)(1+o_\PP(1))$. More precisely, for every $\varepsilon\in(0,1/2)$ and all sufficiently large $d$, an event of probability at least $1-4\nout e^{-\varepsilon^2d/3}$ satisfies both
\begin{equation}
\label{eq:gap_statement}
 \begin{aligned}
 (1-4\varepsilon)\frac d{\nout}
 &\le\lambda_{\max}(\frameOp_{\bs\mu})\le(1+4\varepsilon)\frac d{\nout},\\
 \frac1{\sqrt d}\left\|\frameOp_{\bs\mu}\bs v_{I_d}
       -\frac d{\nout}\bs v_{I_d}\right\|_2
 &\le4\varepsilon\sqrt{1+4\varepsilon}\,\frac d{\nout}.
 \end{aligned}
\end{equation}
Almost surely, $\operatorname{rank}\mathbf M_{\bs\mu}=\min(d^2,\nout)$, so the frame and Gram matrix have $(d^2-\nout)_+$ and $(\nout-d^2)_+$ zero eigenvalues, respectively.
\end{theorem}
\begin{proof}
\parTitle{Centered Gaussian bulk} Set $\mathbf W=\nout^{-1}\sum_a\mathbf m_a\mathbf m_a^\dagger$ and $\bar{\mathbf m}=\nout^{-1}\sum_a\mathbf m_a$. Expanding the Gaussian lift gives
\begin{equation}
\label{eq:A_decomposition}
 \mathbf A=\mathbf W+\bs v_{I_d}\bs v_{I_d}^\dagger
 +\bs v_{I_d}\bar{\mathbf m}^\dagger
 +\bar{\mathbf m}\bs v_{I_d}^\dagger,\qquad
 \bs P_0\mathbf A\bs P_0=\bs P_0\mathbf W\bs P_0.
\end{equation}
The difference $\mathbf A-\mathbf W$ has rank at most two. The quadratic-form estimate~\cref{lem:yaskov_condition_chaos} says that $\mathbf m_a^\dagger\bs H\mathbf m_a/d^2$ concentrates around $\Trace(\bs H)/d^2$ for each deterministic bounded Hermitian $\bs H$. That is the condition in~\cref{thm:yaskov_mp} giving the MP law despite dependence among lift coordinates. A rank-two change shifts eigenvalue counts across any threshold by at most two (interlacing, \cref{lem:congruence_spectral}), so $\mathbf A$ has the same limiting histogram. By~\cref{lem:gaussian_lift_rank,lem:congruence_spectral}, every positive eigenvalue changes by a uniform multiplicative factor $1+o_\PP(1)$ under $\mathbf A\mapsto\mathbf T\mathbf A\mathbf T^\dagger=\nout\frameOp_{\bs\mu}$, so the full frame has this limit in probability. Cauchy interlacing~\cref{lem:congruence_spectral} changes its empirical distribution function by at most $1/(d^2-1)$ under traceless compression, proving~\cref{eq:mp_convergence}.

\parTitle{Identity outlier from Gram row sums} The matrix $\mathfrak G_{\bs\mu}$ has nonnegative entries and row sums $\omega_a$ because $\sum_b\Trace(\mu_a\mu_b)=\Trace(\mu_a)$. Its Rayleigh quotient on the uniform vector is $\nout^{-1}\sum_a\omega_a=d/\nout$, while its norm is at most its largest row sum. Consequently,
\begin{equation}
\label{eq:gram_outlier_loc_simple}
 0\le\lambda_{\max}(\mathfrak G_{\bs\mu})-\frac d{\nout}
 \le\max_a\left\lvert\omega_a-\frac d{\nout}\right\rvert.
\end{equation}
Moreover, $\mathbf M_{\bs\mu}^\dagger\bs v_{I_d}=\bs\omega$ and $\mathbf M_{\bs\mu}\mathbf1_{\nout}=\bs v_{I_d}$ imply
\begin{equation}
\label{eq:Fu_identity_decomposition}
 \begin{aligned}
 \frameOp_{\bs\mu}\bs v_{I_d}-\frac d{\nout}\bs v_{I_d}
 &=\mathbf M_{\bs\mu}\!\left(\bs\omega-\frac d{\nout}\mathbf1_{\nout}\right),\\
 \frac{\left\|\frameOp_{\bs\mu}\bs v_{I_d}-\frac d{\nout}\bs v_{I_d}\right\|}{\sqrt d}
 &\le\frac{\|\mathbf M_{\bs\mu}\|_{\rm op}}{\sqrt d}
 \left\|\bs\omega-\frac d{\nout}\mathbf1_{\nout}\right\|.
 \end{aligned}
\end{equation}
On the event in~\cref{lem:trace_conc}, the last two norms are bounded by $\sqrt{(1+4\varepsilon)d/\nout}$ and $4\varepsilon d/\sqrt{\nout}$. These bounds prove~\cref{eq:gap_statement}; taking $\varepsilon\to0$ slowly proves the outlier limit. The rank assertion follows from~\cref{lem:gaussian_lift_rank}.
\end{proof}

\parTitle{Singular-value scales} The MP law controls empirical spectra, allowing a vanishing fraction of eigenvalues outside the bulk. The corresponding bulk singular values of $\mathbf M_{\bs\mu}$ have scale $\nout^{-1/2}$, whereas its leading singular value has scale $\sqrt{d/\nout}$. At finite dimension, the identity direction is only approximately an eigenvector.

\subsection{Finite-dimensional outlier correction}
\label[endsec]{app:thm_POVM_extensions}

The leading outlier scale of~\cref{thm:main} can be sharpened to the additive $d+1$ value shown in the insets of main-text~\cref{fig:spectrum}. Effect-trace fluctuations raise the identity Rayleigh quotient above $d/\nout$; the coupling to traceless directions produces only a vanishing correction after rescaling by $\nout$.

\parTitle{Trace fluctuations shift the quotient} The identity Rayleigh quotient satisfies the exact identity
\begin{equation}
\label{eq:u_rayleigh_exact}
 \frac{\bs v_{I_d}^\dagger\frameOp_{\bs\mu}\bs v_{I_d}}d
 =\frac1d\sum_a\omega_a^2
 =\frac d{\nout}+\frac1d\sum_a\left(\omega_a-\frac d{\nout}\right)^2.
\end{equation}
Its expectation is $(d+1)/(\nout+1)$ by the Beta second moment. Controlling its coupling to the traceless sector gives the following refinement.
\begin{corollary}[Additive identity-outlier correction]
\label{cor:outlier_additive}
In the proportional limit of~\cref{thm:main},
\begin{equation}
\label{eq:outlier_additive}
 \nout\lambda_{\max}(\frameOp_{\bs\mu})=d+1+o_\PP(1).
\end{equation}
The scaled identity Rayleigh quotient has exact mean $\nout(d+1)/(\nout+1)$ and variance $O(\nout^{-1})$; its difference from $\nout\lambda_{\max}(\frameOp_{\bs\mu})$ is $O_\PP(\log d/d)$.
\end{corollary}
\begin{proof}
\parTitle{Concentration of the identity quotient} Let $\bs H$ be the Haar rank-$d$ projection in $\CC^{\nout}$ and $\delta_a=H_{aa}-d/\nout=\omega_a-d/\nout$. The scaled quotient is $\upsilon=(\nout/d)\bs v_{I_d}^\dagger\frameOp_{\bs\mu}\bs v_{I_d}=d+(\nout/d)\sum_a\delta_a^2$, with the stated expectation. The logarithmic Sobolev inequality on $U(\nout)$ implies the variance bound $\Var\upsilon\le(6/\nout)\EE\|\nabla\upsilon\|_F^2$~\cite[Theorem 15]{meckes2013spectral}. Differentiating along $H\mapsto e^{tX}He^{-tX}$, with $\bs D_\delta=\diag\bs\delta$, gives
\begin{equation}
 \begin{aligned}
 \|\nabla\upsilon\|_F^2
 &\le4(\nout/d)^2\|[\bs H,\bs D_\delta]\|_F^2
 \le16(\nout/d)^2\sum_a\delta_a^2\omega_a,\\
 \EE[\delta_a^2\omega_a]
 &=\frac{d(\nout-d)[(d+2)\nout-2d]}{\nout^3(\nout+1)(\nout+2)}
 \le\frac{d(d+2)}{\nout^3}.
 \end{aligned}
\end{equation}
The last expression uses the first three Beta moments. Hence $\EE\|\nabla\upsilon\|_F^2\le16(d+2)/d$, $\Var\upsilon\le C/\nout$, and $\upsilon=d+1+o_\PP(1)$.

\parTitle{Coupling to traceless directions} Put $\bs C=\nout\bs P_0\frameOp_{\bs\mu}\bs P_0$ on the traceless sector and $\bs b=\nout\bs P_0\frameOp_{\bs\mu}\bs v_{I_d}/\sqrt d$. The Gaussian lift yields
$\|\bs C\|_{\rm op}\le2\|\mathbf T\|_{\rm op}^2\|\mathbf W\|_{\rm op}+2\|\bs P_0\mathbf T\bs v_{I_d}\|^2=O_\PP(\log d)$.
Indeed, $\|\mathbf T-I_{d^2}\|_{\rm op}=O_\PP(d^{-1/2})$. Truncating the independent centered-lift outer products to $\|\hat{\mathbf g}_a\|^2\le2d$ changes no column with probability tending to one; each summand of $\mathbf W$ then has norm at most $Cd^2/\nout$, and their summed expectation is at most $I_{d^2}$. The matrix Chernoff bound for sums of independent positive matrices~\cite{tropp2012user} therefore gives $\|\mathbf W\|_{\rm op}=O_\PP(\log d)$. Finally,~\cref{eq:Fu_identity_decomposition} and $\bs P_0\mathbf M_{\bs\mu}\mathbf1_{\nout}=0$ give $\|\bs b\|^2\le\|\bs C\|_{\rm op}(\nout/d)\sum_a\delta_a^2=\|\bs C\|_{\rm op}(\upsilon-d)$. With probability tending to one, $\upsilon>\|\bs C\|_{\rm op}$, so the Schur complement for $\Lambda=\nout\lambda_{\max}(\frameOp_{\bs\mu})$ gives $0\le\Lambda-\upsilon\le\frac{\|\bs b\|^2}{\upsilon-\|\bs C\|_{\rm op}} =O_\PP(\log d/d)=o_\PP(1)$. This proves~\cref{eq:outlier_additive}.
\end{proof}

\parTitle{Identity direction in outcome space} In Gram space, the identity direction maps to $\mathbf M_{\bs\mu}^\dagger\bs v_{I_d}=\bs\omega$. Trace concentration gives $\|\bs\omega/\sqrt d-(\sqrt d/\nout)\mathbf1_{\nout}\|=O_\PP(\sqrt{\log\nout/\nout})$. Since $\sum_a\omega_a=d$, the squared overlap of $\bs\omega/\|\bs\omega\|$ with the uniform unit vector is $d^2/(\nout\sum_a\omega_a^2)$. By~\cref{eq:u_rayleigh_exact},
\begin{equation}
\label{eq:gram_overlap_prob}
 \PP\!\left(\left\lvert\left\langle\frac{\bs\omega}{\|\bs\omega\|},
 \frac{\mathbf1_{\nout}}{\sqrt{\nout}}\right\rangle\right\rvert^2
 \ge\frac1{1+16\varepsilon^2}\right)
 \ge1-4\nout e^{-\varepsilon^2d/3}.
\end{equation}
Thus the identity-induced Gram direction approaches the uniform vector.

\subsection{Fixed-rank states and measurement effects}
\label[appsec]{app:thm_random_states}

The fixed-rank ensembles below extend the covariance input to the quantum peak of~\cref{sec:quantum_ic}. Mixing $k$ orthogonal pure states reduces the traceless training covariance by an asymptotic factor $1/k$; grouping $\ell$ microscopic effects into each POVM outcome gives the measurement factor $1/\ell$. The lemma makes these bulk laws precise and supplies the scales entering the peak amplitudes in~\cref{eq:weak_rank_peak}, with $k$ and $\ell$ fixed as $d\to\infty$. ESD denotes the empirical spectral distribution, or normalized eigenvalue histogram.

\parTitle{Distinct identity scales} Independent training states have no resolution-of-identity constraint across samples. For pure states, $\EE\sum_i\rho_i=(\ntr/d)I_d$, so their frame has an identity spike of order $\ntr/d$, rather than the measurement scale $d/\nout$.

\begin{lemma}[Bulk laws for fixed-rank states and measurement effects]
\label{lem:fixed_rank_bulk_bridge}
Let $d\to\infty$. Fix positive integers $k$ and $\ell$, independent of $d$, and assume $(d^2-1)/\ntr\to\gamma_{\rm tr}\in(0,\infty)$ and $(d^2-1)/\nout\to\gamma_{\rm out}\in(0,\infty)$. All empirical spectral distributions below are normalized by $(d^2-1)$, include zero eigenvalues, and act on the real $(d^2-1)$-dimensional space of traceless Hermitian operators. Convergence is weak in probability.

\begin{enumerate}
\item Let $\rho_i=U_iP_kU_i^\dagger/k$, where $P_k$ is a deterministic rank-$k$ orthogonal projector and the $U_i\in U(d)$ are independent Haar unitaries. With $\bs\Pi_{\ntr}=I_{\ntr}-\mathbf1_{\ntr}\mathbf1_{\ntr}^T/\ntr$ and the convention of~\cref{eq:general_training_frame},
\begin{equation}
 \operatorname{ESD}\!\left(
 \frac{kd(d+1)}{\ntr}\bs P_0\bs M_{\bs\rho}\bs\Pi_{\ntr}
 \bs M_{\bs\rho}^T\bs P_0\big\vert_{\calL(\calH)_0}\right)
 \Rightarrow\mu_{{\rm MP},\gamma_{\rm tr}}.
 \label{eq:fixed_rank_state_bulk_bridge}
\end{equation}
For every deterministic symmetric matrix $\bs A$ on this sector, the exactly isotropic column $\bs x=a_d\bs v_{\rho-I_d/d}$, with $a_d^2=kd(d^2-1)/(d-k)=kd(d+1)(d-1)/(d-k)$, satisfies $\EE\bs x=0$, $\EE\bs x\bs x^T=I_{d^2-1}$, and, for $d>k$,
\begin{equation}
\label{eq:fixed_rank_state_qf_bridge}
 \Var(\bs x^T\bs A\bs x)
 \le\frac{C(d^2-1)^2}{k(d-k)}\|\bs A\|_{\rm op}^2
 =O_k\!\left(\frac{(d^2-1)^2}{d}\right)\|\bs A\|_{\rm op}^2.
\end{equation}
\item Let $\mathbf V\in\CC^{(\nout\ell)\times d}$ be a Haar isometry, partitioned into blocks $\mathbf V_a\in\CC^{\ell\times d}$, and set $\mu_a=\mathbf V_a^\dagger\mathbf V_a$. Then
\begin{equation}
 \operatorname{ESD}\!\left(
 \nout\ell\,\bs P_0\frameOp_{\bs\mu}\bs P_0
 \big\vert_{\calL(\calH)_0}\right)
 \Rightarrow\mu_{{\rm MP},\gamma_{\rm out}}.
 \label{eq:fixed_rank_povm_bulk_bridge}
\end{equation}
The state list and POVM may be drawn independently when both statements are used together.
\end{enumerate}
\end{lemma}

\begin{proof}
\parTitle{Normalized projector states} The Haar moment formula~\cref{eq:moment_2_operator}, applied to $P_k/k-I_d/d$, gives
$\EE\bs v_{\rho-I_d/d}\bs v_{\rho-I_d/d}^T=(1/k-1/d)I_{d^2-1}/(d^2-1)$.
Thus $\bs x$ is centered and isotropic, with $\|\bs x\|^2=d^2-1$ deterministically. The unitary Poincar\'e inequality following from~\cite[Theorem 15]{meckes2013spectral} bounds variance by $C/d$ times the mean squared Hilbert--Schmidt gradient. To apply it, let $F(U)=\bs x^T\bs A\bs x$ and differentiate along $U\mapsto e^{tX}U$, where $X^\dagger=-X$. Since $\dot\rho=[X,\rho]$ and $\|\rho\|_{\rm op}=1/k$,
$\lvert\dot F\rvert\le2a_d\|\bs A\|_{\rm op}\sqrt{d^2-1}\,\|[X,\rho]\|_F\le4a_d\sqrt{d^2-1}\|\bs A\|_{\rm op}\|X\|_F/k$.
The resulting gradient bound proves~\cref{eq:fixed_rank_state_qf_bridge}. In particular, the variance divided by $(d^2-1)^2$ tends to zero uniformly for $\|\bs A\|_{\rm op}\le1$. The MP theorem~\cref{thm:yaskov_mp} therefore applies to $\ntr^{-1}\sum_i\bs x_i\bs x_i^T$. Subtracting the empirical mean changes this matrix by rank at most one, so~\cref{lem:congruence_spectral} gives the same limit for the sample-centered frame. Finally, $kd(d+1)/a_d^2=(d-k)/(d-1)\to1$, proving~\cref{eq:fixed_rank_state_bulk_bridge}. Equivalently, generate each $\rho_i$ as $G_i(G_i^\dagger G_i)^{-1}G_i^\dagger/k$ with $G_i\in\CC^{d\times k}$ standard Gaussian. The proof uses the normalized-projector law itself.

\parTitle{Independent block lifts} Let $\mathbf G\in\CC^{(\nout\ell)\times d}$ have independent $\mathcal{CN}(0,1)$ entries and put $\mathbf S=\mathbf G^\dagger\mathbf G/(\nout\ell)$. Write the conjugate-transposed rows of its $a$th block as $g_{at}\in\CC^d$, $1\le t\le\ell$, and define
$Y_a=\sum_{t=1}^{\ell}g_{at}g_{at}^\dagger$ and $\bs z_a=\bs v_{Y_a-\ell I_d}/\sqrt\ell$.
The $d^2$-dimensional real vectors $\bs z_a$ are independent across $a$, centered, and isotropic by the Wick identity~\cref{eq:wicks_identity}. They also satisfy the quadratic-form concentration needed for the MP law. Indeed, for symmetric $\bs H$ on the full Hermitian space and $Q=\bs z_a^T\bs H\bs z_a$, Gaussian Poincar\'e and differentiation with respect to the $\ell$ Gaussian vectors give
\begin{equation}
 \Var Q\le\frac{C}{\ell}\|\bs H\|_{\rm op}^2
 \EE\!\left[\left(\sum_{t=1}^{\ell}\|g_{at}\|^2\right)
                   \|\bs z_a\|^2\right]
 =\frac{C}{\ell}(\ell d^3+2d^2)\|\bs H\|_{\rm op}^2.
 \label{eq:fixed_rank_lift_qf_bridge}
\end{equation}
For the last equality, expand the centered sum: cross terms vanish by independence, $\EE\|\bs v_{gg^\dagger-I_d}\|^2=d^2$, and $\EE[\|g\|^2\|gg^\dagger-I_d\|_F^2]=d^3+2d^2$. Hence $d^{-4}\Var Q=O_\ell(d^{-1})\|\bs H\|_{\rm op}^2$, as in~\cref{lem:yaskov_condition_chaos}. Consequently $\mathbf W=\nout^{-1}\sum_a\bs z_a\bs z_a^T$ has MP limit with aspect $d^2/\nout\to\gamma_{\rm out}$.

\parTitle{Restoring measurement normalization} The exact Haar normalization is $\mathbf V=\mathbf G\mathbf S^{-1/2}/\sqrt{\nout\ell}$, so $\mu_a=\mathbf S^{-1/2}Y_a\mathbf S^{-1/2}/(\nout\ell)$. Let $\mathbf T$ represent the linear map $A\mapsto\mathbf S^{-1/2}A\mathbf S^{-1/2}$ in the same orthonormal Hermitian coordinates, and let $\mathbf Y=[\bs v_{Y_1},\ldots,\bs v_{Y_{\nout}}]$. Then
\begin{equation}
 \nout\ell\,\frameOp_{\bs\mu}
 =\mathbf T\left(\frac{\mathbf Y\mathbf Y^T}{\nout\ell}\right)\mathbf T^T,
 \qquad
 \operatorname{rank}\!\left(\frac{\mathbf Y\mathbf Y^T}{\nout\ell}
                              -\mathbf W\right)\le2.
\end{equation}
The rank bound follows by writing $\mathbf Y/\sqrt\ell=[\bs z_1,\ldots,\bs z_{\nout}]+\sqrt\ell\,\bs v_{I_d}\mathbf1_{\nout}^T$. By~\cref{lem:wishart_conc,eq:S_inv_sqrt_bound},
$\|\mathbf T-I_{d^2}\|_{\rm op}=O_\PP(\sqrt{d/(\nout\ell)})=o_\PP(1)$.
The finite-rank, near-identity congruence, and codimension-one compression bounds in~\cref{lem:congruence_spectral} transfer the MP limit to the exact traceless POVM frame, proving~\cref{eq:fixed_rank_povm_bulk_bridge}.
\end{proof}

\parTitle{Covariance scales for peak amplitudes} The bulk statements supply the fixed-$k$ and fixed-$\ell$ covariance scales used in~\cref{eq:weak_rank_peak}: the frame $\frameOp_{\bs\rho}^{\rm c}$ has the MP law rescaled by $1/k$, and $\nout\bs P_0\frameOp_{\bs\mu}\bs P_0\vert_{\calL(\calH)_0}$ has the MP law rescaled by $1/\ell$. The MP zero atom is included once. The lemma does not identify an additive outlier correction, an extremal spectral edge, or an inverse spectral moment.

\parTitle{Separate control of the largest eigenvalues} The upper-frame hypotheses used in the fixed-rank peak examples follow separately from the rank-one bounds. Write $\rho_i=k^{-1}\sum_{t=1}^k\rho_{it}$ as its constituent orthogonal pure states. For each fixed $t$, the list $\bs\rho_t=(\rho_{it})_i$ consists of independent Haar pure states. Applying the outer-product Cauchy inequality to the centered coordinates gives $\frameOp_{\bs\rho}^{\rm c}\preceq k^{-1}\sum_t\frameOp_{\bs\rho_t}^{\rm c}$. Each term is bounded by the Haar fourth moment~\cref{eq:joined_Haar_fourth} and~\cref{lem:rm_sample_upper}; fixed $k$ makes their sum bounded with probability tending to one. Likewise, the rank-$\ell$ effects group a single Haar rank-one POVM $\bs\nu$ with $\nout\ell$ microscopic outcomes. Cauchy--Schwarz for each operator response gives $\nout\bs P_0\frameOp_{\bs\mu}\bs P_0\preceq \nout\ell\bs P_0\frameOp_{\bs\nu}\bs P_0$. The right side is bounded by~\cref{lem:alt_frame_upper} at the fixed microscopic aspect ratio $(d^2-1)/(\nout\ell)\to\gamma_{\rm out}/\ell$. These upper bounds do not rely on weak spectral convergence or assert an extremal-edge limit.

\section{Classical peak in test variance}
\label[appsec]{app:bounds}

The classical calculation uses the main-text reconstruction weights $\bs w_{\rm ex}=\bs M_{\bs\mu}^+\bs v_\calO$, which reproduce the measurement-visible part of the target with minimum coefficient norm. An outcome $a$ is assigned the value $w_{{\rm ex},a}$. At fixed test state, averaging $M$ independent outcomes divides the single-shot variance by $M$. Small positive frame eigenvalues correspond to weak probability responses; inverting them enlarges these weights and amplifies test-shot fluctuations. The resulting variance peaks at $\nout=d^2$, as stated in~\cref{sec:classical_peak}. \Cref{nav:inverse_weight} controls the inverse weight of small eigenvalues, \cref{nav:classical_variance} proves the variance profile, and~\cref{nav:exact_training} identifies when exact-feature training returns the same coefficients, as discussed in~\cref{sec:classical_training}.

\parTitle{Positive modes and their inverse weight} Define the positive spectral measure and its inverse weight,
\begin{equation}
\label{eq:positive_spectral_measure}
 \nu_d^+=\frac1{d^2-1}\sum_{j:\lambda_j>0}\delta_{\nout\lambda_j},\qquad
 S_d=\int_{(0,\infty)}\frac{d\nu_d^+(x)}x
 =\frac{\Trace(\frameOp_{\bs\mu}^+)}{\nout(d^2-1)},
\end{equation}
where $\lambda_j$ are the frame eigenvalues. If $d^2/\nout\to\gamma$, then $\nu_d^+$ has total mass $\min(d^2,\nout)/(d^2-1)\to\min(1,\gamma^{-1})$. The remaining MP mass comes from the exact zero eigenvalues in~\cref{thm:main}. For test functions vanishing at zero, replacing the full empirical measure by $\nu_d^+$ changes only the asymptotically immaterial factor $d^2/(d^2-1)$. Weak spectral convergence alone does not control $S_d$, because $x^{-1}$ is unbounded at zero.

\subsection{Inverse weight near zero}
\label[appsec]{nav:inverse_weight}

A vanishing fraction of exceptionally small positive eigenvalues can dominate $S_d$ without affecting the MP limit. Above informational completeness, the estimate below rules out a finite contribution from such modes by bounding their total inverse weight, rather than the smallest eigenvalue. The bound is the extra input needed for the overcomplete branch of the main-text classical variance law, through~\cref{eq:proof_tail}.

\begin{lemma}[Inverse moments without an extremal-edge bound]
\label{lem:lift_inverse_tail}
For independent standard complex Gaussian $g_a\in\CC^d$, let $\mathbf x_a=\vect{g_ag_a^\dagger}$ and $\mathbf K=\sum_{a=1}^{\nout}\mathbf x_a\mathbf x_a^\dagger$. If $d^2/\nout\to\gamma\in(0,1)$, then $\Trace(\mathbf K^{-1})=O_\PP(1)$ and $\nout\Trace(\mathbf K^{-2})=O_\PP(1)$. Consequently,
$\frac{\nout}{d^2-1}\sum_{\lambda_j(\mathbf K)\le\theta\nout}\lambda_j(\mathbf K)^{-1}=O_\PP(\sqrt\theta)$ uniformly for sufficiently large $d$. Thus the eigenvalues below $\theta\nout$ carry vanishing inverse weight without a lower bound on the smallest eigenvalue.
\end{lemma}
\begin{proof}
\parTitle{Removing one independent column} Remove column $a$'s outer product so that the remaining matrix is independent of $\mathbf x_a$. Put $\mathbf K_{(a)}=\mathbf K-\mathbf x_a\mathbf x_a^\dagger$, $\xi_a=\mathbf x_a^\dagger\mathbf K_{(a)}^{-1}\mathbf x_a$, $\eta_a=\mathbf x_a^\dagger\mathbf K_{(a)}^{-2}\mathbf x_a$, and $\mathcal T_j=\Trace(\mathbf K^{-j})$ for $j=1,2$. For sufficiently large $d$, $\nout-1\ge d^2$, so~\cref{lem:gaussian_lift_rank} makes every $\mathbf K_{(a)}$ invertible almost surely. Sherman--Morrison gives $\mathbf K^{-1}\mathbf x_a=\mathbf K_{(a)}^{-1}\mathbf x_a/(1+\xi_a)$. Consequently, $\mathbf x_a^\dagger\mathbf K^{-1}\mathbf x_a=\xi_a/(1+\xi_a)$ and $\mathbf x_a^\dagger\mathbf K^{-2}\mathbf x_a=\eta_a/(1+\xi_a)^2$. Summing over columns yields the two exact sum rules
\begin{equation}
\label{eq:proof_leverage}
 \sum_{a=1}^{\nout}\frac{\xi_a}{1+\xi_a}=d^2
 =\gamma_{\rm out}\nout,\qquad
 \sum_{a=1}^{\nout}\frac{\eta_a}{(1+\xi_a)^2}=\mathcal T_1.
\end{equation}
Since $\mathbf K_{(a)}\preceq\mathbf K$, $\Trace(\mathbf K_{(a)}^{-j})\ge\mathcal T_j$ for $j=1,2$. The small-ball bound~\cref{lem:smallball_lift} says that an independent lift is unlikely to have quadratic weight much smaller than $\Trace(\mathbf B)$, even for an ill-conditioned $\mathbf B$. Specifically,
\begin{equation}
\label{eq:proof_smallball}
 \PP\!\left(\mathbf x^\dagger\mathbf B\mathbf x
 \le\varepsilon\Trace(\mathbf B)\mid\mathbf B\right)
 \le C_0\varepsilon^{1/4}
 \quad\text{for every nonzero }\mathbf B\succeq0
 \text{ independent of }\mathbf x.
\end{equation}
Fix $\gamma'\in(\gamma,1)$ and set $R=2\gamma'/(1-\gamma')$, $\beta=(1-\gamma')/2$. We show, for all sufficiently large $d$ with $\gamma_{\rm out}\le\gamma'$ and all $\varepsilon\in(0,1)$, that
\begin{equation}
\label{eq:T1_bound}
 \PP\!\left(\mathcal T_1>\frac R\varepsilon\right)
 \le\frac{C_0\varepsilon^{1/4}}\beta,\qquad
 \PP\!\left(\nout\mathcal T_2>
 \frac{2R(1+R)^2}{\varepsilon^2\beta}\right)
 \le\frac{3C_0\varepsilon^{1/4}}\beta.
\end{equation}
These bounds imply both asserted inverse-moment estimates.

\parTitle{First inverse moment} Take $t>0$ and $N_1=\#\{a:\xi_a\le\varepsilon t\}$. The event $\{\mathcal T_1\ge t,\ \xi_a\le\varepsilon t\}$ is contained in $F_a=\{\xi_a\le\varepsilon\Trace(\mathbf K_{(a)}^{-1})\}$. Independence of $\mathbf x_a$ and $\mathbf K_{(a)}$ gives $\PP(F_a)\le C_0\varepsilon^{1/4}$. Summing and applying Markov's inequality gives $\PP(\mathcal T_1\ge t,N_1\ge\beta\nout)\le C_0\varepsilon^{1/4}/\beta$. On $\{N_1<\beta\nout\}$, the first sum rule instead implies $\gamma_{\rm out}\ge(1-\beta)\varepsilon t/(1+\varepsilon t)$, hence $\varepsilon t\le\gamma'/(1-\beta-\gamma')=R$. Thus $\{\mathcal T_1\ge t,N_1<\beta\nout\}$ is empty for $t>R/\varepsilon$, proving the first bound in~\cref{eq:T1_bound}.

\parTitle{Second inverse moment} The first sum rule bounds the number of large $\xi_a$ deterministically:
\begin{equation}
\label{eq:many_small_xi}
 \#\{a:\xi_a\ge R\}\le\gamma'\nout\frac{1+R}{R}
 =\frac{1+\gamma'}2\nout,\qquad
 \lvert\mathcal A\rvert\ge\beta\nout,\quad
 \mathcal A=\{a:\xi_a<R\}.
\end{equation}
Let $N_2=\#\{a:\eta_a\le\varepsilon t\}$. Apply the same small-ball argument to $\mathbf K_{(a)}^{-2}$ to obtain $\PP(\mathcal T_2\ge t,N_2\ge\beta\nout/2)\le2C_0\varepsilon^{1/4}/\beta$. On $\{N_2<\beta\nout/2\}$, at least $\beta\nout/2$ indices in $\mathcal A$ satisfy $\eta_a>\varepsilon t$. The second sum rule then gives $\mathcal T_1\ge\beta\nout\varepsilon t/[2(1+R)^2]$. Intersecting with $\{\mathcal T_1\le R/\varepsilon\}$ forces $t\le2R(1+R)^2/(\varepsilon^2\beta\nout)$. Adding the two failure probabilities proves the second bound in~\cref{eq:T1_bound}.

\parTitle{Total weight of small eigenvalues} Every eigenvalue below $\theta\nout$ contributes at least $(\theta\nout)^{-1}$ to $\mathcal T_1$. Cauchy--Schwarz gives
\begin{equation}
\label{eq:proof_CS}
 \sum_{\lambda_j(\mathbf K)\le\theta\nout}\frac1{\lambda_j(\mathbf K)}
 \le\sqrt{\#\{j:\lambda_j(\mathbf K)\le\theta\nout\}}\sqrt{\mathcal T_2}
 \le\sqrt\theta\sqrt{\mathcal T_1(\nout\mathcal T_2)}.
\end{equation}
The inverse-moment estimates and $\nout/(d^2-1)\to\gamma^{-1}$ prove the tail bound uniformly in $d$.
\end{proof}

\subsection{Classical variance profile}
\label[appsec]{nav:classical_variance}

The theorem below gives the precise variance law stated in~\cref{sec:classical_peak}: test states and target rotations are averaged at fixed POVM, and convergence in probability is over the POVM. Nearly uniform mean outcome probabilities reduce the normalized variance to $S_d$. Its limit diverges at $\gamma=1$ because the positive bulk reaches zero; inaccessible directions have zero pseudoinverse weight and account for the factor $\min\{1,\gamma^{-1}\}$. Under the stated target bound, the identity component contributes asymptotically vanishing normalized variance.

\begin{theorem}[Classical variance peak]
\label{thm:classical_peak}
Let $\bs\mu$ be a Haar-random rank-one POVM with $\nout$ outcomes in dimension $d$, $\sigma$ an independent Haar pure state, and $\calO_U=U^\dagger\calO U$ for fixed Hermitian $\calO$ and Haar $U$ independent of the POVM and test state. Assume $\Delta_{\calO}^2>0$ and $\lvert\Trace(\calO)\rvert/(d\sqrt{d\Delta_{\calO}^2})\le C$. Define the conditional Haar average, which depends on the POVM through $\bs M_{\bs\mu}$,
\begin{equation}
\label{eq:conditional_classical_variance}
 V_{\rm ex}
 =\EE_{U,\sigma}\!\left[
 \big\|\bs D_\sigma^{1/2}\bs M_{\bs\mu}^+\bs v_{\calO_U}\big\|^2
 -\bigl\lvert\bs v_{\calO_U}^\dagger\Pi_{\bs\mu}\bs v_\sigma\bigr\rvert^2
 \;\middle\vert\;\bs\mu\right].
\end{equation}
As $d\to\infty$ with $\gamma_{\rm out}\to\gamma\in(0,\infty)$, convergence in probability over the POVM gives
\begin{equation}
\label{eq:statement_classical_peak}
 \frac{V_{\rm ex}}{d\Delta_{\calO}^2}
 \xrightarrow{\PP}
 \begin{cases}
 \mathcal J_{\rm cl}(\gamma)
 =\dfrac{\min\{1,\gamma^{-1}\}}{\lvert1-\gamma\rvert},&\gamma\ne1,\\[1ex]
 +\infty,&\gamma=1.
 \end{cases}
\end{equation}
\end{theorem}
\begin{proof}
\parTitle{Identity target contribution} Write $\calO_U=c_\calO I_d+U^\dagger\calO_0U$. By~\cref{eq:moment_2_operator}, the traceless term has mean zero and covariance $d\Delta_{\calO}^2\bs P_0/(d^2-1)$, so its mixed variance term with the identity vanishes after the $U$ average. The identity shadow coefficients are $\bs w_{\rm shad}(I_d)=\Pi_{\im\bs M_{\bs\mu}^T}\mathbf1_{\nout}$. Since $\nout\bsp_{I/d}$ lies in this outcome-space range, $\Delta_d=\max_a\lvert\nout p_{I/d,a}-1\rvert$ gives $\|\bs w_{\rm shad}(I_d)-\mathbf1_{\nout}\|\le\sqrt{\nout}\Delta_d$ and $V_{\rm ex}(I_d)\le(1+\Delta_d)\Delta_d^2$. The latter uses $\barSigma_{\rm te}\mathbf1_{\nout}=0$ and $\barSigma_{\rm te}\preceq\bs D_{I/d}\preceq(1+\Delta_d)I_{\nout}/\nout$. Trace concentration and the target assumption therefore imply $c_\calO^2V_{\rm ex}(I_d)/(d\Delta_{\calO}^2)=o_\PP(1)$, without requiring uniform minimum-norm identity coefficients.

\parTitle{Reduction to inverse spectral weight} The independent target and test-state moments give the exact average
\begin{equation}
\label{eq:exact_classical_average}
 \begin{aligned}
 V_{\rm ex}
 &=\frac{d\Delta_{\calO}^2}{d^2-1}
 \big\|\bs D_{I/d}^{1/2}\bs M_{\bs\mu}^+\bs P_0\big\|_F^2
 -\mathcal B_d(\bs\mu)+c_\calO^2V_{\rm ex}(I_d),\\
 \mathcal B_d(\bs\mu)
 &=\frac{\Delta_{\calO}^2}{(d^2-1)(d+1)}
 \Trace\!\left(\bs P_0\Pi_{\bs\mu}
 (I_{d^2}+\bs v_{I_d}\bs v_{I_d}^\dagger)\Pi_{\bs\mu}\right).
 \end{aligned}
\end{equation}
Because the two projectors are contractions and $\Trace(I_{d^2}+\bs v_{I_d}\bs v_{I_d}^\dagger)=d^2+d$, the subtracted term obeys $0\le\mathcal B_d\le d\Delta_{\calO}^2/(d^2-1)$. By~\cref{lem:trace_conc}, $\Delta_d=\|\nout\bs D_{I/d}-I_{\nout}\|_{\rm op}=O_\PP(\sqrt{\log\nout/d})=o_\PP(1)$, so on $\{\Delta_d<1\}$,
\begin{equation}
\label{eq:diagonal_sandwich}
 \frac{1-\Delta_d}{\nout}\|\bs M_{\bs\mu}^+\bs P_0\|_F^2
 \le\|\bs D_{I/d}^{1/2}\bs M_{\bs\mu}^+\bs P_0\|_F^2
 \le\frac{1+\Delta_d}{\nout}\|\bs M_{\bs\mu}^+\bs P_0\|_F^2.
\end{equation}
The identity contribution is removed exactly by
\begin{equation}
\label{eq:identity_sector_exact}
 \begin{aligned}
 \|\bs M_{\bs\mu}^+\bs P_0\|_F^2
 &=\Trace(\frameOp_{\bs\mu}^+)
 -\frac{\bs v_{I_d}^\dagger\frameOp_{\bs\mu}^+\bs v_{I_d}}d,\\
 0&\le\frac{\bs v_{I_d}^\dagger\frameOp_{\bs\mu}^+\bs v_{I_d}}d
 \le\frac{\nout}d.
 \end{aligned}
\end{equation}
For the upper bound, $\bs M_{\bs\mu}\mathbf1_{\nout}=\bs v_{I_d}$ makes $\mathbf1_{\nout}$ a feasible solution, so $\|\bs M_{\bs\mu}^+\bs v_{I_d}\|^2\le\nout$. The normalized subtracted term is at most $1/(d^2-1)$, and the removed identity contribution is at most $1/[d(d^2-1)]$. Together with trace concentration and the identity-target estimate, these bounds give
\begin{equation}
\label{eq:proof_sandwich}
 \frac{V_{\rm ex}}{d\Delta_{\calO}^2}
 =(1+o_\PP(1))S_d+o_\PP(1).
\end{equation}

\parTitle{Transferring positive eigenvalues} By~\cref{lem:gaussian_lift_rank}, $\nout\frameOp_{\bs\mu}=\mathbf T\mathbf A\mathbf T^\dagger$, with $\delta_d=\|\mathbf T-I_{d^2}\|_{\rm op}=O_\PP(d^{-1/2})$. Both matrices have rank $\min(d^2,\nout)$. The congruence estimate~\cref{lem:congruence_spectral} bounds their ordered positive eigenvalues between $(1-\delta_d)^2\lambda_j(\mathbf A)$ and $(1+\delta_d)^2\lambda_j(\mathbf A)$. Inverting gives
\begin{equation}
\label{eq:proof_transfer}
 \begin{aligned}
 S_d&=(1+O_\PP(d^{-1/2}))\frac{\Trace(\mathbf A^+)}{d^2-1},\\
 \int_{(0,\theta)}\frac{d\nu_d^+(x)}x
 &\le\frac4{d^2-1}\sum_{0<\lambda_j(\mathbf A)\le4\theta}
 \frac1{\lambda_j(\mathbf A)}\qquad(\delta_d\le1/2).
 \end{aligned}
\end{equation}
Indeed, an eigenvalue of $\nout\frameOp_{\bs\mu}$ below $\theta$ corresponds to one of $\mathbf A$ below $4\theta$, with reciprocal enlarged by at most four.

\parTitle{Critical divergence from bounded tests} For $a>0$, $f_a(x)=x/(x+a)^2$ is bounded and continuous, vanishes at zero, and increases to $x^{-1}$ for $x>0$ as $a\downarrow0$. The MP law and monotone convergence give
\begin{equation}
\label{eq:bounded_inverse_test}
 S_d\ge\int f_a\,d\nu_d^+
 \xrightarrow{\PP}\int f_a\,d\mu_{{\rm MP},\gamma}
 \nearrow\int_{(0,\infty)}\frac{d\mu_{{\rm MP},\gamma}(x)}x
 \qquad(a\downarrow0).
\end{equation}
At $\gamma=1$, the density behaves as $x^{-1/2}$ near zero, so its inverse moment diverges. Given $K$, choose $a$ with $\int f_a\,d\mu_{{\rm MP},1}>K+1$; then $\PP(S_d>K)\to1$. Away from $\gamma=1$, the same argument gives $\PP(S_d\ge\mathcal J_{\rm cl}(\gamma)-\delta)\to1$ for every $\delta>0$.

\parTitle{Below informational completeness, $\gamma>1$} Here $\mathbf X$ has independent columns and full column rank. The negative second moment identity~\cref{lem:neg_second_moment} gives
\begin{equation}
\label{eq:proof_dist_identity}
 \Trace(\mathbf A^+)=\nout\sum_{a=1}^{\nout}\frac1{D_a},\qquad
 D_a=\operatorname{dist}\!\left(\mathbf x_a,
 \operatorname{span}\{\mathbf x_b:b\ne a\}\right)^2.
\end{equation}
The orthogonal complement of the other columns has projector $\mathbf Q_a$, independent of $\hat{\mathbf g}_a$, of rank $d^2-\nout+1$. Thus $D_a=\mathbf x_a^\dagger\mathbf Q_a\mathbf x_a$ has conditional mean $d^2-\nout+1+\|\mathbf Q_a\bs v_{I_d}\|^2=d^2-\nout+1+O(d)$. Apply~\cref{lem:lift_distance} at scale $d^{-1/4}$ and take a union bound over $\nout\le d^2$ columns to obtain $\max_a\lvert D_a/d^2-(1-\nout/d^2)\rvert=O_\PP(d^{-1/4})$. Since $1-\nout/d^2\to1-\gamma^{-1}>0$, these distances can be inverted uniformly. Therefore,
\begin{equation}
\label{eq:proof_Sd_under}
 S_d=(1+O_\PP(d^{-1/4}))
 \frac{\nout^2}{(d^2-1)(d^2-\nout)}
 \xrightarrow{\PP}\frac1{\gamma(\gamma-1)}
 =\mathcal J_{\rm cl}(\gamma).
\end{equation}

\parTitle{Above informational completeness, $\gamma<1$} The rows of $\mathbf X$ are dependent quadratic functions, so the column-distance argument cannot be applied to them. Instead choose $0<\theta<\lambda_-=(1-\sqrt\gamma)^2$ and split $S_d$ into integrals on $[\theta,\infty)$ and $(0,\theta)$. The first integrand is bounded and its only discontinuity lies outside the MP support. Hence~\cref{thm:main} gives
\begin{equation}
\label{eq:proof_bulk}
 \int_{[\theta,\infty)}\frac{d\nu_d^+(x)}x
 \xrightarrow{\PP}\int_{[\theta,\infty)}
 \frac{d\mu_{{\rm MP},\gamma}(x)}x
 =\frac1{1-\gamma}=\mathcal J_{\rm cl}(\gamma).
\end{equation}
For $\mathbf K=\mathbf X\mathbf X^\dagger=\nout\mathbf A$,~\cref{lem:lift_inverse_tail,eq:proof_transfer} give
\begin{equation}
\label{eq:proof_tail}
 \int_{(0,\theta)}\frac{d\nu_d^+(x)}x
 \le\frac{4\nout}{d^2-1}
 \sum_{\lambda_j(\mathbf K)\le4\theta\nout}\frac1{\lambda_j(\mathbf K)}
 =O_\PP(\sqrt\theta)
\end{equation}
uniformly in $d$. Given $\varepsilon,\varrho>0$, choose $\theta$ below $\lambda_-$ so that the limiting upper probability of a tail exceeding $\varepsilon/2$ is at most $\varrho$. For this fixed $\theta$,~\cref{eq:proof_bulk} puts the bulk within $\varepsilon/2$ of $\mathcal J_{\rm cl}(\gamma)$ with probability tending to one. Letting $\varrho\downarrow0$ proves $S_d\to_\PP1/(1-\gamma)$. All three regimes, together with~\cref{eq:proof_sandwich}, prove the theorem.
\end{proof}

\subsection{Relation to exact-feature training}
\label[appsec]{nav:exact_training}

Exact-feature regression uses $P_{ai}=\Trace(\mu_a\rho_i)$ and $y_i=\Trace(\calO\rho_i)$, choosing the minimum-norm least-squares readout $\bs w_0=(\bs P^T)^+\bs y$. The question in~\cref{sec:classical_training} is when this training fit returns the reconstruction coefficients used above. Full-rank measurement and training designs give those coefficients exactly. Below informational completeness, finite training sets can couple accessible and inaccessible operator directions through their empirical state covariance; sufficiently many Haar training states suppress that coupling and recover the projected target in the relevant test-noise norm.

\begin{corollary}[Relation to exact training]
\label{cor:classical_training}
Under the measurement and target assumptions of~\cref{thm:classical_peak}, if both measurement and training designs have rank $d^2$, exact minimum-norm regression gives $\bs w_0=\bs M_{\bs\mu}^+\bs v_\calO$, and the theorem applies to its test variance. If $\nout<d^2$ and the training states are independent Haar pure states with $d^2\log d/\ntr\to0$, the learned coefficients approach the shadow coefficients in the Haar-averaged test-noise norm normalized by $\sqrt{d\Delta_{\calO}^2}$. At fixed $\gamma_{\rm out}>1$, the averaged variance consequently has the limit in~\cref{eq:statement_classical_peak}.
\end{corollary}
\begin{proof}
\parTitle{Full-rank designs} If both designs have rank $d^2$, the reverse-order identity gives $\bs P^+=\bs M_{\bs\rho}^+(\bs M_{\bs\mu}^T)^+$. Transpose this identity and apply it to $\bs y=\bs M_{\bs\rho}^T\bs v_\calO$. Since $\bs M_{\bs\rho}\bs M_{\bs\rho}^+=I_{d^2}$, the result is $\bs w_0=\bs M_{\bs\mu}^+\bs v_\calO$.

\parTitle{Whitened training covariance} The raw state frame is $\frameOp_{\bs\rho}=\bs M_{\bs\rho}\bs M_{\bs\rho}^T$. For $\nout<d^2$, multiply state coordinates by $\bs C_\rho^{-1/2}$, where $\bs C_\rho=\EE[\bs v_\rho\bs v_\rho^T]$. This whitening equalizes the population second moments without subtracting the mean: put $\bs H_\rho=\bs C_\rho^{-1/2}\frameOp_{\bs\rho}\bs C_\rho^{-1/2}/\ntr$ and $e_{\ntr}=\|\bs H_\rho-I_{d^2}\|_{\rm op}$. Each whitened state has squared norm $d^2$ and second moment $I_{d^2}$. Matrix Bernstein~\cite{tropp2012user} gives $e_{\ntr}=o_\PP(1)$ when $d^2\log d/\ntr\to0$.

\parTitle{Mixing of accessible and inaccessible directions} The measurement range $L=\im\bs M_{\bs\mu}$ contains $\bs v_{I_d}$, so $L^\perp$ is traceless and $\Pi_{\bs\mu}$ commutes with $\bs C_\rho$. Write $\bs w_0=\bs M_{\bs\mu}^+\bs w$ with $\bs w\in L$. The normal equations give
\begin{equation}
 \bs w-\Pi_{\bs\mu}\bs v_\calO
 =\bs C_{\rho,L}^{-1/2}
 (\Pi_{\bs\mu}\bs H_\rho\Pi_{\bs\mu}\vert_L)^{-1}
 \Pi_{\bs\mu}(\bs H_\rho-I_{d^2})(I_{d^2}-\Pi_{\bs\mu})
 \bs C_\rho^{1/2}\bs v_\calO.
\end{equation}
The eigenvalue of $\bs C_\rho$ on $L^\perp$ is $1/[d(d+1)]$, and its eigenvalues on $L$ are no smaller. On $\{e_{\ntr}<1\}$, the map on the right has norm at most $e_{\ntr}/(1-e_{\ntr})$ and annihilates $\bs v_{I_d}$. For the learned weights corresponding to $\calO_U$, the target second moment therefore gives
\begin{equation}
 \EE_U\|\bs D_{I/d}^{1/2}(\bs w_0-\bs M_{\bs\mu}^+\bs v_{\calO_U})\|^2
 \le\frac{d\Delta_{\calO}^2}{d^2-1}
 \left(\frac{e_{\ntr}}{1-e_{\ntr}}\right)^2
 \|\bs D_{I/d}^{1/2}\bs M_{\bs\mu}^+\|_F^2.
\end{equation}
By~\cref{thm:classical_peak,eq:diagonal_sandwich}, the right side is $o_\PP(d\Delta_{\calO}^2)$ at fixed $\gamma_{\rm out}>1$. Since $\barSigma_{\rm te}\preceq\bs D_{I/d}$, Cauchy--Schwarz bounds the variance cross term by $o_\PP(d\Delta_{\calO}^2)$ as well. The limiting shadow observable is $\Pi_{\bs\mu}\calO$.
\end{proof}

\subsection{Shot-averaged least-squares problem}
\label[appsec]{nav:shot_averaged}
The noise-induced regularization discussed in~\cref{sec:classical_training} compares the least-squares problem of a measured record with its average over the training shots. The lemma below collects the exact identities behind this comparison. The training states are held fixed throughout; only the multinomial counts are averaged.

\begin{lemma}[Shot-averaged least-squares problem]
\label{lem:shot_averaged_ls}
Fix training states $\rho_1,\ldots,\rho_{\ntr}$ with exact probability vectors $\bs p_i$ and labels $\bs y$, and let $\hatbs n_i\sim\operatorname{Mult}(N,\bs p_i)$ be independent. Write $\EE_{\hatbs n}$ for the expectation over the counts, $\barSigma_{\rm tr}=\ntr^{-1}\sum_i\bs\Sigma_i$ with $\bs\Sigma_i=\diag\bs p_i-\bs p_i\bs p_i^T$, and $\bar{\bs p}=\ntr^{-1}\sum_i\bs p_i$.
\begin{enumerate}
\item For every $\bs w\in\RR^{\nout}$ independent of the counts,
\begin{equation}
\frac1{\ntr}\EE_{\hatbs n}\|\hatbs P_N^T\bs w-\bs y\|^2
=\frac1{\ntr}\|\bs P^T\bs w-\bs y\|^2+\frac1N\bs w^T\barSigma_{\rm tr}\bs w.
\label{eq:shot_averaged_loss_lemma}
\end{equation}
\item The coefficient matrices of the normal equations $\hatbs P_N\hatbs P_N^T\bs w=\hatbs P_N\bs y$ have the averages
\begin{equation}
\frac1{\ntr}\EE_{\hatbs n}[\hatbs P_N\hatbs P_N^T]
=\frac1{\ntr}\bs P\bs P^T+\frac1N\barSigma_{\rm tr},
\qquad
\EE_{\hatbs n}[\hatbs P_N\bs y]=\bs P\bs y,
\label{eq:shot_averaged_normal}
\end{equation}
which are the normal equations of the right side of~\cref{eq:shot_averaged_loss_lemma}.
\item $\barSigma_{\rm tr}\mathbf1_{\nout}=0$ and $\barSigma_{\rm tr}=\bs\Sigma(\bar{\bs p})-\ntr^{-1}\sum_i(\bs p_i-\bar{\bs p})(\bs p_i-\bar{\bs p})^T$, where $\bs\Sigma(\bar{\bs p})=\diag\bar{\bs p}-\bar{\bs p}\bar{\bs p}^T$.
\item If $\max_a\lvert\nout\bar p_a-1\rvert\le\varepsilon$, $\|\frameOp_{\bs\rho}^{\rm c}\|_{\rm op}\le K$ and $\|\nout\bs P_0\frameOp_{\bs\mu}\bs P_0\|_{\rm op}\le K$, with the centered training frame of~\cref{eq:general_training_frame}, then
\begin{equation}
\|\nout\barSigma_{\rm tr}-\bs\Pi_{\nout}\|_{\rm op}\le3\varepsilon+\varepsilon^2+\frac{K^2}{d(d+1)}.
\label{eq:shot_averaged_isotropy}
\end{equation}
\end{enumerate}
\end{lemma}
\begin{proof}
\parTitle{Items 1 and 2} Each frequency vector $\hatbs n_i/N$ has mean $\bs p_i$ and covariance $\bs\Sigma_i/N$, and different training states are independent. For fixed $\bs w$, the prediction $\bs w^T\hatbs n_i/N$ therefore has mean $\bs w^T\bs p_i$ and variance $\bs w^T\bs\Sigma_i\bs w/N$, and $\EE_{\hatbs n}(\bs w^T\hatbs n_i/N-y_i)^2=(\bs w^T\bs p_i-y_i)^2+\bs w^T\bs\Sigma_i\bs w/N$. Summing over $i$ and dividing by $\ntr$ gives~\cref{eq:shot_averaged_loss_lemma}. Likewise $\EE_{\hatbs n}[(\hatbs n_i/N)(\hatbs n_i/N)^T]=\bs p_i\bs p_i^T+\bs\Sigma_i/N$ and $\EE_{\hatbs n}[\hatbs n_i/N]=\bs p_i$; summing over $i$ gives~\cref{eq:shot_averaged_normal}. Setting the gradient of the right side of~\cref{eq:shot_averaged_loss_lemma} to zero gives $(\bs P\bs P^T/\ntr+\barSigma_{\rm tr}/N)\bs w=\bs P\bs y/\ntr$.

\parTitle{Item 3} Since $\mathbf1_{\nout}^T\bs p_i=1$, $\bs\Sigma_i\mathbf1_{\nout}=\bs p_i-\bs p_i=0$. The decomposition follows from $\ntr^{-1}\sum_i\bs p_i\bs p_i^T=\bar{\bs p}\bar{\bs p}^T+\ntr^{-1}\sum_i(\bs p_i-\bar{\bs p})(\bs p_i-\bar{\bs p})^T$.

\parTitle{Item 4} Put $\delta_a=\nout\bar p_a-1$, so $\|\bs\delta\|\le\varepsilon\sqrt{\nout}$. Then $\nout\bs\Sigma(\bar{\bs p})-\bs\Pi_{\nout}=\diag\bs\delta-(\bs\delta\mathbf1_{\nout}^T+\mathbf1_{\nout}\bs\delta^T+\bs\delta\bs\delta^T)/\nout$, whose operator norm is at most $\varepsilon+2\varepsilon+\varepsilon^2$. For the between-state term, $\bs p_i-\bar{\bs p}=\bs M_{\bs\mu}^T\bs P_0\bs v_{\rho_i-\bar\rho}$ because $\rho_i-\bar\rho$ is traceless, so that, with~\cref{eq:general_training_frame},
$\nout\ntr^{-1}\sum_i(\bs p_i-\bar{\bs p})(\bs p_i-\bar{\bs p})^T=[d(d+1)]^{-1}\nout\bs M_{\bs\mu}^T\bs P_0\frameOp_{\bs\rho}^{\rm c}\bs P_0\bs M_{\bs\mu}$. Its norm is at most $\|\frameOp_{\bs\rho}^{\rm c}\|_{\rm op}\|\nout\bs P_0\frameOp_{\bs\mu}\bs P_0\|_{\rm op}/[d(d+1)]\le K^2/[d(d+1)]$.
\end{proof}

\parTitle{Scope} Item 4 makes the averaged objective a ridge objective on $\mathbf1_{\nout}^\perp$ with $\tau_N=1/(N\nout)$, i.e., scaled ridge $\eta=1$ in the units of~\cref{nav:def_limit}. The lemma concerns averages at fixed $\bs w$. It describes the fitted weights of a single record only if the realized matrices in~\cref{eq:shot_averaged_normal} concentrate around their averages, which we do not prove; near $\nout=\ntr$ they do not (\cref{sec:uc_finite_shots}).

\section{Finite-shot regression and physical readout}
\label[appsec]{app:quantum_profiles}

The interpolation peaks of~\cref{sec:quantum,sec:uc_main} occur near $\nout=\ntr$, where the measured feature matrix is square and small singular directions can require large fitted weights. They concern the actual multinomial records, both above and below informational completeness. Their analysis begins with a common calculation at fixed positive scaled ridge, valid above and below the outcome-count threshold $\nout=d^2$. \Cref{nav:def_geometry} specifies the empirical geometry and averages, \cref{nav:def_centered} evaluates the centered costs, and~\cref{app:proof_physical_mean_profile} restores the empirical mean. That restoration is essential because the physical fit has no separately fitted intercept.

\subsection{Geometry, targets, and asymptotic regime}
\label{nav:def_geometry}

The hypotheses below express which operator directions the training states explore, which directions the apparatus resolves, and how strongly shot noise perturbs each measured feature. \Cref{nav:def_readout} defines the physical estimator and its two costs;~\cref{nav:def_training_geometry} separates state variation from categorical noise;~\cref{nav:def_limit} fixes the large-dimension regime used for the main-text peak laws.

\subsubsection{Readout and target ensembles}
\label{nav:def_readout}

We first specify the physical ridge readout of~\cref{eq:main_ridge_estimator} and the averages used to compare it with minimum norm. The conventions are those of~\cref{app:conventions}: $d(d+1)$ normalizes Haar expectation-value fluctuations, while $d^2$ bounds the rank of the Born matrix. Haar test averaging separates reconstruction of the traceless target from the constant prediction, which must be tracked independently.

\parTitle{Physical ridge and minimum norm} For each training state $\rho_i$, independently sample $\hatbs n_i\sim\operatorname{Mult}(N,\bsp_i)$ and collect frequencies in $\hatbs P_N=[\hatbs n_i/N]$. The physical ridge readout is
\begin{equation}
 \hatbs w_{N,\tau}=(\hatbs P_N\hatbs P_N^T/\ntr+\tau I_{\nout})^{-1}
          \hatbs P_N\bs y/\ntr,\qquad
 y_i=\Trace(\calO\rho_i),\qquad
 \eta=\nout N\tau.
 \label{eq:alt_physical_ridge}
\end{equation}
Here $\tau$ is the coefficient of $\|\bs w\|^2$ added to the averaged fitting loss $\ntr^{-1}\|\hatbs P_N^T\bs w-\bs y\|^2$; the equivalent penalty for the unnormalized sum of squares is $\ntr\tau$. For a measured singular value $s$, ridge replaces $1/s$ by $s/(s^2+\ntr\tau)$, limiting the weight assigned to features with small singular values. At zero penalty, $\hatbs w_{N,0}=(\hatbs P_N^T)^+\bs y$ is the readout $\hatbs{w}_N$ of~\cref{eq:qelm_exact_linear_rule}: among all least-squares minimizers, it has the smallest Euclidean weight norm.

\parTitle{Target law and conditional averages} The target is drawn from one of the ensembles of~\cref{def:target_ensembles}, whose isotropic moments~\cref{eq:centered_full_target_covariance} are all that the target-averaged costs require.
Unless stated otherwise, states and POVMs are held fixed, target averages condition on the measured record, and probability limits concern the counts and the independent target. For random training states or apparatuses, probability statements also include that randomness; an ensemble expectation is taken only when explicitly indicated.

\parTitle{Physical error decomposition} For the fixed fitted record, the ideal-test error of~\cref{eq:hat_MSE}, averaged over Haar test states, is $\calE_{N,\tau}(\calO)=\EE_\sigma[(\bs p_\sigma^T\hatbs w_{N,\tau}-\Trace(\calO\sigma))^2]$. By~\cref{lem:haar_error_split}, this average separates traceless reconstruction from the error in the constant prediction:
\begin{equation}
 {d(d+1)}\calE_{N,\tau}(\calO)
 =\|\bs P_0\bs M_{\bs\mu}\hatbs w_{N,\tau}-\bs v_{\calO_0}\|^2
       +{d(d+1)}(\bsp_{I/d}^T\hatbs w_{N,\tau}-c_\calO)^2.
 \label{eq:general_physical_split}
\end{equation}
The second term evaluates the mean prediction on $I_d/d$ and can retain the empirical training mean even when normalized spectral traces do not. The variance numerator $V_{N,\tau}(\calO)=\hatbs w_{N,\tau}^T\barSigma_{\rm te}\hatbs w_{N,\tau}$ contributes $V_{N,\tau}(\calO)/M$ to the error of $M$ independent test shots, with covariance~\cref{eq:barSigma_exact}. Neither cost here is averaged over targets or training records.

\subsubsection{Training geometry and measurement-outcome noise}
\label{nav:def_training_geometry}

The peak depends on two sources of variability in the measured features: the variation of the exact probabilities across training states, which is the signal to be learned, and the shot fluctuations within each state. All fixed-ridge formulas of~\cref{nav:def_centered} and all peak coefficients of~\cref{app:interpolation_mechanisms} depend on the training states, the measurement, and the counts only through three matrices, which we define in this subsection: the centered training frame $\frameOp_{\bs\rho}^{\rm c}$, which measures the first source in operator coordinates; the rescaled traceless measurement matrix $\tilde{\bs M}_{\bs\mu}$, which transfers it to outcome space in units of shot noise; and the training-averaged single-shot covariance $\barSigma_{\rm tr}$, which measures the second source. Keeping these matrices distinct identifies which small eigenvalues are supplied by noise in~\cref{sec:quantum_ic} and which already occur in the exact training matrix in~\cref{sec:uc_main}.

\parTitle{Empirical means and shot covariance} Write $\bar\rho=\ntr^{-1}\sum_i\rho_i$ and $\bar{\bsp}=\bs M_{\bs\mu}^T\bs v_{\bar\rho}$ for the empirical means. The single-shot covariance is $\bs\Sigma(\rho_i)=\diag\bsp_i-\bsp_i\bsp_i^T$, $\bsp_i\equiv \bsp(\rho_i)$, as in~\cref{eq:main_counts}, and $\barSigma_{\rm tr}=\ntr^{-1}\sum_i\bs\Sigma(\rho_i)$ averages it over the actual training list, as in~\cref{eq:noise_induced_regularization}. A frequency vector has covariance $\bs\Sigma(\rho_i)/N$. We define the projectors
$\bs\Pi_{\nout}=I_{\nout}-\mathbf1_{\nout}\mathbf1_{\nout}^T/\nout$ and $\bs\Pi_{\ntr}=I_{\ntr}-\mathbf1_{\ntr}\mathbf1_{\ntr}^T/\ntr$, which remove the constant outcome direction and the mean training coordinate, respectively. They play for outcome and training coordinates the role that the projector $\bs P_0$ of~\cref{eq:common_projectors} plays for operator coordinates.

\parTitle{Explored operator directions} We define the normalized centered training frame as the sample covariance of the traceless coordinates of the training states, rescaled by the Haar variance $1/[d(d+1)]$:
\begin{equation}
 \frameOp_{\bs\rho}^{\rm c}
 =\frac{d(d+1)}{\ntr}\bs P_0\bs M_{\bs\rho}\bs\Pi_{\ntr}\bs M_{\bs\rho}^T\bs P_0,
 \qquad
 \frac1{d^2-1}\Trace(\frameOp_{\bs\rho}^{\rm c})
 =\frac{d(d+1)}{d^2-1}\left[\frac1{\ntr}\sum_i\Trace(\rho_i^2)-\Trace(\bar\rho^2)\right].
 \label{eq:general_training_frame}
\end{equation}
It is the analogue, for the training states, of the measurement frame operator $\frameOp_{\bs\mu}$ of~\cref{sec:classical_frame}~\cite{innocenti2023shadow,scott2006TightInformationallyComplete}, with the empirical mean removed. For a traceless observable $A$ with $\Trace(A^2)=1$, $\bs v_A^T\frameOp_{\bs\rho}^{\rm c}\bs v_A=({d(d+1)}/\ntr)\sum_i[\Trace(A(\rho_i-\bar\rho))]^2$. The frame eigenvalues therefore measure the empirical variances of expectation values across the training list in units of the Haar value $1/[d(d+1)]$, so that independent Haar training states give eigenvalues of order one (\cref{lem:fixed_rank_bulk_bridge}); unexplored operator directions give zero eigenvalues. The raw frame differs by the empirical-mean contribution:
\begin{equation}
 \frac{d(d+1)}{\ntr}\bs P_0\frameOp_{\bs\rho}\bs P_0
 =\frameOp_{\bs\rho}^{\rm c}
       +{d(d+1)}\bs v_{\bar\rho-I_d/d}\bs v_{\bar\rho-I_d/d}^T.
 \label{eq:general_raw_frame}
\end{equation}
The rank-one difference leaves the limiting eigenvalue distribution unchanged but can affect the constant prediction in~\cref{eq:general_physical_split}.

\parTitle{Signal in units of shot noise} We define the rescaled traceless measurement matrix
\begin{equation}
 \tilde{\bs M}_{\bs\mu}=\sqrt{\frac{\nout N}{d(d+1)}}\,\bs P_0\bs M_{\bs\mu},\qquad
 \tilde{\bs M}_{\bs\mu}\mathbf1_{\nout}=0,\qquad
 \tilde{\bs M}_{\bs\mu}\tilde{\bs M}_{\bs\mu}^T
       =\frac{\nout N}{d(d+1)}\bs P_0\frameOp_{\bs\mu}\bs P_0.
 \label{eq:alt_scaled_measurement}
\end{equation}
The squared singular values of $\tilde{\bs M}_{\bs\mu}$ measure the apparatus response along traceless operator directions; the normalization is chosen so that they are signal-to-noise ratios. An operator direction with rescaled frame eigenvalue $x$ gives the squared singular value $\alpha_dx$, with $\alpha_d=N/[d(d+1)]$, and~\cref{lem:snr_direction} below identifies $\alpha_dx$ as the ratio between the variation of that direction across Haar-random states and the noise of its $N$-shot estimate. Combining the two frames gives the exact signal covariance
\begin{equation}
 \bs S\equiv\tilde{\bs M}_{\bs\mu}^T\frameOp_{\bs\rho}^{\rm c}
                  \tilde{\bs M}_{\bs\mu}
 =\frac{\nout N}{\ntr}\sum_{i=1}^{\ntr}
 (\bsp_i-\bar{\bsp})(\bsp_i-\bar{\bsp})^T,
 \label{eq:general_ridge_signal_noise}
\end{equation}
whereas the categorical covariances obey
\begin{equation}
 \barSigma_{\rm tr}=\diag\bar{\bsp}
 -\frac1{\ntr}\bs M_{\bs\mu}^T\frameOp_{\bs\rho}\bs M_{\bs\mu},\qquad
 \nout N[\bs\Sigma(\bar{\bsp})-\barSigma_{\rm tr}]
 =\tilde{\bs M}_{\bs\mu}^T\frameOp_{\bs\rho}^{\rm c}\tilde{\bs M}_{\bs\mu},
 \label{eq:general_covariance_identity}
\end{equation}
where $\bs\Sigma(\bar{\bsp})=\diag\bar{\bsp}-\bar{\bsp}\bar{\bsp}^T$. Thus~\cref{eq:general_covariance_identity} identifies $\bs S$ as $\nout N$ times the covariance of exact training probabilities, distinct from within-state noise $\nout\barSigma_{\rm tr}$. These matrices enter all subsequent formulas: the deterministic equivalent~\cref{eq:general_profile_Q} contains $\bs S$ and $\nout\barSigma_{\rm tr}$, the costs~\cref{eq:general_centered_costs,eq:general_centered_ideal_cost} are weighted traces involving $\tilde{\bs M}_{\bs\mu}$, and the peak coefficients~\cref{eq:general_peak_coefficients} involve all three.

\parTitle{Signal-to-noise ratio of an operator direction} The following lemma justifies the normalization of $\tilde{\bs M}_{\bs\mu}$ and supplies the interpretation of the shot ratio $\alpha$ used in~\cref{sec:classical_training,sec:quantum_ic}.
\begin{lemma}[Signal-to-noise ratio of a frame direction]
\label{lem:snr_direction}
Let $A$ be a traceless Hermitian operator with $\Trace(A^2)=1$ and $\frameOp_{\bs\mu}(A)=\lambda A$ for some $\lambda>0$, put $x=\nout\lambda$, and let $\bs w_A=\bs M_{\bs\mu}^T\bs v_A/\sqrt\lambda$, a unit vector of outcome weights. Then $\bs w_A^T\bs p(\rho)=\sqrt\lambda\,\Trace(A\rho)$ for every state $\rho$. For a Haar-random pure state $\rho$,
\begin{equation}
\Var_\rho[\bs w_A^T\bs p(\rho)]=\frac{\lambda}{d(d+1)},\qquad
\bs w_A^T\barSigma_{\rm te}\bs w_A=\bs w_A^T\bs D_{I/d}\bs w_A-\frac{\lambda}{d(d+1)}.
\label{eq:snr_direction_moments}
\end{equation}
If $\max_a\lvert\nout\omega_a/d-1\rvert\le\varepsilon$, the ratio $R_A$ of the across-state variance to the Haar-averaged variance of an $N$-shot estimate satisfies
\begin{equation}
R_A\equiv\frac{N\Var_\rho[\bs w_A^T\bs p(\rho)]}{\bs w_A^T\barSigma_{\rm te}\bs w_A}
=\frac{\alpha x}{1+\delta-x/[d(d+1)]},\qquad\lvert\delta\rvert\le\varepsilon,
\label{eq:snr_direction}
\end{equation}
with $\alpha=N/[d(d+1)]$. Moreover, $\alpha x$ is the squared singular value of $\tilde{\bs M}_{\bs\mu}$ along $\bs v_A$.
\end{lemma}
\begin{proof}
The vector $\bs w_A$ has unit norm because $\|\bs M_{\bs\mu}^T\bs v_A\|^2=\bs v_A^T\frameOp_{\bs\mu}\bs v_A=\lambda$. Since $\bs M_{\bs\mu}\bs w_A=\frameOp_{\bs\mu}\bs v_A/\sqrt\lambda=\sqrt\lambda\,\bs v_A$, the weights reconstruct $\sqrt\lambda A$, so $\bs w_A^T\bs p(\rho)=\sqrt\lambda\Trace(A\rho)$. As $A$ is traceless,~\cref{lem:haar_error_split} gives $\EE_\rho[\Trace(A\rho)]=0$ and $\EE_\rho[\Trace(A\rho)^2]=1/[d(d+1)]$, hence the first identity. For the second, insert~\cref{eq:barSigma_exact}: $\bs\omega^T\bs w_A=\bs v_{I_d}^T\bs M_{\bs\mu}\bs w_A=\sqrt\lambda\Trace(A)=0$ and $\bs w_A^T\mathfrak G_{\bs\mu}\bs w_A=\|\bs M_{\bs\mu}\bs w_A\|^2=\lambda$. The trace condition gives $\bs w_A^T\bs D_{I/d}\bs w_A=(1+\delta)/\nout$ with $\lvert\delta\rvert\le\varepsilon$, and dividing $N\lambda/[d(d+1)]$ by $(1+\delta)/\nout-\lambda/[d(d+1)]$ gives~\cref{eq:snr_direction}. Finally, $\tilde{\bs M}_{\bs\mu}\tilde{\bs M}_{\bs\mu}^T\bs v_A=[\nout N/(d(d+1))]\bs P_0\frameOp_{\bs\mu}\bs v_A=\alpha x\,\bs v_A$ by~\cref{eq:alt_scaled_measurement}.
\end{proof}

\subsubsection{Proportional limit and empirical bounds}
\label{nav:def_limit}

The main-text interpolation laws keep the numbers of operator directions, outcomes, training states, and shots comparable as dimension grows. The ratios below locate the measurement and interpolation thresholds separately; the empirical bounds exclude collapse of the training signal and anomalously large or vanishing noise. All fixed-ridge comparisons take the dimensional limit at positive scaled penalty before any zero-penalty limit.

\parTitle{Dimension and shot ratios} Take
\begin{equation}
 \kappa_d=\frac{d^2-1}{\nout-1}\to\kappa\in(0,\infty),\qquad
 \gamma_d=\frac{\nout-1}{\ntr}\to\gamma\in(0,\infty),\qquad
 \alpha_d=\frac N{d(d+1)}\to\alpha\in(0,\infty).
 \label{eq:alt_dimensions}
\end{equation}
The ratio $\kappa$ places the outcome count above ($\kappa<1$) or below ($\kappa>1$) the informational-completeness threshold; outcome count alone does not assert full operator rank. The separate peak asymptotics below take $\kappa\ne1$; the fixed-positive-ridge comparison itself applies at every fixed $\kappa>0$. The ratio $\gamma$ locates interpolation, and $\alpha$ compares shot precision with Haar expectation-value fluctuations (\cref{lem:snr_direction}). The physical ratio $\nout/\ntr$ has the same limit as $\gamma_d$. At finite dimension the centered matrix has $\nout-1$ outcome and $\ntr-1$ training coordinates, so it is exactly square at $\nout=\ntr$.

\parTitle{Bounds on geometry and noise} Assume uniform empirical bounds
\begin{align}
 &0\preceq\frameOp_{\bs\rho}^{\rm c}\preceq K\bs P_0,
 \qquad
 \nout\bs P_0\frameOp_{\bs\mu}\bs P_0\preceq K\bs P_0\label{eq:alt_geometry},\\
 &c\bs\Pi_{\nout}\preceq \nout\barSigma_{\rm tr}\preceq K\bs\Pi_{\nout},
 \qquad
 c\bs\Pi_{\nout}\preceq \nout\barSigma_{\rm te}\preceq K\bs\Pi_{\nout},\label{eq:general_noise_bounds}
\end{align}
with dimension-independent $c>0$ and $K<\infty$. The frame bounds exclude unbounded geometric eigenvalues, while the covariance bounds ensure controlled, nonzero shot noise in every nonconstant outcome direction. A nonzero variance amplitude additionally requires
\begin{equation}
 \liminf\frac1{d^2-1}\Trace(\frameOp_{\bs\rho}^{\rm c})>0.
 \label{eq:general_noncollapse}
\end{equation}
For random states or apparatuses, these empirical conditions are imposed on events of probability tending to one.

\subsection{Centered auxiliary fit and fixed-ridge costs}
\label{nav:def_centered}

Centering makes the state-dependent fit accessible to spectral analysis, but changes how the training mean is fitted. \Cref{nav:def_centering} derives that change exactly, and~\cref{app:fixed_ridge_reduction_statement} gives the centered variance and traceless reconstruction cost at fixed positive ridge. Their physical interpretation requires the mean restoration in~\cref{app:proof_physical_mean_profile}; neither the centered calculation nor its Gaussian comparison alone controls the physical minimum-norm inverse.

\subsubsection{Uniform weights and the mean-fitting penalty}
\label{nav:def_centering}

Probability normalization allows a uniform outcome weight to predict a constant. Eliminating that weight reveals a centered auxiliary regression plus a single mean-fitting correction. The correction disappears from normalized traces with bounded weights at fixed positive ridge, but can survive in the constant prediction and is therefore retained exactly below.

\parTitle{Eliminating the uniform weight} The physical readout minimizes $\mathcal L_{N,\tau}(\bs w)=\ntr^{-1}\|\hatbs P_N^T\bs w-\bs y\|^2+\tau\|\bs w\|^2$. Since $\hatbs P_N^T\mathbf1_{\nout}=\mathbf1_{\ntr}$, a uniform outcome weight predicts a constant. Write $\bs w=w_0\mathbf1_{\nout}+\bs w_\perp$, with $\mathbf1_{\nout}^T\bs w_\perp=0$, and set $\bar y=\ntr^{-1}\mathbf1_{\ntr}^T\bs y$ and $\overline{\hat{\bsp}}=\ntr^{-1}\hatbs P_N\mathbf1_{\ntr}$. Orthogonal mean and fluctuation coordinates give
\begin{equation}
\mathcal L_{N,\tau}(w_0\mathbf1_{\nout}+\bs{w}_\perp)
=\frac1{\ntr}\|\bs\Pi_{\ntr}(\hatbs P_N^T\bs{w}_\perp-\bs y)\|^2
+\tau\|\bs{w}_\perp\|^2+(w_0+\overline{\hat{\bsp}}{}^T\bs{w}_\perp-\bar y)^2
+\nout\tau w_0^2.
\label{eq:general_intercept_main}
\end{equation}
Minimizing over $w_0$ leaves
\begin{equation}
 \min_{w_0}\mathcal L_{N,\tau}
 =\frac1{\ntr}\|\bs\Pi_{\ntr}(\hatbs P_N^T\bs w_\perp-\bs y)\|^2
 +\tau\|\bs w_\perp\|^2
 +\frac{\nout\tau}{1+\nout\tau}
 (\overline{\hat{\bsp}}{}^T\bs w_\perp-\bar y)^2.
\end{equation}
The centered labels $(\bs\Pi_{\ntr}\bs y)_i=\Trace(\calO_0(\rho_i-\bar\rho))$ depend only on the traceless target. Dropping the final mean-fitting penalty defines the auxiliary centered fit
\begin{equation}
\hatbs w^{\rm c}_{N,\tau}(\calO_0)
\equiv \operatorname*{arg\,min}_{\bs{w}\in\mathbf1_{\nout}^\perp}
\left\{
\frac1{\ntr}\|\bs\Pi_{\ntr}(\hatbs P_N^T\bs{w}-\bs y)\|^2
+\tau\|\bs{w}\|^2
\right\}.
\label{eq:general_centered_fit}
\end{equation}
At $\tau=0$ choose the minimum-Euclidean-norm solution in $\mathbf1_{\nout}^\perp$; at $\tau>0$ its normal equation is
\begin{equation}
\label{eq:reduced_normal}
(\hatbs P_N\bs\Pi_{\ntr}\hatbs P_N^T/\ntr+\tau\bs\Pi_{\nout})\hatbs w^{\rm c}_{N,\tau}
=\frac1{\ntr}\hatbs P_N\bs\Pi_{\ntr}\bs y.
\end{equation}
The physical nonconstant coefficients instead satisfy
\begin{equation}
 \left(\hatbs P_N\bs\Pi_{\ntr}\hatbs P_N^T/\ntr+\tau\bs\Pi_{\nout}+\beta_\eta\bs q\bs q^T\right)\bs w_\perp
 =\frac1{\ntr}\hatbs P_N\bs\Pi_{\ntr}\bs y+\beta_\eta\bar y\,\bs q,
 \label{eq:general_physical_rank_one_normal}
\end{equation}
where $\bs q=\bs\Pi_{\nout}\overline{\hat{\bsp}}$ and $\beta_\eta=\eta/(N+\eta)=\nout\tau/(1+\nout\tau)$. The extra matrix and source fit the empirical mean. At fixed positive $\eta$, the resulting rank-one map correction vanishes from normalized traces with bounded weights. Its constant prediction in~\cref{eq:general_physical_split} must still be restored exactly.

\parTitle{Comparable signal and noise scales} Introduce the rescaled matrices
\begin{equation}
\bs M^{\rm c}_{\bs\rho}=\sqrt{d(d+1)}\,\bs P_0\bs M_{\bs\rho}\bs\Pi_{\ntr},
\qquad
\hatbs P_N^{\rm c}=\sqrt{\nout N/\ntr}\,\hatbs P_N\bs\Pi_{\ntr},
\qquad
\frac1{\ntr}\bs M^{\rm c}_{\bs\rho}(\bs M^{\rm c}_{\bs\rho})^T
=\frameOp_{\bs\rho}^{\rm c}.
\label{eq:general_proof_matrices}
\end{equation}
For $\eta>0$, define on $\mathbf1_{\nout}^\perp$
\begin{equation}
\hat{\bs Q}
=[\hatbs P_N^{\rm c}(\hatbs P_N^{\rm c})^T
+\eta\bs\Pi_{\nout}]^{-1},\qquad
\hat{\bs K}=\hat{\bs Q}\hatbs P_N^{\rm c}
(\bs M^{\rm c}_{\bs\rho})^T/\sqrt{\ntr}.
\label{eq:general_centered_map}
\end{equation}
The normal equation~\cref{eq:reduced_normal} gives
\begin{equation}
\hatbs w^{\rm c}_{N,\tau}(\calO_0)
=(\hatbs P_N\bs\Pi_{\ntr}\hatbs P_N^T/\ntr+\tau\bs\Pi_{\nout})^{-1}
\hatbs P_N\bs\Pi_{\ntr}\bs y/\ntr=\sqrt{\frac{\nout N}{d(d+1)}}\,\hat{\bs K}\bs v_{\calO_0}.
\label{eq:general_centered_weights}
\end{equation}
The factors $\sqrt{d(d+1)}$ and $\sqrt{\nout N/\ntr}$ place state fluctuations and frequency noise on comparable scales. The map $\hat{\bs K}$ depends on the training record; a new target enters only through $\bs v_{\calO_0}$.

\subsubsection{Centered costs and their fixed-ridge equivalent}
\label{app:proof_centered_fixed_ridge}
\label{app:fixed_ridge_reduction_statement}

The centered fit has two distinct costs: amplification of independent test-shot noise and failure to reconstruct the traceless target. \Cref{prop:general_centered_fixed_ridge} evaluates both at fixed positive scaled ridge for every fixed $\kappa>0$. The proof replaces count fluctuations only inside normalized regularized traces, so the result does not yet justify substituting a pseudoinverse at zero penalty.

\parTitle{Centered costs} Define $V^{\rm c}_{N,\tau}(\calO)=(\hatbs w^{\rm c}_{N,\tau})^T\barSigma_{\rm te}\hatbs w^{\rm c}_{N,\tau}$ and the traceless reconstruction cost ${d(d+1)}\calE^{\rm c}_{N,\tau}(\calO)=\|\bs P_0\bs M_{\bs\mu}\hatbs w^{\rm c}_{N,\tau}-\bs v_{\calO_0}\|^2$. Their conditional target averages $V^{\rm c}_{N,\tau}$ and $\calE^{\rm c}_{N,\tau}$ give
\begin{align}
 &\widehat F_{\eta,d}
\equiv \frac{V^{\rm c}_{N,\tau}}{d\Delta_{\calO}^2}
=\frac{\alpha_d}{d^2-1}
\Trace(\hat{\bs K}^T\nout\barSigma_{\rm te}\hat{\bs K})\label{eq:general_centered_costs},\\
 &\widehat G_{\eta,d}
\equiv \frac{{d(d+1)}\calE^{\rm c}_{N,\tau}}{d\Delta_{\calO}^2}
=\frac1{d^2-1}\|\tilde{\bs M}_{\bs\mu}\hat{\bs K}-\bs P_0\|_F^2.\label{eq:general_centered_ideal_cost}
\end{align}
The first cost measures amplification of test-shot noise; the second measures traceless reconstruction. At strong ridge the centered prediction vanishes, giving $F\to0$ and $G\to1$.

\parTitle{Count comparison at fixed ridge} For every fixed $\kappa>0$, the centered measured matrix~\cref{eq:general_proof_matrices} has conditional mean
$\EE[\hatbs P_N^{\rm c}\mid\bs\rho,\bs\mu] =\frac1{\sqrt{\ntr}}\tilde{\bs M}_{\bs\mu}^T\bs M_{\bs\rho}^{\rm c}.$
The fluctuations are categorical and independent before centering, with average outcome covariance $\nout\barSigma_{\rm tr}/\ntr$. At fixed $\eta>0$,~\cref{lem:general_count_comparison} permits a Gaussian replacement inside the normalized traces of regularized inverse covariance matrices, called resolvents:
\begin{equation}
 \hatbs P_{N,{\rm G}}^{\rm c}\equiv
 \frac1{\sqrt{\ntr}}\left[
 \tilde{\bs M}_{\bs\mu}^T\bs M_{\bs\rho}^{\rm c}
 +(\nout\barSigma_{\rm tr})^{1/2}\bs G\right],
 \qquad G_{ai}\overset{\rm iid}{\sim}\mathcal N(0,1).
 \label{eq:general_ridge_gaussian_model}
\end{equation}
The replacement preserves these averaged inverse-covariance quantities at fixed positive penalty; it does not control the smallest measured singular values or the zero-penalty pseudoinverse. The matrix $\bs{\mathcal R}$ below is a deterministic equivalent: with the geometry fixed, its normalized weighted traces approximate those of the random inverse, rather than every matrix entry. In its inverse, $a^{-1}$ scales the signal covariance $\bs S$, and $b$ scales the shot covariance $\nout\barSigma_{\rm tr}$. The scalar $b$ describes the mean eigenvalue of the training residual filter. All outcome matrices here are restricted to $\mathbf1_{\nout}^{\perp}$; $I_{\nout-1}$ is the identity in orthonormal coordinates on that subspace. For $z$ near zero and a bounded positive weight matrix $\bs W$, select the positive solution of

\begin{align}
 &\bs{\mathcal R}=[\eta I_{\nout-1}+z\bs W+b\,\nout\barSigma_{\rm tr}
                                   +a^{-1}\bs S]^{-1},\qquad a=1+\frac1{\ntr}\Trace(\nout\barSigma_{\rm tr}\bs{\mathcal R})\label{eq:general_profile_Q},\\
 &b=\frac1a-\frac1{\ntr a^2}\Trace(\bs S\bs{\mathcal R}).\label{eq:general_profile_scalars}
\end{align}
Define
\begin{equation}
 \mathcal L_d(\eta,z;\bs W)\equiv
 \frac1{a(d^2-1)}\Trace\left(\frameOp_{\bs\rho}^{\rm c}
 -a^{-1}\frameOp_{\bs\rho}^{\rm c}
 \tilde{\bs M}_{\bs\mu}\bs{\mathcal R}
 \tilde{\bs M}_{\bs\mu}^T\frameOp_{\bs\rho}^{\rm c}\right).
 \label{eq:general_profile_loss}
\end{equation}

\parTitle{Extracting quadratic costs} Fix $\eta>0$ and a bounded positive semidefinite outcome metric $\bs W$. The derivative of a minimized loss with respect to a quadratic penalty measures the fitted coefficient norm weighted by that penalty. Temporarily varying the matrix penalty by $z\bs W$ therefore extracts test variance or the quadratic term of the traceless reconstruction cost, according to the choice of $\bs W$. Set
\begin{equation}
\begin{aligned}
 \hat{\bs Q}_z&\equiv
 [\hatbs P_N^{\rm c}(\hatbs P_N^{\rm c})^T
                  +\eta I_{\nout-1}+z\bs W]^{-1},\\
 \hat{\bs K}_z&\equiv
 \hat{\bs Q}_z\hatbs P_N^{\rm c}
             (\bs M_{\bs\rho}^{\rm c})^T/\sqrt{\ntr},\\
 \widehat{\mathcal L}(\eta,z;\bs W)&\equiv
 \frac1{\ntr (d^2-1)}\Trace\!\left(
 \bs M_{\bs\rho}^{\rm c}
 (I_{\ntr}-(\hatbs P_N^{\rm c})^T\hat{\bs Q}_z
                    \hatbs P_N^{\rm c})
 (\bs M_{\bs\rho}^{\rm c})^T\right).
\end{aligned}
\label{eq:general_sample_loss}
\end{equation}
All outcome-space inverses here and below act on $\mathbf1_{\nout}^\perp$. The quantity $\widehat{\mathcal L}$ is the minimum of the centered fitting loss including the temporary penalty $z\bs W$, averaged over unit traceless directions. At the minimizing weights, derivatives through the optimizer cancel by stationarity; differentiating the minimum therefore returns the explicit penalty term (the envelope identity). Using $\partial_z\hat{\bs Q}_z=-\hat{\bs Q}_z\bs W\hat{\bs Q}_z$ gives
\begin{equation}
 \partial_z\widehat{\mathcal L}(\eta,z;\bs W)
 =\frac1{d^2-1}\Trace(\hat{\bs K}_z^T\bs W\hat{\bs K}_z).
 \label{eq:general_sample_envelope}
\end{equation}
For $z=0$, $\hat{\bs K}_z=\hat{\bs K}$. Thus the $z$-derivative with $\bs W=\nout\barSigma_{\rm te}$ gives the centered test variance; with $\bs W=\tilde{\bs M}_{\bs\mu}^T\tilde{\bs M}_{\bs\mu}$ it gives the quadratic term in the traceless reconstruction error. 
Covariance bounds on $\mathbf1_{\nout}^\perp$ force individual mean probabilities to have order $\nout^{-1}$, as needed for the shotwise replacement.

\begin{lemma}[Balanced mean outcome probabilities]
\label{lem:general_probability_bounds}
Under~\cref{eq:alt_dimensions,eq:alt_geometry,eq:general_noise_bounds}, the numbers $\nout p_{I/d,a}$ and $\nout\bar p_a$ are bounded above and below by fixed positive constants. In particular $p_{ai}\le\Trace(\mu_a)\le Kd/\nout$ for every state and outcome.
\end{lemma}

\begin{proof}
\parTitle{Balanced outcome probabilities} Equation~\cref{eq:general_covariance_identity} shows that $\nout\bs\Sigma(\bar{\bsp})$ differs from $\nout\barSigma_{\rm tr}$ by a positive matrix of norm $O(N^{-1})$. The Haar test identity gives
\begin{equation}
 \nout\barSigma_{\rm te}=\nout\bs\Sigma(\bsp_{I/d})
       -\tilde{\bs M}_{\bs\mu}^T\tilde{\bs M}_{\bs\mu}/N.
 \label{eq:general_test_categorical}
\end{equation}
For either mean vector $\bs p$, the diagonal bounds imply $c/2\le \nout p_a(1-p_a)\le K$ for large dimension. Thus $p_a\ge c/(2\nout)$; summing all other entries bounds $1-p_a$ below by a constant. The upper bound then gives $\nout p_a\le K'$. Finally $\rho_i\preceq I_d$ gives $p_{ai}\le\Trace(\mu_a)=dp_{I/d,a}$.
\end{proof}

\begin{lemma}[Multinomial comparison with the empirical categorical matrix]
\label{lem:general_count_comparison}
Under the same hypotheses, normalized bounded-weight fixed positive penalty traces of the count matrix $\hatbs P_N^{\rm c}$ in~\cref{eq:general_sample_loss,eq:general_trace_transfer} agree in probability with those of the Gaussian design~\cref{eq:general_ridge_gaussian_model}, whose entries are standard real Gaussian in orthonormal outcome-range coordinates. The convergence is local on positive-penalty intervals and transfers first penalty derivatives.
\end{lemma}

\begin{proof}
\parTitle{Shotwise count replacement} Condition on the complete list of training states. Start with $\hatbs P_{N,*}^{\rm c}=\sqrt{\nout N/\ntr}(\hatbs P_N-\bar{\bsp}\mathbf1_{\ntr}^T)$. It has independent centered-noise columns and mean $\tilde{\bs M}_{\bs\mu}^T\bs M_{\bs\rho}^{\rm c}/\sqrt{\ntr}$; centering its training columns changes it by rank one. A single block inverse contains the regularized feature inverse, the fitted prediction map, and the training residual filter. Controlling its traces therefore controls all three quantities needed for the costs. For $\bs\Lambda\succeq\eta I/2$ use
\begin{equation}
\begin{aligned}
 \mathsf R(\hatbs P_{N,*}^{\rm c})
 &=\begin{pmatrix}
     \bs\Lambda&\hatbs P_{N,*}^{\rm c}\\
     (\hatbs P_{N,*}^{\rm c})^T&-I_{\ntr}
   \end{pmatrix}^{-1}=\begin{pmatrix}
     \hat{\bs Q}&\hat{\bs Q}\hatbs P_{N,*}^{\rm c}\\
     (\hatbs P_{N,*}^{\rm c})^T\hat{\bs Q}
       &-I_{\ntr}+(\hatbs P_{N,*}^{\rm c})^T\hat{\bs Q}\hatbs P_{N,*}^{\rm c}
   \end{pmatrix},\\
 \hat{\bs Q}&\equiv
   (\hatbs P_{N,*}^{\rm c}(\hatbs P_{N,*}^{\rm c})^T+\bs\Lambda)^{-1}.
\end{aligned}
 \label{eq:general_linearization}
\end{equation}
Intermediate diagonal-noise matrices act on the full outcome space. There extend the range penalty by $\eta(I_{\nout}-\bs\Pi_{\nout})$. It is uniformly positive, and at the two categorical endpoints the data lie in $\im\bs\Pi_{\nout}$, so their restricted traces agree with the original ones. Removing or projecting the extra uniform coordinate is a bounded-rank change. The ridge filter bounds all four blocks by $K_\eta$ for every possible feature matrix. For a deterministic bounded weight $\bs W_0$, the third directional derivative of $(d^2-1)^{-1}\Trace(\bs W_0\mathsf R)$ in a rank-one feature perturbation is bounded by $K_\eta\|\bs W_0\|_{\rm op}\|\Delta\hatbs P_{N,*}^{\rm c}\|_{\rm op}^3/(d^2-1)$. Each shot replacement has matching mean and covariance, so the first two Taylor terms cancel. Replacing all $\ntr N$ independent categorical shots by these Gaussian increments gives total error
$K_\eta\|\bs W_0\|_{\rm op}\frac{\nout^{3/2}}{(d^2-1)\sqrt{N\ntr}}=o(1).$
The Gaussian third moments obey the same estimate because each increment has covariance trace at most $\nout/(N\ntr)$. Bounding the effect of resampling one independent shot and summing its squared change (the Efron--Stein variance inequality) gives variance at most $K_\eta\|\bs W_0\|_{\rm op}^2\nout/(d^2-1)^2=o(1)$.

\parTitle{State-dependent Gaussian covariances} The matched Gaussian column with covariance $\nout\bs\Sigma(\rho_i)$ is
\begin{equation}
 \sqrt{\nout}\,\diag(\sqrt{\bsp_i})\bs g_i
    -\sqrt{\nout}\,\bsp_i(\sqrt{\bsp_i}^{\,T}\bs g_i),
 \qquad \bs g_i\sim\mathcal N(0,I_{\nout}).
 \label{eq:general_gaussian_bridge}
\end{equation}
After division by $\sqrt{\ntr}$, the second term has a rank-one mean part plus a remainder $\bs E$ with
$\EE\|\bs E\|_F^2 =\frac \nout \ntr\|\bs P-\bar{\bsp}\mathbf1_{\ntr}^T\|_F^2 \le K\Trace(\frameOp_{\bs\rho}^{\rm c})/[d(d+1)]=O(1).$
Bounded rank changes normalized weighted inverse traces by $O_\eta((d^2-1)^{-1})$; the remainder changes them by $O_\eta(\|\bs E\|_F/\sqrt{d^2-1})=o_{\PP}(1)$.

\parTitle{A common Gaussian covariance} For the remaining independent Gaussian entries, interpolate their variances $\nout p_{ai}$ to the row variances $\nout\bar p_a$. Write $\Delta=\nout(\bs P-\bar{\bsp}\mathbf1_{\ntr}^T)$, where $\bs P$ is the exact probability matrix. Gaussian integration by parts bounds the weighted trace error by
\begin{equation}
 K_\eta\|\bs W_0\|_{\rm op}\left(\|\Delta\|_{\rm op}/\ntr+\|\Delta\|_{\max}/\ntr\right)
 =O_\eta({[d(d+1)]}^{-1/2}+d/\ntr)=o(1).
 \label{eq:general_variance_error}
\end{equation}
Indeed $\|\Delta\|_{\rm op}\le K\sqrt{\nout \ntr/[d(d+1)]}$ by the two frame bounds, and the maximum entry is at most $Kd$ by~\cref{lem:general_probability_bounds}. For clarity, the second derivative of the block inverse trace is a sum of two products of diagonal entries and one product of off-diagonal entries. The first two sums use $\|\Delta\|_{\rm op}$ and Euclidean diagonal-vector norms. The last uses $\|\Delta\|_{\max}$ and the two block Frobenius norms. This gives the two terms in~\cref{eq:general_variance_error}.

\parTitle{Categorical covariance and derivatives} The resulting common diagonal covariance can be changed, by a rank-one Gaussian bridge, to $\nout\bs\Sigma(\bar{\bsp})$. Its difference from $\nout\barSigma_{\rm tr}$ is $N^{-1}\tilde{\bs M}_{\bs\mu}^T\frameOp_{\bs\rho}^{\rm c} \tilde{\bs M}_{\bs\mu}$, of norm $O(N^{-1})$. The square-root inequality for positive matrices bounds the difference of Gaussian noise matrices in operator norm by $O_{\PP}(N^{-1/2})$. Thus the endpoint is~\cref{eq:general_ridge_gaussian_model}. Outcome and training projections change bounded rank. Concavity and local positive-penalty difference quotients give the first derivatives, as used in the main proof.
\end{proof}

\begin{lemma}[Evaluation of the common-covariance Gaussian traces]
\label{lem:general_gaussian}
Under the empirical bounds, the Gaussian design~\cref{eq:general_ridge_gaussian_model} has the unique positive resolvent solution~\cref{eq:general_profile_Q,eq:general_profile_scalars}. Its weighted residual and mixed prediction traces satisfy
\begin{equation}
 \widehat{\mathcal L}-\mathcal L_d\xrightarrow{\PP}0,
 \qquad
 \frac1{d^2-1}\Trace(\tilde{\bs M}_{\bs\mu}\hat{\bs K})
 -\frac1{a(d^2-1)}\Trace(\frameOp_{\bs\rho}^{\rm c}
 \tilde{\bs M}_{\bs\mu}\bs{\mathcal R}
 \tilde{\bs M}_{\bs\mu}^T)\xrightarrow{\PP}0.
 \label{eq:general_trace_transfer}
\end{equation}
where $\widehat{\mathcal L},\hat{\bs K}$ are evaluated with the Gaussian design. The equivalents are smooth locally at positive penalty, and the comparison transfers first penalty derivatives.
\end{lemma}

\begin{proof}
\parTitle{Gaussian resolvent equations} The deterministic mean and its covariance are given by~\cref{eq:general_ridge_gaussian_model,eq:general_ridge_signal_noise}. For a positive penalty $\bs\Lambda=\eta\bs\Pi_{\nout}+z\bs W$, apply the weighted Gaussian inverse-trace theorem~\cite{hachem2013bilinear} after multiplication by $\bs\Lambda^{-1/2}$. It supplies deterministic approximations to the normalized weighted inverse and residual traces, with the unique positive solution displayed below. Diagonalize $\bs\Lambda^{-1/2}(\nout\barSigma_{\rm tr})\bs\Lambda^{-1/2}$ by a real orthogonal matrix. The resulting row covariance is bounded above and below, the column covariance is the identity, and the deterministic mean has bounded norm. Real Gaussian independence is preserved. The positive deterministic matrices are
\begin{equation}
\begin{aligned}
 \bs{\mathcal R}&=[\bs\Lambda+b\,\nout\barSigma_{\rm tr}
                        +a^{-1}\bs S]^{-1},\\ \bs U&=\biggl[aI_{\ntr}+\frac1{\ntr}(\bs M_{\bs\rho}^{\rm c})^T
 \tilde{\bs M}_{\bs\mu}
 (\bs\Lambda+b\,\nout\barSigma_{\rm tr})^{-1}\tilde{\bs M}_{\bs\mu}^T
                  \bs M_{\bs\rho}^{\rm c}\biggr]^{-1},\\
 a&=1+\frac1{\ntr}\Trace(\nout\barSigma_{\rm tr}\bs{\mathcal R}),
 \qquad b=\frac1{\ntr}\Trace(\bs U).
\end{aligned}
 \label{eq:general_gaussian_equations}
\end{equation}
Woodbury's identity gives
\begin{equation}
 \bs U=\frac1a I_{\ntr}-\frac1{\ntr a^2}
 (\bs M_{\bs\rho}^{\rm c})^T\tilde{\bs M}_{\bs\mu}
 \bs{\mathcal R}\tilde{\bs M}_{\bs\mu}^T
 \bs M_{\bs\rho}^{\rm c}.
\end{equation}
Tracing it gives~\cref{eq:general_profile_scalars}; weighting it by $\bs M_{\bs\rho}^{\rm c}/\sqrt{\ntr}$ gives~\cref{eq:general_profile_loss}. The same pair also gives the exact deterministic trace identity
\begin{equation}
 b=1-\gamma_d+\Trace(\bs\Lambda\bs{\mathcal R})/\ntr.
 \label{eq:general_schur_trace}
\end{equation}
This follows by tracing $I=\bs\Lambda\bs{\mathcal R}+b(\nout\barSigma_{\rm tr})\bs{\mathcal R}+\bs S\bs{\mathcal R}/a$ and using $a-1=\Trace((\nout\barSigma_{\rm tr})\bs{\mathcal R})/\ntr$. Existence, positivity, uniqueness of the resolvent branch and its local parameter convergence are supplied by the positive Gaussian resolvent equations; all input covariance and penalty bounds are uniform here.

\parTitle{Mixed prediction trace} For the mixed prediction trace, use the Gaussian design~\cref{eq:general_ridge_gaussian_model} and its resolvent $\hat{\bs Q}=(\hatbs P_{N,{\rm G}}^{\rm c} (\hatbs P_{N,{\rm G}}^{\rm c})^T+\bs\Lambda)^{-1}$. Gaussian integration by parts gives, with $\delta=\Trace((\nout\barSigma_{\rm tr})\hat{\bs Q})/\ntr$,
\begin{equation}
 \EE[\hat{\bs Q}(\nout\barSigma_{\rm tr})^{1/2}\bs G/\sqrt{\ntr}]
 =-\EE[\delta\hat{\bs Q}\hatbs P_{N,{\rm G}}^{\rm c}]
          -\ntr^{-1}\EE[\hat{\bs Q}(\nout\barSigma_{\rm tr})\hat{\bs Q}\hatbs P_{N,{\rm G}}^{\rm c}].
 \label{eq:general_gaussian_cross}
\end{equation}
Poincar\'e and the filter give $\Var\delta=O_\eta(\ntr^{-2})$ and bounded norms of the two filtered matrices. Hence, in operator norm,
\begin{equation}
 \EE(\hat{\bs Q}\hatbs P_{N,{\rm G}}^{\rm c})
 =\frac{(\EE\hat{\bs Q})\tilde{\bs M}_{\bs\mu}^T
              \bs M_{\bs\rho}^{\rm c}}{\sqrt{\ntr}(1+\EE\delta)}
   +O_\eta(\ntr^{-1}).
\end{equation}
The weighted inverse theorem replaces the normalized traces of $\hat{\bs Q}$ by those of $\bs{\mathcal R}$, while concentration makes the corresponding fluctuations vanish. Substituting the displayed mixed matrix mean gives the second limit in~\cref{eq:general_trace_transfer}.

\parTitle{Uniform derivative transfer} The derivative comparison can be made uniform in the dimension even when the empirical spectra have no limits. For $\bs\Lambda=\eta I_{\nout-1}+z\bs W\succeq\eta I_{\nout-1}/2$, differentiating the sample envelope gives
$\partial_z^2\widehat{\mathcal L}=-2(d^2-1)^{-1}\Trace(\hat{\bs K}_z^T\bs W\hat{\bs Q}_z\bs W\hat{\bs K}_z)$ and hence $\lvert\partial_z^2\widehat{\mathcal L}\rvert\le2\|\bs W\|_{\rm op}^2\Trace(\frameOp_{\bs\rho}^{\rm c})/((d^2-1)\eta^2)$ by the ridge filter. To obtain the same bound for $\mathcal L_d$, fix its finite deterministic geometry. Write its Gaussian mean as $\bs A=\tilde{\bs M}_{\bs\mu}^T\bs M_{\bs\rho}^{\rm c}/\sqrt{\ntr}$, its response map as $\bs R=\bs M_{\bs\rho}^{\rm c}/\sqrt{\ntr}$, and its noise covariance as $\bs C=\nout\barSigma_{\rm tr}$. Replicate this fixed geometry as an auxiliary mathematical construction: replace $\bs A,\bs R,\bs C,\bs\Lambda,\bs W$ by their tensor products with $I_k$ and use independent Gaussian noise scaled by $1/\sqrt{\ntr k}$. The enlarged dimensions are $(\nout-1)k,\ntr k,(d^2-1)k$. The scalar equations, normalized deterministic loss, and curvature bound are unchanged. At fixed original geometry, the same Gaussian weighted inverse-trace theorem makes the enlarged sample loss converge to $\mathcal L_d$ as $k\to\infty$. The sample loss is bounded by $\Trace(\frameOp_{\bs\rho}^{\rm c})/(d^2-1)$, so the convergence also holds in expectation. Its uniform second-difference bound passes to $\mathcal L_d$; smoothness on the positive resolvent branch then gives the same curvature bound. Centered difference quotients therefore approximate both sample and deterministic derivatives with an error $O_\eta(h)$ uniformly in $d$. Apply the value comparison at fixed step $h$ and then let $h\downarrow0$ to transfer first penalty derivatives.
\end{proof}

\begin{proposition}[Fixed-ridge centered costs]
\label{prop:general_centered_fixed_ridge}
Under~\cref{eq:alt_dimensions,eq:alt_geometry,eq:general_noise_bounds}, for any fixed $\kappa\in(0,\infty)$, condition on the states and POVM and fix $\eta>0$. Let $a,b,\bs{\mathcal R}$ and $\mathcal L_d$ be defined by~\cref{eq:general_profile_Q,eq:general_profile_scalars,eq:general_profile_loss}. The deterministic centered costs are
\begin{align}
 &F_{\eta,d}\equiv\left.\alpha_d\partial_z
 \mathcal L_d(\eta,z;\nout\barSigma_{\rm te})\right\vert_{z=0}\label{eq:general_profile_F},\\
 &G_{\eta,d}\equiv 1-\frac2{a(d^2-1)}
 \Trace(\frameOp_{\bs\rho}^{\rm c}\tilde{\bs M}_{\bs\mu}
 \bs{\mathcal R}\tilde{\bs M}_{\bs\mu}^T)
 +\left.\partial_z\mathcal L_d
 (\eta,z;\tilde{\bs M}_{\bs\mu}^T
             \tilde{\bs M}_{\bs\mu})\right\vert_{z=0}.\label{eq:general_complete_profiles}
\end{align}
with all $z$-dependence included in the derivatives. Then $\widehat F_{\eta,d}-F_{\eta,d}\xrightarrow{\PP}0$ and $\widehat G_{\eta,d}-G_{\eta,d}\xrightarrow{\PP}0$.
\end{proposition}
\begin{proof}
\parTitle{Variance and reconstruction derivatives} The envelope identity~\cref{eq:general_sample_envelope} with $\bs W=\nout\barSigma_{\rm te}$ gives $\widehat F_{\eta,d}/\alpha_d$. With $\bs W=\tilde{\bs M}_{\bs\mu}^T\tilde{\bs M}_{\bs\mu}$ it gives the quadratic term in
\begin{equation}
 \widehat G_{\eta,d}
 =1-\frac2{d^2-1}\Trace(\tilde{\bs M}_{\bs\mu}\hat{\bs K})
 +\frac1{d^2-1}\Trace(\hat{\bs K}^T\tilde{\bs M}_{\bs\mu}^T
              \tilde{\bs M}_{\bs\mu}\hat{\bs K}).
 \label{eq:general_fixed_ridge_cost_expansion}
\end{equation}
The count comparison~\cref{lem:general_count_comparison} transfers these normalized positive-penalty traces to the Gaussian design. Exact empirical centering changes the independent-column model by bounded rank and has no effect on these limits. The Gaussian trace identities~\cref{eq:general_trace_transfer} then replace the sample loss and mixed prediction trace by their deterministic counterparts.

\parTitle{Recovering the penalty responses} To transfer derivatives, the sample residual is concave in $z$, being a minimum of affine penalty functions. Its derivative lies between positive and negative difference quotients. The dimension-uniform curvature bound in~\cref{lem:general_gaussian} controls the secant error for both losses. Apply the value comparison at fixed step size, take the dimensional limit, and only then let the step decrease. The two envelope derivatives and the mixed-trace limit prove~\cref{eq:general_profile_F,eq:general_complete_profiles}, without assuming a limiting empirical spectrum.
\end{proof}

\parTitle{Limiting centered profiles} At fixed geometry, $F_{\eta,d}$ and $G_{\eta,d}$ are deterministic approximations to the record-dependent, target-averaged centered costs. Write their dimensional limits as $F_\eta(\gamma)$ and $G_\eta(\gamma)$ when those limits exist. The comparison itself needs no limiting empirical spectrum; the peak and complete-curve results later impose the trace limits they use.

\subsection{Restoring the physical readout}
\label{app:proof_physical_mean_profile}

The physical readout must fit the training mean with its own outcome weights. A one-dimensional Schur complement restores that constraint exactly, including at zero penalty on count matrices of any rank. \Cref{lem:general_centering} gives the finite-record identities;~\cref{thm:alt_ridge_profiles} combines them with the centered calculation to identify the physical costs used in~\cref{sec:quantum_ic,sec:ridge_risk}. The ideal-test error can retain one directional mean contribution even when normalized bulk traces have deterministic limits.

\parTitle{Mean displacement and residual label} Define
\begin{equation}
 \bar{\bs h}=\sqrt{d(d+1)}\bs v_{\bar\rho-I_d/d},\qquad
 \bs b_N=\sqrt{\nout N}\bs\Pi_{\nout}\overline{\hat{\bsp}},\qquad
 \bs b_0=\sqrt{\nout N}\bs\Pi_{\nout}\bsp_{I/d}.
 \label{eq:general_mean_vectors}
\end{equation}
The difference $\bs b_N-\bs b_0$ is the measured displacement of the training mean from the Haar test mean, and $\bs r_\eta=\bar{\bs h}-\hat{\bs K}^T\bs b_N$ is the mean label not supplied by the centered weights. With $\beta_\eta=\eta/(N+\eta)$ as above, set
\begin{equation}
 J_\eta=1+\beta_\eta\bs b_N^T\hat{\bs Q}\bs b_N,\qquad
 A_\eta=\frac{1-\beta_\eta+\beta_\eta\bs b_0^T\hat{\bs Q}\bs b_N}{J_\eta}.
 \label{eq:general_mean_scalars}
\end{equation}
The factor $J_\eta$ normalizes the rank-one correction, and $A_\eta$ is the Haar mean prediction for constant unit labels. Define the identity-label readout $\hatbs w^I_{N,\tau}=\hatbs w_{N,\tau}(I_d)$ and the mean-error vector $\bs t_\eta=\hat{\bs K}^T\bs b_0+A_\eta\bs r_\eta$. The latter singles out the target direction affected by the mean prediction.

\parTitle{Exact finite-record restoration} The formulas below recover the original weights, identity-label readout, and target-averaged costs from the centered fit. Their zero-penalty form includes the nullspace correction selected by the physical minimum-norm criterion; the subsequent bounds identify which terms survive the dimensional limit.

\begin{lemma}[Exact restoration of the empirical mean]
\label{lem:general_centering}
For every count record and $\eta>0$, the quantities in~\cref{eq:general_mean_vectors,eq:general_mean_scalars} give the original readout as follows. Write
\begin{equation}
 \bs z_\eta=\frac{\beta_\eta\hat{\bs Q}\bs b_N}{J_\eta},\qquad
 \bs Z_\eta=\hat{\bs K}+\bs z_\eta\bs r_\eta^T.
 \label{eq:general_exact_mean_map}
\end{equation}
If $\bs{w}=w_0\mathbf1_{\nout}+\bs{w}_c$, with $\mathbf1_{\nout}^T\bs{w}_c=0$, then exactly
\begin{equation}
 \bs{w}_c=\sqrt{\frac{\nout N}{d(d+1)}}
       [\bs Z_\eta\bs v_{\calO_0}+\sqrt{d(d+1)} c_\calO\bs z_\eta],
 \qquad w_0=\frac{1-\beta_\eta}{\sqrt{d(d+1)} J_\eta}
       (\bs r_\eta^T\bs v_{\calO_0}+\sqrt{d(d+1)} c_\calO).
 \label{eq:general_exact_schur}
\end{equation}
The identity-label readout and mean prediction are exactly
\begin{equation}
 \hatbs w^I_{N,\tau}
 =\frac{1-\beta_\eta}{J_\eta}\mathbf1_{\nout}
 +\frac{\sqrt{\nout N}\beta_\eta}{J_\eta}\hat{\bs Q}\bs b_N.
 \label{eq:general_mean_readout}
\end{equation}
\begin{equation}
 \bsp_{I/d}^T\hatbs w^I_{N,\tau}=A_\eta,\qquad
 \sqrt{d(d+1)}(\bsp_{I/d}^T\hatbs w_{N,\tau}-c_\calO)
 =\bs t_\eta^T\bs v_{\calO_0}+\sqrt{d(d+1)}c_\calO(A_\eta-1).
 \label{eq:general_mean_exact}
\end{equation}
Conditional on the record and target invariants, all stated target laws have the exact cost averages
\begin{align}
 &\frac{V_{N,\tau}}{d\Delta_{\calO}^2}
 =\frac{\alpha_d}{d^2-1}\Trace(\bs Z_\eta^T\nout\barSigma_{\rm te}\bs Z_\eta)
       +\frac{Nc_\calO^2}{d\Delta_{\calO}^2}
                         \bs z_\eta^T\nout\barSigma_{\rm te}\bs z_\eta\label{eq:general_exact_average_V},\\
 &\frac{{d(d+1)}\calE_{N,\tau}}{d\Delta_{\calO}^2}
 =\frac1{d^2-1}\|\tilde{\bs M}_{\bs\mu}\bs Z_\eta-\bs P_0\|_F^2
 +\frac{{d(d+1)}c_\calO^2}{d\Delta_{\calO}^2}\|\tilde{\bs M}_{\bs\mu}\bs z_\eta\|^2
 +\frac{\|\bs t_\eta\|^2}{d^2-1}
 +\frac{{d(d+1)}c_\calO^2}{d\Delta_{\calO}^2}(A_\eta-1)^2.\label{eq:general_exact_average_G}
\end{align}
These identities also hold at zero penalty, on every rank, with $\hat{\bs K}_0=((\hatbs P_N^{\rm c})^T)^+(\bs M^{\rm c}_{\bs\rho})^T/\sqrt{\ntr}$ and $\bs\Pi_N=\bs\Pi_{\nout}-\hatbs P_N^{\rm c}(\hatbs P_N^{\rm c})^+$, after setting
\begin{equation}
 \bs z_0=\frac{\bs\Pi_N\bs b_N}{N+\|\bs\Pi_N\bs b_N\|^2},
 \qquad A_0=\frac{N+\bs b_0^T\bs\Pi_N\bs b_N}
                         {N+\|\bs\Pi_N\bs b_N\|^2},
 \qquad w_0=\frac{\bs r_0^T\bs v_{\calO_0}+\sqrt{d(d+1)} c_\calO}
                  {\sqrt{d(d+1)}[1+\|\bs\Pi_N\bs b_N\|^2/N]}.
 \label{eq:general_zero_mean_map}
\end{equation}
Here $\bs r_0,\bs Z_0,\bs t_0$ have the same definitions using $\hat{\bs K}_0,\bs z_0,A_0$. The physical identity readout at zero penalty is
\begin{equation}
 \hatbs w^I_{N,0}=\frac{N\mathbf1_{\nout}+\sqrt{\nout N}\bs\Pi_N\bs b_N}
                          {N+\|\bs\Pi_N\bs b_N\|^2}.
 \label{eq:general_zero_identity_readout}
\end{equation}
\end{lemma}

\begin{proof}
\parTitle{Positive-ridge Schur complement} Eliminate $w_0$ in the exact objective~\cref{eq:general_intercept_main}. The centered normal matrix acquires $\beta_\eta\bs b_N\bs b_N^T$, and its right side acquires $\beta_\eta\bs b_N(\bar{\bs h}^T\bs v_{\calO_0}+\sqrt{d(d+1)} c_\calO)$ in the scaled coordinates. Sherman--Morrison gives~\cref{eq:general_exact_mean_map,eq:general_exact_schur}. Since $\sqrt{d(d+1)}\bsp_{I/d}^T\bs{w}=\sqrt{d(d+1)} w_0+ \bs b_0^T[\bs Z_\eta\bs v_{\calO_0}+\sqrt{d(d+1)} c_\calO\bs z_\eta]$, substitution gives~\cref{eq:general_mean_exact}. The target second moment is $d\Delta_{\calO}^2\bs P_0/(d^2-1)$ and its first moment is zero. Applying these identities to the variance and~\cref{eq:general_physical_split} proves~\cref{eq:general_exact_average_V,eq:general_exact_average_G}.

\parTitle{Minimum-norm nullspace correction} At zero ridge every centered least-squares solution is the minimum-norm centered solution plus a vector in $\ker(\hatbs P_N^{\rm c})^T$. Write its scaled correction as $\bs z$. The part of the original coefficient norm that depends on $\bs z$ is proportional to
\begin{equation}
 N\|\bs z\|^2+
 (\bs r_0^T\bs v_{\calO_0}+\sqrt{d(d+1)} c_\calO-\bs b_N^T\bs z)^2,
 \qquad \bs z\in\im\bs\Pi_N.
 \label{eq:general_zero_norm_objective}
\end{equation}
Its minimum gives~\cref{eq:general_zero_mean_map} and the same exact physical costs. Taking constant labels gives~\cref{eq:general_zero_identity_readout}.
\end{proof}

\parTitle{Descriptors of the empirical mean} For the bounds used in the asymptotic arguments, introduce the descriptors
$q_d=\|\bar\rho-I_d/d\|_F^2,\qquad t_d^2=\nout\|\bsp_{I/d}-\mathbf1_{\nout}/\nout\|^2.$
They measure mean-state purity above $1/d$ and imbalance of effect traces. Both are bounded under the basic empirical assumptions; no vanishing condition is imposed on them.

\begin{lemma}[Fixed-ridge size of the mean correction]
\label{lem:general_mean_bounds}
Under the empirical bounds, at each fixed $\eta>0$, $\|\hat{\bs K}\|_{\rm op},\|\bs Z_\eta\|_{\rm op},\lvert A_\eta\rvert=O_{\PP,\eta}(1)$, $\|\bs z_\eta\|=O_{\PP,\eta}({[d(d+1)]}^{-1/2})$ and $\|\bs r_\eta\|=O_{\PP,\eta}(\sqrt{d(d+1)})$. The difference $\bs Z_\eta-\hat{\bs K}$ has rank at most one and bounded norm, while
\begin{equation}
 \|\bs t_\eta\|^2/(d^2-1)=O_{\PP,\eta}(q_d+(d^2-1)^{-1}).
 \label{eq:general_mean_projection_bound}
\end{equation}
\end{lemma}

\begin{proof}
\parTitle{Sizes of the mean vectors} The bounded mean probabilities and independence of the records give
\begin{equation}
 \|\bs b_N\|^2=O_{\PP}[N(q_d+t_d^2)+1],\qquad
 \|\bar{\bs h}\|^2={d(d+1)}q_d,
 \qquad
 \bs b_N-\bs b_0=\tilde{\bs M}_{\bs\mu}^T\bar{\bs h}
                                      +O_{\PP}(1).
 \label{eq:general_mean_sizes}
\end{equation}
The filter $\|\hat{\bs Q}\hatbs P_N^{\rm c}\|_{\rm op}\le1/(2\sqrt\eta)$ and $\|\bs M^{\rm c}_{\bs\rho}/\sqrt{\ntr}\|_{\rm op}\le\sqrt K$ bound $\hat{\bs K}$. Since $J_\eta\ge1$ and $\|\hat{\bs Q}\|_{\rm op}\le1/\eta$, the definitions give the sizes of $\bs z_\eta,\bs r_\eta,\bs Z_\eta$. They also bound $A_\eta$, using $\|\bs b_0\|^2=Nt_d^2$.

\parTitle{Directional mean bound} For the sharper bound on the mean query, use the original uniform normal equation to write its coefficient for a traceless target as
\begin{equation}
 \bs t_\eta=\bar{\bs h}
 -\left(\bs Z_\eta\right)^T(\bs b_N-\bs b_0)
 -\frac{\beta_\eta}{J_\eta}\bs r_\eta.
 \label{eq:general_mean_query_bound_identity}
\end{equation}
The last term is $O_{\PP,\eta}({[d(d+1)]}^{-1/2})$. Together with~\cref{eq:general_mean_sizes} and the bounded map $\bs Z_\eta$, this bounds $\|\bs t_\eta\|^2$ by $O_{\PP,\eta}({d(d+1)}q_d+1)$, proving the claim.
\end{proof}

\begin{lemma}[Diffuse mean for a pure-state training frame]
\label{lem:general_pure_mean}
For pure training states, the upper frame bound and~\cref{eq:general_noncollapse} imply $\|\bar\rho\|_{\rm op}=O(d^{-1})$ and $q_d=O(d^{-1})$.
\end{lemma}

\begin{proof}
\parTitle{Purity and a diffuse training mean} For an eigenvector of $\bar\rho$ with eigenvalue $\lambda$, purity gives the mean squared length of $(\rho_i-\bar\rho)$ applied to that vector as $\lambda-\lambda^2$. Sum its $d$ matrix-entry variances and use the upper centered-frame bound to obtain
$\lambda(1-\lambda)\le Kd/[d(d+1)].$
On the other hand~\cref{eq:general_training_frame} and the positive average frame trace bound $\Trace(\bar\rho^2)$ away from one. Thus $\lambda_{\max}(\bar\rho)$ stays away from one; the bound on $\lambda(1-\lambda)$ forces every eigenvalue to be $O(d^{-1})$. Finally $q_d=\Trace(\bar\rho^2)-1/d\le\|\bar\rho\|_{\rm op}$.
\end{proof}

\begin{lemma}[Identity-label readout]
\label{lem:identity_readout}
For constant unit labels, every multinomial record and every $\eta\ge0$ satisfy
\begin{equation}
 \|\hatbs w_{N,\eta/(\nout N)}(I_d)\|\le\sqrt{\nout},\qquad
 \|\bs\Pi_{\nout}\hatbs w_{N,\eta/(\nout N)}(I_d)\|
       \le\sqrt{\nout/N}\|\bs b_N\|.
 \label{eq:identity_coefficient_bound}
\end{equation}
Under the proportional-dimensional assumptions, uniformly for bounded $\eta\ge0$,
\begin{equation}
 \begin{aligned}
 V_{N,\eta/(\nout N)}(I_d)&=O_{\PP}(q_d+t_d^2+[d(d+1)]^{-1}),\\
 d(d+1)\calE_{N,\eta/(\nout N)}(I_d)
 &=O_{\PP}[d(d+1)q_d(q_d+t_d^2)+q_d+t_d^2+[d(d+1)]^{-1}].
 \end{aligned}
 \label{eq:identity_error_bound}
\end{equation}
The assertions include the actual minimum-norm solution on deficient records.
\end{lemma}

\begin{proof}
\parTitle{Identity-label coefficient bounds} Probability normalization gives $\hatbs P_N^T\mathbf1_{\nout}=\mathbf1_{\ntr}$. For $\tau>0$, the identity readout is $(\hatbs S+\tau I)^{-1}\hatbs S\mathbf1_{\nout}$ with $\hatbs S=\hatbs P_N\hatbs P_N^T/\ntr$; this is a contraction applied to $\mathbf1_{\nout}$. At zero ridge it is the orthogonal projection onto $\im\hatbs P_N$. This proves the first bound. For positive ridge,~\cref{eq:general_mean_readout} gives $\bs w_c=\sqrt{\nout N}\,\beta_\eta\hat{\bs Q}\bs b_N/J_\eta$. The bounds $J_\eta\ge1$ and $\|\hat{\bs Q}\|_{\rm op}\le1/\eta$ prove the second coefficient estimate. At zero ridge,~\cref{eq:general_zero_identity_readout} gives the same bound with the existing nullspace projector $\bs\Pi_N$, for every rank. Equation~\cref{eq:general_mean_sizes} now implies $\|\bs{w}_c\|^2=O_{\PP}[\nout(q_d+t_d^2)+1]$. The upper bound on $\nout\barSigma_{\rm te}$ proves the variance estimate.

\parTitle{Identity prediction error} The original normal equation in the uniform direction gives exactly $\overline{\hat{\bsp}}{}^T\bs{w}(I_d)-1 =-\eta\mathbf1_{\nout}^T\bs{w}(I_d)/(\nout N)$. Also $\|\bar{\bsp}-\bsp_{I/d}\|^2\le Kq_d/\nout$ and the record mean has mean-square fluctuation at most $1/(N\ntr)$. These facts and the coefficient bounds show that ${d(d+1)}\lvert\bsp_{I/d}^T\bs{w}(I_d)-1\rvert^2$ is bounded in probability by the right side of~\cref{eq:identity_error_bound}. The Haar split~\cref{eq:general_physical_split} adds the squared traceless prediction norm, bounded by $K\|\bs w_c\|^2/\nout$. This proves the ideal-test bound.
\end{proof}

\parTitle{From averaged to realized costs} The target bounds promote conditional quadratic traces to typical realized centered costs. The squared error of the mean prediction, which probes a single direction, remains separate.

\begin{lemma}[Concentration for spherical and sign target directions]
\label{lem:centered_full_sphere_sign_concentration}
Let $v$ be an independent uniform unit vector on the traceless operator sphere of dimension $d^2-1$, or let $v_j=\varepsilon_j/\sqrt{d^2-1}$ for independent symmetric signs $\varepsilon_j$. For every deterministic real symmetric operator $\bs A$ on this space,
\begin{equation}
\EE(v^T\bs A v)=\frac1{d^2-1}\Trace(\bs A),\qquad
\Var(v^T\bs A v)\le\frac2{(d^2-1)^2}\Trace(\bs A^2).
\label{eq:centered_full_sphere_sign_concentration}
\end{equation}
The same conclusions hold conditionally on any independent training record and apparatus.
\end{lemma}
\begin{proof}
\parTitle{Spherical and sign fourth moments} For a sphere vector, $\EE[v_iv_jv_kv_l]=(\delta_{ij}\delta_{kl}+\delta_{ik}\delta_{jl}+\delta_{il}\delta_{jk})/[(d^2-1)(d^2+1)]$, so $\Var(v^T\bs A v)=2[\Trace(\bs A^2)-(\Trace(\bs A))^2/(d^2-1)]/[(d^2-1)(d^2+1)]$. For signs,
$v^T\bs A v=(d^2-1)^{-1}\Trace(\bs A)+2(d^2-1)^{-1}\sum_{i<j}A_{ij}\varepsilon_i\varepsilon_j$.
Distinct unordered pairs have zero cross covariance, giving $\Var(v^T\bs A v)=4(d^2-1)^{-2}\sum_{i<j}A_{ij}^2\le2(d^2-1)^{-2}\Trace(\bs A^2)$. Both laws have mean $\Trace(\bs A)/(d^2-1)$.
\end{proof}

\begin{lemma}[Concentration for a target on a unitary orbit]
\label{lem:identity_target_concentration}
For a fixed traceless Hermitian $\Omega$ with $\Trace(\Omega^2)=1$, an independent Haar unitary $U$, and real symmetric $\bs A$ on the traceless operator space,
\begin{equation}
 \EE[\bs v_{U^\dagger\Omega U}\bs v_{U^\dagger\Omega U}^T]
 =\bs P_0/(d^2-1),\qquad
 \Var(\bs v_{U^\dagger\Omega U}^T\bs A\bs v_{U^\dagger\Omega U})
 \le\frac{K\|\Omega\|_{\rm op}^2}{d(d^2-1)}\Trace(\bs A^2)
 \le\frac{Kd}{(d^2-1)^2}\Trace(\bs A^2).
 \label{eq:identity_orbit_variance}
\end{equation}
\end{lemma}

\begin{proof}
\parTitle{Concentration on a unitary orbit} The second moment is~\cref{eq:moment_2_operator}. Regard $\bs A$ as a self-adjoint map on Hermitian operators. Differentiation along a unitary tangent gives a gradient bounded in squared norm by $16\|\Omega\|_{\rm op}^2\|\bs A\bs v_{U^\dagger\Omega U}\|^2$. The unitary Poincar\'e inequality with constant $K/d$ ~\cite{meckes2013spectral}, followed by the second moment, gives the first variance bound. Use $\|\Omega\|_{\rm op}\le1$ and $(d^2-1)\le d^2$ for the last one.
\end{proof}

\begin{proposition}[Fixed-ridge physical costs]
\label{thm:alt_ridge_profiles}
Assume~\cref{eq:alt_dimensions,eq:alt_geometry,eq:general_noise_bounds}, independent multinomial records, and a random target from~\cref{def:target_ensembles}. For fixed $\eta>0$,
\begin{equation}
 \frac{V_{N,\tau}(\calO)}{d\Delta_{\calO}^2}
 =F_{\eta,d}+\frac{c_\calO^2}{d\Delta_{\calO}^2}
       (\hatbs w^{I}_{N,\tau})^T\barSigma_{\rm te}\hatbs w^{I}_{N,\tau}
       +o_{\PP}(1),\qquad \tau=\eta/(\nout N),
 \label{eq:general_profile_equivalence}
\end{equation}
\begin{equation}
 \frac{{d(d+1)}\calE_{N,\tau}(\calO)}{d\Delta_{\calO}^2}
 =G_{\eta,d}+\frac{c_\calO^2}{d\Delta_{\calO}^2}
       \|\bs P_0\bs M_{\bs\mu}\hatbs w^{I}_{N,\tau}\|^2
 +\frac{[\bs t_\eta^T\bs v_{\calO_0}+\sqrt{d(d+1)} c_\calO(A_\eta-1)]^2}{d\Delta_{\calO}^2}
       +o_{\PP}(1).
 \label{eq:general_complete_mean_G}
\end{equation}
The empirical mean is arbitrary. Its error square is retained exactly, even when it grows with dimension; the remainders concern the bounded normalized bulk contributions at fixed ridge.
\end{proposition}
\begin{proof}
\parTitle{Bulk correction and target concentration} By~\cref{lem:general_centering}, the physical nonconstant map is $\bs Z_\eta=\hat{\bs K}+\bs z_\eta\bs r_\eta^T$, and its constant prediction is exactly~\cref{eq:general_mean_exact}. The mean-size estimate~\cref{lem:general_mean_bounds} makes $\bs Z_\eta-\hat{\bs K}$ a bounded rank-one map at each fixed positive ridge. The ridge filter bounds $\hat{\bs K}$, and the apparatus and categorical metrics are bounded. Expanding the quadratic traces therefore gives
\begin{equation}\begin{aligned}
 \frac1{d^2-1}\Trace(\bs Z_\eta^T\nout\barSigma_{\rm te}\bs Z_\eta)
 -\frac1{d^2-1}\Trace(\hat{\bs K}^T\nout\barSigma_{\rm te}\hat{\bs K})
 &=O_{\PP,\eta}((d^2-1)^{-1}),\\
 \frac1{d^2-1}\|\tilde{\bs M}_{\bs\mu}\bs Z_\eta-\bs P_0\|_F^2
 -\widehat G_{\eta,d}&=O_{\PP,\eta}((d^2-1)^{-1}).
\end{aligned}\label{eq:centered_full_bulk_comparison}\end{equation}
The fixed-ridge centered proposition identifies their limits. The target concentration lemmas transfer the normalized quadratic traces to realized costs. The identity-label readout has bounded state-dependent prediction norm by~\cref{lem:identity_readout}; the target second moment makes its cross term with the independent traceless target vanish. Keeping the squared mean-prediction error exactly proves~\cref{eq:general_profile_equivalence,eq:general_complete_mean_G}. These remainder estimates hold at fixed positive ridge, with no uniformity asserted when $\eta$ decreases with dimension.
\end{proof}

\parTitle{One directional error can survive} For a traceless target, the normalized physical variance equals $F_{\eta,d}+o_{\PP}(1)$, while the ideal-test error includes one additional target projection. Normalized bulk traces average over many operator directions; this mean term probes a single direction and need not concentrate. Conditional on the record and scalar target invariants, the mean-error square averages to $\|\bs t_\eta\|^2/(d^2-1)+d(d+1)c_\calO^2(A_\eta-1)^2/(d\Delta_{\calO}^2)$.
For a spherical traceless target, the extra square has distribution $(\|\bs t_\eta\|^2/(d^2-1))Z_{d^2-1}^2$, where $Z_{d^2-1}$ is $\sqrt{d^2-1}$ times a coordinate of a uniform unit sphere and tends to a standard normal variable. If $\|\bs t_\eta\|^2/(d^2-1)\xrightarrow{\PP}\sigma_\eta^2$ and $G_{\eta,d}\to G_\eta$ have finite deterministic limits, the ideal-test cost converges in distribution to $G_\eta+\sigma_\eta^2\chi_1^2$; its target average and median generally differ. The other direction laws share the second moment but need not share the projection distribution.

\parTitle{When the mean term vanishes} For pure training states with a bounded, noncollapsed frame,~\cref{lem:general_pure_mean} gives $\|\bar\rho\|_{\rm op}=O(d^{-1})$, and~\cref{eq:general_mean_projection_bound} makes the traceless-target mean term vanish at fixed positive ridge. Mixed-state lists can retain it. A deterministic full profile therefore requires limits of the explicit mean terms as well as of the centered traces; replacing the measured mean terms by spectral integrals would require additional directional inverse control.

\section{Interpolation mechanisms and peaks}
\label[appsec]{app:interpolation_mechanisms}

As in the main text, the interpolation ratio is $\gamma=\lim\nout/\ntr$, the shot ratio is $\alpha=\lim N/[d(d+1)]$, and the scaled ridge is $\eta=\nout N\tau$. Interpolation at $\gamma=1$ has two origins of weak training features. Above the outcome-count threshold, shot noise fills exact Born null modes and creates the quantum peak of~\cref{sec:quantum_ic}; below it, a nearly square exact training matrix can amplify the inaccessible part of the target, as in~\cref{sec:uc_main}. The fixed-ridge equations of~\cref{app:quantum_profiles} connect both mechanisms to the same rounded denominator, under different conditions for its existence. \Cref{nav:def_above_peak} gives the universal peak above the threshold, \cref{app:proof_peak_amplitudes} evaluates its geometry-dependent coefficients, and~\cref{sec:uc_interpolation} states the additional spectral criterion below it.

\subsection{Quantum peak above informational completeness}
\label{nav:def_above_peak}

For $\kappa<1$, exact centered probabilities span at most $d^2-1$ of the $\nout-1$ nonconstant outcome directions, while shot noise acts on all of them. The signal-free directions guarantee the positive critical solution behind the quantum peak in~\cref{sec:quantum_ic}. When the exact Born matrix has rank $d^2$, projecting away its row and column spaces leaves a shot-noise block of size $(\nout-d^2)\times(\ntr-d^2)$. Its becoming square at $\nout=\ntr$ explains the small measured singular values in the mechanism of~\cref{sec:quantum_mechanism}, whose exact block algebra is given in~\cref{prop:block_interpolation}; the theorems below establish the peak through the critical equations, without assuming a Gaussian smallest-singular-value law for the counts. \Cref{app:proof_quantum_peak} determines the ridge-rounded profile and its coefficients;~\cref{app:proof_centered_zero_penalty} proves divergence of the actual minimum-norm variance and states the extra inverse-tail condition needed away from interpolation.

\begin{proposition}[Block form of the interpolating weights]
\label{prop:block_interpolation}
Let the POVM be informationally complete and the training states span the operator space, so that $\rank\bs P=d^2$, and let $\bs U_1,\bs U_2,\bs V_1,\bs V_2$ and the blocks $\bs P_{11}$, $\hatbs\Xi_{kl}$ be those of~\cref{eq:alt_companion_blocks}. Suppose $\nout=\ntr$ and that $\hatbs P_N$ and $\hatbs\Xi_{22}$ are invertible. Then the least-squares weights interpolate, $\hatbs P_N^T\hatbs w_N=\bs y$, and their components $\hatbs w_k=\bs U_k^T\hatbs w_N$ satisfy~\cref{eq:main_null_weights} and
\begin{equation}
 \bigl[\bs P_{11}+N^{-\frac12}\bigl(\hatbs\Xi_{11}-\hatbs\Xi_{12}\hatbs\Xi_{22}^{-1}\hatbs\Xi_{21}\bigr)\bigr]^T\hatbs w_1=\bs V_1^T\bs y,
 \label{eq:main_signal_weights}
\end{equation}
whereas the exact coefficients satisfy $\bs P_{11}^T\bs U_1^T\bs w_{\rm ex}=\bs V_1^T\bs y$. The reconstructed observable depends only on $\hatbs w_1$, $\widetilde\calO_N=\sum_a(\bs U_1\hatbs w_1)_a\mu_a$, whereas the test variance $V_N(\sigma,\calO)=\hatbs w_N^T\bs\Sigma_\sigma\hatbs w_N$ depends on both components.
\end{proposition}
\begin{proof}
Since $\hatbs P_N$ is square and invertible, $\hatbs P_N^T\bs w=\bs y$ has a unique solution, which is the least-squares minimizer. Let $\bs B$ denote the block matrix on the right side of~\cref{eq:alt_companion_blocks}. Multiplying the interpolation conditions by $[\bs V_1\ \bs V_2]^T$ and inserting $[\bs U_1\ \bs U_2][\bs U_1\ \bs U_2]^T=I_{\nout}$ gives $\bs B^T(\hatbs w_1;\hatbs w_2)=(\bs V_1^T\bs y;\bs V_2^T\bs y)$. The columns of $\bs V_2$ span $\ker\bs P=\ker(\bs M_{\bs\mu}^T\bs M_{\bs\rho})$; because the POVM is informationally complete, $\bs M_{\bs\mu}^T$ is injective, so $\bs M_{\bs\rho}\bs V_2=0$ and $\bs V_2^T\bs y=(\bs M_{\bs\rho}\bs V_2)^T\bs v_\calO=0$. The second block row then reads $N^{-1/2}(\hatbs\Xi_{12}^T\hatbs w_1+\hatbs\Xi_{22}^T\hatbs w_2)=0$, which is~\cref{eq:main_null_weights}. Substituting it into the first block row, $(\bs P_{11}+N^{-1/2}\hatbs\Xi_{11})^T\hatbs w_1+N^{-1/2}\hatbs\Xi_{21}^T\hatbs w_2=\bs V_1^T\bs y$, gives~\cref{eq:main_signal_weights}. The exact coefficients obey $\bs P^T\bs w_{\rm ex}=\bs M_{\bs\rho}^T\bs M_{\bs\mu}\bs w_{\rm ex}=\bs M_{\bs\rho}^T\bs v_\calO=\bs y$; multiplying by $\bs V_1^T$ and using $\bs P^T\bs U_2=0$ gives $\bs P_{11}^T\bs U_1^T\bs w_{\rm ex}=\bs V_1^T\bs y$. Finally, the columns of $\bs U_2$ span $\ker\bs P^T=\ker(\bs M_{\bs\rho}^T\bs M_{\bs\mu})$, and $\bs M_{\bs\rho}^T$ is injective because the training states span the operator space; hence $\bs M_{\bs\mu}\bs U_2=0$, i.e., $\sum_a(\bs U_2\hatbs w_2)_a\mu_a=0$.
\end{proof}

\subsubsection{Critical coefficients and universal profile}
\label{app:proof_quantum_peak}

After the dimensional limit, each centered cost with positive amplitude grows as $1/\lvert\gamma-1\rvert$ near interpolation. Positive ridge rounds this divergence over a width of order $\sqrt\eta$, with height of order $\eta^{-1/2}$. \Cref{cor:alt_rounding} states that profile under the empirical bounds and critical trace regularity. The coefficients depend on how much signal remains unresolved when the fitted noise strength and inverse covariance are determined together. The trace-regularity assumption requires the spectral averages and penalty responses used below to have compatible limits near interpolation; the fixed-ridge comparison alone does not ensure this. The definitions below separate training, apparatus, and shot-noise contributions.

\parTitle{Critical inverse and signal spectrum} At interpolation, define the inverse on $\mathbf1_{\nout}^\perp$ by
\begin{equation}
 \bs{\mathcal R}_*\equiv
  [\zeta_*\nout\barSigma_{\rm tr}+\bs S]^{-1},
 \qquad
 \frac1{\nout-1}\Trace(\nout\barSigma_{\rm tr}\bs{\mathcal R}_*)=1.
 \label{eq:general_critical_inverse}
\end{equation}
To determine $\zeta_*$ in~\cref{eq:general_critical_inverse}, let $\lambda_1,\ldots,\lambda_{\nout-1}$ include all eigenvalues of
\begin{equation}
 (\nout\barSigma_{\rm tr})^{-1/2}\bs S
 (\nout\barSigma_{\rm tr})^{-1/2}
 \quad\text{on }\mathbf1_{\nout}^\perp.
 \label{eq:general_critical_spectrum}
\end{equation}
The positive spectrum also equals that of the training Gram matrix
\begin{equation}
 \frac N{\ntr}\bs\Pi_{\ntr}\bs P^T\barSigma_{\rm tr}^{+}
                  \bs P\bs\Pi_{\ntr}.
 \label{eq:general_critical_gram}
\end{equation}
Both matrices are Gram products of $\sqrt{N/\ntr}\,(\barSigma_{\rm tr}^{+})^{1/2}\bs P\bs\Pi_{\ntr}$. At $\nout=\ntr$, their restrictions to $\mathbf1_{\nout}^\perp$ and $\mathbf1_{\ntr}^\perp$ have equal dimension, so zero multiplicities agree as well; away from interpolation they must be counted separately.

\parTitle{Signal-to-noise eigenvalues} The whitening compares between-state signal with within-state noise. Indeed,
\begin{equation}
 \frac1{\ntr}\bs P\bs\Pi_{\ntr}\bs P^T
 =\bs\Sigma(\bar{\bsp})-\barSigma_{\rm tr},
 \label{eq:general_critical_between_covariance}
\end{equation}
and the $\lambda_j$ are the stationary values over nonzero $\bs w\in\mathbf1_{\nout}^\perp$ of
\begin{equation}
 \frac{N\,\ntr^{-1}\sum_i
       [\bs w^T(\bsp_i-\bar{\bsp})]^2}
 {\ntr^{-1}\sum_i\bigl[\sum_a p_{ai}w_a^2
                    -(\sum_a p_{ai}w_a)^2\bigr]}.
 \label{eq:general_critical_signal_to_noise}
\end{equation}
The numerator is the variance of exact predictions across training states; the denominator is their mean single-shot variance. The factor $N$ converts the ratio to the signal-to-noise ratio of an $N$-shot record. When $\nout\barSigma_{\rm tr}\to\bs\Pi_{\nout}$ in operator norm, the eigenvalues approach those of $(\nout N/\ntr)\bs\Pi_{\ntr}\bs P^T\bs P\bs\Pi_{\ntr}$, apart from zero multiplicities.

\parTitle{Existence from signal-free directions} The critical equation becomes
\begin{equation}
 \frac1{\nout-1}\sum_{j=1}^{\nout-1}\frac1{\zeta_*+\lambda_j}=1,
 \qquad 1-\kappa_d\le\zeta_*\le1.
 \label{eq:general_critical_scalar}
\end{equation}
The left-hand side decreases strictly, is at most one at $\zeta=1$, and is at least one at $\zeta=1-\kappa_d$ because at least $\nout-d^2$ eigenvalues vanish. Hence the positive root is unique and bounded away from zero for $\kappa<1$. Below completeness, existence requires a separate spectral condition.

\parTitle{Residual signal and peak amplitudes} The residual signal at this effective noise scale is
\begin{equation}
 \mathcal H_*\equiv\frac1{d^2-1}\min_{\bs W}\Big\{
 \|(\bs P_0-\bs W\tilde{\bs M}_{\bs\mu}^T)
   (\frameOp_{\bs\rho}^{\rm c})^{1/2}\|_F^2
 +\zeta_*\Trace(\bs W(\nout\barSigma_{\rm tr})\bs W^T)\Big\},
 \label{eq:general_H_variational}
\end{equation}
Here $\bs W$ estimates centered operator coordinates from the rescaled outcome coordinates. The first term is its training-frame-weighted reconstruction error; the second is its variance for outcome-noise covariance $\zeta_*\nout\barSigma_{\rm tr}$. Completing the square gives $\bs W_*=\frameOp_{\bs\rho}^{\rm c}\tilde{\bs M}_{\bs\mu}\bs{\mathcal R}_*$ and
\begin{equation}
\begin{aligned}
 \mathcal H_*
 &=\frac1{d^2-1}\Trace(\frameOp_{\bs\rho}^{\rm c}
 -\frameOp_{\bs\rho}^{\rm c}\tilde{\bs M}_{\bs\mu}
 \bs{\mathcal R}_*\tilde{\bs M}_{\bs\mu}^T
 \frameOp_{\bs\rho}^{\rm c}),\\
 C_V&=\frac{\alpha\mathcal H_*}{\nout-1}
             \Trace(\nout\barSigma_{\rm te}\bs{\mathcal R}_*),\\
 C_G&=\frac{\mathcal H_*}{\nout-1}
             \Trace(\tilde{\bs M}_{\bs\mu}^T\tilde{\bs M}_{\bs\mu}
                 \bs{\mathcal R}_*).
\end{aligned}
\label{eq:general_peak_coefficients}
\end{equation}
The coefficients in~\cref{eq:general_peak_coefficients} share the unresolved-signal factor $0\le\mathcal H_*\le(d^2-1)^{-1}\Trace(\frameOp_{\bs\rho}^{\rm c})$. The two traces weight that residual by test-shot noise and the traceless POVM response. Under~\cref{eq:general_noncollapse}, $C_V>0$; $C_G>0$ additionally requires a positive limiting average scaled measurement-frame eigenvalue. The ridge width is
\begin{equation}
 h\equiv\frac{\zeta_*}{\nout-1}\Trace(\bs{\mathcal R}_*).
 \label{eq:general_ridge_width}
\end{equation}
Flat noise $\nout\barSigma_{\rm tr}=\bs\Pi_{\nout}$ gives $h=\zeta_*$; the covariance bounds give $(1-\kappa)/K\le h\le1/c$ in the limit.

\begin{remark}[Trace regularity for the critical expansion]
\label{rem:critical_traces}
The rounded-profile statements assume limits, at each fixed $\eta>0$, of the normalized traces in~\cref{eq:general_profile_Q,eq:general_profile_scalars,eq:general_profile_loss} and the mixed trace $(d^2-1)^{-1}\Trace(\frameOp_{\bs\rho}^{\rm c}\tilde{\bs M}_{\bs\mu}\bs{\mathcal R}\tilde{\bs M}_{\bs\mu}^T)$, including first derivatives in the two penalty directions $\nout\barSigma_{\rm te}$ and $\tilde{\bs M}_{\bs\mu}^T\tilde{\bs M}_{\bs\mu}$. They also require local convergence, including first penalty derivatives, of the corresponding weighted normalized traces of $[uI_{\nout-1}+v\nout\barSigma_{\rm tr}+w\bs W+\bs S]^{-1}$ near $(u,v,w)=(0,\zeta_*,0)$, with limits continuous as $\gamma\to1$. Here $u,v,w$ are scalar arguments and $\bs W$ is either of the two penalty metrics above. Mere existence of fixed-ridge limits does not supply this additional critical regularity. The mechanisms guaranteeing a positive critical root are stated separately.
\end{remark}

\begin{theorem}[Quantum peak and ridge rounding]
\label{cor:alt_rounding}
Assume~\cref{eq:alt_dimensions,eq:alt_geometry,eq:general_noise_bounds,eq:general_noncollapse} with $\kappa<1$. Assume the critical trace regularity of~\cref{rem:critical_traces}. After the proportional limit at each fixed $\eta>0$, the joint limit $(\gamma,\eta)\to(1,0)$ gives
\begin{equation}
 F_\eta(\gamma)\sim
 \frac{C_V}{\sqrt{(\gamma-1)^2+4h\eta}},\qquad
 G_\eta(\gamma)\sim
 \frac{C_G}{\sqrt{(\gamma-1)^2+4h\eta}}
 \quad\text{if }C_G>0,
 \label{eq:alt_rounding}
\end{equation}
with coefficients evaluated at interpolation.
\end{theorem}
\begin{proof}
\parTitle{Critical scalar balances} The rescaling below separates a vanishing scalar $e=1/a$ from a finite inverse $\bs{\mathcal R}^{\sharp}$. The loss becomes $\mathcal L=e\mathcal H$; its singular penalty response is isolated in the derivative of $e$. Work with the deterministic equations from~\cref{prop:general_centered_fixed_ridge}, and take the local limits of their normalized traces, assumed continuous in $\gamma$ at one. The signal matrix $\bs S$ is defined in~\cref{eq:general_ridge_signal_noise}. Let $e=1/a$, $u=\eta/e$, $v=ab$ and $w=z/e$, and write
\begin{equation}
\begin{aligned}
 \bs{\mathcal R}^{\sharp}&\equiv
 [u I_{\nout-1}+v\,\nout\barSigma_{\rm tr}+w\bs W+\bs S]^{-1},\\
 f&\equiv\frac1{\nout-1}\Trace(\nout\barSigma_{\rm tr}\bs{\mathcal R}^{\sharp}),
 \qquad j\equiv\frac1{\nout-1}\Trace(\bs{\mathcal R}^{\sharp}),
 \qquad k\equiv\frac1{\nout-1}\Trace(\bs W\bs{\mathcal R}^{\sharp}).
\end{aligned}
 \label{eq:general_central_variables}
\end{equation}
Multiplying the first equation of~\cref{eq:general_profile_Q} by $e=1/a$, and using the trace identity $b=1-\gamma+\Trace((\eta I_{\nout-1}+z\bs W)\bs{\mathcal R})/\ntr$ proved in~\cref{eq:general_schur_trace}, yields the following \emph{equivalent}, rather than additional, scalar equations:
\begin{equation}\begin{aligned}
 1-e&=\gamma f,\\
 ve^2&=(1-\gamma)e+\eta\gamma j+z\gamma k,\\
 \mathcal L&=e\mathcal H,\quad
 \mathcal H=\frac1{d^2-1}\Trace(\frameOp_{\bs\rho}^{\rm c}
 -\frameOp_{\bs\rho}^{\rm c}\tilde{\bs M}_{\bs\mu}\bs{\mathcal R}^{\sharp}
                 \tilde{\bs M}_{\bs\mu}^T\frameOp_{\bs\rho}^{\rm c}).
\end{aligned}\label{eq:general_central_equations}\end{equation}

\parTitle{Positive critical root} We first identify the critical values of $e,u,v$. The trace identity above implies $0<v=1-\Trace(\bs S\bs{\mathcal R}^{\sharp})/\ntr\le1$. Because $\rank\bs S\le d^2-1$ and $(d^2-1)/(\nout-1)\to\kappa<1$, $\ker\bs S$ has dimension at least $(1-\kappa+o(1))(\nout-1)$. For $\nout\barSigma_{\rm tr}\succeq cI_{\nout-1}$ and $\nout\barSigma_{\rm tr}\preceq KI_{\nout-1}$, the inverse restricted to this kernel obeys the lower trace bound below. Indeed, if $P$ projects onto $\ker\bs S$, the Schur-complement inequality gives $P\bs{\mathcal R}^{\sharp} P\succeq[P(uI_{\nout-1}+v\nout\barSigma_{\rm tr})P]^{-1}$ on $\operatorname{im}P$. Using $\nout\barSigma_{\rm tr}\succeq cI_{\nout-1}$ and $\nout\barSigma_{\rm tr}\preceq KI_{\nout-1}$ then yields
\begin{equation}
 f=\frac1{\nout-1}\Trace(\nout\barSigma_{\rm tr}\bs{\mathcal R}^{\sharp})
 \ge \frac{c(1-\kappa+o(1))}{u+Kv}.
\end{equation}
Suppose $e$ were bounded below along a sequence with $(\gamma,\eta)\to(1,0)$. Then $u=\eta/e\to0$, while the second equation of~\cref{eq:general_central_equations} forces $v\to0$. The displayed lower bound would give $f\to\infty$, contrary to $1-e=\gamma f\le1$. Hence $e\to0$. At $z=0$, division of the second scalar equation by $e$ gives $ve=1-\gamma+\gamma uj$, so $uj\to0$. Since $\bs S$ is bounded and $v\le1$, $uj\ge u/(u+K')$ for a fixed $K'$; consequently $u\to0$. The same lower bound for $f$ now prevents $v\to0$. Any limit point $v_*$ therefore lies in $(0,1]$ and satisfies the critical normalization~\cref{eq:general_critical_inverse} with $v_*$ in place of $\zeta_*$. After whitening $\nout\barSigma_{\rm tr}$, the left side is strictly decreasing in $v_*$; the nullity of $\bs S$ and its nonnegativity give a unique solution $v_*=\zeta_*\in[1-\kappa,1]$. This proves~\cref{eq:general_critical_inverse,eq:general_critical_scalar}. For bounded positive $\bs W$, the derivatives of the normalized inverse traces with respect to $u,v,w$ are uniformly bounded near this point. In particular $\partial_v f=-(\nout-1)^{-1}\Trace(\nout\barSigma_{\rm tr}\bs{\mathcal R}^{\sharp}\nout\barSigma_{\rm tr}\bs{\mathcal R}^{\sharp})$ is bounded away from zero in absolute value, because $\nout\barSigma_{\rm tr}\succeq cI_{\nout-1}$ and $\bs{\mathcal R}^{\sharp}$ has uniformly bounded spectrum above and a positive normalized trace below.

\parTitle{Singular penalty response} The partial resolvent derivatives with respect to $u,v,w$ remain bounded, but the envelope formula contains the potentially singular factor $e_z$. Set $\delta=\gamma-1$ and $\Delta=\sqrt{\delta^2+4\eta\gamma jv}$, at $z=0$. The scalar equation gives exactly
\begin{equation}
 e=\frac{\Delta-\delta}{2v},\qquad
 u=\frac{\Delta+\delta}{2\gamma j},\qquad
 2ve+\delta=\Delta,
 \label{eq:general_discriminant}
\end{equation}
so $e+u=O(\Delta)$. The identity $ve^2+\delta e-\eta\gamma j-z\gamma k=0$ shows explicitly why a quadratic discriminant appears. Differentiate it and $1-e=\gamma f$ with respect to $z$, keeping $u=\eta/e$ and $w=z/e$ dependent on $z$. Differentiating $1-e=\gamma f$ gives $\lvert v_z\rvert\le K[(1+\eta/e^2)\lvert e_z\rvert+e^{-1}]$. The second gives $\Delta e_z=\gamma k+O(\Delta)+O(\Delta^2\lvert e_z\rvert)$; here $(e^2+\eta)(1+\eta/e^2)=(e+u)^2$ controls the derivative errors. Consequently
\begin{equation}
 e_z=\gamma k/\Delta+O(1),\qquad
 e\mathcal H_z=O(1),\qquad
 \partial_z\mathcal L=\mathcal H\gamma k/\Delta+O(1).
 \label{eq:general_central_derivative}
\end{equation}
For the middle bound, use $eu_z=-ue_z$ and $ew_z=1$ at $z=0$; the bound on $v_z$ gives $ev_z=O(1)$ because $e+u=O(\Delta)$. Take $\bs W=\nout\barSigma_{\rm te}$ or $\tilde{\bs M}_{\bs\mu}^T\tilde{\bs M}_{\bs\mu}$. The mixed prediction trace in~\cref{eq:general_complete_profiles} is bounded by positive-ridge and frame bounds. In~\cref{eq:general_central_derivative}, the factor $\mathcal H$ tends to the residual $\mathcal H_*$ of~\cref{eq:general_H_variational}, whereas $k$ tends to $(\nout-1)^{-1}\Trace(\bs W\bs{\mathcal R}_*)$. Also $jv\to (\zeta_*/(\nout-1))\Trace(\bs{\mathcal R}_*)=h$. Thus $\Delta\sim\sqrt{(\gamma-1)^2+4h\eta}$. Inserting the two indicated choices of $\bs W$ into the normalized derivative, and restoring the prefactor $\alpha$ for test-shot variance, gives exactly~\cref{eq:general_peak_coefficients,eq:alt_rounding}.

\parTitle{Positive amplitudes and physical variance} The whitened residual identity is
\begin{equation}
 \mathcal H_* =\frac{\zeta_*}{d^2-1}\Trace\left(\frameOp_{\bs\rho}^{\rm c}
 \left\{\zeta_*\bs P_0+(\frameOp_{\bs\rho}^{\rm c})^{1/2}
 \tilde{\bs M}_{\bs\mu}(\nout\barSigma_{\rm tr})^{-1}
 \tilde{\bs M}_{\bs\mu}^T(\frameOp_{\bs\rho}^{\rm c})^{1/2}
 \right\}^{-1}\right).
 \label{eq:general_positive_residual}
\end{equation}
The frame and categorical bounds make this uniformly positive under~\cref{eq:general_noncollapse}. They also bound $h$ above and below by positive constants. The test categorical lower bound makes $C_V$ positive; a positive average measurement-frame eigenvalue makes $C_G$ positive. The identity-label variance bound is uniform over bounded $\eta$, and $q_d,t_d$ are bounded. Its variance contribution is $O_{\PP}(1)$ for the full target class, and its mixed term is at most of the order of the square root of the divergent variance. Both are lower order than the centered variance pole. The ideal-test cost retains its explicit mean contribution in~\cref{eq:general_complete_mean_G}.
\end{proof}

\parTitle{Width, height, and physical costs} The singular tails scale as $1/\lvert\gamma-1\rvert$. Ridge rounds them over $\lvert\gamma-1\rvert\asymp2\sqrt{h\eta}$, with heights $C_V/(2\sqrt{h\eta})$ and $C_G/(2\sqrt{h\eta})$ when the coefficients are positive. The nonsingular background retains the full geometry. The physical variance has the same leading singularity by~\cref{thm:alt_ridge_profiles,lem:identity_readout}; its complete ideal-test error retains the mean terms of~\cref{eq:general_complete_mean_G}.

\subsubsection{Minimum norm and small measured singular values}
\label{app:proof_centered_zero_penalty}
\label{app:proof_actual_central_min_norm}

The unregularized estimator in~\cref{sec:quantum_ic} inverts each positive measured singular value at finite dimension, whereas~\cref{cor:alt_rounding} first takes a dimensional limit at positive ridge. \Cref{thm:alt_central} bridges that distinction at interpolation: the physical minimum-norm variance diverges without an inverse-tail assumption, even on deficient records. Away from interpolation, \cref{cor:alt_fixed_gap} identifies the centered zero-penalty profiles only when the total label-weighted contribution of measured modes with arbitrarily small singular values vanishes.

\begin{theorem}[Minimum-norm variance at interpolation]
\label{thm:alt_central}
Assume~\cref{eq:alt_dimensions,eq:alt_geometry,eq:general_noise_bounds,eq:general_noncollapse} with $\kappa<1$, independent multinomial records, and a random target from~\cref{def:target_ensembles}. If $\nout/\ntr\to1$, the physical minimum-norm readout satisfies
\begin{equation}
V_{N,0}(\calO)/(d\Delta_{\calO}^2)\xrightarrow{\PP}+\infty,
\qquad
V_{N,0}/(d\Delta_{\calO}^2)\xrightarrow{\PP}+\infty.
\label{eq:alt_central}
\end{equation}
The median diverges. The empirical mean is arbitrary, and deficient count matrices are included.
\end{theorem}
\begin{proof}
\parTitle{Whitened lower bound for minimum norm} We lower-bound the physical minimum-norm variance by a positive-ridge coefficient norm. A uniform square-root bound on the minimized loss supplies the divergence without assuming convergence of the empirical spectra. Write $\bs C_{\rm tr}=\nout\barSigma_{\rm tr}$ and $\bs C_{\rm te}=\nout\barSigma_{\rm te}$ on $\mathbf1_{\nout}^\perp$, and whiten the centered design as $\bs X=\bs C_{\rm tr}^{-1/2}\hatbs P_N^{\rm c}$. Set $\bs R=\bs M_{\bs\rho}^{\rm c}/\sqrt{\ntr}$ and $\bs K^{\rm w}_\eta=(\bs X\bs X^T+\eta I_{\nout-1})^{-1}\bs X\bs R^T$. For a unit target direction $\bs v=\bs v_{\calO_0}/\sqrt{d\Delta_{\calO}^2}$, the nonconstant part of every physical least-squares solution solves the centered problem. After whitening, its coefficient norm is at least the centered minimum-norm coefficient norm, which dominates $\|\bs K^{\rm w}_\eta\bs v\|$ by singular-value calculus. The covariance bounds give, record by record,
\begin{equation}
 \frac{V_{N,0}(\calO)}{d\Delta_{\calO}^2}
 \ge\alpha_d\frac cK\|\bs K^{\rm w}_\eta\bs v\|^2.
 \label{eq:central_whitened_lower_bridge}
\end{equation}
The constant target component disappears from the centered labels, so this bound applies to the full target class.

\parTitle{Uniform square-root loss bound} For the whitened fixed-ridge equivalent, let $\bs B=\tilde{\bs M}_{\bs\mu}\bs C_{\rm tr}^{-1/2}$, $\bs F=\frameOp_{\bs\rho}^{\rm c}$, and $\bs S=\bs B^T\bs F\bs B$. Choose fixed constants $q_0>0$, $L<\infty$, and $\zeta_0>0$ such that eventually $(d^2-1)^{-1}\Trace(\bs F)\ge q_0$, $\|\bs F\|_{\rm op}\|\bs B\|_{\rm op}^2\le L$, and $1-(d^2-1)/\ntr\ge\zeta_0$; these exist by noncollapse, the frame and covariance bounds, and $\kappa<1$, $\gamma_d\to1$. The rescaled equations~\cref{eq:general_central_variables,eq:general_central_equations} now have identity noise. In particular $j=f=(1-e)/\gamma_d$, and
\begin{equation}
 v\ge1-(d^2-1)/\ntr\ge\zeta_0,\qquad v\le1,\qquad
 ve^2=(1-\gamma_d)e+\eta(1-e).
 \label{eq:central_uniform_balance_bridge}
\end{equation}
The lower bound on $v$ follows from $v=1-\ntr^{-1}\Trace(\bs S(uI_{\nout-1}+vI_{\nout-1}+\bs S)^{-1})$ and $\rank\bs S\le d^2-1$. For every fixed sufficiently small $\eta>0$, taking $d\to\infty$ in the last balance bounds every scalar limit point by $\sqrt{\eta/2}\le e\le\sqrt{\eta/\zeta_0}$. Hence $u=\eta/e\le\sqrt{2\eta}$. The residual has the exact whitened form
$\mathcal H=(u+v)(d^2-1)^{-1}\Trace(\bs F[(u+v)\bs P_0+\bs F^{1/2}\bs B\bs B^T\bs F^{1/2}]^{-1})$.
For $\eta$ small enough that $u\le1$, it obeys $\zeta_0q_0/(2+L)\le\mathcal H\le K$. Since $\mathcal L_d^{\rm w}=e\mathcal H$, fixed constants $a,b>0$ therefore satisfy
\begin{equation}
 a\sqrt\eta\le\liminf_d\mathcal L_d^{\rm w}(\eta)
 \le\limsup_d\mathcal L_d^{\rm w}(\eta)\le b\sqrt\eta.
 \label{eq:central_uniform_loss_bridge}
\end{equation}
For example, one may use $a=\zeta_0q_0/[\sqrt2(2+L)]$ and $b=K/\sqrt{\zeta_0}$. No matrix spectral limit has been taken.

\parTitle{A divergent coefficient norm} Choose a fixed number $R>(b/a)^2$. The sampled whitened minimized loss is concave, and its envelope derivative is $(d^2-1)^{-1}\|\bs K^{\rm w}_\eta\|_F^2$. Thus that derivative is at least its forward difference quotient between $\eta$ and $R\eta$. The fixed-positive-penalty count comparison transfers the two loss values to their deterministic equivalents. Applying~\cref{eq:central_uniform_loss_bridge} at both penalties gives, for every sufficiently small fixed $\eta$, the lower bound $(a\sqrt R-b)/[(R-1)\sqrt\eta]+o_\PP(1)$ on $(d^2-1)^{-1}\|\bs K^{\rm w}_\eta\|_F^2$.

\parTitle{Typical targets and deficient records} Conditional target concentration transfers this trace bound to $\|\bs K^{\rm w}_\eta\bs v\|^2$: at each fixed $\eta$, the ridge filter and response-frame bound give a dimension-independent operator norm, so the sphere, sign, and unitary-orbit variance estimates vanish. Given any threshold, choose the positive penalty small enough first and then take the dimensional limit. Together with~\cref{eq:central_whitened_lower_bridge}, this proves divergence in probability and hence divergence of the median. Averaging~\cref{eq:central_whitened_lower_bridge} over the target instead of using concentration bounds $V_{N,0}/(d\Delta_\calO^2)$ below by $\alpha_d(c/K)(d^2-1)^{-1}\|\bs K^{\rm w}_\eta\|_F^2$, which diverges by the previous step. Deficient count matrices are included throughout.
\end{proof}

\parTitle{Required control away from interpolation} Let $\sigma_j>0$ be the positive singular values of the measured matrix $\hatbs P_N^{\rm c}$, and let $\bs v_j$ be their right singular vectors in training space. For the scaled response $(\bs M^{\rm c}_{\bs\rho})^T\bs v_{\calO_0}/\sqrt{\ntr}$, the mean squared label in that direction, averaged over the independent target and divided by $d\Delta_{\calO}^2$, is $\|\bs M^{\rm c}_{\bs\rho}\bs v_j\|^2/((d^2-1)\ntr)$. Minimum norm divides this label by $\sigma_j$, so the contribution to its normalized squared coefficient norm is $\|\bs M^{\rm c}_{\bs\rho}\bs v_j\|^2/((d^2-1)\ntr\sigma_j^2)$. At fixed $\gamma\ne1$, impose
\begin{equation}
 \lim_{u\downarrow0}\limsup_d\PP\left\{
 \frac1{(d^2-1)\ntr}\sum_{j:\,0<\sigma_j^2<u}
       \frac{\|\bs M^{\rm c}_{\bs\rho}\bs v_j\|^2}{\sigma_j^2}
            >\varepsilon\right\}=0
 \quad\text{for every }\varepsilon>0.
 \label{eq:general_inverse_condition}
\end{equation}
It requires the total coefficient norm from measured modes with arbitrarily small singular values to vanish, rather than requiring every mode to have a fixed lower singular-value bound. A bounded corresponding weighted inverse moment of order $1+\epsilon$ is sufficient.

\begin{corollary}[Centered zero-penalty profiles]
\label{cor:alt_fixed_gap}
Under~\cref{eq:alt_dimensions,eq:alt_geometry,eq:general_noise_bounds} and the random target laws of~\cref{def:target_ensembles}, fix $\gamma\ne1$ and assume the fixed-ridge trace limits and~\cref{eq:general_inverse_condition}. The normalized realized centered variance and traceless ideal-test error, computed with $\hat{\bs K}_0=((\hatbs P_N^{\rm c})^T)^+(\bs M^{\rm c}_{\bs\rho})^T/\sqrt{\ntr}$ converge in probability to $F_0=\lim_{\eta\downarrow0}F_\eta$ and $G_0=\lim_{\eta\downarrow0}G_\eta$. Their conditional target averages converge to the same limits in probability over the counts; their medians have these limits as well. The apparatus and training list are conditioned on unless a geometry ensemble is specified. Under the continuity hypotheses of~\cref{cor:alt_rounding}, each positive amplitude gives
\begin{equation}
F_0(\gamma)\sim C_V/\lvert\gamma-1\rvert,\qquad
G_0(\gamma)\sim C_G/\lvert\gamma-1\rvert.
\label{eq:general_poles}
\end{equation}
The complete physical minimum-norm costs retain the mean and nullspace restoration in~\cref{eq:general_zero_mean_map}.
\end{corollary}
\begin{proof}
\parTitle{Transfer from ridge to pseudoinverse} Use the weighted inverse-tail condition~\cref{eq:general_inverse_condition}. The exact minimum-norm coefficients, including deficient records, are given in~\cref{eq:general_zero_mean_map}. To identify the centered zero-penalty limits, use $\bs R=\bs M^{\rm c}_{\bs\rho}/\sqrt{\ntr}$ and $\hat{\bs K}_\eta=(\hatbs P_N^{\rm c}(\hatbs P_N^{\rm c})^T+\eta\bs\Pi_{\nout})^{-1} \hatbs P_N^{\rm c}\bs R^T$. Singular-value calculus gives
\begin{equation}
 \frac1{d^2-1}\|\hat{\bs K}_0-\hat{\bs K}_\eta\|_F^2
 \le\frac1{d^2-1}\Trace(\bs R\mathbf1_{(0,u)}((\hatbs P_N^{\rm c})^T\hatbs P_N^{\rm c})
                         ((\hatbs P_N^{\rm c})^T\hatbs P_N^{\rm c})^+\bs R^T)
        +\frac{\eta^2}{u^3(d^2-1)}\Trace(\bs R\bs R^T).
 \label{eq:general_inverse_transfer}
\end{equation}
The normalized response trace is bounded. After the dimensional limit, take $\eta\downarrow0$ at fixed $u$, then $u\downarrow0$. The inverse-tail condition makes the maps converge in normalized Frobenius norm. The upper categorical and apparatus bounds transfer their two quadratic costs; target second moments transfer realized costs in probability. Fixed-positive-ridge limits then identify $F_0,G_0$. Their central poles follow from~\cref{eq:alt_rounding}.

\parTitle{Restoring the zero-penalty physical fit} For the original solution, the additional nullspace coefficient and mean query remain in~\cref{eq:general_zero_mean_map}. Their exact physical and target-averaged costs are~\cref{eq:general_physical_split,eq:general_exact_average_V,eq:general_exact_average_G}. In particular, if $\hatbs P_N^{\rm c}$ has full nonconstant row rank, then $\bs\Pi_N=0$ and $\bs{w}^I_{N,0}=\mathbf1_{\nout}$ exactly. Constant labels then have zero variance and zero ideal-test error. A nonzero mean query for a traceless target is still evaluated by $\bs t_0$. This proves~\cref{cor:alt_fixed_gap} without a restriction on the empirical mean.
\end{proof}

\subsection{Peak coefficients above informational completeness}
\label{app:proof_peak_amplitudes}

The universal denominator does not fix how strongly the two physical costs grow. The examples below evaluate its coefficients for $\kappa<1$, clarifying the geometry and shot-budget dependence discussed in~\cref{sec:ridge_risk}. \Cref{nav:def_tight_frames} gives an equal-strength benchmark;~\cref{nav:def_haar_apparatus,nav:def_haar_training} isolate apparatus and training randomness;~\cref{nav:def_fixed_rank_peaks} includes fixed-rank states and effects. All examples assume asymptotically flat categorical covariances:
\begin{equation}
\nout\barSigma_{\rm tr},\nout\barSigma_{\rm te}\to\bs\Pi_{\nout}
\quad\text{in operator norm},\qquad
\frameOp_{\bs\mu}^{\rm s}
\equiv \tilde{\bs M}_{\bs\mu}\tilde{\bs M}_{\bs\mu}^T.
\label{eq:general_flat_categorical}
\end{equation}
Then~\cref{eq:general_peak_coefficients,eq:general_ridge_width} gives $h=\zeta_*$ and $C_V=\alpha\mathcal H_*$. The coefficients describe the centered singularity; the full physical ideal-test error follows~\cref{eq:general_complete_mean_G}.

\subsubsection{Tight empirical training and measurement frames}
\label{nav:def_tight_frames}

Equal-strength empirical frames remove directional variation from both the state exploration and apparatus response, leaving a solvable benchmark for the peak coefficients. At interpolation, suppose $\frameOp_{\bs\rho}^{\rm c}\to\beta\bs P_0$ and $\frameOp_{\bs\mu}^{\rm s}\to\lambda_\mu\bs P_0$, with $\beta,\lambda_\mu>0$. The empirical expectation variance of every unit traceless observable is $\beta/[d(d+1)]$, and $\lambda_\mu$ is the scaled measurement-frame eigenvalue. Writing $x=\beta\lambda_\mu$, the signal covariance has $d^2-1$ eigenvalues $x$ and $\nout-d^2$ zeros, which gives
\begin{align}
 &\zeta_*=\frac{1-x+\sqrt{(1-x)^2+4(1-\kappa)x}}2,\qquad
 h=\zeta_*,\qquad
 \mathcal H_*=\frac{\beta \zeta_*}{\zeta_*+x}\label{eq:general_tight_peak},\\
 &C_V=\frac{\alpha\beta \zeta_*}{\zeta_*+x},\qquad
 C_G=\frac{\kappa \lambda_\mu\beta \zeta_*}{(\zeta_*+x)^2}.\label{eq:general_tight_amplitudes}
\end{align}
The root tends to one as $x\downarrow0$ and to $1-\kappa$ as $x\to\infty$; at $\kappa=1/2$, $x=1$, it equals $1/\sqrt2$.

\parTitle{Evaluation for tight frames} The normalization of the critical inverse is $(1-\kappa)/\zeta_*+\kappa/(\zeta_*+x)=1$. Multiplication by the two denominators gives the quadratic whose positive root is~\cref{eq:general_tight_peak}. The residual and the two weighted inverse traces then give~\cref{eq:general_tight_amplitudes}.

\subsubsection{Haar apparatus and an empirical training frame}
\label{nav:def_haar_apparatus}

For an independent Haar rank-one apparatus, the covariance-resolvent calculation reduces the peak coefficients to an integral over the empirical training spectrum. The integral shows how unequal training-frame eigenvalues change the unresolved signal. For a centered empirical training-frame law $\mu_1$ at interpolation, define
\begin{equation}
 \phi=\int\frac{\dd\mu_1(x)}{1+\alpha x},\qquad
 \mathcal H_*=\int\frac{x\,\dd\mu_1(x)}{1+\alpha x}
             =\frac{1-\phi}{\alpha}.
 \label{eq:weak_haar_peak}
\end{equation}
Then
\begin{equation}
 C_V=1-\phi,\qquad C_G=\kappa\phi(1-\phi),\qquad
 h=\zeta_*=1-\kappa+\kappa\phi.
 \label{eq:general_haar_empirical_amplitudes}
\end{equation}
Each eigenvalue $x$ is ${d(d+1)}$ times an empirical expectation variance. Its residual contribution $x/(1+\alpha x)$ increases with $x$ and decreases with $\alpha$; concavity in $x$ implies that a tight frame maximizes $\mathcal H_*$ at fixed mean eigenvalue.

\parTitle{Tight-training specialization} For $\frameOp_{\bs\rho}^{\rm c}\to\beta\bs P_0$, put $q=\alpha\beta$. The coefficients reduce to
\begin{equation}
 C_V=\frac{q}{1+q},\qquad
 C_G=\frac{\kappa q}{(1+q)^2},\qquad
 h=\zeta_*=1-\kappa+\frac{\kappa}{1+q}.
 \label{eq:general_haar_tight_amplitudes}
\end{equation}
At $q=1$, $C_V=1/2$ and $C_G=\kappa/4$. The variance amplitude increases toward one with $q$, while the ideal-test amplitude is maximal at $q=1$.

\parTitle{Covariance-resolvent evaluation} For the independent Haar apparatus, the covariance resolvent for its Gaussian operator lifts gives, at positive $\xi$ and with critical training-frame law $\mu_1$,
\begin{equation}
 w=\frac{\alpha}{1+\alpha\kappa\int x/(\xi+wx)\,\dd\mu_1(x)},
 \qquad
 g=\int\frac{\dd\mu_1(x)}{\xi+wx},\qquad
 \mathcal H(\xi)=\xi\int\frac{x\,\dd\mu_1(x)}{\xi+wx}.
 \label{eq:general_haar_critical_integrals}
\end{equation}
On the traceless sector, put $\bs Q_\xi=[\xi\bs P_0+(\frameOp_{\bs\mu}^{\rm s})^{1/2}\frameOp_{\bs\rho}^{\rm c}(\frameOp_{\bs\mu}^{\rm s})^{1/2}]^{-1}$. The normalized traces $(d^2-1)^{-1}\Trace(\bs Q_\xi)$ and $(d^2-1)^{-1}\Trace(\frameOp_{\bs\mu}^{\rm s}\bs Q_\xi)$ have limits $g$ and $wg$, respectively. The bounded empirical training-state frame and quadratic-form concentration of the independent effect lifts justify the covariance resolvent~\cref{lem:weak_population_resolvent}. Haar normalization changes the bounded traces by a near-identity congruence and bounded-rank identity coupling. The equation also reads $w=\alpha(1-\kappa+\kappa\xi g)$. At the critical value $\xi=\zeta_*=h$, the inverse normalization is $h=1-\kappa+\kappa h g$. Thus $w=\alpha h$, $hg=\phi$ and $\mathcal H(h)=(1-\phi)/\alpha$. The two weighted metrics give~\cref{eq:general_haar_empirical_amplitudes}. A tight training-frame law $\mu_1=\delta_\beta$ yields~\cref{eq:general_haar_tight_amplitudes}.

\subsubsection{Haar training states and a fixed apparatus spectrum}
\label{nav:def_haar_training}

For independent Haar pure training states, every unit traceless observable has ensemble expectation-value variance $1/[d(d+1)]$. A finite training list still has unequal frame eigenvalues when $\ntr$ is comparable with $d^2$; the coefficients below include those sampling fluctuations. As in~\cref{sec:ridge_risk}, the apparatus enters through its spectrum. Let $\nu$ be the limiting traceless spectrum of $\nout\bs P_0\frameOp_{\bs\mu}\bs P_0$. Define
\begin{equation}
 \phi=\int\frac{\dd\nu(\lambda)}{1+\alpha\lambda},\qquad
 h=\zeta_*=1-\kappa+\kappa\phi.
 \label{eq:weak_haar_training_moments}
\end{equation}
For a fresh Haar pure state measured with $N$ shots, $\phi$ is the limiting minimum linear state-reconstruction error normalized by its mean squared Hilbert--Schmidt distance from $I_d/d$. The empirical residual is $\mathcal H_*=\phi h$, giving
\begin{equation}
 C_V=\alpha\phi h,\qquad C_G=\kappa\phi(1-\phi).
 \label{eq:general_haar_training_amplitudes}
\end{equation}
Complete centered curves, including the regular background, follow in~\cref{app:haar_centered_full_results}.

\parTitle{Critical expansion for Haar training} The outcome covariance eigenvalue associated with apparatus eigenvalue $\lambda$ is $t=1+\alpha\lambda$: unit shot noise plus signal. Its limiting law is $H=(1-\kappa)\delta_1+\kappa(1+\alpha\lambda)_*\nu$, where the mapped measure replaces $\lambda$ by $1+\alpha\lambda$ with the same weight. Thus $h=\int t^{-1}\dd H=1-\kappa+\kappa\phi$. The scalar $c>0$ solves $1=\eta c+\gamma\int tc/(1+tc)\,\dd H$, the equation used for the complete curves in~\cref{eq:alt_scalar}. As $(\gamma,\eta)\to(1,0)$, it gives $c\to\infty$ and $\eta c^2+(\gamma-1)c-h=o(1)$ at leading order. Differentiating this balance in the two physical penalty metrics gives the amplitudes in~\cref{eq:general_haar_training_amplitudes}.

\subsubsection{Independent Haar states and effects of fixed rank}
\label{nav:def_fixed_rank_peaks}

The independent fixed-rank ensembles of~\cref{lem:fixed_rank_bulk_bridge} give explicit shot-dependent coefficients and distinguish the large-shot limit from exact-feature training. With state rank $k$ and effect rank $\ell$ fixed as $d\to\infty$, put $q=\alpha/(k\ell)$. The spectral integral is
\begin{align}
 &\phi=\frac{2}{1+q(1-\kappa)
          +\sqrt{[1+q(1-\kappa)]^2+4\kappa q}}\label{eq:alt_MP_a},\\
 &C_V=\ell(1-\phi),\qquad
 C_G=\kappa\phi(1-\phi),\qquad
 h=\zeta_*=1-\kappa+\kappa\phi.\label{eq:weak_rank_peak}
\end{align}
For the pure-state, rank-one-POVM case $k=\ell=1$, fixed $0<\kappa<1$ gives
\begin{equation}
 \begin{aligned}
 C_V&=1-\frac1{(1-\kappa)\alpha}+O(\alpha^{-2}),\\
 C_G&=\frac{\kappa}{(1-\kappa)\alpha}+O(\alpha^{-2}),
 \end{aligned}
 \quad(\alpha\to\infty).
 \label{eq:ic_haar_large_shot_amplitudes}
\end{equation}
Increasing training shots suppresses the traceless ideal-test singular coefficient but leaves the test-variance coefficient nonzero. A weight component in $\ker\bs M_{\bs\mu}$ reconstructs no observable but still carries categorical test-shot variance. Finite-shot fitting can use these directions; exact-feature minimum norm removes them. These parameter limits follow the dimensional limit and cannot be interchanged with setting $N=\infty$ before fitting.

\parTitle{Fixed-rank covariance scales} For Haar conjugates of a fixed state $\rho_*$, the traceless covariance scale is $\beta_d={d(d+1)}(\Trace(\rho_*^2)-1/d)/(d^2-1)$. At fixed projector rank $k$, $\beta_d\to1/k$. For the exact normalized rank-$k$ projector states,~\cref{lem:fixed_rank_bulk_bridge} gives the empirical covariance law $(x\mapsto x/k)_*\mu_{\rm MP,\kappa}$ at interpolation. The same lemma gives the independent rank-$\ell$ apparatus law scaled by $1/\ell$. These scales replace $\alpha$ inside the spectral integral by $\alpha/\ell$, while leaving its physical variance prefactor $\alpha$. Evaluating $\phi=\int(1+qx)^{-1}\,\dd\mu_{{\rm MP},\kappa}(x)$, including the zero atom, gives~\cref{eq:alt_MP_a,eq:weak_rank_peak}.

\subsection{Interpolation below informational completeness}
\label{sec:uc_interpolation}

At $\nout=\ntr<d^2$, the inaccessible part of a generic target cannot be reconstructed on all states, yet its labels can be fitted on the finite training list when the square exact feature matrix is nonsingular. Near interpolation, small singular values of the exact training matrix can make that fit require large weights even when the accessible POVM directions are well conditioned. \Cref{sec:uc_empirical_spectrum} establishes this exact-feature mechanism, and~\cref{sec:uc_finite_shots} explains why finite shots need not remove it. \Cref{sec:uc_theorem} gives the positive-root condition and physical interpolation theorem supporting~\cref{sec:uc_main};~\cref{sec:uc_coefficients} evaluates coefficients and distinguishes conditions that do and do not force a peak.

\subsubsection{Exact probabilities and the spectral edge at zero}
\label{sec:uc_empirical_spectrum}

The exact-feature example below separates near-singularity of the training matrix from poor conditioning of the apparatus. Haar training creates a spectrum reaching zero at interpolation even for a flat accessible POVM frame; an error peak additionally requires label strength in those small singular directions. At the Haar fluctuation scale, the exact centered training matrix is
\begin{equation}
 \bs P^{\rm c}\equiv\sqrt{\nout {d(d+1)}/\ntr}\,\bs P\bs\Pi_{\ntr}
 =\sqrt{\nout/\ntr}\,\bs M_{\bs\mu}^T\bs P_0
                         \bs M^{\rm c}_{\bs\rho}.
 \label{eq:uc_exact_training_matrix}
\end{equation}
It maps $\ntr-1$ centered training coordinates to $\nout-1$ nonconstant outcomes and becomes square at $\nout=\ntr$.

\parTitle{Hard edge from training-state sampling} The accessible measurement Gram matrix can stay gapped while sampling a nearly square training matrix creates a lower spectral edge at zero, called a hard edge.
\begin{lemma}[Exact-feature hard edge]
\label{lem:uc_hard_edge}
Let the training states be independent Haar pure states, and let the POVM satisfy $\nout\bs M_{\bs\mu}^T\bs P_0\bs M_{\bs\mu}=x_0\bs\Pi_{\nout}$ for some fixed $x_0>0$, i.e., its rescaled traceless frame has all its nonzero eigenvalues equal to $x_0$, which requires $\nout\le d^2$. As $d\to\infty$ with $(d^2-1)/(\nout-1)\to\kappa\in[1,\infty)$ and $(\nout-1)/\ntr\to\gamma\in(0,\infty)$, the empirical eigenvalue distribution of $\bs P^{\rm c}(\bs P^{\rm c})^T$ on $\mathbf1_{\nout}^\perp$ converges weakly in probability to the law of $x_0X$ with $X\sim\mu_{{\rm MP},\gamma}$ of~\cref{eq:MP_density}. Its continuous part has edges $x_\pm=x_0(1\pm\sqrt\gamma)^2$, and at $\gamma=1$ its density is
\begin{equation}
 \rho_{x_0}(x)=
 \frac{\sqrt{4x_0-x}}{2\pi x_0\sqrt x}
 \mathbf1_{(0,4x_0)}(x)
 \sim\frac1{\pi\sqrt{x_0 x}}
 \quad(x\downarrow0).
 \label{eq:uc_exact_mp_hard_edge}
\end{equation}
For $\gamma\ne1$, the inverse moment of the positive part of the limiting law is
\begin{equation}
\int_{(0,\infty)}\frac{\dd\mu_{x_0}(x)}{x}=\frac{1}{x_0}\,\frac{\min\{1,\gamma^{-1}\}}{\lvert1-\gamma\rvert},
\label{eq:uc_exact_inverse_moment}
\end{equation}
where $\mu_{x_0}$ is the limiting law; it diverges as $\gamma\to1$.
\end{lemma}
\begin{proof}
Because the columns of $\bs P\bs\Pi_{\ntr}$ are traceless state differences, $\bs P^{\rm c}=\ntr^{-1/2}[\bs x_1\ \cdots\ \bs x_{\ntr}]\bs\Pi_{\ntr}$ with $\bs x_i=\sqrt{\nout}\,\bs M_{\bs\mu}^T\bs P_0\bs h_i$ and $\bs h_i=\sqrt{d(d+1)}\bs P_0\bs v_{\rho_i}$. The vectors $\bs h_i$ are independent, centered, and isotropic on the traceless sector, with the quadratic-form concentration $\Var(\bs h_i^T\bs A\bs h_i)\le K(d^2-1)^2\|\bs A\|_{\rm op}^2/d$ used in the proof of~\cref{prop:general_haar_full_curves}. Hence $\EE\bs x_i\bs x_i^T=x_0\bs\Pi_{\nout}$, and for symmetric $\bs A$ on $\mathbf1_{\nout}^\perp$, $\Var(\bs x_i^T\bs A\bs x_i)\le Kx_0^2(d^2-1)^2\|\bs A\|_{\rm op}^2/d=o((\nout-1)^2)\|\bs A\|_{\rm op}^2$, because $\nout-1\asymp d^2$. \Cref{thm:yaskov_mp} therefore gives the stated Marchenko--Pastur limit for $\ntr^{-1}\sum_i\bs x_i\bs x_i^T$, and the rank-one centering by $\bs\Pi_{\ntr}$ does not change it (\cref{lem:congruence_spectral}); equivalently,~\cref{lem:weak_population_resolvent} applies. The density~\cref{eq:uc_exact_mp_hard_edge} is~\cref{eq:MP_density} at $\gamma=1$, rescaled by $x_0$. The inverse moment of the positive part of $\mu_{{\rm MP},\gamma}$ is $\min\{1,\gamma^{-1}\}/\lvert1-\gamma\rvert$~\cite{bai2010spectral}, the function $\mathcal J_{\rm cl}$ of~\cref{eq:statement_classical_peak}, and rescaling by $x_0$ gives~\cref{eq:uc_exact_inverse_moment}. These are statements about the limiting law; they do not control the smallest empirical eigenvalues.
\end{proof}
Its divergent inverse moment satisfies the limiting positive-root criterion at every fixed finite $\alpha>0$. Centering leaves the spectral law unchanged, but that law alone does not control finite-dimensional pseudoinverses.

\parTitle{Label strength in small singular directions} Small singular directions produce an error peak only when their labels carry nonnegligible weight. Representable labels inherit the feature singular-value factors, whereas an inaccessible target component need not supply a factor that cancels the inverse singular value. In this subsection use an independent isotropic \emph{traceless} target with the second moment~\cref{eq:centered_full_target_covariance}, $\Trace(\calO^2)=d\Delta_{\calO}^2>0$. Omitting the target argument denotes its average conditional on the apparatus, training states, and any counts. If $\sigma_j>0$ and $\bs v_j$ are the positive singular values and right singular vectors of $\bs P^{\rm c}$, then
\begin{equation}
 \frac{V^{\rm c}_{\infty,0}}{d\Delta_{\calO}^2}
 =(1+o(1))\frac1{(d^2-1)\ntr}\sum_{\sigma_j>0}
       \frac{\|\bs M^{\rm c}_{\bs\rho}\bs v_j\|^2}{\sigma_j^2},
 \label{eq:uc_exact_inverse_trace}
\end{equation}
provided $\|\nout\barSigma_{\rm te}-\bs\Pi_{\nout}\|_{\rm op}\to0$. Here $N=\infty$ means exact training probabilities, while $V$ remains the numerator of the test-shot variance. Each summand divides target-averaged label strength by the squared feature response.

\parTitle{Physical minimum-norm lower bound} For the physical fit, eliminating the uniform weight $w_0$ leaves the centered least-squares equations. Its nonconstant coefficients therefore have at least the centered minimum norm. If $\nout\barSigma_{\rm te}\succeq c\bs\Pi_{\nout}$, then
\begin{equation}
 \frac{V_{\infty,0}}{d\Delta_{\calO}^2}
 \ge\frac c{(d^2-1)\ntr}\sum_{\sigma_j>0}
       \frac{\|\bs M^{\rm c}_{\bs\rho}\bs v_j\|^2}{\sigma_j^2}.
 \label{eq:uc_physical_exact_lower_bound}
\end{equation}
The inaccessible target component can supply the amplified residual; a square matrix alone does not force a peak for every target. A well-conditioned accessible measurement also converts large coefficients into a large reconstruction error, as follows.

\begin{lemma}[Accessible-frame lower bound on ideal-test error]
\label{lem:accessible_error_bound}
Suppose
\begin{equation}
 a\bs\Pi_{\nout}\preceq \nout\bs M_{\bs\mu}^T\bs P_0\bs M_{\bs\mu}
 \preceq K\bs\Pi_{\nout},\qquad
 \nout\barSigma_{\rm te}\preceq K\bs\Pi_{\nout},
\end{equation}
with dimension-independent $a,K>0$. For any physical readout, with $\calE$ and $V$ defined as above,
\begin{equation}
 \frac{{d(d+1)}\calE_{N,\tau}(\calO)}{d\Delta_{\calO}^2}
 \ge\left[\sqrt{\frac aK\frac{V_{N,\tau}(\calO)}{d\Delta_{\calO}^2}}-1\right]_+^2,
 \qquad [x]_+=\max\{x,0\}.
 \label{eq:uc_ideal_variance_bound}
\end{equation}
The same bound holds for target-averaged costs.
\end{lemma}
\begin{proof}
\parTitle{Accessible-frame comparison} By~\cref{eq:general_physical_split}, the reconstructed traceless norm is at least $\sqrt{a/\nout}\|\bs w_\perp\|$, whereas $V_{N,\tau}(\calO)\le K\|\bs w_\perp\|^2/\nout$. The reverse triangle inequality gives~\cref{eq:uc_ideal_variance_bound}; applying it in target $L^2$ gives the averaged version.
\end{proof}

\subsubsection{Finite shots and residual amplification}
\label{sec:uc_finite_shots}

Finite shots penalize large weights in the loss averaged at a fixed readout, but the fitted weights depend on the realized noise. Near interpolation the fit can follow that noise, so the averaged penalty does not establish a spectral gap. The residual and singular-vector identities below make this distinction precise for the mechanism of~\cref{sec:uc_main}.

\parTitle{Averaged loss and realized fit} By~\cref{lem:shot_averaged_ls}, averaging the training loss over independent multinomial shots at a fixed readout gives
\begin{equation}
 \frac1{\ntr}\EE_{\hatbs n}\|\hatbs P_N^T\bs w-\bs y\|^2
 =\frac1{\ntr}\|\bs P^T\bs w-\bs y\|^2
 +\frac1N\bs w^T\barSigma_{\rm tr}\bs w.
 \label{eq:uc_shot_averaged_loss}
\end{equation}
The final term penalizes large weights and can suppress the classical peak when the averaged objective approximates the realized one uniformly over the readouts relevant to the fit. Near interpolation, however, the fit can follow the particular observed noise: if $\nout=\ntr$ and $\hatbs P_N$ is invertible, every label is fitted exactly. Averaging at fixed $\bs w$ does not constrain this record-dependent solution.

\parTitle{Inaccessible labels and shot residual} Let $\bs w_{\rm ex}$ represent the Hilbert--Schmidt projection of $\calO$ onto the effects, and write $\calO_{\bs w_{\rm ex}}=\sum_a w_{{\rm ex},a}\mu_a$. The residual is
\begin{equation}
 (\bs y-\hatbs P_N^T\bs w_{\rm ex})_i
 =\Trace((\calO-\calO_{\bs w_{\rm ex}})\rho_i)
 -\bs w_{\rm ex}^T(\hatbsp_i-\bsp_i).
 \label{eq:uc_training_mismatch}
\end{equation}
Its inaccessible-target term persists with exact probabilities; changing $N$ does not generally cancel it.

\parTitle{Why shot noise need not open a gap} An isotropic Gaussian surrogate illustrates why finite-shot noise need not open a gap. Replace the rescaled exact signal and noise by $\sqrt{\alpha x_0/\ntr}\,\bs G_1$ and $\bs G_2/\sqrt{\ntr}$, respectively, with independent standard Gaussian $(\nout-1)\times(\ntr-1)$ matrices. Their sum satisfies
\begin{equation}
 \frac{\sqrt{\alpha x_0}\,\bs G_1+\bs G_2}{\sqrt{\ntr}}
 \ \overset{\rm law}=\
 \sqrt{\frac{1+\alpha x_0}{\ntr}}\,\bs G.
 \label{eq:uc_gaussian_surrogate}
\end{equation}
The limiting squared-singular-value edges are $(1+\alpha x_0)(1\pm\sqrt\gamma)^2$, so the lower edge remains zero at $\gamma=1$. An independent isotropic label residual contributes the inverse moment
\begin{equation}
 \int\frac{x}{(x+\eta)^2}\rho_{1+\alpha x_0}(x)\,\dd x
 \sim\frac1{2\sqrt{(1+\alpha x_0)\eta}}
 \quad(\eta\downarrow0).
 \label{eq:uc_gaussian_inverse_moment}
\end{equation}
The inverse-moment asymptotic concerns the stated Gaussian surrogate; the physical multinomial amplitudes require the categorical comparison and interpolation theorem.

\parTitle{Residual amplification in measured modes} For the positive singular values and singular vectors $\sigma_j,\bs u_j,\bs v_j$ of $\hatbs P_N^{\rm c}$, the zero-penalty normal equations give
\begin{equation}
 \bs u_j^T(\hatbs w_N-\bs w_{\rm ex})
 =\sqrt{\frac{\nout N}{\ntr}}\,
 \frac{\bs v_j^T\bs\Pi_{\ntr}
 (\bs y-\hatbs P_N^T\bs w_{\rm ex})}{\sigma_j}.
 \label{eq:uc_mismatch_amplification}
\end{equation}
Small measured singular values amplify residual components that do not decay sufficiently fast. Finite shots preserve the nearly square geometry; ridge instead replaces each inverse singular value by $\sigma_j/(\sigma_j^2+\eta)$ in the observed record.

\subsubsection{Positive critical root and interpolation theorem}
\label{app:proof_uc_interpolation}
\label{sec:uc_theorem}

Below completeness, no exact signal-free outcome direction is guaranteed, so a peak requires sufficiently many weak signal directions relative to shot noise. The positive critical root below expresses that requirement. Under the stated geometry, mean, and trace assumptions, \cref{thm:uc_interpolation} gives both the rounded physical target-averaged profiles and divergence of the actual minimum-norm variance, supplying the precise version of~\cref{thm:informal_uc_interpolation}.

\parTitle{Geometry and mean assumptions} The fixed-ridge equivalent of~\cref{prop:general_centered_fixed_ridge} applies to either ordering of $d^2-1$ and $\nout-1$. Use~\cref{eq:alt_dimensions} with $\kappa>1$, $\gamma$ near one, and $\kappa\gamma>1$, so both $\nout,\ntr<d^2$. Retain~\cref{eq:alt_geometry,eq:general_noise_bounds,eq:general_noncollapse}, and additionally require
\begin{equation}
 \liminf\frac1{\nout-1}\Trace(\tilde{\bs M}_{\bs\mu}^T
\tilde{\bs M}_{\bs\mu})>0,\qquad
 q_d=\|\bar\rho-I_d/d\|_F^2\longrightarrow0.
 \label{eq:uc_mean_and_measurement}
\end{equation}
The first condition ensures positive ideal-test amplitude; the second removes the mean-prediction contribution. For random states or apparatuses, these empirical conditions must hold with probability tending to one.

\parTitle{A positive critical root} For the centered Born covariance $\bs S$ of~\cref{eq:general_ridge_signal_noise}, require at $\gamma=1$
\begin{equation}
 \bs{\mathcal R}_*=[\zeta_*\nout\barSigma_{\rm tr}+\bs S]^{-1},\qquad
 \frac1{\nout-1}\Trace(\nout\barSigma_{\rm tr}\bs{\mathcal R}_*)=1,
 \qquad \zeta_*\ge\zeta_0>0,
 \label{eq:uc_critical_root}
\end{equation}
with inverses on $\mathbf1_{\nout}^\perp$. Using the generalized eigenvalues of~\cref{eq:general_critical_spectrum,eq:general_critical_gram,eq:general_critical_signal_to_noise}, write
$f_d(\zeta)\equiv\frac1{\nout-1}\sum_{j=1}^{\nout-1} \frac1{\zeta+\lambda_j}.$
The function decreases strictly on $\zeta>0$, and $f_d(1)\le1$. A strictly positive finite-dimensional root exists if and only if
\begin{equation}
 \lim_{\zeta\downarrow0}f_d(\zeta)>1.
 \label{eq:uc_positive_root_criterion}
\end{equation}
For positive eigenvalues this is $(\nout-1)^{-1}\sum_j\lambda_j^{-1}>1$; a zero eigenvalue makes the inverse trace infinite. Equality at zero is a boundary case. The proportional theorem further requires $\zeta_*\ge\zeta_0>0$ and regular trace limits.

\parTitle{Resolved and unresolved directions} An equivalent interpretation uses the effective number of signal-bearing directions:
\begin{equation}
 d_{\rm eff}(\zeta)\equiv
 \sum_{j=1}^{\nout-1}\frac{\lambda_j}{\zeta+\lambda_j},
 \qquad
 \zeta_*=1-\frac{d_{\rm eff}(\zeta_*)}{\nout-1},
 \quad\zeta_*>0.
 \label{eq:uc_effective_dimension}
\end{equation}
The identity is equivalent to the critical equation only for $\zeta_*>0$. Strong modes $\lambda_j\gg\zeta_*$ contribute approximately one to $d_{\rm eff}$, while modes with $\lambda_j\ll\zeta_*$ contribute little. Thus $\zeta_*$ is a self-consistent unresolved fraction of outcome directions, not the physical ridge penalty. Above completeness exact null modes guarantee that fraction; below completeness sufficiently many weak modes can sustain it.

\parTitle{Trace assumptions and coefficients} Impose the critical trace regularity of~\cref{rem:critical_traces}, in addition to the uniformly positive root. The amplitudes and ridge width retain the common form~\cref{eq:general_peak_coefficients,eq:general_ridge_width}; the theorem below concerns target-averaged physical costs for traceless targets, with the empirical-mean contribution removed by~\cref{eq:uc_mean_and_measurement}.

\begin{theorem}[Interpolation below informational completeness]
\label{thm:uc_interpolation}
Assume $\kappa>1$ and $\kappa\gamma>1$,~\cref{eq:alt_dimensions,eq:alt_geometry,eq:general_noise_bounds,eq:general_noncollapse,eq:uc_mean_and_measurement,eq:uc_critical_root} and the trace regularity just specified. Use independent multinomial records and an independent isotropic traceless target; the training set and POVM are otherwise arbitrary.

At fixed $\eta>0$, with $\tau=\eta/(\nout N)$, the target-averaged physical costs $V_{N,\tau}/(d\Delta_{\calO}^2)$ and ${d(d+1)}\calE_{N,\tau}/(d\Delta_{\calO}^2)$ converge in probability to $F_\eta(\gamma)$ and $G_\eta(\gamma)$. After this dimensional limit,
\begin{equation}
 F_\eta(\gamma)\sim
 \frac{C_V}{\sqrt{(\gamma-1)^2+4h\eta}},\qquad
 G_\eta(\gamma)\sim
 \frac{C_G}{\sqrt{(\gamma-1)^2+4h\eta}},
 \qquad (\gamma,\eta)\to(1,0),
 \label{eq:uc_rounded_peak}
\end{equation}
with positive $C_V,C_G,h$ from~\cref{eq:general_peak_coefficients,eq:general_ridge_width}.

At $\nout/\ntr\to1$, the physical minimum-norm readout also satisfies $V_{N,0}/(d\Delta_{\calO}^2)\xrightarrow{\PP}+\infty$. Its target-averaged ideal-test error diverges as well if $\nout\bs M_{\bs\mu}^T\bs P_0\bs M_{\bs\mu}\succeq a\bs\Pi_{\nout}$ for a dimension-independent $a>0$, by~\cref{eq:uc_ideal_variance_bound}.
\end{theorem}
\begin{proof}
\parTitle{Excluding a noncritical limit} The fixed-ridge comparison and scalar balances~\cref{eq:general_central_variables,eq:general_central_equations} also hold for $\kappa>1$. We identify their critical solution from the positive-root assumption~\cref{eq:uc_critical_root}, since the kernel argument for $\kappa<1$ no longer applies. Write $e=a^{-1}$, $u=\eta/e$ and $v=ab$, as in~\cref{eq:general_central_variables}, and initially set $z=0$. The balances~\cref{eq:general_central_equations} hold with $0<v\le1$. Suppose that along a sequence $(\gamma,\eta)\to(1,0)$ we had $e\ge\varepsilon_0>0$. Then $u\to0$ and the second equation forces $v\to0$, since $j$ is bounded by the categorical lower bound and $f=(1-e)/\gamma$. Let $\lambda_1,\ldots,\lambda_{\nout-1}$ denote the eigenvalues of the noise-whitened signal, as in~\cref{eq:general_critical_spectrum}. Choose $0<\varepsilon<\zeta_0$ so that, for all sufficiently large dimensions, $uI_{\nout-1}+v \nout\barSigma_{\rm tr}\preceq \varepsilon \nout\barSigma_{\rm tr}$. Operator monotonicity of inversion gives $f\ge\frac1{\nout-1}\sum_{j=1}^{\nout-1}\frac1{\varepsilon+\lambda_j}>1$. The last inequality holds uniformly in a neighborhood of the critical geometry: the root condition equals one at $\zeta_*\ge\zeta_0$, the whitened signal is bounded, and the assumed local trace convergence preserves the strictly positive gap between its values at $\varepsilon$ and $\zeta_*$. This contradicts $f=(1-e)/\gamma$. Hence $e\to0$.

\parTitle{Identification of the positive limit} At $z=0$ the second balance divided by $e$ reads $ve=1-\gamma+\gamma uj$; it implies $uj\to0$. The covariance and frame bounds imply $j\ge(u+K')^{-1}$ for a dimension-independent $K'$, so $u\to0$. If $v$ had a subsequential limit of zero, the same spectral comparison with $\varepsilon=\zeta_0/2$ would contradict $f\to1$. Thus $v$ stays positive. Taking limits in $1-e=\gamma f$ makes the normalized trace in~\cref{eq:uc_critical_root} tend to one. The normalized inverse trace is strictly decreasing in $v$, so the positive-root assumption identifies $v\to\zeta_*$ and $\bs{\mathcal R}^{\sharp}\to\bs{\mathcal R}_*$ in the required normalized traces. We have established exactly the critical properties $e\to0$, $u\to0$, $v\to\zeta_*>0$ used after the root-identification step in the complete-regime proof.

\parTitle{Rounded peak below completeness} Since $\zeta_*\ge\zeta_0$, the inverse and its $u,v,w$ trace derivatives remain bounded. The discriminant and derivative calculation~\cref{eq:general_discriminant,eq:general_central_derivative} therefore apply unchanged. Using the two penalty directions $\nout\barSigma_{\rm te}$ and $\tilde{\bs M}_{\bs\mu}^T\tilde{\bs M}_{\bs\mu}$ gives the centered profiles~\cref{eq:uc_rounded_peak} with precisely the coefficients~\cref{eq:general_peak_coefficients,eq:general_ridge_width}. The residual identity~\cref{eq:general_positive_residual}, together with the positive-root, frame and covariance hypotheses, makes $\mathcal H_*>0$. The test covariance lower bound makes $C_V>0$; the assumed average traceless POVM strength also makes $C_G>0$. The same assumptions bound $h$ away from zero.

\parTitle{Physical fixed-ridge costs} The identity~\cref{lem:general_centering} is exact for each count record, while~\cref{lem:general_mean_bounds} makes its rank-one bulk correction negligible at fixed positive ridge. For a traceless independent isotropic target, the target-averaged constant-prediction term is $\|\bs t_\eta\|^2/(d^2-1)$. Its fixed-ridge bound is $O_{\PP,\eta}(q_d+(d^2-1)^{-1})$, which tends to zero under the present $q_d\to0$ assumption. This proves that the physical target-averaged costs have the centered limits used above. We do not claim this replacement for a general target with a nonzero identity component or a nonvanishing empirical mean correction.

\parTitle{Minimum-norm divergence} For the minimum-norm statement, take an arbitrary count record and eliminate the constant label direction. The nonconstant part of the physical least-squares solution solves the centered equations. After whitening by $(\nout\barSigma_{\rm tr})^{-1/2}$, its coefficient norm is no smaller than that of the whitened minimum-norm centered solution, and the latter dominates its fixed-positive-ridge norm. The uniform covariance bounds give the same lower bound, up to a positive constant, for the physical test variance. At $\gamma=1$, the preceding centered critical calculation makes the regularized lower bound diverge as its fixed penalty tends to zero. For any finite threshold, choose that positive penalty first and then take the proportional limit. The target-averaged physical minimum-norm variance consequently diverges in probability even when the count matrix is deficient. Finally, under the stated uniform lower bound on the accessible measurement Gram matrix,~\cref{eq:uc_ideal_variance_bound} transfers this divergence to the target-averaged ideal-test error. This completes the proof.
\end{proof}

\parTitle{Order of limits and physical scope} Dimensions must tend to infinity at fixed ridge before $(\gamma,\eta)\to(1,0)$. At fixed $\gamma\ne1$, identifying a full zero-penalty profile additionally requires weighted inverse-tail control, as in~\cref{eq:general_inverse_condition}. Minimum-norm divergence at interpolation needs no such condition. Without $q_d\to0$, the centered profiles remain valid but the physical ideal-test error retains its mean contribution; nontraceless targets also require the identity restoration of~\cref{lem:general_centering}.

\subsubsection{Peak amplitudes and limits of the root criterion}
\label{app:proof_uc_no_root}
\label{sec:uc_coefficients}

The positive-root criterion makes the incomplete peak conditional on geometry. The examples below show how training shots change its amplitudes and width, including the nonzero large-shot ideal-test coefficient highlighted in~\cref{sec:uc_main}. They also identify the scope of the result: strict failure of the root condition bounds centered regularized costs, while the threshold case and physical minimum norm require separate control.

\parTitle{Small shot parameter with general geometry} Hold the apparatus and training geometry fixed while varying $\alpha=N/[d(d+1)]$, and evaluate critical traces at interpolation. Since $\bs S=O(\alpha)$, $\zeta_*=1+O(\alpha)$ and $\bs{\mathcal R}_*=(\nout\barSigma_{\rm tr})^{-1}+O(\alpha)$. Hence
\begin{equation}
 \begin{aligned}
 C_V&=\alpha\frac{\Trace(\frameOp_{\bs\rho}^{\rm c})}{d^2-1}
 \frac{\Trace(\barSigma_{\rm te}\barSigma_{\rm tr}^{-1})}{\nout-1}
 +O(\alpha^2),\\
 C_G&=\alpha\frac{\Trace(\frameOp_{\bs\rho}^{\rm c})}{d^2-1}
 \frac{\Trace(\bs M_{\bs\mu}^T\bs P_0\bs M_{\bs\mu}
                         \barSigma_{\rm tr}^{-1})}{\nout-1}
 +O(\alpha^2).
 \end{aligned}
 \label{eq:uc_small_alpha_coefficients}
\end{equation}
Both amplitudes vanish linearly as $\alpha\downarrow0$, yet remain positive at every fixed $\alpha>0$ within the theorem's regime. These limits follow the dimensional limit and do not describe fixed $N$.

\parTitle{Haar apparatus and general training} Assume flat categorical covariances and an independent normalized Haar rank-one POVM. Let $\mu_1$ be the centered training-frame law on all $d^2-1$ traceless directions at interpolation, including zeros. The transform~\cref{eq:general_haar_critical_integrals} allows singular frames and gives
\begin{equation}
 \begin{gathered}
 \phi=\int\frac{\dd\mu_1(x)}{1+\alpha x},\qquad
 C_V=1-\phi,\qquad C_G=\kappa\phi(1-\phi),\\
 h=\zeta_*=1-\kappa+\kappa\phi.
 \end{gathered}
 \label{eq:uc_haar_apparatus_coefficients}
\end{equation}
The rank bound $\rank\frameOp_{\bs\rho}^{\rm c}\le \ntr-1$ gives zero mass at least $1-1/\kappa$. If the spectrum is bounded by $K$, then $h\ge(1+\alpha K)^{-1}>0$; noncollapse ensures positive amplitudes at finite $\alpha$. The first two expansions below are for $\alpha\downarrow0$ and the last two limits for $\alpha\to\infty$, after the proportional limit:
\begin{equation}
 \begin{aligned}
 C_V&=\alpha\int x\,\dd\mu_1(x)+O(\alpha^2),\\
 C_G&=\kappa\alpha\int x\,\dd\mu_1(x)+O(\alpha^2),\\
 C_V&\longrightarrow1-\mu_1(\{0\}),\\
 C_G&\longrightarrow\kappa\mu_1(\{0\})[1-\mu_1(\{0\})].
 \end{aligned}
 \label{eq:uc_haar_apparatus_limits}
\end{equation}
If the zero mass equals $1-1/\kappa$, the large-shot amplitudes are $1/\kappa$ and $1-1/\kappa$, while $h\to0$. The shot and ridge limits are then nonuniform; identifying these coefficients with an exact-training calculation requires inverse-spectrum control.

\parTitle{A frame flat on its sampled subspace} For $\mu_1=(1-1/\kappa)\delta_0+\kappa^{-1}\delta_\beta$, with $0<\beta\le\kappa$, direct integration gives
\begin{equation}
 C_V=\frac{\alpha\beta}{\kappa(1+\alpha\beta)},\qquad
 C_G=\frac{\alpha\beta[\kappa+(\kappa-1)\alpha\beta]}
{\kappa(1+\alpha\beta)^2},\qquad
 h=\frac1{1+\alpha\beta}.
\label{eq:uc_supported_flat_coefficients}
\end{equation}
A physical realization in prime-power dimensions uses all $\ntr=bd$ projectors from $b\sim d/\kappa$ mutually unbiased orthonormal bases, with $\nout=\ntr$~\cite{wootters1989optimal}. Their mean is $I_d/d$, and their centered state Gram matrix has $b$ blocks $I_d-\mathbf1_d\mathbf1_d^T/d$. The frame therefore has $b(d-1)$ positive eigenvalues ${d(d+1)}/\ntr\to\kappa$, realizing $\beta=\kappa$; uniform depolarization yields $0<\beta\le\kappa$. Balanced effect traces and the bounded frame give flat categorical covariances for the independent Haar apparatus.

\parTitle{Haar training and a fixed apparatus} Retain flat categorical covariances and a bounded apparatus frame. For independent Haar pure training states, let $\nu$ be the traceless spectral law of $\nout\bs P_0\frameOp_{\bs\mu}\bs P_0$, including zeros. Then
\begin{equation}
 \begin{gathered}
 \phi=\int\frac{\dd\nu(\lambda)}{1+\alpha\lambda},\qquad
 h=1-\kappa+\kappa\phi,\\
 C_V=\alpha\phi h,\qquad C_G=\kappa\phi(1-\phi).
 \end{gathered}
 \label{eq:uc_haar_training_coefficients}
\end{equation}
The outcome covariance law is $H=(1-\kappa)\delta_1+\kappa(\lambda\mapsto1+\alpha\lambda)_*\nu$. The zero atom of $\nu$, of mass at least $1-1/\kappa$, cancels the formally negative first term, so $H$ is a probability law with harmonic moment $h=\int t^{-1}\dd H(t)>0$. The positive-ridge scalar obeys $1=\eta c+\gamma\int tc/(1+tc)\,\dd H(t)$. Its large-$c$ expansion gives $\eta c^2+(\gamma-1)c\sim h$, from which the penalty derivatives yield the amplitudes. Haar training gives $q_d\to0$, so physical and centered fixed-ridge limits agree.

\parTitle{Equal accessible apparatus strengths} For a flat accessible apparatus spectrum $\nu=(1-1/\kappa)\delta_0+\kappa^{-1}\delta_{x_0}$, with $0<x_0\le\kappa$ for a balanced physical POVM,
\begin{equation}
 C_V=\frac{\alpha[\kappa+(\kappa-1)\alpha x_0]}
                {\kappa(1+\alpha x_0)^2},\qquad
 C_G=x_0 C_V.
\label{eq:uc_flat_apparatus_coefficients}
\end{equation}
At small $\alpha$, $C_V\sim\alpha$ and $C_G\sim\alpha x_0$; at large $\alpha$ they approach $(\kappa-1)/(\kappa x_0)$ and $1-1/\kappa$. Both increase monotonically for $\kappa\ge2$, whereas for $1<\kappa<2$ they can exceed their large-shot limits.

\parTitle{Independent Haar states and apparatus} For independent Haar training states and a normalized Haar rank-one POVM, $\nu=\mu_{{\rm MP},\kappa}$ includes a zero atom of mass $1-1/\kappa$. The transform in~\cref{eq:alt_MP_a} with $k=\ell=1$ gives $\phi$. At fixed $\kappa>1$, after the proportional limit,
\begin{align}
 &\begin{aligned}
 C_V&=\alpha+O(\alpha^2)\label{eq:uc_haar_small_shot_amplitudes},\\
 C_G&=\kappa\alpha+O(\alpha^2)
 \end{aligned}
 \qquad(\alpha\downarrow0),\\
\begin{aligned}
 C_V&\longrightarrow\frac1\kappa,\\
 C_G&\longrightarrow1-\frac1\kappa,\\
 h&\sim\frac1{(\kappa-1)\alpha}
 \end{aligned}
 \qquad(\alpha\to\infty).\label{eq:uc_haar_shot_amplitude_limits}
\end{align}
The variance amplitude $C_V=1-\phi$ increases monotonically with $\alpha$. For $1<\kappa<2$, however, $C_G=\kappa\phi(1-\phi)$ reaches
$C_G^{\rm max}=\frac\kappa4 \quad\text{at}\quad \alpha=\frac2{2-\kappa}, \qquad 1<\kappa<2,$
exceeding its large-shot limit by $(2-\kappa)^2/(4\kappa)$; for $\kappa\ge2$ it increases monotonically. Thus shot noise changes the undercomplete peak's amplitude and width rather than creating its exact-data mechanism. The variance reduction at smaller budgets is illustrated in~\cref{fig:undercomplete_interpolation}(b), but monotonicity need not hold for every geometry or both costs. The limit $h\to0$ makes large-shot and zero-ridge limits nonuniform. With $M$ independent test shots, $\MSE_{N,M,\tau}=\calE_{N,\tau}+V_{N,\tau}/M$.

\parTitle{Strict failure of the root condition} Failure of~\cref{eq:uc_positive_root_criterion}, interpreted by~\cref{eq:uc_effective_dimension}, excludes the positive-root expansion~\cref{eq:uc_rounded_peak}; it does not alone establish bounded costs. Strict failure gives the following sufficient condition.

\begin{proposition}[Bounded centered costs under strict failure of the root condition]
\label{prop:strict_no_root}
Under~\cref{eq:alt_dimensions,eq:alt_geometry,eq:general_noise_bounds} with $\kappa>1$, suppose $\bs S$ is invertible on $\mathbf1_{\nout}^\perp$ at $\nout=\ntr$ for all sufficiently large dimensions and
\begin{equation}
 \limsup_{d\to\infty}\frac1{\nout-1}\sum_{j=1}^{\nout-1}\frac1{\lambda_j}<1.
 \label{eq:uc_strict_no_root}
\end{equation}
Then the centered deterministic costs at interpolation remain bounded as $\eta\downarrow0$ after the proportional limit.
\end{proposition}
\begin{proof}
\parTitle{Strict root failure keeps the filter finite} Work at $\nout=\ntr$ and use the existing fixed-point variables $e,u,v,f,j$ and $\bs{\mathcal R}^{\sharp}$ from~\cref{eq:general_central_variables,eq:general_central_equations}, with $z=0$ and finite-dimensional ratio $\gamma_d=\frac{\nout-1}{\ntr}=1-1/\ntr$. Since $u,v\ge0$, invertibility of $\bs S$ gives $\bs{\mathcal R}^{\sharp}\preceq\bs S^{-1}$. Consequently,
\begin{equation}
 1-e=\gamma_d f\le\gamma_d\frac1{\nout-1}
 \Trace(\nout\barSigma_{\rm tr}\bs S^{-1})
 =\gamma_d\frac1{\nout-1}\sum_{j=1}^{\nout-1}\lambda_j^{-1}.
\end{equation}
Since $\gamma_d=\frac{\nout-1}{\ntr}<1$ at $\nout=\ntr$, the strict upper limit in~\cref{eq:uc_strict_no_root} makes $e$ uniformly positive. The categorical lower bound also gives $j\le f/c$. The second scalar balance therefore implies $u=\eta/e=O(\eta)$ and $v=O(\eta+\ntr^{-1})$, uniformly in dimension.

\parTitle{Bounding the centered residual} At $\nout=\ntr$, the centered training frame has rank at most $\ntr-1=(\nout-1)$. Invertibility of $\bs S$ forces that rank to equal $\nout-1$, and $\frameOp_{\bs\rho}^{\rm c}\tilde{\bs M}_{\bs\mu}$ spans its range. The corresponding projection identity is
\begin{equation}
 \frameOp_{\bs\rho}^{\rm c}
 =\frameOp_{\bs\rho}^{\rm c}\tilde{\bs M}_{\bs\mu}
 \bs S^{-1}\tilde{\bs M}_{\bs\mu}^T
 \frameOp_{\bs\rho}^{\rm c}.
\end{equation}
Using $0\preceq\frameOp_{\bs\rho}^{\rm c}\preceq K\bs P_0$, the residual in~\cref{eq:general_central_equations} satisfies
\begin{equation}
\begin{aligned}
 0\le\mathcal H
 &=\frac1{d^2-1}\Trace\!\left(
 \frameOp_{\bs\rho}^{\rm c}\tilde{\bs M}_{\bs\mu}
 (\bs S^{-1}-\bs{\mathcal R}^{\sharp})
 \tilde{\bs M}_{\bs\mu}^T\frameOp_{\bs\rho}^{\rm c}\right)\\
 &\le\frac K{d^2-1}\Trace(I_{\nout-1}-\bs S\bs{\mathcal R}^{\sharp})
 =\frac{K(\nout-1)}{d^2-1}(uj+vf)
 =O(\eta+\ntr^{-1}).
\end{aligned}
\label{eq:uc_no_root_loss_bound}
\end{equation}
Here the trace inequality uses $(\frameOp_{\bs\rho}^{\rm c})^2\preceq K\frameOp_{\bs\rho}^{\rm c}$, and the equality follows from $(uI_{\nout-1}+v \nout\barSigma_{\rm tr}+\bs S) \bs{\mathcal R}^{\sharp}=I_{\nout-1}$. Since $\mathcal L_d=e\mathcal H$ and $0<e\le1$, the minimized deterministic loss obeys $\mathcal L_d\le C(\eta+\ntr^{-1})$. Gaussian replication as in~\cref{lem:general_gaussian} makes this loss nonnegative, nondecreasing and concave, just as each sampled penalized loss. Hence $0\le\partial_\eta\mathcal L_d\le\mathcal L_d/\eta\le C[1+(\ntr\eta)^{-1}]$. Taking the dimensional upper limit at fixed $\eta>0$, then $\eta\downarrow0$, bounds the derivative without requiring a unique dimensional loss limit. For any bounded positive penalty direction $\bs W$, the envelope identity~\cref{eq:general_sample_envelope} bounds its penalty derivative by $\|\bs W\|_{\rm op}\partial_\eta\mathcal L_d$. Both derivative terms defining the centered costs are therefore bounded. The mixed term in $G_{\eta,d}$ is bounded as well, because $0\le\Trace(\bs S\bs{\mathcal R}^{\sharp})\le (\nout-1)$. This proves the stated absence of a divergent centered fixed-ridge cost at interpolation, without asserting an unregularized inverse limit.
\end{proof}

\parTitle{Boundary cases and physical scope} The bounded centered result does not automatically extend to the physical Moore--Penrose readout or its full ideal-test error: measured modes with small singular values and the empirical mean still require control. At the boundary where the inverse trace approaches one, $\zeta_*$ can approach zero and different singular scaling is possible. A uniformly gapped \emph{measured} Gram matrix, together with response and mean bounds, also yields bounded costs, but a fixed-positive-ridge comparison does not establish such a gap.

\parTitle{Targets and geometries outside the peak regime} Label structure can remove a peak despite a hard edge. With exact features, a target represented by weights satisfying $\|\bs w_{\rm repr}\|^2/(\nout\,d\Delta_{\calO}^2)=O(1)$ has no inaccessible residual; its labels already contain the singular-value factors. Such a target is outside the independent isotropic target average. A zero traceless Born map gives $C_G=0$ even if noise fitting produces $C_V>0$, violating measurement strength; collapsed training fluctuations violate noncollapse and need not produce either pole. Outside the positive-root, noncollapse regime, persistence of an exact-data instability requires separate spectral analysis.

\section{Complete profiles and finite-resource limits}
\label[appsec]{app:complete_profiles_resources}

The peak laws of~\cref{app:interpolation_mechanisms} determine the singular contribution near interpolation, but comparisons with the full curves in~\cref{sec:ridge_risk} also need the regular background. \Cref{app:haar_centered_full_results} evaluates that complete centered curve for Haar training and identifies the assumptions for its physical interpretation. \Cref{app:finite_resource_results} then treats fixed resources and fixed training-shot budgets, which lie outside the proportional regime $N/[d(d+1)]\to\alpha>0$ and need not display its divergent limit.

\subsection{Complete centered curves for Haar training states}
\label{app:proof_haar_centered_profiles}
\label{app:shared_haar_profiles_intro}
\label{app:haar_centered_full_results}

For independent Haar pure training states, \cref{prop:general_haar_full_curves} gives the full centered test-variance and traceless-reconstruction curves, including the finite background omitted by the main-text peak expansion. The input is the apparatus eigenvalue law $\nu$ on all traceless directions, including zeros. The quantities $\phi$ and $h$ below measure the unresolved fraction in minimum linear single-state reconstruction after $N$ shots and set the ridge-rounding width, respectively. Under the stated trace balance and frame bounds, physical mean and identity corrections vanish at fixed positive ridge. Minimum-norm limits additionally require measured modes with small singular values to carry vanishing label-weighted inverse contribution, as specified in~\cref{eq:general_inverse_condition}; exact physical mean restoration is retained at zero ridge.

\begin{proposition}[Complete centered Haar-training profiles]
\label{prop:general_haar_full_curves}
Let the training states be independent Haar pure states, independent of an apparatus with bounded rescaled frame as in~\cref{eq:alt_geometry} and balanced traces $\max_a\lvert\nout p_{I/d,a}-1\rvert\xrightarrow{\PP}0$. Let $\nu$ be the limiting scaled apparatus law, including its zero modes on the $d^2-1$-dimensional traceless operator space. Let $\kappa\in(0,\infty)$ be fixed away from the measurement threshold $\kappa=1$. Write $\phi=\int(1+\alpha\lambda)^{-1}\,\dd\nu(\lambda)$ and $h=1-\kappa+\kappa\phi$. In the proportional regime~\cref{eq:alt_dimensions}, with fixed $\alpha>0$ and $\eta>0$, the conditional target-averaged centered profiles are determined by
\begin{equation}
 H=(1-\kappa)\delta_1+\kappa(1+\alpha\lambda)_*\nu.
 \label{eq:general_haar_outcome_law}
\end{equation}
For $\kappa<1$, the first term counts the extra outcome directions, whose categorical covariance is asymptotically one; the second counts the measured operator directions. For $\kappa>1$ the expression is still a nonnegative probability measure: the rank of the Born map is at most $\nout-1$, so $\nu$ has an atom at zero of mass at least $1-1/\kappa$. Its image at $t=1$ cancels the formally negative $(1-\kappa)\delta_1$; the remaining modes have total covariance $1+\alpha\lambda$. The measure $H$ in~\cref{eq:general_haar_outcome_law} is therefore the limiting covariance law on nonconstant outcome space. The mapped measure $(1+\alpha\lambda)_*\nu$ assigns $t=1+\alpha\lambda$ the same weight as $\lambda$. At fixed $\eta>0$, $c>0$ parametrizes the regularized training inverse: $\eta c$ is the normalized trace of its residual filter. The derivative $c'=\partial_\eta c$ describes the response to ridge, and $v$ the response of the residual-filter trace to a penalty weighted by the apparatus strength $t-1$. They determine the profiles through
\begin{equation}
\begin{aligned}
 1&=\eta c+\gamma\int\frac{tc}{1+tc}\,\dd H(t),\\
 c'&=-\frac{c}{\eta+\gamma\int t(1+tc)^{-2}\,\dd H(t)},\\
 v&=-\gamma c'\int\frac{t(t-1)}{(1+tc)^2}\,\dd H(t).
\end{aligned}
 \label{eq:alt_scalar}
\end{equation}
With $t=1+\alpha\lambda$ in integrals over $\nu$,
\begin{equation}
 F_\eta=\alpha\left\{\int\frac{t-1}{t^2}
  \left[\frac{tc}{1+tc}+\frac{\eta tc'}{(1+tc)^2}\right]
                     \dd\nu(\lambda)+\phi(c+\eta c')\right\},
 \label{eq:alt_F}
\end{equation}
\begin{equation}
 G_\eta=\int\left(\frac{1+c}{1+tc}\right)^2\dd\nu(\lambda)
       +v\left[\phi+\int\frac{t-1}{t(1+tc)^2}\dd\nu(\lambda)\right].
 \label{eq:alt_G}
\end{equation}

For the conditional target averages, convergence in probability is over the Haar training states and counts, and the apparatus draw when specified. The independent sphere, sign, and unitary-orbit target laws also give convergence of the realized centered costs to these profiles. The first equation in~\cref{eq:alt_scalar} has a unique positive solution for every fixed $\eta,\gamma>0$. With $\phi_2=\int(1+\alpha\lambda)^{-2}\,\dd\nu(\lambda)$, the weighted inverse-tail condition~\cref{eq:general_inverse_condition} gives the centered minimum-norm limits, for $\gamma<1$,

\begin{equation}
 F_0=\alpha(\phi-\phi_2)+\alpha\phi h\frac{\gamma}{1-\gamma},
 \qquad
 G_0=\phi_2+\kappa\phi(1-\phi)\frac{\gamma}{1-\gamma}.
 \label{eq:alt_left}
\end{equation}

For $\gamma>1$, let $c_0>0$ solve $1=\gamma\int tc_0/(1+tc_0)\,\dd H(t)$ and put $v_0=c_0\int t(t-1)(1+tc_0)^{-2}\dd H/ \int t(1+tc_0)^{-2}\dd H$. Then
\begin{equation}
 F_0=\alpha\left[\int\frac{t-1}{t^2}\frac{tc_0}{1+tc_0}
              \dd\nu(\lambda)+\phi c_0\right],\qquad
 G_0=\int\left(\frac{1+c_0}{1+tc_0}\right)^2\dd\nu(\lambda)+v_0\left[\phi+\int\frac{t-1}{t(1+tc_0)^2}
                       \dd\nu(\lambda)\right].
 \label{eq:alt_right}
\end{equation}

\end{proposition}
\begin{proof}
\parTitle{Haar training-frame bounds} We first check the hypotheses for the Haar examples. For independent pure Haar states the vectors $\sqrt{d(d+1)}\bs P_0\bs v_{\rho_i}$ are centered and isotropic on the $d^2-1$-dimensional traceless space, have squared norm $d^2-1$, and have bounded directional fourth moments. The fourth-moment bounds follow by integrating $[\Trace(A\rho)]^4$ for a traceless Hermitian $A$:
\begin{equation}
 \EE_\rho[\Trace(A\rho)^4]
 =\frac{3(\Trace(A^2))^2+6\Trace(A^4)}{d(d+1)(d+2)(d+3)}.
 \label{eq:joined_Haar_fourth}
\end{equation}
The sample upper bound~\cref{lem:rm_sample_upper}, with independent columns appended if necessary, therefore bounds the centered empirical frame at every fixed proportional ratio. Also $q_d=\|\bar\rho-I_d/d\|_F^2=O_\PP(\ntr^{-1})$, by its exact mean $(1-1/d)/\ntr$. The normalized frame trace tends to one.

\parTitle{Balanced covariance and physical mean} For an independent normalized Haar rank-one POVM,~\cref{lem:trace_conc,lem:alt_frame_upper} give asymptotically equal effect traces and a bounded rescaled traceless frame. Conditional scalar Bernstein concentration of the independent Haar probabilities gives
\begin{equation}
 \max_a\lvert\nout\bar p_a-\nout p_{I/d,a}\rvert
 =O_\PP\!\left(\sqrt{\frac{\log \nout}{\ntr}}+
                         \frac{d\log \nout}{\ntr}\right)=o_\PP(1).
 \label{eq:joined_Haar_mean_balance}
\end{equation}
Here $\nout p_{ai}\le Kd$ and $\Var(\nout p_{ai})\le K$ on the trace event, by the Haar state moments. Equations~\cref{eq:general_covariance_identity,eq:general_test_categorical} then imply asymptotically flat training and test categorical covariances. The same conclusions hold for a fixed apparatus with balanced traces and a bounded rescaled frame. Its empirical mean descriptors obey $q_d=O_\PP(\ntr^{-1})$ and $t_d=o(1)$. Consequently~\cref{lem:general_mean_bounds,lem:identity_readout} make all normalized mean and identity corrections vanish at every fixed $\eta>0$, including bounded $\lvert c_\calO\rvert/\sqrt{d\Delta_{\calO}^2}$. This argument does not remove the separate inverse-tail requirement at zero penalty.

\parTitle{Joint feature and response covariance} Let $\bs B=\tilde{\bs M}_{\bs\mu}$, $\bs T=I_{\nout-1}+\bs B^T\bs B$, and $\bs C=\bs P_0-\bs B\bs T^{-1}\bs B^T$. The eigenvalues of $\bs B\bs B^T$ have limiting law $(\lambda\mapsto\alpha\lambda)_*\nu$, so the eigenvalues of $\bs T$ have law $H$ and $(d^2-1)^{-1}\Trace(\bs C)\to\phi$. Before empirical centering, a training column and its operator response have the form $\bs x_i=\bs B^T\bs h_i+\bs g_i$ and $\bs h_i=\sqrt{d(d+1)}\bs P_0\bs v_{\rho_i}$, where $\bs g_i$ is an independent standard real Gaussian vector. Their joint covariance is
\begin{equation}
 \EE\begin{pmatrix}\bs x_i\\\bs h_i\end{pmatrix}
       \begin{pmatrix}\bs x_i^T&\bs h_i^T\end{pmatrix}
 =\begin{pmatrix}\bs T&\bs B^T\\\bs B&\bs P_0\end{pmatrix}.
 \label{eq:haar_joint_covariance_bridge}
\end{equation}
Haar spherical concentration gives $\Var(\bs h_i^T\bs A\bs h_i)\le K (d^2-1)^2\|\bs A\|_{\rm op}^2/d$: the map from a unit state vector to this quadratic form has Lipschitz constant $O((d^2-1)\|\bs A\|_{\rm op})$. Bounded transformations by $\bs B$, together with the Gaussian quadratic and cross-term moment bounds, give the same vanishing normalized quadratic-form variance for the joint vector. Together with the sample upper bounds just verified, these estimates establish the hypotheses of~\cref{lem:joint_response_reduction}. The lemma permits evaluating the feature inverse, mixed response trace, and minimized response loss using Gaussian pairs with the same joint covariance. The original Haar regression residual is not assumed independent.

\parTitle{Loss from a scalar residual} For a commuting outcome penalty $\ell_z(t)>0$, let $r_z>0$ solve the scalar balance below. The feature inverse has deterministic diagonal $[\ell_z(t)+tr_z]^{-1}$, while $r_z$ is the normalized trace of the training residual filter. Completing the square in the joint Gaussian covariance gives the normalized loss
\begin{equation}
\begin{aligned}
 1&=r_z+\gamma\int\frac{tr_z}{\ell_z(t)+tr_z}\,\dd H(t),\\
 \mathcal L_z
 &=\phi r_z+\int\frac{t-1}{t}
          \frac{\ell_z(t)r_z}{\ell_z(t)+tr_z}\,\dd\nu(\lambda).
\end{aligned}
 \label{eq:haar_deformed_loss_bridge}
\end{equation}
The first term is the response covariance $\bs C$ left after its linear prediction from the features. The second term is the residual of that linear prediction. In integrals over $\nu$, $t=1+\alpha\lambda$; zero apparatus modes are included. At $\ell_0(t)=\eta$, put $r_0=\eta c$. The first balance becomes~\cref{eq:alt_scalar}, and the loss is $\eta c[\phi+\int(t-1)/(t(1+tc))\,\dd\nu]$. Its derivative with respect to a uniform penalty is $F_\eta/\alpha$ by the envelope identity; direct differentiation gives precisely~\cref{eq:alt_F}, including the term $\phi(c+\eta c')$.

\parTitle{Traceless reconstruction cost} For the Born metric choose $\ell_z(t)=\eta+z(t-1)$. Equivalently, writing $r_z=\eta c_z$, its deformed scalar equation is $1=\eta c_z+\gamma\int tc_z/[1+tc_z+z(t-1)/\eta]\,\dd H$. Differentiation at zero gives $\partial_z r_z\vert_0=-\gamma c'\int t(t-1)(1+tc)^{-2}\,\dd H=v$. The mixed prediction trace and the Born-metric quadratic trace are consequently
\begin{equation}
\begin{aligned}
 \frac1{d^2-1}\Trace(\bs B\hat{\bs K})
 &\longrightarrow\int\frac{(t-1)c}{1+tc}\,\dd\nu,\\
 \left.\partial_z\mathcal L_z\right\vert_0
 &=\int\left(\frac{(t-1)c}{1+tc}\right)^2\dd\nu
 +v\left[\phi+\int\frac{t-1}{t(1+tc)^2}\,\dd\nu\right].
\end{aligned}
 \label{eq:haar_mixed_quadratic_bridge}
\end{equation}
Combining the mixed and quadratic terms in~\cref{eq:general_complete_profiles} leaves $[1-(t-1)c/(1+tc)]^2=[(1+c)/(1+tc)]^2$, proving~\cref{eq:alt_G}. The formulas include the regular background as well as the critical singularity. Their singular coefficients agree with~\cref{cor:alt_rounding} for $\kappa<1$ and with~\cref{thm:uc_interpolation} under its positive-root hypothesis for $\kappa>1$.

\parTitle{Zero-penalty limits on either side} For $\gamma<1$, the ridge equation gives $c\sim(1-\gamma)/\eta$, and the limiting finite terms after the cancellation of the leading divergences yield~\cref{eq:alt_left}. For $\gamma>1$, it admits the finite limit $c_0$ determined by the equation stated before~\cref{eq:alt_right}, and direct substitution yields that formula. Passing from these positive-ridge limits to the actual centered minimum-norm costs requires~\cref{eq:general_inverse_condition}, as proved in~\cref{app:proof_centered_zero_penalty}.
\end{proof}

\subsection{Finite resources and fixed training shots}
\label{app:proof_finite_shot_resources}
\label{app:finite_resource_results}

The divergent proportional peak does not imply infinite error at fixed resources. \Cref{prop:alt_finite_N} proves that finite multinomial records have finite error moments and that keeping $N$ fixed while dimension grows leaves a positive fraction of outcomes unobserved. Unobserved outcomes leave zero rows in the measured matrix, so the fixed-shot behavior discussed in~\cref{sec:ridge_risk} requires a separate limit in either measurement regime.

\parTitle{Combining the two shot budgets} For $M$ independent test shots the physical costs always combine as $\MSE_{N,M,\tau}=\calE_{N,\tau}+V_{N,\tau}/M.$ Both costs include the physical empirical-mean contribution. The proposition distinguishes a fixed finite record, the limit with fixed $N$ and $\nout,\ntr\asymp d^2$, and a separate Haar absence calculation with $d,N$ fixed.

\begin{proposition}[Finite resources and fixed training shots]
\label{prop:alt_finite_N}
At fixed $d,\nout,\ntr,N$, bounded labels give finite prediction-error moments for the multinomial minimum-norm readout. If $N$ is fixed, $\nout,\ntr\asymp d^2$, $\nout/\ntr\to\gamma>0$, the frame bounds hold and $\nout^{-1}\sum_a\lvert\nout\bar p_a-1\rvert^2\to0$, the number $Z_0$ of outcome rows absent from all training records satisfies
\begin{equation}
 Z_0/\nout\xrightarrow{\PP}e^{-N/\gamma},\qquad
 \rank\hatbs P_N/\nout\le1-e^{-N/\gamma}+o_{\PP}(1).
 \label{eq:alt_empty_rows}
\end{equation}
For $N=1$ and $\Trace(\calO^2)=1$, every record has $\MSE_{1,M,0}(\calO)\le4$ and $V_{1,0}(\calO)\le1$.

For an independent Haar pure-state training list and a normalized rank-one Haar POVM, the expected fraction of unobserved rows can also be evaluated exactly at finite $d,\nout,\ntr,N$:

\begin{equation}
 \EE\frac{Z_0}{\nout}
 =\EE_W\left[\left\{\EE_U(1-WU)^N\right\}^{\ntr}\right].
 \label{eq:finite_Haar_absence}
\end{equation}

Here $W\sim\operatorname{Beta}(d,\nout-d)$ is an effect trace and, independently, $U\sim\operatorname{Beta}(1,d-1)$ is its overlap with a Haar state; $WU$ is the outcome probability. Writing $\bar u_N=\EE[Z_0/\nout]$, one finds in the separate limit $\nout,\ntr\to\infty$ at fixed $d,N$ and $\nout/\ntr\to\gamma$,
\begin{equation}
 \bar u_N\longrightarrow\left(1+\frac{N}{d\gamma}\right)^{-d}.
 \label{eq:general_haar_empty_row_limit}
\end{equation}
\end{proposition}

\begin{proof}
\parTitle{Finite records and unobserved outcomes} At fixed resources the count matrix has finite support and its pseudoinverse takes finitely many values. For the fixed-shot limit, the probability row $a$ is unobserved is $u_a=\prod_i(1-p_{ai})^N$. The frame bounds and the balanced empirical mean imply $\nout^{-1}\sum_{a,i}p_{ai}^2=O(\nout^{-1})$. Discard the vanishing fraction of rows where the squared-probability sum or mean-probability imbalance is large. On the remaining rows, expanding $\log(1-p_{ai})$ gives $\log u_a=-N\sum_i p_{ai}+o(1)=-N\ntr\bar p_a+o(1)\to-N/\gamma$. Hence $\nout^{-1}\sum_a u_a\to e^{-N/\gamma}$. Distinct absence indicators have nonpositive covariance, since $1-p_{ai}-p_{bi}\le(1-p_{ai})(1-p_{bi})$ in every independent record. Thus $\Var Z_0\le \nout$ and~\cref{eq:alt_empty_rows} follows. At $N=1$, each observed outcome weight is the average of its assigned labels, and each unseen outcome receives zero. Targets with $\Trace(\calO^2)=1$ have $\|\calO\|_{\rm op}\le1$, so all these values lie in $[-1,1]$, giving the stated bounds.

\parTitle{Exact Haar absence probabilities} A row of a normalized Haar isometry has squared norm $W\sim\operatorname{Beta}(d,\nout-d)$. Conditional on that row, a Haar pure training state has overlap $U\sim\operatorname{Beta}(1,d-1)$ with its normalized direction, independently of $W$. Hence its outcome probability is $WU$. Conditional on $W$, the training states are independent, so the probability that no one of the $\ntr$ records with $N$ shots visits the row is $[\EE_U(1-WU)^N]^{\ntr}$. Taking the expectation over $W$ and using exchangeability of the $\nout$ POVM rows proves~\cref{eq:finite_Haar_absence}.

\parTitle{Fixed-dimension absence limit} In the limit $\nout,\ntr\to\infty$ at fixed $d,N$, $X=\nout W/d$ converges to $\operatorname{Gamma}(d,\text{rate }d)$. Because $\EE U=1/d$ and $\ntr/\nout\to1/\gamma$, the conditional absence probability converges to $e^{-NX/\gamma}$. Its expectation is the Gamma Laplace transform $(1+N/(d\gamma))^{-d}$, proving~\cref{eq:general_haar_empty_row_limit}.
\end{proof}

\parTitle{Effect imbalance slows outcome discovery} Taking logarithms of~\cref{eq:general_haar_empty_row_limit} gives the absence rate $-N^{-1}\log\bar u_N\simeq(d/N)\log[1+N/(d\gamma)]$. For $N/(d\gamma)\ll1$, the rate is $1/\gamma-N/(2d\gamma^2)+O(N^2/(d^2\gamma^3))$. Weaker effects remain unseen longer, reducing the decay rate by $N/(2d\gamma^2)$ at first order. For a small relative error in $\bar u_N$, truncating $\log\bar u_N$ after its correction $+N^2/(2d\gamma^2)$ also requires $N^3/(d^2\gamma^3)\ll1$. The unexpanded expression~\cref{eq:general_haar_empty_row_limit} retains all orders in $N/(d\gamma)$;~\cref{eq:finite_Haar_absence} additionally retains finite-$\nout$ corrections.

\section{Shared matrix and covariance estimates}
\label[appsec]{app:rm_prelim}

The shared estimates justify three distinct steps in the physical argument: transferring Gaussian spectra to exact Haar frames, controlling the inverse weights that amplify classical test noise, and evaluating finite-shot regression at positive penalty. The spectral comparisons in~\cref{nav:g_spectral_comparisons}, lift and frame estimates in~\cref{nav:g_lift_frame_bounds}, and covariance formulas in~\cref{nav:g_covariance_resolvents} provide these steps. Norms, compressions and vectorization follow~\cref{app:conventions}; matrix and sample dimensions below are generic unless the physical POVM is specified.

\subsection{Spectral comparisons and Gaussian normalization}
\label{nav:g_spectral_comparisons}

The spectrum theorem~\cref{thm:main} starts from independent Gaussian lifts, then restores Haar normalization and removes the identity direction. Near-identity congruence controls each positive eigenvalue, while interlacing controls the empirical spectrum after compression or a finite-rank change. The same multiplicative comparison transfers the inverse spectral weight in the classical variance proof~\cref{eq:proof_transfer}.

\begin{lemma}[Spectral comparisons]
\label{lem:congruence_spectral}
Let $\bs A\succeq0$ have rank $r$, and let $\bs T$ be invertible. The ordered positive eigenvalues of $\bs A$ and $\bs T\bs A\bs T^\dagger$ satisfy $s_{\min}(\bs T)^2\lambda_i\le\widetilde\lambda_i\le\|\bs T\|_{\rm op}^2\lambda_i$. In particular, if $\|\bs T-I\|_{\rm op}\le\delta<1$, then
\begin{equation}
\label{eq:congruence_interval}
 (1-\delta)^2\lambda_i\le\widetilde\lambda_i\le(1+\delta)^2\lambda_i,
 \qquad 1\le i\le r.
\end{equation}
For a Hermitian $(p+1)\times(p+1)$ matrix, compression to a $p$-dimensional subspace gives $\lambda_{j+1}(\bs A)\le\widetilde\lambda_j\le\lambda_j(\bs A)$, $1\le j\le p$. More generally, for Hermitian $p\times p$ matrices with $\operatorname{rank}(\bs A-\bs B)\le k$,
\begin{equation}\label{eq:rank_interlace}
 \lambda_{j+k}(\bs A)\le\lambda_j(\bs B)\le\lambda_{j-k}(\bs A),
\end{equation}
where missing end indices are ignored. Their empirical distribution functions differ by at most $k/p$.
\end{lemma}
\begin{proof}
\parTitle{Congruence bounds}
The positive eigenvalues of $\bs T\bs A\bs T^\dagger$ are the squared nonzero singular values of $\bs T\bs A^{1/2}$. Singular-value inequalities give the first claim, and $1-\delta\le s_{\min}(\bs T)\le\|\bs T\|_{\rm op}\le1+\delta$ gives~\cref{eq:congruence_interval}.

\parTitle{Compression and finite-rank interlacing}
The compression is a principal submatrix in a basis adapted to the subspace, so Cauchy interlacing applies. For the finite-rank claim, intersect the subspaces in the Courant--Fischer principle with $\ker(\bs A-\bs B)$, whose codimension is at most $k$. The dimension shift gives~\cref{eq:rank_interlace}; counting eigenvalues above each threshold gives the distribution-function bound.
\end{proof}

\parTitle{Inverse weight from column distances}
Below informational completeness, the independent lift columns have full column rank. The identity below expresses their inverse spectral weight as a sum of inverse squared distances from each column to the others; concentration of those distances gives the undercomplete branch of the classical peak in~\cref{eq:proof_Sd_under}.

\begin{lemma}[Negative second moment identity]
\label{lem:neg_second_moment}
Let $\mathbf Y\in\CC^{P\times n}$ with $n\le P$ have linearly independent columns $\mathbf y_1,\dots,\mathbf y_n$ and let $H_a\equiv\operatorname{span}\{\mathbf y_b:b\ne a\}$. Then
\begin{equation}
\label{eq:neg_second_moment}
\sum_{a=1}^{n}\frac{1}{s_a(\mathbf Y)^2}
=
\Trace\!\left((\mathbf Y^\dagger\mathbf Y)^{-1}\right)
=
\sum_{a=1}^{n}\frac{1}{\operatorname{dist}(\mathbf y_a,H_a)^2},
\end{equation}
where $s_1,\dots,s_n$ are the nonzero singular values of $\mathbf Y$.
\end{lemma}

\begin{proof}
\parTitle{Column-distance identity}
The singular-value equality is immediate. Let $\mathfrak G=\bs Y^\dagger\bs Y$ and let $\bs Y_{-a}$ omit column $a$. Its Schur complement gives $(\mathfrak G^{-1})_{aa}=[\|\bs y_a\|^2-\bs y_a^\dagger\bs Y_{-a}(\bs Y_{-a}^\dagger\bs Y_{-a})^{-1}\bs Y_{-a}^\dagger\bs y_a]^{-1}=\operatorname{dist}(\bs y_a,H_a)^{-2}$, since the matrix between $\bs y_a^\dagger$ and $\bs y_a$ is the projector onto $H_a$. Sum over $a$.
\end{proof}

\parTitle{Haar normalization}
The exact Gaussian construction~\cref{lem:gaussian_lift_rank} multiplies the lifted columns by a map induced by $\bs S^{-1/2}$. The bounds below make that map close to the identity when $n\gg d$, allowing~\cref{lem:congruence_spectral} to transfer the Gaussian spectrum to the normalized POVM.

\begin{lemma}[Gaussian singular values]\label{lem:wishart_conc}
Let $\hat{\mathbf{G}}\in\CC^{n\times d}$ have i.i.d.\ $\mathcal{CN}(0,1)$ entries. Assume $n\ge d$. Let $s_{\min}(\hat{\mathbf{G}})$ and $s_{\max}(\hat{\mathbf{G}})$ be respectively the smallest and largest nonzero singular values. Then for every $t\ge 0$,
\begin{equation}\label{eq:gauss_sval_bound}
\PP\!\left(
\sqrt{n}-\sqrt{d}-t
\;\le\;
s_{\min}(\hat{\mathbf{G}})
\le s_{\max}(\hat{\mathbf{G}})
\;\le\;
\sqrt{n}+\sqrt{d}+t
\right)
\;\ge\;
1-2e^{-t^2/2}.
\end{equation}

For $\bs S=n^{-1}\hat{\bs G}^\dagger\hat{\bs G}$, the same event gives, when $\sqrt{d/n}+t/\sqrt n\le1$,
\begin{equation}\label{eq:bound_lambda_S}
\lambda_{\min}(\mathbf S)\ge \Big(1-\sqrt{d/n}-\tfrac{t}{\sqrt{n}}\Big)^2,
\qquad
\lambda_{\max}(\mathbf S)\le \Big(1+\sqrt{d/n}+\tfrac{t}{\sqrt{n}}\Big)^2.
\end{equation}
In particular, if $n\gg d$ then
$\|\mathbf S-I\|_{\rm op}=O_\PP(\sqrt{d/n})$.

If the eigenvalues of $\mathbf{S}$ lie in $[1-\varepsilon,1+\varepsilon]$ with $\varepsilon\in(0,1)$, then
\begin{equation}\label{eq:S_inv_sqrt_bound}
\|\mathbf{S}^{-1/2}-\mathbf{I}\|_{\mathrm{op}}
\le
\max\!\left\{\frac{1}{\sqrt{1-\varepsilon}}-1,\ 1-\frac{1}{\sqrt{1+\varepsilon}}\right\}
\le
\frac{\varepsilon}{1-\varepsilon}.
\end{equation}

\end{lemma}
\begin{proof}
\parTitle{Concentration and spectral calculus}
The singular-value estimate is Gaussian concentration~\cite{vershynin2018high}. Square the singular values of $n^{-1/2}\hat{\bs G}$ under the stated nonnegativity condition to obtain the eigenvalue bounds. Spectral calculus gives the inverse-square-root estimate.
\end{proof}

\subsection{Gaussian lifts and empirical frame bounds}
\label{nav:g_lift_frame_bounds}

The MP criterion below, verified for Gaussian rank-one lifts in~\cref{lem:yaskov_condition_chaos}, yields the bulk spectrum in~\cref{thm:main} despite dependence among coordinates of each lift. The distance and small-ball estimates then control inverse weights in~\cref{thm:classical_peak}, which the bulk law alone does not determine. Separate fourth-moment bounds control the largest eigenvalues of the centered training frame and traceless-compressed POVM frame, verifying the empirical geometry assumptions~\cref{eq:alt_geometry} for the finite-shot Haar examples.

\begin{theorem}[MP law under quadratic-form concentration~\cite{yaskov2016short}]\label{thm:yaskov_mp}
Let $\mathbf{x}_1,\dots,\mathbf{x}_n\in\CC^p$ be i.i.d.\ with $\EE\mathbf{x}_i=0$ and $\EE[\mathbf{x}_i\mathbf{x}_i^\dagger]=\mathbf{I}_p$. Assume that for every deterministic Hermitian $\mathbf{A}$ with $\|\mathbf{A}\|_{\mathrm{op}}\le 1$,
$p^{-1}(\bs x_1^\dagger\bs A\bs x_1-\Trace(\bs A))\xrightarrow{\PP}0$.
If $p,n\to\infty$ with $p/n\to\gamma_\infty\in(0,\infty)$, the empirical spectrum of $\bs W=n^{-1}\sum_{a=1}^n\bs x_a\bs x_a^\dagger$ converges almost surely to $\mu_{\mathrm{MP},\gamma_\infty}$.
\end{theorem}

\begin{lemma}[Quadratic-form concentration for Gaussian rank-one lifts]
\label{lem:yaskov_condition_chaos}
For $\bs g\sim\mathcal{CN}(0,I_d)$ and $\bs m=\vect{\bs g\bs g^\dagger-I_d}$, every deterministic Hermitian $\bs H$ with $\|\bs H\|_{\rm op}\le1$ satisfies $d^{-2}(\bs m^\dagger\bs H\bs m-\Trace(\bs H))\xrightarrow{\PP}0$ as $d\to\infty$.
\end{lemma}
\begin{proof}
\parTitle{Variance of the lifted quadratic form}
The coordinate map is an isometry, so $\bs H$ induces a self-adjoint operator $\Phi_{\bs H}$ on matrices with $\|\Phi_{\bs H}\|_{\rm op}\le1$. Set $F(\bs g)=\langle\bs g\bs g^\dagger-I_d,\Phi_{\bs H}(\bs g\bs g^\dagger-I_d)\rangle$. Wick's identity gives $\EE F=\Trace(\bs H)$. Differentiating along $\bs h\in\CC^d$ gives $DF(\bs g)[\bs h]=2\Re\langle\bs g\bs h^\dagger+\bs h\bs g^\dagger,\Phi_{\bs H}(\bs g\bs g^\dagger-I_d)\rangle$. Consequently, Gaussian Poincar\'e in the $2d$ real coordinates yields
\begin{equation}
\label{eq:poincare}
 \|\nabla F(\bs g)\|_2\le4\|\bs g\|_2\|\bs g\bs g^\dagger-I_d\|_F,
 \qquad
 \Var F\le C\EE[\|\bs g\|_2^2\|\bs g\bs g^\dagger-I_d\|_F^2]=O(d^3).
\end{equation}
The last estimate uses $\|\bs g\bs g^\dagger-I_d\|_F^2=\|\bs g\|^4-2\|\bs g\|^2+d$ and $\EE\|\bs g\|^k=O(d^{k/2})$ for each fixed $k$. Chebyshev's inequality now bounds the probability of a normalized deviation exceeding $\varepsilon$ by $O(\varepsilon^{-2}d^{-1})$.
\end{proof}

\parTitle{Distances for undercomplete measurements}
Taking $\bs Q$ to project onto the orthogonal complement of the other lift columns turns $\bs x^\dagger\bs Q\bs x$ into the squared distance in~\cref{lem:neg_second_moment}. The uniform tail below permits a union bound over the columns in~\cref{eq:proof_dist_identity}; the resulting squared distances stay of order $d^2-\nout$ at fixed $\gamma_{\rm out}>1$.

\begin{lemma}[Uniform concentration of the lifted distances]
\label{lem:lift_distance}
Let $\hat{\mathbf g}\sim\calC\calN(0,\bfI_d)$, let $\mathbf x=\vect{\hat{\mathbf g}\hat{\mathbf g}^\dagger}$, and let $\mathbf Q$ be an orthogonal projector on $\CC^{d^2}$ of rank $R$, independent of $\hat{\mathbf g}$. Then
\begin{equation}
\label{eq:dist_mean}
\EE\big[\mathbf x^\dagger\mathbf Q\mathbf x\,\bigm\vert\,\mathbf Q\big]
=R+\big\|\mathbf Q\bs v_{I_d}\big\|^2\in[R,R+d],
\end{equation}
and there are universal constants $c,C>0$ such that, for every $t\in(0,1/2)$,
\begin{equation}
\label{eq:dist_conc}
\PP\!\left(
\left\lvert\frac{\mathbf x^\dagger\mathbf Q\mathbf x}{d^2}-\frac{R}{d^2}\right\rvert
\ge C\left(t+\frac{1}{\sqrt d}\right)
\;\middle\vert\;\mathbf Q\right)
\le C\exp\!\big(-c\,d\,t^2\big).
\end{equation}

\end{lemma}

\begin{proof}
\parTitle{Conditional mean}
Write $\hat{\bs g}=\sqrt u\,\bs h$, with $u=\|\hat{\bs g}\|^2\sim\Gamma(d,1)$ independent of the uniform unit vector $\bs h\in\CC^d$. Then $\bs x^\dagger\bs Q\bs x=u^2\varphi(\bs h)$, where $\varphi(\bs h)=\|\bs Q\vect{\bs h\bs h^\dagger}\|^2\in[0,1]$. By~\cref{eq:moment_2_state}, $\EE[\varphi\mid\bs Q]=(R+\|\bs Q\bs v_{I_d}\|^2)/[d(d+1)]$. Since $\EE u^2=d(d+1)$ and $\|\bs Q\bs v_{I_d}\|^2\le d$, the conditional mean is~\cref{eq:dist_mean}.

\parTitle{Angular and radial concentration}
The estimate $\|\bs h\bs h^\dagger-\bs h'\bs h'^\dagger\|_F\le2\|\bs h-\bs h'\|$ makes $\varphi$ $4$-Lipschitz on $S^{2d-1}$. L\'evy's inequality therefore gives, uniformly in $\bs Q$,
\begin{equation}\label{eq:levy}
 \PP(\lvert\varphi-\EE\varphi\rvert\ge t+C'd^{-1/2}\mid\bs Q)\le2e^{-cdt^2}.
\end{equation}
The $O(d^{-1/2})$ mean--median gap follows by integrating the concentration tail. Meanwhile~\cref{eq:gamma_two_sided} gives $\PP(\lvert u-d\rvert\ge td)\le2e^{-t^2d/3}$. On the intersection of these events, $\bs x^\dagger\bs Q\bs x=d^2(1+O(t))[R/d^2+O(t)+O(d^{-1/2})]$, using $R\le d^2$ and $\|\bs Q\bs v_{I_d}\|^2/[d(d+1)]\le(d+1)^{-1}$. A union bound proves~\cref{eq:dist_conc}.
\end{proof}

\parTitle{Inverse weights for overcomplete measurements}
Above informational completeness, the lift rows are dependent, so the column-distance argument cannot be reused for the inverse frame. The small-ball estimate below instead applies to independent columns and leave-one-column inverse matrices, regardless of their conditioning. It supplies the inverse-moment tail bound~\cref{lem:lift_inverse_tail} needed for the overcomplete branch of~\cref{thm:classical_peak}.

\begin{lemma}[Anticoncentration for the rank-one lift]
\label{lem:smallball_lift}

Let $\hat{\mathbf g}\sim\mathcal{CN}(0,\bfI_d)$ and $\mathbf x=\vect{\hat{\mathbf g}\hat{\mathbf g}^\dagger}$. For every nonzero $\mathbf B\succeq0$ independent of $\hat{\mathbf g}$ and $\varepsilon\in(0,1)$, a universal constant $C$ gives $\PP(\mathbf x^\dagger\mathbf B\mathbf x\le\varepsilon\Trace(\mathbf B)\mid\mathbf B)\le C\varepsilon^{1/4}$. 
\end{lemma}

\begin{proof}
\parTitle{Polynomial small-ball estimate}
Condition on $\bs B=\sum_l\beta_l\bs w_l\bs w_l^\dagger$ and define $W_l$ by $\vect{W_l}=\bs w_l$. The nonnegative polynomial $Q(\hat{\bs g})=\bs x^\dagger\bs B\bs x=\sum_l\beta_l\lvert\hat{\bs g}^\dagger W_l^\dagger\hat{\bs g}\rvert^2$ has degree at most four in the $2d$ real Gaussian coordinates. Wick's identity~\cref{eq:wicks_identity} gives $\EE[Q\mid\bs B]=\Trace(\bs B)+\bs v_{I_d}^\dagger\bs B\bs v_{I_d}\ge\Trace(\bs B)$, hence $\|Q\|_{L^2}\ge\Trace(\bs B)>0$. The Carbery--Wright inequality~\cite{carbery2001distributional}, $\PP(\lvert Q\rvert\le\varepsilon\|Q\|_{L^2})\le Ck\varepsilon^{1/k}$ for a degree-$k$ polynomial under a log-concave law, which includes the Gaussian law here, gives the claim with $k=4$.
\end{proof}

\parTitle{Upper frame bounds beyond the bulk law}
Weak spectral convergence permits a small number of large eigenvalues and therefore does not verify~\cref{eq:alt_geometry}. The fourth-moment estimate below supplies the upper training-frame bound used in~\cref{prop:general_haar_full_curves} and, after Haar normalization, the measurement-frame bound in~\cref{lem:alt_frame_upper}. The fixed-rank examples of~\cref{lem:fixed_rank_bulk_bridge} inherit these upper bounds.

\begin{lemma}[Upper sample covariance from fourth moments]
\label{lem:rm_sample_upper}
Let $\bs v_i\in\RR^k$ be iid centered isotropic vectors with $\EE\lvert\bs a^{T}\bs v_i\rvert^4\le C\|\bs a\|^4$ uniformly in dimension, and $2k\le t\asymp k$. Then, with failure probability tending to zero,
\begin{equation}
 \left\|t^{-1}\sum_{i=1}^t\bs v_i\bs v_i^{T}\right\|_{\rm op}
 \le C'\left(1+t^{-1}\max_i\|\bs v_i\|^2\right).
 \label{eq:rm_sample_upper}
\end{equation}
For a sample size comparable to $k$ but below $2k$, the same conclusion follows by appending independent columns and removing their positive outer products, provided the same maximum-norm bound holds for the appended sample. The constants depend only on the moment bound and the fixed sample-to-dimension ratios.
\end{lemma}

\begin{proof}
\parTitle{Fourth moments and sample enlargement}
Apply~\cite[Theorem 1]{tikhomirov2016sample} with moment exponent four. At fixed $t/k$, its logarithmic ratio factor is bounded and its failure probability tends to zero. This gives~\cref{eq:rm_sample_upper}. Adding independent vectors to reach $2k$ changes normalization by a bounded factor. Removing their positive outer products cannot increase the operator norm.
\end{proof}

\parTitle{Upper bound after Haar normalization}
Haar normalization couples all effects through the same inverse covariance. The following estimate carries the independent-lift upper bound through that normalization and verifies the measurement part of~\cref{eq:alt_geometry} on either side of informational completeness, as required by the Haar finite-shot profiles~\cref{prop:general_haar_full_curves}.

\begin{lemma}[Upper frame bound for the random POVM]
\label{lem:alt_frame_upper}
For a normalized Haar rank-one POVM independent of the data, with $(d^2-1)/(\nout-1)\to \kappa\in(0,\infty)$, $\|\nout\bs P_0\frameOp_{\bs\mu}\bs P_0\|_{\rm op}\le C$ with probability tending to one for a dimension-independent constant, which may depend on the fixed limiting ratio $\kappa$.
\end{lemma}

\begin{proof}
\parTitle{Independent-lift upper bound}
Use the exact Gaussian representation of~\cref{app:concentration_POVM}: $\bs V=\nout^{-1/2}\bs G\bs C^{-1/2}$, $\bs C=\bs G^\dagger\bs G/\nout$, with independent standard complex Gaussian rows $\bs g_a$. The centered real Hermitian-coordinate lifts $\bs z_a=\vect{\bs g_a\bs g_a^\dagger-I_d}$ are isotropic in dimension $d^2$. Their coordinates are quadratic Gaussian polynomials, so every directional fourth moment is bounded by a dimension-independent constant times the squared second moment (Gaussian hypercontractivity). Gaussian radial tails also give $\max_a\|\bs z_a\|^2\le C d^2$ with probability tending to one. The sample bound~\cref{lem:rm_sample_upper} consequently bounds $\nout^{-1}\sum_a\bs z_a\bs z_a^{T}$ uniformly. If $\nout<2d^2$, append independent lifts to reach that sample size and then remove their positive outer products. The normalization changes by at most a fixed factor depending on $\kappa$; the Gaussian radial-tail bound holds for the appended list as well. At every such fixed ratio, $\|\bs C-I\|_{\rm op}=O_\PP(\sqrt{d/\nout})=O_\PP(d^{-1/2})$, so the common-shift bound $d\|\bs C^{-1}-I\|_{\rm op}^2=O_\PP(d^2/\nout)$ remains bounded.

\parTitle{Normalization and common shift}
Let $\bs L_C$ denote the coordinate map induced by $A\mapsto\bs C^{-1/2}A\bs C^{-1/2}$ on Hermitian operators. The normalized traceless lift is $\bs P_0\bs L_C\bs z_a+\bs P_0\vect{\bs C^{-1}}$. Its sample second moment is bounded by twice the transformed centered second moment plus twice the common-shift outer product. The Wishart estimate~\cref{lem:wishart_conc} gives $\|\bs C-I\|_{\rm op}\le C d^{-1/2}$ with probability tending to one. Thus $\|\bs L_C\|_{\rm op}$ is bounded and $\|\bs P_0\vect{\bs C^{-1}}\|^2 \le d\|\bs C^{-1}-I\|_{\rm op}^2\le C$. These estimates bound $\nout\bs P_0\frameOp_{\bs\mu}\bs P_0$ with probability tending to one. The argument uses no lower frame bound and holds on both sides of informational completeness.
\end{proof}

\subsection{Covariance resolvents at positive penalty}
\label{nav:g_covariance_resolvents}

Positive penalty keeps the inverse covariance bounded, so quadratic-form concentration can determine its normalized weighted traces without controlling the smallest sample eigenvalue. The population covariance is the second moment before drawing a finite training list; its eigenvalues need not equal those of that list's sample covariance. The integral below converts the population spectrum into weighted inverse-covariance averages. It supplies the Haar-apparatus peak coefficients in~\cref{eq:general_haar_critical_integrals} and the spectrum of exact training features in~\cref{eq:uc_exact_mp_hard_edge}. The joint extension~\cref{lem:joint_response_reduction} supplies the response traces needed for the complete Haar-training profiles~\cref{prop:general_haar_full_curves}; the argument uses the joint feature--response covariance without assuming independence of the original Haar residual and the features.

\begin{lemma}[Population covariance integrals]
\label{lem:weak_population_resolvent}
Let $\bs x_1,\ldots,\bs x_n$ be independent copies of a centered real vector in dimension $s$, with covariance $\bs T\succeq0$ and $\|\bs T\|_{\rm op}\le K$. Assume $s/n\to y\in(0,\infty)$ and, uniformly over deterministic symmetric $\bs A$,
\begin{equation}
 \EE\lvert\bs x_i^T\bs A\bs x_i-\Trace(\bs T\bs A)\rvert^2
       \le\epsilon_s s^2\|\bs A\|_{\rm op}^2,\qquad\epsilon_s\to0.
 \label{eq:weak_population_concentration}
\end{equation}
If the spectrum of $\bs T$ converges to $H$, then, for $\xi>0$ and bounded continuous $f$, with $\bs W=n^{-1}\sum_i\bs x_i\bs x_i^T$,
\begin{equation}
 \frac1s\Trace(f(\bs T)(\bs W+\xi I_s)^{-1})
 \xrightarrow{\PP}\int\frac{f(t)}{\xi+wt}\,\dd H(t),\qquad
 w=\left[1+y\int\frac{t}{\xi+wt}\,\dd H(t)\right]^{-1}.
 \label{eq:weak_population_resolvent}
\end{equation}
The scalar $w>0$ is unique. The statement also holds with the finite empirical measure of $\bs T$ in the integral, as a deterministic approximation. Locally at positive $\xi$, derivatives of the normalized traces converge.
\end{lemma}

\begin{proof}
\parTitle{Leave-one-column reduction}
Let $\bs Q=(\bs W+\xi I)^{-1}$ and $\bs Q_i$ be the same inverse with column $i$ deleted, retaining normalization by $n$. Condition~\cref{eq:weak_population_concentration}, conditional on the other columns, gives
$n^{-1}\sum_i\lvert\bs x_i^T\bs Q_i\bs x_i/n-\Trace(\bs T\bs Q_i)/n\rvert=O_\PP(\sqrt{\epsilon_s})$.
A rank-one update changes $\Trace(\bs T\bs Q_i)$ by $O_\xi(1)$. Thus the denominators in Sherman--Morrison's identity can be replaced in normalized traces by $1+\Trace(\bs T\bs Q)/n$.

\parTitle{Weighted covariance integral}
Denote the inverse of the common denominator by $w_s$. For a spectral projector $\Pi$ of $\bs T$, the traced normal equation then gives
$\Trace(\Pi)/s=\xi\Trace(\Pi\bs Q)/s+w_s\Trace(\Pi\bs T\bs Q)/s+o_\PP(1)$.
The numerator uses the symmetric part of $\bs Q_i\Pi$ in~\cref{eq:weak_population_concentration}. The error tends to zero in mean absolute value; the same second-moment bound controls the products with leverage errors by Cauchy--Schwarz. In particular, $\EE_i\lvert\bs x_i^T\bs Q_i\Pi\bs x_i/n\rvert^2\le K_\xi$ uniformly in the exposed column and spectral projector, since $\|\bs Q_i\Pi\|_{\rm op}\le\xi^{-1}$ and $s/n$ stays bounded. On a spectral interval of width $\delta$ about $t$, positivity gives $\lvert\Trace(\Pi(\bs T-tI)\bs Q)\rvert\le\delta\Trace(\Pi\bs Q)$. Partition the bounded spectrum into finitely many such intervals, use this traced identity, and let $\delta\downarrow0$. This yields the weighted integral with denominator $\xi+w_st$.

\parTitle{Scalar uniqueness and penalty derivatives}
The defining trace for $w_s$ gives the displayed equation for $w$. Equivalently $1=w+y\int wt/(\xi+wt)\dd H(t)$; its right side is strictly increasing from zero to infinity, giving existence and uniqueness. The same argument retains the finite empirical measure. Resolvent derivatives are uniformly bounded on compact positive $\xi$ intervals. Difference quotients with positive weights, and decomposition into positive weights for general $f$, prove derivative convergence. All assertions are local in the strictly positive penalty.
\end{proof}

\parTitle{Response traces from the joint covariance}
The feature spectrum alone does not determine reconstruction error: the fitted labels and operator response share the same training states. For the joint covariance~\cref{eq:haar_joint_covariance_bridge}, the result below transfers the feature inverse, mixed prediction, and minimized loss to Gaussian pairs with that covariance. In the Haar-training application, $p=\nout-1$, $s=d^2-1$, and $n=\ntr$: $\bs x_i$ is a feature column in the Gaussian-noise comparison and $\bs h_i$ its scaled traceless state response. Those three traces yield the scalar loss and complete centered profiles in~\cref{eq:haar_deformed_loss_bridge,eq:alt_F,eq:alt_G} without assuming an independent Haar regression residual.

\begin{lemma}[Joint feature--response traces at positive penalty]
\label{lem:joint_response_reduction}
Let $(\bs x_i,\bs h_i)\in\RR^p\times\RR^s$ be independent copies of a centered pair, with $p/n$ and $s/n$ bounded above and below by positive constants. Write $\bs z_i=(\bs x_i^T,\bs h_i^T)^T$, assume $\|\EE\bs z_i\bs z_i^T\|_{\rm op}\le K$, and suppose
$\EE\lvert\bs z_i^T\bs A\bs z_i-\Trace((\EE\bs z_i\bs z_i^T)\bs A)\rvert^2\le\epsilon_d(p+s)^2\|\bs A\|_{\rm op}^2$, with $\epsilon_d\to0$, uniformly over deterministic symmetric $\bs A$. Set $\bs X=[\bs x_i]/\sqrt n$ and $\bs R=[\bs h_i]/\sqrt n$, and assume $\|[\bs X^T\ \bs R^T]^T\|_{\rm op}=O_\PP(1)$. For deterministic $\eta I_p\preceq\bs\Lambda\preceq KI_p$, with fixed $\eta>0$, put $\bs Q=(\bs X\bs X^T+\bs\Lambda)^{-1}$. The normalized feature inverse $p^{-1}\Trace(\bs W\bs Q)$, mixed response $s^{-1}\Trace(\bs B\bs Q\bs X\bs R^T)$, and minimized response loss $s^{-1}\Trace(\bs R(I_n-\bs X^T\bs Q\bs X)\bs R^T)$ differ by $o_\PP(1)$ from those of Gaussian pairs with the same joint covariance. Here $\bs W$ and $\bs B$ are arbitrary deterministic matrices of compatible dimensions and uniformly bounded norms. The claim concerns fixed positive penalty, not pseudoinverses.
\end{lemma}
\begin{proof}
\parTitle{Weighted joint resolvents}
The leave-one-column argument of~\cref{lem:weak_population_resolvent} also gives bounded deterministic weights that need not commute with the population covariance. To see this, whiten a positive deterministic penalty and let $\bs Q_i$ be the whitened inverse with column $i$ deleted. In the traced resolvent identity, the numerator is a quadratic form with the symmetric part of $\bs Q_i\bs A\bs R_w$, where $\bs R_w=(I+w\bs T)^{-1}$ and $\bs T$ is the whitened covariance. The stated concentration bound replaces it by its conditional trace, just as for the scalar denominator. The rank-one change in $n^{-1}\Trace(\bs T\bs Q_i)$ is $O(n^{-1})$. A finite grid in the bounded scalar $w$, followed by the uniform positive-penalty bound, handles its dependence on the exposed column. The resulting identity is $\Trace(\bs A(\bs Q-\bs R_w))/(p+s)=o_\PP(1)$; taking $\bs A=\bs T$ identifies the same unique scalar equation as in~\cref{eq:weak_population_resolvent}. It depends only on the covariance. This proves equality of the weighted joint resolvent limits for the two pair ensembles.

\parTitle{Recovering the three regression traces}
An auxiliary large penalty suppresses the response block, so expanding the joint inverse recovers the three regression traces. Apply the argument to the joint sample covariance with penalty $\operatorname{diag}(\bs\Lambda,tI_s)$, initially at fixed $t>0$. With $\bs A=\bs X\bs X^T$, $\bs E=\bs X\bs R^T$, and $\bs C=\bs R\bs R^T$, write
$\bs Q_t=\left(\begin{smallmatrix}\bs A+\bs\Lambda&\bs E\\\bs E^T&\bs C+tI_s\end{smallmatrix}\right)^{-1}$.
Schur complements give, as $t\to\infty$,
\begin{equation}
 (\bs Q_t)_{xx}=\bs Q+O_\PP(t^{-1}),\qquad
 -t(\bs Q_t)_{xh}=\bs Q\bs E+O_\PP(t^{-1}),\qquad
 (\bs Q_t)_{hh}=t^{-1}I_s-t^{-2}(\bs C-\bs E^T\bs Q\bs E)+O_\PP(t^{-3}).
 \label{eq:joint_response_expansion}
\end{equation}
The errors are in operator norm, uniformly in dimension on events where the joint sample norm is bounded. Indeed, $0\preceq\bs C-\bs E^T\bs Q\bs E\preceq\bs C$, and $\|\bs Q\|_{\rm op}\le\eta^{-1}$; the Gaussian comparison has the same sample-norm bound. The first two block limits recover the feature inverse and mixed trace. The third gives the minimized response loss as $t-t^2\Trace((\bs Q_t)_{hh})/s+O_\PP(t^{-1})$. Take the dimensional limit at fixed $t$, then $t\to\infty$. The uniform errors transfer all three traces without assuming independence of the non-Gaussian regression residual.
\end{proof}

\clearpage
\twocolumngrid

\makeatletter
\long\def\addtocontents#1#2{%
  \SMoriginaladdtocontents{#1}{#2}%
  \def\SMtocfile{#1}%
  \def\SMstandardtoc{toc}%
  \ifx\SMtocfile\SMstandardtoc
    \SMoriginaladdtocontents{smt}{#2}%
  \fi
}
\def\toclevel@section{0}
\makeatother

\bibliography{bibliography}

\end{document}